\documentclass[fleqn,final]{unmpandathesis}
\pdfoutput=1
\usepackage[bookmarks = true, pdfpagemode = None, pdfstartview = FitH, colorlinks = true, urlcolor = blue,hyperfootnotes=false]{hyperref}
\usepackage{hypcap}
\usepackage{mathrsfs}
\usepackage{graphicx}
\usepackage{amssymb}
\usepackage{amsthm}
\usepackage{pgf}
\usepackage{amsmath}

\newcommand{\leftexp}[2]{{\vphantom{#2}}^{#1}\!{#2}}
\newcommand{\C}[1]{\leftexp{C}{#1}}
\newcommand{\CZ}{\leftexp{C}{Z}}
\newcommand{\CX}{\leftexp{C}{X}}
\newcommand{\mat}{\mathbf}
\newcommand{\set}{\mathcal}
\newcommand{\po}[1]{p_{\scriptscriptstyle\hspace{-.04em}{#1}}}
\newcommand{\pt}[2]{p_{\scriptscriptstyle\hspace{-.04em}{#1}\hspace{-.06em}{#2}}}

\newcommand{\qt}[2]{q_{\scriptscriptstyle\hspace{-.04em}{#1}\hspace{-.06em}{#2}}}
\newcommand{\loc}{\mathcal}
\newcommand{\source}{\circ}
\newcommand{\sink}{\bullet}
\newcommand{\braket}[2]{\left\langle\left.#1\vphantom{#2}\right|#2\right\rangle}
\newcommand{\brakket}[3]{\left\langle#1\left|\vphantom{#1}#2\vphantom{#3}\right|#3\right\rangle}
\newcommand{\abs}[1]{\left\vert#1\right\vert}
\newcommand{\tr}{\text{tr}}
\newcommand{\group}{\mathcal}
\newtheorem{lem}{Lemma}
\newtheorem{thm}{Theorem}
\newtheorem{defi}{Definition}
\newcommand{\fun}{\text}
\newcommand{\op}{\mathcal}
\newcommand{\trans}{\textrm{T}}
\newcommand{\field}{\mathbb}
\newcommand{\code}{\mathscr}
\newcommand{\prove}{\vspace{-1em}\noindent}
\newcommand{\wt}{\text{wt}}
\newcommand{\e}{\text{e}}
\renewcommand{\i}{\text{i}}
\newcommand{\leftsub}[2]{{\vphantom{#2}}_{#1}\!{#2}}
\newcommand{\command}{\texttt}
\newcommand{\subsubsubsection}[1]{\noindent\textit{#1}\newline}
\definecolor{uglyBlue}{rgb}{.172,.188,.547}
\definecolor{uglyGreen}{rgb}{.07,.64,.3}

\newcommand{\SWAP}{{
  \begin{pgfpicture}{0cm}{0cm}{.33cm}{.275cm}
  \pgfsetlinewidth{.8pt}
  \pgfxyline(0,.274)(.088,.274)
  \pgfxyline(0,.015)(.088,.015)
  \pgfxyline(.242,.274)(.33,.274)
  \pgfxyline(.242,.015)(.33,.015)
  \pgfxyline(.076,.2806)(.254,.0083)
  \pgfxyline(.076,.0083)(.254,.2806)
  \end{pgfpicture}
  }}

\usepackage{rotating}

\input{Qcircuit}
\xyoption{curve}

\begin{document}

\frontmatter

\title{Error Channels and the Threshold for Fault-tolerant Quantum Computation}

\author{Bryan Eastin}

\degreesubject{Ph.D., Physics}

\degree{Doctor of Philosophy \\ Physics}

\documenttype{Dissertation}

\previousdegrees{B.S., Physics, California Institute of Technology, 2001}

\date{July, 2007}

\maketitle

\begin{dedication}
   To my parents, who taught me to dream unreasonable dreams.
\end{dedication}

\begin{acknowledgments}

\renewcommand{\baselinestretch}{1.45}\selectfont
   Foremost, I would like to recognize Carlton Caves and Ivan Deutsch for advising me over the years.  They have been great friends and mentors, and have always given selflessly of their time and knowledge.  I cannot thank them enough.

   My research has also benefitted from the suggestions and criticisms of many people, including, but not limited to, Andrew Landahl, Matthew Elliott, Steven Flammia, Jim Harrington, Andrew Silberfarb, Anil Shaji, John Preskill, JM Geremia, Joseph Renes, Cristopher Moore, Ben Reichardt, Emanuel Knill, Sergio Boixo, Aaron Denney, Seth Merkel, Animesh Datta, Rene Stock, Kiran Manne, Iris Reichenbach, David Hayes, and Shohini Ghose.

\end{acknowledgments}

\maketitleabstract

\begin{abstract}
The threshold for fault-tolerant quantum computation depends on the available resources, including  knowledge about the error model.  I investigate the utility of such knowledge by designing a fault-tolerant procedure tailored to a restricted stochastic Pauli channel and studying the corresponding threshold for quantum computation.  Surprisingly, I find that tailoring yields, at best, modest gains in the threshold, while substantial losses occur for error models only marginally different from the assumed channel.  This result is shown to derive from the fact that the ancillae used in threshold estimation are of exceedingly high quality and, thus, difficult to improve upon.  Motivated by this discovery, I propose a tractable algebraic algorithm for predicting the outcome of threshold estimates, one which approximates ancillae as having independent and identically distributed errors on their constituent qubits.  In the limit of an infinitely large code, the algorithm simplifies tremendously,  yielding a rigorous threshold bound given the availability of ancillae with i.i.d. errors.  I use this bound as a metric to judge the relative performance of various fault-tolerant procedures in combination with different error models.  Modest gains in the threshold are observed for certain restricted error models, and, for the assumed ancillae, Knill's fault-tolerant method is found to be superior to that of Steane.  My algorithm generally yields high threshold bounds, reflecting the computational value of large, low-error ancillae.  In an effort to render these bounds achievable, I develop a novel procedure for directly constructing large ancillae.  Numerically, the scaling and average error properties of this procedure are found to be encouraging, and, though it is not fault-tolerant, I prove that each error can spread to only one additional location.  Promising means of improving the ancillae are proposed, and I discuss briefly the challenges associated with preparing the cat states necessary for my procedure.
\clearpage 
\end{abstract}

\tableofcontents
\listoffigures
\listoftables

\mainmatter

\chapter{Introduction}

The field of quantum computation seeks to harness the processing power
implicit in the structure of quantum mechanics.  To do so, however, requires
the ability to create and precisely control quantum mechanical states on large numbers of subsystems.
This is a difficult task for the same reasons that macroscopic quantum effects such as superpositions are exceedingly rare.  In order to manipulate a quantum system, it must be made to interact with its environment, but these interactions inevitably expose it to corruption from environmental factors over which we have imperfect control.  Thus, the production of a complex quantum state spanning many subsystems is unlikely to proceed flawlessly.

Flaws need not be fatal, however.  Through quantum coding, quantum information can be made robust against many kinds of error.  Quantum codes store data in a distributed fashion over multiple quantum systems so that damage caused by errors on a small number of the systems is fully reversible.  Because unencoded data is at the mercy of the elements, methods have been developed for applying operations and correcting accumulated errors without ever decoding the stored information.

Not all ways of manipulating encoded data are equal.  Encoded operations that spread errors between different parts of an encoded state are likely to cause irrevocable damage since quantum coding is based on the fact that errors affecting many subsystems are improbable.  Consequently, an important topic in quantum computing is the construction of encoded operations that avoid propagating errors, a property known as fault tolerance.

Fault-tolerant design minimizes the impact of errors, but it does not guarantee that a computation will succeed.  It is possible for the probability of error to be so high that an encoded computation with error correction is more likely to fail than an unencoded computation.  If the unencoded error probability is below a certain threshold, however, encoding and error correction provide a means to implement an arbitrary quantum algorithm using resources that scale efficiently in the size and desired accuracy of the computation.  This error probability is known, aptly enough, as the threshold for quantum computation.

Knowledge of
thresholds is clearly a crucial design criterion for use in the engineering of quantum
computing architectures, but no simple, unified scheme exists
for determining them.
Information regarding the threshold for quantum computation is obtained from explicit constructions of fault-tolerant procedures, and, as such, is strongly dependent on the particular procedure utilized.  Moreover, to permit the broadest possible applicability, threshold calculations are generally performed for generic or worst case error models using fault-tolerant procedures designed to match.  By contrast, actual implementations of a quantum computer are likely to suffer from errors that possess more structure.  The initial impetus behind the work contained in this dissertation was the desire to determine whether quantum computing might be rendered more feasible by taking advantage of that structure.

It is to this end that I embark in Chapter~\ref{chap:channelDependencyOfTheThreshold} on a program of tailoring the fault-tolerant procedure of Steane to a specialized, though somewhat unrealistic, error model which would seem to hold major promise for improving the threshold.  There I develop procedures for constructing ancillary states (henceforth ancillae) and implementing error correction that suit the error model adopted, and I investigate the performance of my tailored procedure by analytically bounding the threshold as well as by estimating its value.  I also estimate the threshold using Steane's method, and, comparing the two, I find that, contrary to expectations, the advantage of my approach over that of Steane turns out to be quite small.  Moreover, I show through further numerical estimates that my tailored procedure is not robust against small perturbations in the error model, thereby severely restricting its applicability.  To clarify the origin of this uninspiring achievement, I estimate the threshold assuming that ancillae with no errors whatsoever are available as a resource.  Ancillae are singled out because I expect the biggest impact of my modified fault-tolerant procedure to be in reducing ancillary errors, but, in fact, I find minimal improvement in the threshold even when ancillae are perfect.

Motivated by the relatively minor role that errors on ancillae seem to play, I devote Chapter~\ref{chap:thresholdsForHomogeneousAncillae} to developing a method for determining the threshold given ancillae with simplified error distributions.
For such ancillae, I am able to derive quite high bounds on the threshold for fault-tolerant
quantum computation in the limit that the size of the code becomes large.  I say ``bounds'' rather than ``bound'' because the technique is sufficiently simple that I apply it to a selection of different error models and fault-tolerant procedures.  As in Chapter~\ref{chap:channelDependencyOfTheThreshold}, I see only a small improvement in the threshold for substantially restricted error models.
Comparing fault-tolerant procedures, I find that the method of Knill always performs best for ancillae of the form considered.
In addition to their interpretation as bounds for idealized resources, I discuss the merit of these threshold results as a means of approximating the outcome of threshold estimation.  In that mode, agreement with prior work is found to be tolerable, and a modified version of the method better suited to small codes is explained and vetted.

Having found in Chapter~\ref{chap:thresholdsForHomogeneousAncillae} that large ancillae with simple error properties are a sufficient resource for computing at high rates of error, I devote Chapter~\ref{chap:ancillaConstruction} to addressing the question of where ancillae with such nice error properties might come from.  I do so by proposing a scheme for preparing ancillae in a broad class of quantum states known as graph states.  This is done through a completely novel technique that tracks the locations of some errors during construction and infers the existence and locations of others.  I advance three different variants on my routine for interpreting error information, and investigate the fault-tolerance properties of two of them analytically.  Neither is found to be fault-tolerant, but the error spread associated with each is small.  I also perform numerical studies on the error distributions of states constructed via this method which show that the average number of surviving errors is less than the average number of failures that occurred.
The resource requirements of this approach are found, with some caveats, to compare favorably with those of more traditional procedures, and possible elaborations to deal with correlated errors are discussed.

This dissertation only includes research that I have done which is pertinent to the themes of thresholds, fault tolerance, and atypical error models.  Chapter~\ref{chap:thresholdsForHomogeneousAncillae} covers material published in Reference~\cite{Eastin07}, and Chapter~\ref{chap:ancillaConstruction} deals with a body of work that should eventually coalesce into a paper on ancilla construction.  Topics that I have collaborated on with other researchers are not included, but some of these have resulted in papers that are available online.  Two papers on non-local hidden variables are published in PRA~\cite{Tessier05,Barrett07}, and a paper on graphical representations of stabilizer states is currently available in preprint form~\cite{Elliott07}.

\chapter{Background\label{chap:background}}

\section{Quantum States\label{sec:quantumStates}}

The fundamental difference between classical and quantum physics is that quantum mechanics is incompatible with a local, realistic description of the world.  Classically, the state of a system may be unknown, and widely separated systems may be correlated in complex ways, but there always exists a description in terms of incomplete information about local, objective states.  By contrast, quantum mechanics can be shown both in theory~\cite{Bell64,Mermin90} and in practice~\cite{Pan00,Rowe01} to encompass situations in which either locality or realism must be abandoned to be consistent with observed measurement results.  Thus, while a classical system, such as a coin, must possess a single well-defined classical state, e.g. \textit{heads}, the quantum mechanical analog of a coin can be in any superposition of allowed states, e.g.
\begin{align}
  \ket{\textit{coin}} = \frac{1}{\sqrt{2}}\ket{\textit{heads}} + \frac{1}{\sqrt{2}}\ket{\textit{tails}}
\end{align}
by which we mean that the state of the coin is actually $\ket{\textit{heads}}$ and $\ket{\textit{tails}}$ in equal parts.
Superposition is subtle, measuring whether a quantum coin is in the state \textit{heads} or \textit{tails} always yields one result or the other, but
clever combinations of measurements on multiple quantum coins can be used to show that superposition differs fundamentally from classical uncertainty.
Coins with these bizarre properties exist in nature in the form of spin-$\frac{1}{2}$ particles and in theory in the form of qubits.

Qubits are idealized two-state quantum systems for which, in analogy with classical bits, the standard basis states are labeled $\ket{0}$ and $\ket{1}$ rather than $\ket{\textit{heads}}$ and $\ket{\textit{tails}}$.  The term basis is appropriate here because, mathematically, the state of a qubit exists in a two-dimensional Hilbert space $\group{H}$, that is, a two-dimensional complex vector space with a Hermitian inner product.  Distinct classical states are orthogonal under this inner product, so $\braket{0}{1}=0$.  For simplicity, we additionally assume that $\ket{0}$ and $\ket{1}$ are normalized, $\braket{0}{0}=\braket{1}{1}=1$.  Thus, the states $\ket{0}$ and $\ket{1}$ form a orthonormal basis for single-qubit states, meaning that an arbitrary pure\footnote{The term pure basically excludes any classical uncertainty regarding the state, the state vector is known.  More will be said about purity momentarily.} state $\ket{\psi}$ of a single qubit can be written as
\begin{align}
  \ket{\psi} = \alpha \ket{0} + \beta \ket{1} \label{eq:genericState}
\end{align}
where $\alpha$ and $\beta$ are complex numbers such that $|\alpha|^2$ and $|\beta|^2$ are the probabilities of a measurement in the standard basis finding the states $\ket{0}$ and $\ket{1}$ respectively.  These being the only allowed outcomes, conservation of probability requires that $|\alpha|^2+|\beta|^2=1$.

The states $\ket{0}$ and $\ket{1}$ do not constitute the only possible basis for $\group{H}$ of course.  The result of applying any invertible linear map to a basis is another basis.  Thus, for example,
\begin{align}
  \ket{+}=\frac{1}{\sqrt{2}}(\ket{0}+\ket{1}) & & \ket{-}=\frac{1}{\sqrt{2}}(\ket{0}-\ket{1})
\end{align}
is an equally valid basis for single-qubit states.

For the purpose of measuring in this and other bases, it is convenient to introduce the notion of projectors.  A projector $\Pi$ projects onto a subspace of the Hilbert space, annihilating components of states outside of that subspace.  Since projecting a second time onto the same subspace has no additional effect, projectors satisfy $\Pi\cdot\Pi=\Pi$.  Given a normalized state $\ket{\phi}$ the projector onto $\ket{\phi}$ is defined as $\Pi_\ket{\phi}=\ket{\phi}\bra{\phi}$.  The probability of finding the state $\ket{\phi}$ given the initial state $\ket{\psi}$ is
\begin{align}
  \brakket{\psi}{\Pi_\ket{\phi}}{\psi} = \braket{\psi}{\phi}\braket{\phi}{\psi} = |\braket{\psi}{\phi}|^2
\end{align}
where I have used the property of the Hermitian inner product that $\braket{\phi}{\psi} = \braket{\psi}{\phi}^*$.

The outcome of the measurement of an arbitrary Hermitian operator can be expressed in terms of projectors as well.  For an observable $O$ with normalized eigenvectors $\ket{\theta_1}$ and $\ket{\theta_2}$ corresponding to eigenvalues $o_1$ and $o_2$, we say that, for the initial state $\ket{\psi}$, the outcome $o_1$ is obtained with probability $\brakket{\psi}{\Pi_\ket{\theta_1}}{\psi}$ and the outcome $o_2$ is obtained with probability $\brakket{\psi}{\Pi_\ket{\theta_2}}{\psi}$.  As before, and for any projective measurement, the outcomes can also be regarded as finding the state vector corresponding to the projector.  Conveniently, the expected value of a measurement on such a Hermitian operator is simply
\begin{align}
  \begin{split}
    \fun{E}\left(O\left|\ket{\psi}\rule[-.34em]{0em}{1.2em}\right.\right) = o_1\brakket{\psi}{\Pi_{\scriptscriptstyle\ket{\theta_1}}}{\psi} &+ o_2\brakket{\psi}{\Pi_{\ket{\theta_2}}}{\psi} \\
    &= \brakket{\psi}{o_1\Pi_{\ket{\theta_1}}+o_2\Pi_{\ket{\theta_2}}}{\psi} = \brakket{\psi}{O}{\psi}
  \end{split}
\end{align}

It is only with the faculty to measure in other bases that the distinction between superposition and probabilistic combinations becomes clear.  Given the initial state $\ket{+}$, the probabilities of measuring $\ket{+}$ and $\ket{-}$ are
\begin{align}
  p_+ &= \brakket{+}{\Pi_{\scriptscriptstyle\ket{+}}}{+} = \braket{+}{+}\braket{+}{+} = 1  && \text{and}\\
  p_- &= \brakket{+}{\Pi_\ket{-}}{+} = \braket{+}{-}\braket{-}{+} = 0.
\end{align}
By contrast, consider an initial state that is, with equal probability, either $\ket{0}$ or $\ket{1}$.  Such a state is said to be mixed to distinguish it from pure states where no uncertainty exists in our knowledge of the state.  Given the initial state $\ket{0}$, the probabilities of measuring $\ket{+}$ and $\ket{-}$ are
\begin{align}
  p_+ &= \brakket{0}{\Pi_\ket{+}}{0} = \frac{1}{2}\bra{0}(\ket{0}+\ket{1})(\bra{0}+\bra{1})\ket{0} = \frac{1}{2} && \text{and}\\
  p_- &= \brakket{0}{\Pi_\ket{-}}{0} = \frac{1}{2}\bra{0}(\ket{0}-\ket{1})(\bra{0}-\bra{1})\ket{0} = \frac{1}{2}.
\end{align}
Likewise, $p_+=p_-=\frac{1}{2}$ for the initial state $\ket{1}$.  Thus, we find that, for a state initially prepared in either $\ket{0}$ or $\ket{1}$ with equal probability, measuring in the basis $\{\ket{+},\ket{-}\}$ yields the result $\ket{+}$ $50\%$ of the time and the result $\ket{-}$ $50\%$ of the time.  This is very different from the case of the initial state $\ket{+}$, a superposition of $\ket{0}$ and $\ket{1}$, for which such a measurement always finds the state $\ket{+}$.

The manipulation of mixed states such as the one above is greatly simplified by introduction of the density matrix.  Density matrices represent probabilistic mixtures of states by convex combinations of the projectors corresponding to each state.  The weight of each projector is determined by the probability of the associated state.  Thus, the density matrix $\rho$ for the equal mixture of $\ket{0}$ and $\ket{1}$ described in the preceding paragraph is
\begin{align}
  \rho = \frac{1}{2}\ket{0}\bra{0} + \frac{1}{2}\ket{1}\bra{1}\label{eq:maxMixedState}.
\end{align}
In addition to being compact, this notation has the advantage of lumping together all mixtures with the same measurement statistics.  The maximally mixed state, the situation in which nothing is known about the state of the system, for instance, can be expressed as an equal mixture of any set of basis states.  An equal mixture of $\ket{+}$ and $\ket{-}$ has the same density matrix
\begin{align}
  \begin{split}
    \rho &= \frac{1}{2}\ket{+}\bra{+} + \frac{1}{2}\ket{-}\bra{-} \\
    &= \frac{1}{4}(\ket{0}+\ket{1})(\bra{0}+\bra{1}) + \frac{1}{4}(\ket{0}-\ket{1})(\bra{0}-\bra{1}) \\
    &= \frac{1}{2}\ket{0}\bra{0} + \frac{1}{2}\ket{1}\bra{1}
  \end{split}
\end{align}
as an equal mixture of $\ket{0}$ and $\ket{1}$.

The projective measurements we have discussed so far can also be carried out using density operators.  Given an initial state $\rho$, and a set of projectors $\Pi_a$, the probability of getting result $a$ is given by
\begin{align}
  p_a = \tr(\Pi_a\rho), \label{eq:projectiveMeasurementWithDensityOperators}
\end{align}
which, due to the cyclic nature of the trace and the fact that $\Pi_a^\dag=\Pi_a$ and $(\Pi_a)^2=\Pi_a$, is the same as $\tr(\Pi_a\rho \Pi_a^\dag)$.

This second form of Equation~(\ref{eq:projectiveMeasurementWithDensityOperators}) is of interest because it also applies to more general kinds of measurements.  In fact, for any set of measurement operators $\{E_a\}$ such that $\sum_a E_a^\dagger E_a = I$, the probability of obtaining the measurement result $a$ is
\begin{align}
  p_a = \tr(E_a\rho E_a^\dag) \label{eq:measurementWithDensityOperators}.
\end{align}
where the restriction $\sum_a E_a^\dagger E_a = I$ insures that probability is conserved,
\begin{align}
  \begin{split}
    \sum_a p_a = \sum_a \tr\left(E_a\rho E_a^\dagger\right) &= \sum_a \tr\left(E_a^\dagger E_a\rho\right) \\
    &= \tr\left(\left(\sum_a E_a^\dagger E_a\right)\rho\right) = \tr(I \rho) = \tr(\rho) = 1.
  \end{split}
\end{align}
Like projective measurements, general measurements typically disturb the state of a system.  Subsequent to obtaining the measurement result $a$, the state is
\begin{align}
  \rho^\prime = \frac{E_a\rho E_a^\dagger}{\tr(E_a\rho E_a^\dagger)}.
\end{align}

Until till now, I have only discussed a single qubit, but the richness of quantum mechanics emerges from the properties of composite systems.

A collection of qubits in pure states can be represented by a tensor product of the individual states.  Writing out tensor products is frequently unwieldy, so a variety of shorthand notations are employed.  The equation
\begin{align}
  \ket{0}\otimes\ket{0} = \ket{0}^{\otimes 2} = \ket{0}_1\ket{0}_2 = \ket{0}\ket{0} = \ket{00},
\end{align}
displays five different ways of representing a pair of qubits each in the state $\ket{0}$.

A pure state on $n$ qubits is a vector in a $2^n$-dimensional Hilbert space $\group{H}^n$, so superposition (addition) works the same as in the single-qubit case.  In terms of the component subsystems, any pure state on multiple qubits can be represented by a sum over tensor products of pure states on the individual qubits, e.g.,
\begin{align}
    \ket{\psi} = \frac{1}{\sqrt{2}}(\ket{0}\ket{1}-\ket{1}\ket{0}), \label{eq:singletState}
\end{align}
represents a pair of qubits in a superposition of the joint states $\ket{0}\ket{1}$ and $\ket{1}\ket{0}$.

The tensor product is an appropriate choice for combining quantum systems because it respects the linearity of superposition, e.g., for complex coefficients $\alpha$, $\beta$, $\gamma$, and $\delta$,
\begin{align}
  (\alpha\ket{0}+\beta\ket{1})(\gamma\ket{0}+\delta\ket{1}) = \alpha\gamma\ket{00} + \alpha\delta\ket{01} + \beta\gamma\ket{10} + \beta\delta\ket{11}.
\end{align}
One comforting physical implication of this is that a system which is in the same state in all terms of a superposition is unchanged by the superposition.

The formalism regarding measurement and density operators described earlier carries over directly to collections of qubits, though the additional concept of the partial trace must be introduced.  The partial trace provides a way of ignoring subsystems that we do not wish to consider by tracing over their degrees of freedom.  Tracing over the first qubit of the singlet state, $\ket{\psi}$ from Equation~(\ref{eq:singletState}), for instance, yields the reduced density operator
\begin{align}
  \begin{split}
  \rho_2 = \tr_1(\ket{\psi}\bra{\psi}) = \leftsub{1}{\braket{0}{\psi}}\braket{\psi}{0}_1 + \leftsub{1}{\braket{1}{\psi}}\braket{\psi}{1}_1 = \frac{1}{2}\ket{0}\bra{0} + \frac{1}{2}\ket{1}\bra{1}
  \end{split}
\end{align}
on qubit $2$.  Thus, ignoring qubit $1$ of $\ket{\psi}$, the state of qubit $2$ appears completely random.  An identical situation holds for qubit $1$ when we ignore qubit $2$.

The singlet state is one of an oft-used basis for two-qubit states known as the Bell basis; the complete set is defined by
\begin{align}
  \ket{\beta_{ab}} = \frac{1}{\sqrt{2}}(\ket{0}\ket{b}+(-1)^a\ket{1}\ket{1+b})
\end{align}
where $a,b\in\{0,1\}$ and the symbol $+$ within the state vector represents bitwise \textsc{xor}.
As for the singlet state, measuring a Bell state in the standard basis yields measurement outcomes for the individual qubits that are completely random when considered alone but perfectly correlated between one another.  It is through these sorts of subsystem correlations that it is possible to verify the existence of non-classical effects.  For any orthogonal pair of basis vectors $\ket{\phi}=\alpha\ket{0}+\beta\ket{1}$ and $\ket{\theta}=\beta^*\ket{0}-\alpha^*\ket{1}$, for instance,
\begin{align}
  \begin{split}
    \frac{1}{\sqrt{2}}&(\ket{\phi}\ket{\theta}-\ket{\theta}\ket{\phi}) \\
    &= \frac{1}{\sqrt{2}}((\alpha\ket{0}-\beta\ket{1})(\beta^*\ket{0}+\alpha^*\ket{1})-(\beta^*\ket{0}+\alpha^*\ket{1})(\alpha\ket{0}-\beta\ket{1})) \\
    &= \frac{1}{\sqrt{2}}\left(\abs{\alpha}^2+\abs{\beta}^2\right)(\ket{0}\ket{1}-\ket{1}\ket{0})
    = \frac{1}{\sqrt{2}}(\ket{0}\ket{1}-\ket{1}\ket{0}) = \ket{\beta_{11}},
  \end{split}
\end{align}
so measuring both qubits of the state $\ket{\beta_{11}}$ in any orthogonal basis yields perfectly anti-correlated measurement results.  Classically such correlations are impossible, a fact that was proven by John Bell~\cite{Bell64}.  States that, like the Bell states, cannot be expressed as a product of pure states on the constituent systems are referred to as being entangled.  Entanglement is by no means well understood, but somehow this property divides the realms of quantum and classical physics.

\section{Quantum Gates\label{sec:quantumGates}}
Measurements are not the only way that we can interact with a quantum system.  States evolve over time according to their Hamiltonian.  By modulating that Hamiltonian, it is possible to produce transformations on a state.  In deference to nature, physicists typically treat such evolutions as continuous in time.  In quantum information, however, as in computer science, the focus is on discrete changes of state.  Thus, rather than dealing with Hamiltonians and continuous time, quantum information deals with quantum operations corresponding to discrete time steps.

A general quantum operation looks very much like a general measurement.  In fact, the definition of a general measurement given in Section~\ref{sec:quantumStates} encompasses quantum operations if we allow for the possibility that the results of some measurements are inaccessible.  In such a case, the density operator is given by a sum over the density operators corresponding to possible output states weighted by their probability.  A quantum operation from which no information is learned (and the system is not destroyed) is called trace-preserving.  A general trace-preserving quantum operation $\op{E}$ has the form
\begin{align}
  \op{E}(\rho) = \sum_a E^a\rho {E^a}^\dag
\end{align}
where ${E^a}^\dag E^a=I$.

As in computer science, the allowed gate set is often restricted in quantum information theory since the ability to apply arbitrary gates would make almost anything possible, divorcing the subject of quantum computation completely from reality.  Instead, a basic set of plausibly implementable gates is chosen, generally, single-qubit measurement in the standard basis and a selection of unitary gates that act on one or two qubits at a time.  This discrete, computation-oriented model of quantum mechanics is commonly known as the quantum circuit model.

The gates $X$, $Y$, $Z$, $H$, $P$, $\CX$, $\CZ$, $\SWAP$, and $T$ (defined below) constitute the gate set $\group{U}_\set{G}^\prime$ used in this dissertation.  There is a great deal of redundancy in this set.  Everything but the $T$ gate can be constructed in a straightforward manner using the subset $\group{C}_\set{G}=\{H,P,\CX\}$.  Moreover, the subset $\group{U}_\set{G}=\{H,\CX,T\}$ permits not only a straightforward construction of the gates in $\group{U}_\set{G}^\prime$ but, in a significantly less straightforward fashion that is described in Subsection~\ref{subsec:universality}, the construction of any unitary transformation on qubits.  Nevertheless, it is frequently convenient to refer to the superfluous gates in $\group{U}_\set{G}^\prime$, hence their definition below.

$X$, $Y$, and $Z$ are used to denote the Pauli spin operators.  They are given in the standard ($\{\ket{0},\ket{1}\}$) basis by
\begin{align}
X =
\left[
\begin{array}{cc}
0 & 1 \\
1 & 0
\end{array}
\right],
&&
Y =
\left[
\begin{array}{cc}
0 & -\i \\
\i & 0
\end{array}
\right],
&&
\text{and}
&&
Z =
\left[
\begin{array}{cc}
1 & 0 \\
0 & -1
\end{array}
\right].
\end{align}
In addition to their role as gates, $X$, $Y$, and $Z$ also serve the function of measurement operators with eigenvalues $\pm1$ and eigenstates $\ket{0}\pm\ket{1}$, $\ket{0}\pm \i\ket{1}$, and $\ket{0}$/$\ket{1}$, respectively.  Frequently, the bases corresponding to their eigenvectors are even referenced by the Pauli operator, e.g., the $Z$ basis is the standard basis.

Concordant with convention, I use $H$ to denote the Hadamard gate and $T$ to denote the $\frac{\pi}{4}$ rotation (about the $Z$ axis), also known as the $\frac{\pi}{8}$ gate.  The phase gate, which is the $\frac{\pi}{2}$ rotation about the $Z$ axis, I denote by $P$ ($S$ is frequently used in the literature as well).  These gates are given in the standard basis by
\begin{align}
H=\frac{1}{\sqrt{2}}
\left[
\begin{array}{cc}
1 & 1 \\
1 & -1
\end{array}
\right],
&&
P=
\left[
\begin{array}{cc}
1 & 0 \\
0 & \i
\end{array}
\right],
&&
\text{and}
&&
T=
\left[
\begin{array}{cc}
1 & 0 \\
0 & \e^{\i\pi/4}
\end{array}
\right].
\end{align}

$\CX$ refers to the controlled-NOT gate, a two-qubit gate that applies $X$ to the target qubit conditional on the state of the control qubit being $\ket{1}$.  The controlled-NOT gate is also sometimes called the controlled-$X$ or \textsc{xor} gate.  The controlled-$Z$ gate, $\CZ$, is a similar two-qubit gate which applies $Z$ conditional on the value of the control.  These gates are written in the standard basis as
\begin{align}
\CX =
\left[
\begin{array}{@{\hspace{.2em}}c@{\hspace{.8em}}c@{\hspace{.8em}}c@{\hspace{.8em}}c@{\hspace{.2em}}}
1 & 0 & 0 & 0 \\
0 & 1 & 0 & 0 \\
0 & 0 & 0 & 1 \\
0 & 0 & 1 & 0
\end{array}
\right],
&&
\text{and}
&&
\CZ =
\left[
\begin{array}{@{\hspace{.2em}}c@{\hspace{.8em}}c@{\hspace{.8em}}c@{\hspace{.8em}}c@{\hspace{.2em}}}
1 & 0 & 0 & 0 \\
0 & 1 & 0 & 0 \\
0 & 0 & 1 & 0 \\
0 & 0 & 0 & -1
\end{array}
\right].
\end{align}

I use the remaining symbol in $\group{U}_\set{G}^\prime$, $\SWAP$, to denote the \textsc{swap} gate.  The \textsc{swap} gate exchanges the state of two qubits, effectively relabeling them.  In the standard basis it is
\begin{align}
\SWAP =
\left[
\begin{array}{@{\hspace{.2em}}c@{\hspace{.8em}}c@{\hspace{.8em}}c@{\hspace{.8em}}c@{\hspace{.2em}}}
1 & 0 & 0 & 0 \\
0 & 0 & 1 & 0 \\
0 & 1 & 0 & 0 \\
0 & 0 & 0 & 1
\end{array}
\right].
\end{align}

In addition to limiting the allowed gates, it is necessary to limit the permissible input and output states for a quantum computation.  As described in the previous section, complex measurements command at least as much power as complex gates, so qubits are required to be measured in the standard basis.  Similarly, qubits are required to be initialized in the standard basis partly to avoid complex input states such as ``the solution to my problem''.  These restrictions also prevent us from overlooking a distinctly quantum mechanical problem, the physical difficulty involved in preparing initial states and performing measurements.  Thus, the use of complex initial states or measurements should always be justified.

\section{Classes of Quantum Gates and States\label{sec:gateGroups}}

This section discusses classes of states and gates important to the field of quantum computation, including the gate groups generated by the gate sets $\group{P}_\set{G}$, $\group{C}_\set{G}$, and $\group{U}_\set{G}$.

\subsection{The Pauli Group\label{subsec:PauliGroup}}

The Pauli group is the subgroup of the group of unitaries $\group{U}$ generated by the Pauli gates $\set{P}_\group{G} = \{X,Y,Z\}$.  Explicitly, the Pauli group on $n$-qubits is
\begin{align}
  \group{P}^n = \{\pm1,\pm \i\} \times\{I,X,Y,Z\}^{\otimes n} \label{eq:definePauliGroup}
\end{align}
where the phases arise from products of Pauli operators such as $ZX = \i Y$.  As for the single-qubit case, elements of this group are referred to as Pauli operators.  I occasionally also refer to $X$-type or $Z$-type Pauli operators, by which I mean multi-qubit Pauli operators consisting only of $I$ and the specified single-qubit Pauli operator.

The un-phased single-qubit Pauli operators are Hermitian, so, in addition to being unitaries, many Pauli operators are also observables.  In fact, we can divide the Pauli group into two parts $\group{P}^n = \hat{\set{P}}^n \cup \i\hat{\set{P}}^n$ where
\begin{align}
  \hat{\set{P}}^n = \pm \{I,X,Y,Z\}^{\otimes n}
\end{align}
and all elements of $\hat{\set{P}}^n$ are observables that square to the identity.
The set $\hat{\set{P}}^n$ is not a group since products of elements can yield members of $\i\hat{\set{P}}^n$, e.g. $ZX = \i Y$.  Like $\group{P}^n$, however, $\hat{\set{P}}^n$ does have the property that it is closed under conjugation by elements of $\group{C}_\set{G}$, a fact that will be of interest in the next section.

Another useful partition of the Pauli group is $\group{P}^n = \tilde{\group{P}}^n \cup \i\tilde{\group{P}}^n$ where
\begin{align}
  \tilde{\group{P}}^n = \pm \{I,X,\i Y,Z\}^{\otimes n}.
\end{align}
The Pauli operators in $\tilde{\group{P}}^n$ are all real, as, therefore, are their products, so the set $\tilde{\group{P}}^n$ is closed under multiplication and, consequently, a group.  Unlike $\hat{\set{P}}^n$, however, $\tilde{\group{P}}^n$ is not closed under conjugation by elements of $\group{C}_\set{G}$, e.g. $PXP^\dag = Y$.

Finally, any of the partitions of the Pauli group also forms a basis for operators.  This is most easily seen by showing that any $2^n\times2^n$ dimensional matrix $L^{(n)}(i,j)$ with $L^{(n)}(i,j)^{gh}=\delta_{ig} \delta_{jh}$ can be decomposed into a linear combination of $n$-qubit Pauli operators.

For a $2\times2$ matrix
\begin{align}
  L^{(1)}(0,0)&=(I+Z)/2 & L^{(1)}(0,1)&=(X+\i Y)/2 \nonumber\\
  L^{(1)}(1,0)&=(X-\i Y)/2 & L^{(1)}(1,1)&=(I-Z)/2;
\end{align}
taking tensor products of the single-qubit $L^{(1)}(i,j)$ it is easy to obtain any $L^{(n)}(i,j)$.  The $L^{(n)}(i,j)$ form a basis for $n$-qubit operators with an arbitrary $n$-qubit operator $O$ being written as
\begin{align}
  O = \sum_{i=0}^{n-1} \sum_{j=0}^{n-1} O^{ij} L^{(n)}(i,j)
\end{align}
which demonstrates that any $O\in\group{H}^n\times\group{H}^n$ can be decomposed into elements of $\group{P}^n$ since each $L^{(n)}(i,j)$ can be decomposed into elements of $\group{P}^n$.

\subsection{Stabilizer States\label{subsec:stabilizerStates}}
Previously I specified quantum states by sums of basis vectors with complex coefficients, but a state can also be specified as the eigenvector corresponding to some particular set of eigenvalues of a complete set of commuting observables.  A especially convenient class of observables is the group of $n$-qubit Pauli operators.

Elements of the Pauli group are a desirable choice for constructing complete sets of commuting observables for a number of reasons.  Foremost is the fact that multi-qubit Pauli operators are simply tensor products of single-qubit Pauli operators and thus possess a description efficient in the number of qubits.  Additionally, since the eigenvalues of each of the single-qubit Pauli operators are $\pm1$, the eigenvalues of any $n$-qubit Pauli operator are also $\pm1$.  And, in a similar vein, any two $n$-qubit Pauli operators either commute or anti-commute since the elements of the single-qubit Pauli group all either commute or anti-commute.

Using the Pauli group, we can define a ubiquitous and extremely useful class of quantum states known as the stabilizer states.  The class of stabilizer states is defined as the set of states that can be specified as the simultaneous $+1$ eigenstate of some set $\group{S}_\set{G}=\{G^j\}$ of $n$ independent, commuting Pauli group elements.  The set $\group{S}_\set{G}$ generates a subgroup $\group{S}$ of the Pauli group known as the stabilizer of the state, and the individual elements, $G^j$, are referred to as stabilizer generators.  Stabilizer generator sets are not unique; replacing any generator with the product of itself and another generator yields an equivalent generating set.  Thus, replacing $G^j$ by $G^jG^k$ for $k\neq j$ has no effect on the stabilized state.  An arbitrary product of stabilizer generators, that is, an arbitrary element of $\group{S}$, is called a stabilizer element.

A useful (unnormalized) representation of the stabilized state is
\begin{align}
  2^{-n} \sum_{A\in\group{S}} A\ket{\psi} = 2^{-n} \prod_{D\in\group{S}_\set{G}} (I+D)\ket{\psi}
\end{align}
for any $\ket{\psi}$ whose overlap with the stabilized state is non-zero.
This state satisfies the eigenvalue conditions since
\begin{align}
  B 2^{-n} \sum_{A\in\group{S}} A\ket{\psi} = 2^{-n} \sum_{A\in\group{S}} BA\ket{\psi} = 2^{-n} \sum_{C\in\group{S}} C\ket{\psi}
\end{align}
for any $B\in\group{S}$.

\subsubsection{Binary Generator Matrix\label{subsubsec:binaryGeneratorMatrix}}
An alternative description of Pauli operators, and therefore of stabilizers, is provided by the binary or symplectic representation of the Pauli group. In the binary representation an arbitrary $n$-qubit Pauli operator $A$ is expressed in terms of a pair of length $n$ binary strings $x(A)$ and $z(A)$ such that
\begin{align}
x_j(A) &=
\begin{cases}
1 & \text{if } A^j=X \text{ or } A^j=Y \\
0 & \text{if } A^j=I \text{ or } A^j=Z
\end{cases}
&&
\text{and}\\
z_j(A) &=
\begin{cases}
1 & \text{if } A^j=Y \text{ or } A^j=Z \\
0 & \text{if } A^j=I \text{ or } A^j=X
\end{cases}
\end{align}
where $A^j$ is the $j$th Pauli operator in the tensor product for $A$ and $x_j(A)$ and $z_j(A)$ are the $j$th bits of $x$ and $z$.  The resultant binary strings are typically placed side by side in a matrix or list with a vertical line between them; thus,
\begin{align}
I_1 X_2 Y_3 Z_4 = \left[x(I_1 X_2 Y_3 Z_4)\left|z(I_1 X_2 Y_3 Z_4)\right.\right]
= \left[
\begin{array}{cccc|cccc}
0 & 1 & 1 & 0 & 0 & 0 & 1 & 1
\end{array}
\right].
\end{align}
The natural inner product for such vectors is the symplectic inner product, which satisfies
\begin{align}
  [x(A)|z(A)]\cdot[x(B)|z(B)] = x(A)\cdot z(B) + z(A)\cdot x(B).
\end{align}
where the addition is performed modulo $2$.  It is a simple exercise to show that two Pauli operators commute if and only if their symplectic inner product is $0$.

Binary notation is particularly useful for manipulating stabilizer generators, which are arranged for the purpose in a split matrix where each row is the binary representation of a single generator.  Figure~\ref{fig:stabilizersGHZ} shows a stabilizer generator for the three-qubit GHZ state and the corresponding binary stabilizer generator matrix and set of stabilizers.
\begin{figure}
\capstart
  (a)
  \raisebox{-1.75em}{
    $\left[
    \begin{array}{@{}c@{\hspace{.4em}}c@{\hspace{.4em}}c@{}}
X & X & Y \\
X & Y & X \\
Y & X & X
    \end{array}
    \right]$
  }
  \hspace{3.4em}(b)
  \raisebox{-1.75em}{
    $\left[
    \begin{array}{@{\hspace{.2em}}c@{\hspace{.8em}}c@{\hspace{.8em}}c|c@{\hspace{.8em}}c@{\hspace{.8em}}c@{\hspace{.2em}}}
1 & 1 & 1 & 0 & 0 & 1 \\
1 & 1 & 1 & 0 & 1 & 0 \\
1 & 1 & 1 & 1 & 0 & 0
    \end{array}
    \right]$
  }
  \hspace{3.4em}(c)
  \raisebox{-1.75em}{
    $
    \begin{array}{rr}
XXY, & IZZ \\
XYX, & ZIZ \\
YXX, & ZZI \\
-YYY, & III
    \end{array}
    $
  }
\caption[Representations of the three-qubit GHZ state]{The a) stabilizer generator, b) binary stabilizer generator matrix, and c) complete set of stabilizers for the three-qubit GHZ state.\label{fig:stabilizersGHZ}}
\end{figure}

\subsection{The Clifford Group\label{subsec:CliffordGroup}}
Of the gates introduced in Section~\ref{sec:quantumGates}, all but the $T$ gate have the property that they normalize the Pauli group, which is to say that
\begin{align}
  U \group{P}^n U^\dag = \group{P}^n \label{eq:CliffordsPreservePauli}
\end{align}
for $U\in \group{C}_\set{G}^\prime=\{X,Y,Z,H,P,\CX,\CZ,\SWAP\}$.  This property is easily verified for any $U\in\group{C}_\set{G}=\{H,P,\CX\}$ by explicitly determining the result of conjugating by each $H$ and $P$ for every Pauli operator and by $\CX$ for every pair of Pauli operators.  The result can then be extended to the other gates in $\group{C}_\set{G}^\prime$ or, indeed, an arbitrary sequence of gates in $\group{C}_\set{G}$, by composition of Equation~(\ref{eq:CliffordsPreservePauli}), e.g., for $U,V\in\group{C}_\set{G}$,
\begin{align}
  V U \group{P}^n U^\dag V^\dag = V \group{P}^n V^\dag=\group{P}^n.
\end{align}
Thus, the set $\group{C}_\set{G}$ generates a subgroup\footnote{Recall that $H^{-1}=H^\dag=H$, $\CZ^{-1}=\CZ^\dag=\CZ$, and $P^{-1}=P^\dag=P^3$.} of the group of unitaries $\group{U}$ wherein each element takes Pauli operators to Pauli operators under conjugation.

Compare this group with the Clifford group $\group{C}\subset\group{U}$, which is defined to be the normalizer of the Pauli group, that is, the set of all gates $U$ such that Equation~(\ref{eq:CliffordsPreservePauli}) holds.
It is tempting to simply assert that the group generated by $\group{C}_\set{G}$ is $\group{C}$; using the following lemma it is straightforward to prove that the gate set $\group{C}_\set{G}$
suffices, up to an overall phase of $\i$, to transform any non-identity Pauli operator into any other.

\begin{lem}For any Pauli operator $A$ s.t. $A^j\neq I$, we can construct a unitary, $U$, s.t. $U A U^\dag \propto X_j$ where $U$ is composed exclusively of gates in $\group{C}_\set{G}$ acting on non-identity elements of $A$.\label{lem:CliffordTransformAnyPauliToX}
\end{lem}
\prove Conjugating by Hadamard and phase gates as necessary, transform $A$ to $A^\prime$ such that $A^\prime$ consists only of $X$ and $I$ operators.  Subsequently conjugating $A^\prime$ by $\CX_{jk}$ for all $k\neq j$ such that ${A^\prime}^k=X$ yields a Pauli operator proportional to $X_j$.

Lemma~\ref{lem:CliffordTransformAnyPauliToX} shows that the group of unitaries generated by $\group{C}_\set{G}$ includes, for any choice of $A,B\in\group{P}^n$, unitaries $U$ and $V$ such that $UAU^\dag\propto X_j$ and $VBV^\dag\propto X_k$.  Consequently, since $\group{C}_\set{G}$ generates both $\SWAP$ and $\group{C}_\set{G}^\dag$, it also includes $W=V^\dag \SWAP_{jk} U$, for which $WAW^\dag\propto B$.  The constant of proportionality can only be $\pm1$ or $\pm \i$ since $\group{C}_\set{G}$ preserves $\group{P}^n$, and it can be reduced to either $1$ or $\i$ by allowing the insertion of a conjugation by $Z_j$ after the conjugation by $U$.  While there exists a sequence of gates converting any single Pauli operator into a multiple of any other, however, the process does not transform each Pauli operator independently, so, phases aside, it need not encompass every possible function on $\group{P}$.  In fact, as shown below, unitarity forbids this.

Conjugation by any unitary $U$ is an isomorphism, since the operation is both bijective,
\begin{align}
  UBU^\dag = UAU^\dag && \text{ iff } && B = A, \label{eq:unitaryConjugationIsBijection}
\end{align}
and a homomorphism,
\begin{align}
  UBAU^\dag = UBU^\dag UAU^\dag. \label{eq:unitaryConjugationIsHomomorphism}
\end{align}
Among other things, this implies that commutators are preserved,
\begin{align}
  \begin{split}
    [UBU^\dag,UAU^\dag] &= UBU^\dag UAU^\dag - UAU^\dag UBU^\dag \\
    &= U(BA - AB)U^\dag = U[B,A]U^\dag,
  \end{split}\label{eq:unitaryConjugationPreservesCommutators}
\end{align}
as are eigenvalues,
\begin{align}
  \tr(UBU^\dag U\Pi U^\dag) = \tr(UB\Pi U^\dag) = \tr(B\Pi).
\end{align}
Moreover, unitary conjugation is linear,
\begin{align}
  U(\alpha A + \beta B) U^\dag = \alpha UAU^\dag + \beta UBU^\dag.
\end{align}
Clearly this rules out a variety of functions.  It is, for instance, impossible to take $X_1$ to $X_1$ while at the same time taking $Z_1$ to $X_2$.  Nor can conjugation by a unitary take $X_1$ to $\i X_1$ or any other Pauli operator with imaginary eigenvalues, showing that the formerly described transformations on individual Pauli operators are the most general possible using $\group{C}_\set{G}$.  These restrictions apply equally to $\group{C}$ and the group generated by $\group{C}_\set{G}$, so they simplify rather than settle the question of equality.

In what follows, I show that $\group{C}_\set{G}$ generates the Clifford group $\group{C}$ by giving an explicit routine for constructing a sequence of gates that implements an arbitrary linear isomorphism on $\group{P}^n$.  The equivalence of $\group{C}$ and the group generated by $\group{C}_\set{G}$ was first shown by Gottesman~\cite{Gottesman97,Gottesman98}.  I present a different approach\footnote{It turns out that Aaronson and Gottesman have also proven the equivalence of $\group{C}$ and the group generated by $\group{C}_\set{G}$ in the manner shown here.~\cite{Aaronson04}} developed by Carlton Caves and myself with the assistance of Andrew Silberfarb and Steven Flammia.

An isomorphism is fully described by its effect on a complete basis since, taken together, Equations~(\ref{eq:unitaryConjugationIsBijection}) and~(\ref{eq:unitaryConjugationIsHomomorphism}) imply that applying an isomorphism to a complete basis yields another complete basis.  Thus, by enumerating the elements of the two bases, it is possible to divine the full isomorphism.  Equation~(\ref{eq:unitaryConjugationPreservesCommutators}) shows that commutators are preserved, indicating that it might be wise, for the purpose of enumerating isomorphisms, to choose a form of basis with simple commutation properties.

The Pauli group on $n$-qubits can be expressed in terms of an overall phase (which transforms trivially due to linearity) represented by the Pauli operator $\i I$ and a basis of $2n$ elements of $\set{P}_n$.  With regard to commutation properties, a particularly simple choice of basis would be one in which all basis elements commute.  It is, however, impossible to choose more than $n$ such independent, commuting $n$-qubit Pauli operators, as can be shown using the following lemma.

\begin{lem}
  Given a group $\group{D}$ s.t. $FC=\pm CF$ for every $C,F\in\group{D}$, any element $D\in\group{D}$ either commutes with every $C\in\group{D}$ or it anti-commutes with half of them.\label{lem:halfElementsAnticommute}
\end{lem}
\prove Write $\group{D}$ as $\group{D}=\group{A}\cup\group{B}$ where $\{D,C\}=0$ for all $C\in\group{A}$ and $[D,C]=0$ for all $C\in\group{B}$.  If $\group{A}$ is non-empty (i.e. $D$ anti-commutes with something) then, for any $A\in\group{A}$, $\group{D}=A\group{A}\cup A\group{B}$ is a partition of $\group{D}$ such that the elements of $A\group{A}$ commute with $D$ and the elements $A\group{B}$ anti-commute with $D$.  Thus, $A\group{A}=\group{B}$ and $A\group{B}=\group{A}$, implying that $\group{A}$ and $\group{B}$ have the same size.

Following Preskill~\cite{PreskillNotes}, we can imagine picking independent, commuting Pauli operators from $\set{P}_n$ sequentially.  Given a set of $k-1$ $n$-qubit Pauli operators $\set{A}=\{A^j\}_{j=1}^{k-1}$, the set of Pauli operators that commute with all of them forms a group $\group{A}^\perp$.  Lemma~\ref{lem:halfElementsAnticommute} assures us that any element chosen from $\group{A}^\perp$ either commutes with every element of $\group{A}^\perp$ or anti-commutes with half of them.  An element that commutes with every element of $\group{A}^\perp$ is either proportional to $I$ (a class of elements already chosen) or it is not independent of $\set{A}$.  Thus, the number of available commuting observables decreases by half each time an independent Pauli operator is added to $\set{A}$.  Having chosen $k-1$ independent, commuting Pauli operators from $\set{P}_n$, the number of commuting Pauli operators remaining is $4^{n+1}/2^{k-1}$.  Of these, $4\times2^{k-1}$ are dependent, corresponding to all distinct choices of the four phases and the $k-1$ previously chosen elements of $\set{A}$.  The number of available independent, commuting operators for the $k$th element of $\set{A}$ is thus $4^{n+1}/2^{k-1}-4\times2^{k-1}$ which equals zero when $k=n+1$.

In lieu of a basis of commuting Pauli operators, consider a basis of $2n$ Pauli operators divided into two sets $\set{A}=\{A^j\}_{j=1}^n$ and $\set{B}=\{B^j\}_{j=1}^n$ such that $[A^j,A^k]=0$, $[B^j,B^k]=0$, and $[A^j,B^k]=0$ unless $j=k$ in which case $\{A^j,B^k\}=0$.  Each of these sets stabilizes a state, and each element commutes with every other element except its mate in the other set.  Figure~\ref{fig:matchedStabilizerExample} provides a concrete example of this paired-stabilizer form, which Carlton Caves rediscovered while counting Clifford operations.  An earlier use of the formalism appears in Reference~\cite{Aaronson04}.

To verify that such a pair of stabilizers exists, imagine, after choosing the stabilizer $\set{A}$ by the process outlined in the previous paragraph, that we proceed to pick the elements of $\set{B}$.  Each $B^k$ must commute with all $A^j$ s.t. $j\neq k$ and all previously chosen $B$.  $A^k$ satisfies the same property, so both $A^k$ and $B^k$ are among the $4^{n+1}/2^{n-1+k-1}=2^{n-k+4}$ elements of the group that commutes with the other basis elements.  From Lemma~\ref{lem:halfElementsAnticommute} we know that only half of the elements of this group anti-commute with $A^k$, however, so there are $2^{n-k+3}$ choices for $B^k$, which equals $8$ when $k=n$.

\begin{figure}
\capstart
  \begin{tabular}{ll}
  (a) & (b) \\
  \hspace{.7em}
  \raisebox{-1.75em}{
    $\left[
    \begin{array}{@{}c@{\hspace{.4em}}c@{\hspace{.4em}}c@{\hspace{.4em}}c@{\hspace{.4em}}c@{\hspace{.4em}}c@{\hspace{.4em}}c@{\hspace{.4em}}c@{}}
X & I & I & I & I \\
I & X & I & I & I \\
I & I & X & I & I \\
I & I & I & X & I \\
I & I & I & I & X
    \end{array}
    \right]$
  }
  \raisebox{-1.75em}{
    $\left[
    \begin{array}{@{}c@{\hspace{.4em}}c@{\hspace{.4em}}c@{\hspace{.4em}}c@{\hspace{.4em}}c@{\hspace{.4em}}c@{\hspace{.4em}}c@{\hspace{.4em}}c@{}}
Z & I & I & I & I \\
I & Z & I & I & I \\
I & I & Z & I & I \\
I & I & I & Z & I \\
I & I & I & I & Z
    \end{array}
    \right]$
  }
  &  \hspace{.7em}
  \raisebox{-1.75em}{
    $\left[
    \begin{array}{@{}c@{\hspace{.4em}}c@{\hspace{.4em}}c@{\hspace{.4em}}c@{\hspace{.4em}}c@{\hspace{.4em}}c@{\hspace{.4em}}c@{\hspace{.4em}}c@{}}
X & Z & Z & X & I \\
I & X & Z & Z & X \\
X & I & X & Z & Z \\
Z & X & I & X & Z \\
X & X & X & X & X
    \end{array}
    \right]$
  }
  \raisebox{-1.75em}{
    $\left[
    \begin{array}{@{}c@{\hspace{.4em}}c@{\hspace{.4em}}c@{\hspace{.4em}}c@{\hspace{.4em}}c@{\hspace{.4em}}c@{\hspace{.4em}}c@{\hspace{.4em}}c@{}}
I & I & I & Y & Y \\
Y & I & I & Y & I \\
I & Y & I & Y & I \\
I & I & Y & Y & I \\
Z & Z & Z & Z & Z
    \end{array}
    \right]$
  }
\end{tabular}
\caption[Example stabilizer pairs]{The a) canonical stabilizer pair for 5 qubits and b) a pair of non-canonical 5-qubit stabilizers.  The left stabilizer in a) stabilizes $\ket{+}^{\otimes 5}$.  The left stabilizer in b) stabilizes $\ket{\bar{+}}$ of the 5-qubit code.} \label{fig:matchedStabilizerExample}
\end{figure}

This procedure also provides a way to count the number of linear isomorphisms.  The number of ways to pick $\set{A}$ is
\begin{align}
  \prod_{k=1}^n \frac{4^{n+1}}{2^{k-1}}-2^{k+1} = 2^{3n-\sum_{l=1}^n l} \prod_{k=1}^n 4^n-4^{k-1}
  = 2^{(5n-n^2)/2} \prod_{k=1}^n 4^n-4^{k-1}.
\end{align}
Given $\set{A}$, the number of ways to pick $\set{B}$ is
\begin{align}
  \prod_{k=1}^n 2^{n-k+3} = 2^{n^2 + 3n - \sum_{k=1}^n k} = 2^{(n^2+5n)/2}.
\end{align}
Thus, the total number of linear isomorphisms is
\begin{align}
  2^{5n} \prod_{k=1}^n 4^n-4^{k-1}.
\end{align}
This is not actually the number of elements in the Clifford group, since, as discussed earlier, conjugation by a unitary cannot generate a phase of $\i$.  Half of the possible transformations on each basis element are unobtainable, so the number of elements in the Clifford group is at most (and I show below exactly)
\begin{align}
  \left[2^{5n} \prod_{k=1}^n 4^n-4^{k-1}\right]/2^{2n} = 2^{3n} \prod_{k=1}^n 4^n-4^{k-1}.
\end{align}

The preceding discussion shows that there exist bases for $\group{P}^n$ consisting of $\i I$ and a set of paired stabilizers.  Isomorphisms, due to their preservation of commutation relations, take a basis in paired stabilizer form to another basis in paired stabilizer form.  Since factors of $\i$ are not produced by the transformation in question, it makes sense to restrict to paired stabilizers composed of elements from $\hat{\set{P}}^n$, a group that is without additional phases.  Thus, to show that the gate set $\group{C}_\set{G}$ generates every isomorphism on $\hat{\set{P}}^n$ it is sufficient to show that gates from $\group{C}_\set{G}$ can be used to transform any paired stabilizer basis for $\hat{\set{P}}^n$ into any other paired stabilizer basis for $\hat{\set{P}}^n$.  The primary result we need for this is the following lemma.

\begin{lem}
For any $A,B\in\hat{\set{P}}^n$ s.t. $\{A,B\}=0$ and $B^j\neq I$, we can construct a unitary, $U$, s.t. $U A U^\dag = X_j$ and $U B U^\dag = Z_j$ where $U$ is composed only of gates in $\group{C}_\set{G}$ acting on locations where $A$ and/or $B$ have non-identity elements. \label{lem:CliffordsTakeAntiComPaulisToXZ}
\end{lem}
\prove Having started in $\hat{\set{P}}^n$, Lemma~\ref{lem:CliffordTransformAnyPauliToX} and the subsequent discussion indicate that we can construct a unitary $V$ composed of gates in $\group{C}_\set{G}$ such that $B^\prime=VBV^\dag = X_j$.  From Equation~(\ref{eq:unitaryConjugationPreservesCommutators}), $A^\prime=VAV^\dag$ anti-commutes with $B^\prime$ implying that ${A^\prime}^j = Y\text{ or }Z$. Conjugating by Hadamard and phase gates as necessary, transform $B^\prime$ to $B^{\prime\prime}=Z_j$ and $A^\prime$ to $A^{\prime\prime}$ such that $A^{\prime\prime}$ consists only of $X$ and $I$ operators.  Next conjugate by $\CX_{jk}$ for all $k\neq j$ such that ${A^\prime}^k=X$, converting $A^{\prime\prime}$ to $\pm X_j$.  If $-X_j$ is obtained the sign can be removed by conjugation by $Z_j$.  $B^{\prime\prime}$ is unchanged since none of the gates transform $Z_j$.  As promised, no gates are applied to locations $k$ where $A^k = B^k = I$.

And finally, we see that all possible isomorphisms on paired stabilizers whose elements are taken from $\hat{\set{P}}^n$ are generated by $\group{C}_\set{G}$.
\begin{thm}
For any matched pair of stabilizer generators $\set{A}=\{A^j\}$ and $\set{B}=\{B^j\}$ where $A^j,B^j\in\hat{\set{P}}^n$, $\exists$ a unitary, $U$, s.t. $U A^j U^\dag = X_j\text{ and }U B^j U^\dag = Z_j\ \forall j$ where $U$ is composed only of gates in $\group{C}_\set{G}$.
\end{thm}
\prove Lemma~\ref{lem:CliffordsTakeAntiComPaulisToXZ} guarantees that there exists a unitary $V$ composed of gates in $\group{C}_\set{G}$ such that $V A^j V^\dag = X_k$ and $V B^j V^\dag = Z_k$.  Conjugation by $\SWAP_{jk}$ would then result in ${A^\prime}^j=X_j$ and ${B^\prime}^j=Z_j$.  ${A^\prime}^{hj}={B^\prime}^{hj}=I$ for $h\neq j$ since each of these Pauli operators must commute with $X_j$ and $Z_j$.  This process can be repeated for all values of $j$ without disturbing the previously transformed operators since their non-identity locations are identities for other Pauli operators.

This theorem shows that the Clifford group and the group generated by $\group{C}_\set{G}$ implement the same set of transformations on $\group{P}^n$.  Since $\group{P}^n$ forms a complete basis for operators on $\group{H}^n$ the two groups are, in truth, equal.

\subsection{Universality\label{subsec:universality}}

By all rights this section should contain a proof of the universality of the gate set $\group{U}_\set{G}=\{H,\CX,T\}$.  However, due to time constraints and the fact that I lack anything new to add to the proof, I omit it.  The basic idea is that an arbitrary $n$-qubit unitary can be decomposed into two-level unitaries~\cite{Reck94}, which can then be decomposed into $\CX$ gates and single-qubit unitaries~\cite{Barenco95}.  Arbitrary single-qubit unitaries can be approximated to any precision by judicious sequences of the gates $H$ and $T$~\cite{Boykin99}.  This last fact is shown by identifying a pair of such gate sequences corresponding to an irrational angle of rotation $\theta$ about two orthogonal axes $n_1$ and $n_2$ in the $3$-D rotational representation of single-qubit unitaries.  Euler's decomposition allows the desired rotation to be decomposed into rotations about $n_1$ and $n_2$ each of which can be approximated to any accuracy by a multiple of $\theta$.
Readers desiring a full treatment of this universality proof are referred either to the references above, or to the rendition in the textbook by Nielsen and Chuang~\cite{NielsenChuang}.

\section{Quantum Circuit Diagrams\label{sec:quantumCircuitDiagrams}}

Quantum circuit diagrams provide a pictorial method for representing the application of discrete operations, or gates, to a quantum system.  Diagrams consist of horizontal lines, representing qubits, interrupted by squares and other decorations, representing discrete unitaries applied to the interrupted qubits, and optionally punctuated by any of a variety of symbols representing measurement.  The order of operations is from left to right, with the initial state written to the left of the qubits (lines) it applies to.  Classical data is denoted by double lines where double lines emanating from a measurement are assumed to carry the measurement value and double lines intersecting a unitary denote a classical control.  As with electrical circuits, repeated usage of a small number of standard, simple parts results in quantum circuits that are easier to understand and implement.  Table~\ref{tab:basicCircuitElements} depicts the standard one- and two-qubit quantum gates as rendered in this dissertation.  An example quantum circuit, the teleportation circuit, is given in Figure~\ref{fig:exampleCircuit}.

\begin{table}
  \capstart
\centering{
\begin{tabular}{@{\hspace{3em}}c@{\hspace{4em}}c@{\hspace{3em}}}
Operation & Circuit element\\
\hline
Measurement
&
$
\raisebox{.35em}{
\Qcircuit @R=1em @C=1em {
& \meter
}
\rule[-1.55em]{0em}{3.1em}
}
$
\ or
$
\raisebox{.35em}{
\Qcircuit @R=1em @C=1em {
& \measureD{}
}
\rule[-1.55em]{0em}{3.1em}
}
$
\\
\hline
$U$
&
\raisebox{.35em}{
\Qcircuit @R=1em @C=1em {
& \gate{U} & \qw
}
\rule[-1.55em]{0em}{3.1em}
}
\\
\hline
$\CX_{12}$
&
\raisebox{1.2em}{
\Qcircuit @R=1.3em @C=1em {
& \ctrl{1} & \qw \\
& \targ & \qw
}
}
\ or
\raisebox{1.2em}{
\Qcircuit @R=1em @C=1em {
& \ctrl{1} & \qw \\
& \gate{X} & \qw
}
\rule[-2.6em]{0em}{3.1em}
}
\\
\hline
$\CZ_{12}=\CZ_{21}$
&
\raisebox{1.2em}{
\Qcircuit @R=1.65em @C=1em {
& \ctrl{1} & \qw \\
& \control \qw & \qw
}
}
\ or
\raisebox{1.2em}{
\Qcircuit @R=1em @C=1em {
& \ctrl{1} & \qw \\
& \gate{Z} & \qw
}
\rule[-2.6em]{0em}{3.1em}
}
\\
\hline
$\C{U}_{12}$
&
\raisebox{1.2em}{
\Qcircuit @R=1em @C=1em {
& \ctrl{1} & \qw \\
& \gate{U} & \qw
}
\rule[-2.6em]{0em}{3.1em}
}
\\
\hline
\SWAP
&
\raisebox{1.2em}{
\Qcircuit @R=1.7em @C=1em {
& {\times} \qwx[1] \qw & \qw \\
& {\times} \qw & \qw
}
}
\ or
\raisebox{1.2em}{
\Qcircuit @R=1.7em @C=1em {
& \link{1}{1}  \qw & & \qw \\
& \link{-1}{1} \qw & & \qw
}
\rule[-2.4em]{0em}{3.1em}
}
\\
\hline
\end{tabular}
}
\caption[Basic quantum circuit elements]{A basic set of one- and two-qubit quantum operations and their corresponding circuit elements.  Here $U$ is used to represent an arbitrary single-qubit unitary.  Gates with multiple controls, such as the Tofolli, or a control on $\ket{0}$, indicated by an open dot, are also allowed, and, in general, complex multi-qubit unitaries are indicated by a large labeled box spanning the qubits acted on by the unitary.\label{tab:basicCircuitElements}}
\end{table}

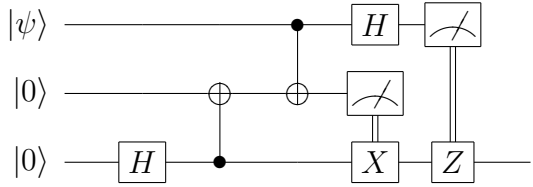
\begin{figure}
\capstart
\centerline{
\Qcircuit @C=.7em @R=.4em @! {
& \lstick{\ket{\psi}} & \qw & \qw & \ctrl{1} & \gate{H} & \meter \cwx[2] \\
& \lstick{\ket{0}} & \qw & \targ & \targ & \meter \\
& \lstick{\ket{0}} & \gate{H} & \ctrl{-1} & \qw & \gate{X} \cwx & \gate{Z} & \qw
}
}
\caption[The teleportation circuit]{An example quantum circuit diagram showing the teleportation circuit.  In this circuit $H_1\protect\CX_{12}\protect\CX_{32}H_3$ is applied to the initial state $\ket{\psi}\ket{0}\ket{0}$ yielding $(\ket{00}\ket{\psi}+\ket{01}X_3\ket{\psi}+\ket{10}Z_3\ket{\psi}+\ket{11}X_3Z_3\ket{\psi})/2$.  The first two qubits are then measured, and, conditional on the measurement results, corrective gates are applied to complete the teleportation, thereby yielding $\ket{\psi}$ on the output.\label{fig:exampleCircuit}}
\end{figure}
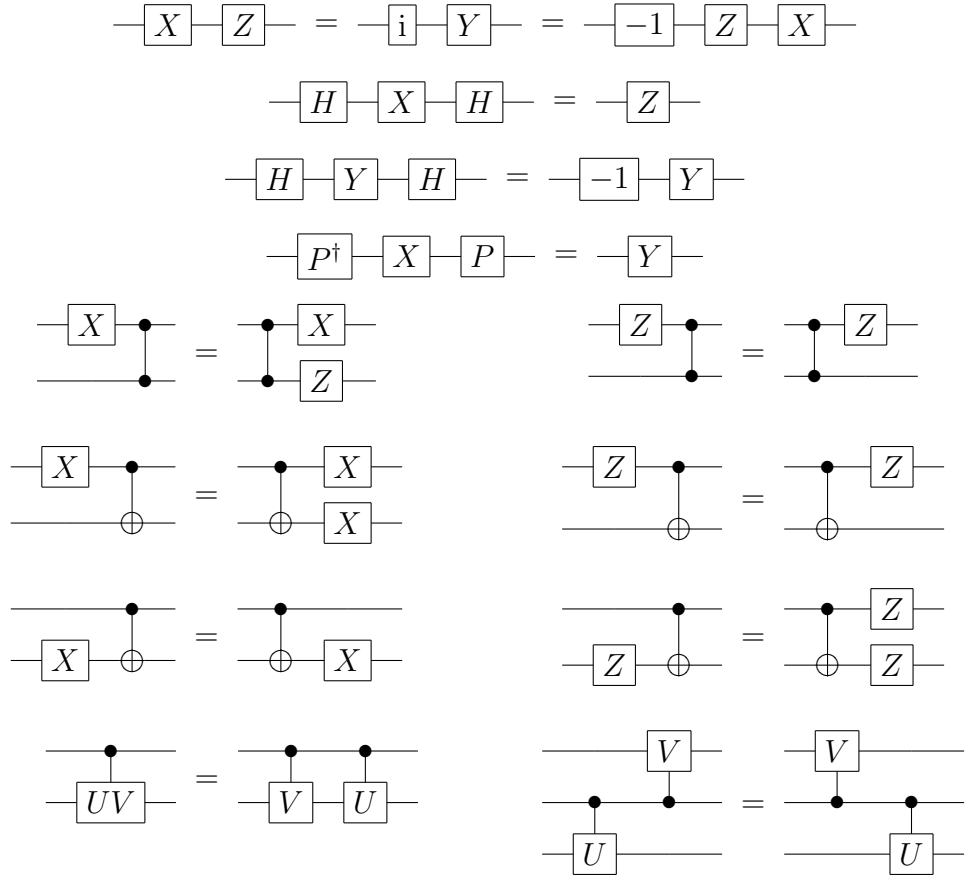
\begin{figure}
\capstart
\centerline{
\Qcircuit @R=.5em @C=1em {
& \gate{X} & \gate{Z} & \qw & {=} & & \gate{\i} & \gate{Y} & \qw & {=} & & \gate{-1} & \gate{Z} & \gate{X} & \qw
}
}
\ \\
\centerline{
\Qcircuit @R=.5em @C=1em {
& \gate{H} & \gate{X} & \gate{H} & \qw & {=} & & \gate{Z} & \qw
}
}
\ \\
\centerline{
\Qcircuit @R=.5em @C=1em {
& \gate{H} & \gate{Y} & \gate{H} & \qw & {=} & & \gate{-1} & \gate{Y} & \qw
}
}
\ \\
\centerline{
\Qcircuit @R=.5em @C=1em {
& \gate{P^\dag} & \gate{X} & \gate{P} & \qw & {=} & & \gate{Y} & \qw
}
}
\ \\
\centerline{
\begin{tabular}{c@{\hspace{4em}}c}
\Qcircuit @R=.5em @C=1em {
& \gate{X} & \ctrl{1} & \qw & \raisebox{-2.3em}{=} & & \ctrl{1} & \gate{X} & \qw \\
& \qw & \control \qw & \qw & & & \control \qw & \gate{Z} & \qw
}
&
\Qcircuit @R=1em @C=1em {
& \gate{Z} & \ctrl{1} & \qw & \raisebox{-2.3em}{=} & & \ctrl{1} & \gate{Z} & \qw \\
& \qw & \control \qw & \qw & & & \control \qw & \qw & \qw
}
\\
\\
\Qcircuit @R=.5em @C=1em {
& \gate{X} & \ctrl{1} & \qw & \raisebox{-2.3em}{=} & & \ctrl{1} & \gate{X} & \qw \\
& \qw & \targ & \qw & & & \targ & \gate{X} & \qw
}
&
\Qcircuit @R=1em @C=1em {
& \gate{Z} & \ctrl{1} & \qw & \raisebox{-2.6em}{=} & & \ctrl{1} & \gate{Z} & \qw \\
& \qw & \targ & \qw & & & \targ & \qw & \qw
}
\\
\\
\Qcircuit @R=1em @C=1em {
& \qw & \ctrl{1} & \qw & \raisebox{-2.3em}{=} & & \ctrl{1} & \qw & \qw \\
& \gate{X} & \targ & \qw & & & \targ & \gate{X} & \qw
}
&
\Qcircuit @R=.5em @C=1em {
& \qw & \ctrl{1} & \qw & \raisebox{-2.3em}{=} & & \ctrl{1} & \gate{Z} & \qw \\
& \gate{Z} & \targ & \qw & & & \targ & \gate{Z} & \qw
}
\\
\\
\hspace{1.3em}
\Qcircuit @R=1em @C=1em {
& \ctrl{1} & \qw & \raisebox{-2.3em}{=} & & \ctrl{1} & \ctrl{1} & \qw \\
& \gate{UV} & \qw & & & \gate{V} & \gate{U} & \qw
}
&
\Qcircuit @R=1em @C=1em {
& \qw & \gate{V} & \qw & & & \gate{V} & \qw & \qw \\
& \ctrl{1} & \ctrl{-1} & \qw & \raisebox{-.6em}{=} & & \ctrl{-1} & \ctrl{1} & \qw \\
& \gate{U} & \qw & \qw & & & \qw & \gate{U} & \qw
}
\end{tabular}
}
\caption[Frequently used circuit identities]{Frequently used circuit identities. $U$ and $V$ here represent arbitrary two-qubit unitaries.  Gates applying a simple phase, such as $\i$ or $-1$, are included in the identities for use in breaking up controlled operations like $\leftexp {C}{Y}$; alone, they merely impart an overall phase and can thus be omitted.\label{fig:simpleCircuitIdentities}}
\end{figure}
In addition to the normal advantages of a graphical depiction for visualization, quantum circuit diagrams provide a powerful tool for proving identities.  Circuit diagrams clearly and concisely indicate the order of operations as well as which qubits each operation acts upon.  This property, when augmented by a selection of simple circuit identities, permits the transformation of many circuits on a grand scale, without ever resorting to either matrices or state vectors.  Figure~\ref{fig:simpleCircuitIdentities} lists the bulk of the simple circuit identities employed during the subsequent chapters, though it omits those that are powers of a gate such as $\CZ_{jk}\CZ_{jk}=I$ and $(P)^2=Z$.

\section{Codes}
At its core, coding is the art of identifying sets of physical states that, under the influence of errors, are unlikely to transition between each other.  A code is nothing more than a mapping between a set of such states, or codewords, and a set of logical (information) states that we wish to protect.  No set of states is robust against every kind of assault, however, so some knowledge about the nature of the errors must be assumed.

Much of coding theory is built around the very reasonable (and not uncommonly true) assumption that independent elements suffer errors independently. In this case, the probability that a pair of errors afflicts two elements is equal to the product of the probabilities of the errors afflicting the individual elements, or, more generally,
\begin{align}
  \fun{P}\left(\bigwedge_j E_j\right) = \prod_j \fun{P}(E_j).
\end{align}
where $E_j$ denotes an error on element $j$.  This assumption ensures that the most probable errors are of those that affect the fewest elements.  Put another way, the most probable error operators are those with the lowest Hamming weight, where the Hamming weight is defined as the number of non-trivial components in a string.

Coding theory, both quantum and classical, are vast subjects, and I will not even begin to cover them in their entirety.  Instead, this section touches upon the basic properties of binary, linear codes relevant to the task of protecting against all errors below a certain weight.

\subsection{Classical Codes}
Protection against low-weight errors is achieved through the use of codes whose logical (encoded) operations have high weight, that is, codes that distribute logical information across many elements.  Perhaps the most straightforward method of constructing a code with this property classically is to employ simple redundancy, as in the class of repetition codes $\code{R}_n$ where logical states are encoded as many copies of themselves.

The smallest of the repetition codes is the two-bit repetition code $\code{R}_2$ for which
\begin{align}
  \bar{0} = 00 && \bar{1}=11
\end{align}
where I have placed a bar over logical states to identify them.  The two-bit repetition code is capable of detecting a single bit error since flipping either the first or the second bit of a codeword yields something that is not a codeword.  It is not, however, able to detect two errors since flipping both bits is the logical (encoded) bit-flip operation.  Nor is the two-bit repetition code capable of correcting one-bit errors since flipping a single bit of either codeword yields a state that might have been produced by flipping a bit of the other codeword.

\begin{figure}
\capstart
  \begin{tabular}{l@{\hspace{4em}}l}
    (a) & (b) \\
    \hspace{1em}
    \Qcircuit[2.5em] @R=1em @C=1.2em {
& \node{01} \\
\node{00} \link{-1}{1} \link{1}{1} & & \node{11} \link{-1}{-1} \link{1}{-1} \\
& \node{10}
    }
  &
    \hspace{1em}
    \Qcircuit[2.5em] @R=1em @C=2.5em {
& \node{100} \link{0}{1} \link{1}{1} & \node{110} \\
\node{000} \link{-1}{1} \link{0}{1} \link{1}{1} & \node{010} \link{1}{1} \link{-1}{1} & \node{101} & \node{111} \link{-1}{-1} \link{0}{-1} \link{1}{-1} \\
& \node{001}  \link{0}{1} \link{-1}{1} & \node{011}
    }
  \end{tabular}
\caption[State diagrams for two and three bits]{State diagrams for a) two and b) three bits.  Connected states differ by a single bit flip.  Consequently, the length of the shortest path between two states is also the smallest number of bit flips that can convert one into the other.  When decoding the three-bit repetition code, the four states on the left side of b) are mapped to $\bar{0}$ while the four states on the right are mapped to $\bar{1}$. \label{fig:classicalCodeStateDiagram}}
\end{figure}
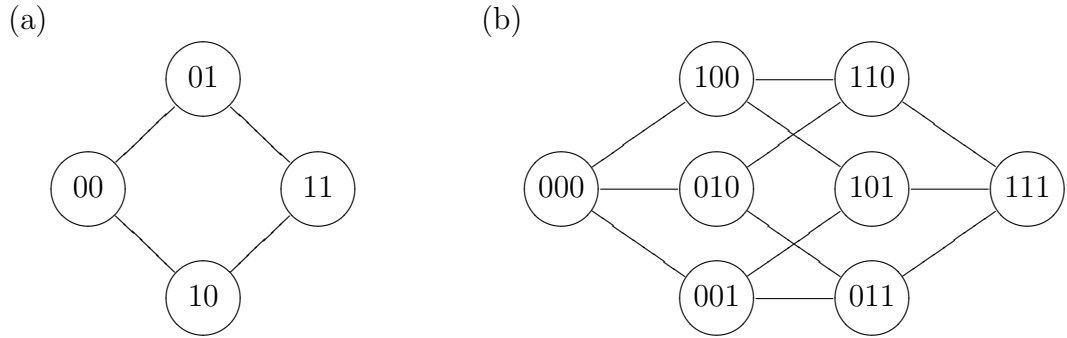

In order to correct errors (while storing information) it is necessary to have three bits.  The three-bit repetition code $\code{R}_3$ is defined by the codewords
\begin{align}
  \bar{0} = 000 && \bar{1}=111.
\end{align}
In addition to detecting any two bit errors, this code can correct any single error since, for a single bit error, the initial value of the logical state can be recovered via majority polling.  Figure~\ref{fig:classicalCodeStateDiagram} shows the state diagrams for the two and three-bit codes.

Implicitly, the preceding example has associated error states with the nearest codeword, where the distance between two states is defined as the minimum weight of any operator that transforms one into the other.  This manner of interpreting error information, known as minimum-distance decoding, is appropriate whenever low-weight errors are more probable than those of high weight.

The number of errors that can be detected or corrected with certainty using minimum-distance decoding is determined by the minimum distance of the code.  The minimum distance of a code is defined as the minimum weight of any of its logical operators.  For linear codes, that is, those for which
\begin{align}
  a + b = c && \text{implies} && \bar{a} + \bar{b} = \bar{c},
\end{align}
the minimum distance is equal to the weight of the lowest-weight codeword since that codeword is generated by applying the lowest-weight logical operator to the zero vector.  A code with distance $d$ can detect $d-1$ errors since all non-trivial errors affecting $d-1$ or fewer bits result in a state that is not a codeword, but at least one error of weight $d$ is a logical operation and so cannot be detected.  The same code can correct $t=\left\lfloor\frac{d-1}{2}\right\rfloor$ errors since all errors of weight $t$ or less yield distinct states, but there exists a pair of errors of weights $\left\lfloor\frac{d}{2}\right\rfloor$ and $\left\lceil\frac{d}{2}\right\rceil$ such that the two errors take two different codewords to the same state and are thus indistinguishable.

Finally, it should be noted that linear codes are frequently referenced by a triplet of parameters $n$, $k$, and $d$ where an $[n,k,d]$ code refers to an $n$-bit code with minimum distance $d$ that encodes $k$ logical bits.  The two and three-bit repetition codes are $[2,1,2]$ and $[3,1,3]$ codes respectively.

\subsection{The Parity Check Matrix\label{subsec:introduceParityCheck}}

An $[n,k,d]$ linear code forms a $k$-dimensional linear subspace of the vector space $\field{Z}_2^n$ and can, consequently, be specified by a set of $k$ basis vectors or by a set of $n-k$ basis vectors of the orthogonal space.  Let $\mat{G}$ be a $k\times n$ matrix whose rows are independent codewords, and let $\mat{H}$ be an $(n-k)\times n$ matrix whose rows are independent vectors orthogonal to every codeword.  $\mat{G}$ and $\mat{H}$ are called the generator matrix and the parity-check matrix respectively and satisfy
\begin{align}
  \mat{H}\cdot\mat{G}^\trans = 0.\label{eq:prodHG}
\end{align}
As is normal for binary linear codes, addition in Equation~(\ref{eq:prodHG}) is performed modulo $2$.  Modular arithmetic engenders a strange kind of orthogonality; any binary vector of even weight, for example, is orthogonal to itself.  Being bases, $\mat{H}$ and $\mat{G}$ are not unique, adding one row to another in either matrix yields an equivalent basis.

The generator matrix provides a very efficient method of encoding logical information.  Given an unencoded data vector $d$ on $\field{Z}_2^k$ the corresponding codeword $\bar{d}$ can be chosen to be
\begin{align}
  \bar{d} = d^\trans\cdot \mat{G}.
\end{align}
Rather than enumerating $2^k$ codewords and associating them with $2^k$ logical states, we enumerate a basis of $k$ codewords and associate them with a basis of $k$ logical states; linearity takes care of the rest.

The parity-check matrix, by contrast, is most useful for what it tells us about states that are not in the code.
The parity-check matrix provides an easy way to separate information about the codeword from information about the error.  For any state $a=\bar{d}+e$ where $\bar{d}$ is a codeword and $e$ is an error string
\begin{align}
  \mat{H}\cdot a = \mat{H}\cdot(\bar{d}+e) = \mat{H}\cdot \bar{d} + \mat{H}\cdot e = \mat{H}\cdot e \label{eq:parityCheckInnerProduct}
\end{align}
since, by definition, codewords have zero inner product with $\mat{H}$.  The bit string $\mat{H}\cdot e$ is known as the syndrome of the error $e$.  The syndrome contains all available information about what error was present since any state with the same syndrome can differ from $e$ only by a codeword, and codewords are undetectable errors.  Each syndrome can thus be associated with an error, reducing the problem of syndrome decoding, that is, finding the most probable causative error, to a single call to a $2^{n-k}$ entry look-up table.  For minimum-distance error correction, the lowest-weight error with a particular syndrome is the one associated with it.

The generator and parity-check matrices are also useful for constructing new codes, either from scratch or by modifying existing codes.  Codes such as the Hamming codes and low-density-parity-check (a.k.a. Gallager) codes result from choosing parity-check matrices that satisfy certain properties.  New codes can be generated from old by a variety of processes including extending, puncturing, concatenation, and taking the dual of the original code.  Each of these topics are covered briefly below.

The Hamming codes $\code{H}_{n}$ are a family of $[2^l-1,2^l-1-l,3]$ codes ($l>1$) invented by Richard Hamming~\cite{Hamming50}.  The parity check matrix for a Hamming code is constructed by choosing the columns to be all non-zero binary strings of length $l$.  This choice is desirable because it ensures that the resulting code has minimum distance $d=3$.  Thinking of error locations as selecting columns of $\mat{H}$, the condition that an error is a codeword, and therefore undetectable, is that it selects columns of $\mat{H}$ that add to zero modulo $2$.  For the Hamming codes, states of weight $1$ cannot be codewords because none of the columns is permitted to be the string of all zeroes.  Nor can states of weight $2$ be orthogonal to $\mat{H}$; no two columns of $\mat{H}$ add to zero since the columns are distinct.  $\mat{H}$ includes all non-zero strings as columns, however, so the sum of any two columns must also be a column, showing that $d=3$.  We have already seen a Hamming code.  The three-bit repetition code introduced in the previous section is the Hamming code for $l=2$.

The term low-density-parity-check (LDPC) code refers to any code defined by a parity check matrix such that the number of $0$'s in each row and column greatly exceeds the number of $1$'s.  The initial work on LDPC codes by Gallager~\cite{Gallager63} concerned parity-check matrices with fixed row and column weights, though, at present, one or both of these constraints are frequently relaxed, with sparse random matrices being a popular choice for parity checks.  The motivation behind this unusual construction is that large LDPC codes have the capacity to achieve very high encoding rates ($k/n$ comparable to Shannon's limit~\cite{Shannon48,MacKay}) while still being efficiently decodable.  The ability to efficiency decode syndrome information rapidly becomes important as the size of a code $n$ increases.  Nearly any randomly chosen linear code achieves the Shannon limit, but the resources required to decode the syndrome for such a code scale exponentially in $n$.   LDPC codes are interesting because there exists an algorithm, the belief propagation or sum-product algorithm, that permits them to be decoded approximately using only order $n$ operations~\cite{MacKay}.

Any code can be extended by adding a column of $0$'s and row of $1$'s to the parity check matrix.  If $\mat{H}$ is the $n\times m$ parity check matrix defining the initial code, then $\mat{H}^\prime$ defines an extended code where
\begin{align}
{\mat{H}^\prime}^{jk}=
\begin{cases}
0 & \text{if } k=n+1\text{ and } j\leq m \\
1 & \text{if } j=m+1 \\
\mat{H}^{jk} & \text{otherwise}
\end{cases}.
\end{align}
Extending a code never reduces the distance since $\mat{H}^\prime\cdot e^\prime = 0$ implies
\begin{align}
\wt(e^\prime)=0 \mod 2 && \text{and} && \mat{H}\cdot e = 0
\end{align}
where $\wt(e^\prime)$ is the weight of $e^\prime$ and $e_j = e^\prime_j$ for all $j\leq n$.  In other words, an error $e^\prime$ that is undetectable by $\mat{H}^\prime$ contains an error $e$ that is undetectable by $\mat{H}$.  A similar argument shows that the distance is increased iff the distance of the original code is odd, as in the case of the Hamming codes.  The extended Hamming codes have parameters $[2^l,2^l-(l+1),4]$; in a slight abuse of notation I subsume them under the label $\code{H}_{n}$.

A code is punctured by deleting one of its bits.  In terms of the generator matrix $\mat{G}$ this is simply the deletion of a column.  Clearly puncturing is likely to reduce the distance since deleting one non-zero bit of a weight $d$ codeword yields a weight $d-1$ codeword.  It is also clear, however, that puncturing cannot reduce the distance by more than $1$.  To express the action of puncturing the $j$th bit of a code using its parity-check matrix, it is helpful to first take products of rows such that the $j$th column of $\mat{H}$ contains at most one $1$.  Puncturing the code then deletes the row containing a $1$ in column $j$ and column $j$.

A very different method of constructing new codes from old is concatenation.  Given two codes $\code{C}_A$ and $\code{C}_B$, we can define a concatenated code $\code{C}_B\circ\code{C}_A$ consisting of the code $\code{C}_B$ where each of the unencoded states composing $\code{C}_B$ is replaced by the corresponding logical state (or codeword) of $\code{C}_A$.  If $\code{C}_A$ is an $[n_A,k_A,d_A]$ code and $\code{C}_B$ is an $[n_B,k_B,d_B]$ code then the code $\code{C}_B\circ\code{C}_A$ is an $[n_An_B,k_B,d_Bd_A]$ code.  The distance $d_Bd_A$ arises from the fact that an undetectable error on $\code{C}_B\circ\code{C}_A$ corresponds to at least $d_B$ errors on the higher level code $\code{C}_B$ each of which, again, if they are to be undetectable, correspond to at least $d_A$ errors in the underlying code $\code{C}_A$.  The primary advantage of using a concatenated code instead of a larger single-layer (or block) code is that the difficulty of decoding grows slowly with the number of layers of concatenation.  The ratio of distance to size, however, decays exponentially; concatenating $\code{C}_A$ $l$ times yields a code with distance to size of $(d_A/n_A)^l$.

The dual of a code $\code{C}$, denoted $\code{C}^\perp$, is the code corresponding to the linear subspace orthogonal to $\code{C}$.  Thus, if $\mat{G}$ and $\mat{H}$ are the generator and parity check matrices of $\code{C}$ then $\mat{G}^\prime=\mat{H}$ and $\mat{H}^\prime=\mat{G}$ are the generator and parity check matrices of $\code{C}^\perp$.

The $7$-bit Hamming code $\code{H}_7$ and its extension $\code{H}_8$ crop up repeatedly through this this work.  $\code{H}_7$ is what is known as a dual-containing code, meaning that $\code{H}_7^\perp\subseteq\code{H}_7$ or, equivalently, that the span of $\mat{G}$ contains the span of $\mat{H}$.  $\code{H}_8$ is self-dual, meaning that $\code{H}_8^\perp=\code{H}_8$ or, equivalently, that the span of $\mat{G}$ equals the span of $\mat{H}$.  Both $\code{H}_7^\perp$ and $\code{H}_8$ also have the property that the weight of all codewords is a multiple of $4$, a property that will prove important in Section~\ref{subsec:transversalGates}.  The generator and parity check matrices of $\code{H}_7$ and $\code{H}_8$ are shown in Figure~\ref{fig:exampleParityCheckMatrices} along with those for the repetition codes introduced in the previous section.

\begin{figure}
\capstart
  \begin{tabular}{l@{\hspace{2.5em}}l@{\hspace{2.5em}}l@{\hspace{2.5em}}l}
    (a) & (b) & (c) & (d)\\
    \hspace{1em}
    $
    \left[
        \begin{array}{@{\hspace{.2em}}c@{\hspace{.8em}}c@{\hspace{.2em}}}
1 & 1
        \end{array}
    \right]$
    &
    \hspace{1em}
    $
    \left[
      \begin{array}{@{\hspace{.2em}}c@{\hspace{.8em}}c@{\hspace{.8em}}c@{\hspace{.2em}}}
1 & 1 & 1
      \end{array}
    \right]$
    &
    \hspace{1em}
    $
    \left[
      \begin{array}{@{\hspace{.2em}}c@{\hspace{.8em}}c@{\hspace{.8em}}c@{\hspace{.2em}}}
1 & 0 & 1 \\
0 & 1 & 1 \\
      \end{array}
    \right]$
    &
    \hspace{1em}
    $\left[
      \begin{array}{@{\hspace{.2em}}c@{\hspace{.8em}}c@{\hspace{.8em}}c@{\hspace{.8em}}c@{\hspace{.8em}}c@{\hspace{.8em}}c@{\hspace{.8em}}c@{\hspace{.2em}}}
1 & 0 & 1 & 0 & 1 & 0 & 1\\
0 & 1 & 1 & 0 & 0 & 1 & 1\\
0 & 0 & 0 & 1 & 1 & 1 & 1
      \end{array}
    \right]$
  \end{tabular}\\\\\\
  \begin{tabular}{l@{\hspace{5em}}l}
    (e) & (f) \\
    \hspace{1em}
    $\left[
      \begin{array}{@{\hspace{.2em}}c@{\hspace{.8em}}c@{\hspace{.8em}}c@{\hspace{.8em}}c@{\hspace{.8em}}c@{\hspace{.8em}}c@{\hspace{.8em}}c@{\hspace{.2em}}}
1 & 0 & 1 & 0 & 1 & 0 & 1\\
0 & 1 & 1 & 0 & 0 & 1 & 1\\
0 & 0 & 0 & 1 & 1 & 1 & 1\\
1 & 1 & 1 & 1 & 1 & 1 & 1
      \end{array}
    \right]$
    &
    \hspace{1em}
    $\left[
      \begin{array}{@{\hspace{.2em}}c@{\hspace{.8em}}c@{\hspace{.8em}}c@{\hspace{.8em}}c@{\hspace{.8em}}c@{\hspace{.8em}}c@{\hspace{.8em}}c@{\hspace{.8em}}c@{\hspace{.2em}}}
0 & 1 & 0 & 1 & 0 & 1 & 0 & 1\\
0 & 0 & 1 & 1 & 0 & 0 & 1 & 1\\
0 & 0 & 0 & 0 & 1 & 1 & 1 & 1\\
1 & 1 & 1 & 1 & 1 & 1 & 1 & 1
      \end{array}
    \right]$
  \end{tabular}
\caption[The generator matrix $\mat{G}$ and the parity check matrix $\mat{H}$ for a selection of codes]{The generator matrix $\mat{G}$ and the parity check matrix $\mat{H}$ for a selection of codes.  The matrices shown correspond to a) $\mat{G}=\mat{H}$ for $\code{R}_2$, b) $\mat{G}$ for $\code{R}_3=\code{H}_3$, c) $\mat{H}$ for $\code{R}_3=\code{H}_3$, d) $\mat{H}$ for $\code{H}_7$, e) $\mat{G}$ for $\code{H}_7$, and f) $\mat{G}=\mat{H}$ for $\code{H}_8$.
The three-bit repetition code $\code{R}_3$ is also the $[3,1,3]$ Hamming code $\code{H}_3$.  The $[7,4,3]$ Hamming code $\code{H}_7$ is dual containing while the two-bit repetition code $\code{R}_2$ and the $[8,4,3]$ extended Hamming code $\code{H}_8$ are both self-dual codes. \label{fig:exampleParityCheckMatrices}}
\end{figure}


\subsection{Quantum Codes\label{subsec:quantumCodes}}
Quantum and classical coding are closely related, so much of the theory of classical coding can be grafted over to the quantum regime.  The classical repetition code $\code{R}_2$, for example, can be used to define a quantum repetition code $\code{R}^X_2$ with encoded states
\begin{align}
\ket{\bar{0}} = \ket{00} && \text{and} && \ket{\bar{1}}=\ket{11}.
\end{align}
Using this encoding, a single bit flip error $X$ no longer interchanges logical states; instead, the operator $X_1X_2$ implements a logical bit flip $\bar{X}$ where, as before, the overhead line indicates a logical or encoded quantity.  This code detects single bit-flip errors just like the classical code, but, in order to maintain superpositions, error checking must be performed in a particular way.  Measuring the bit value, that is, measuring $Z$, of individual qubits of the encoded state is not an option since that would also measure the bit value of the encoded state.  If, for example, the encoded state were initially
\begin{align}
\ket{\bar{\psi}}=\frac{1}{\sqrt{2}}(\ket{\bar{0}}+\ket{\bar{1}})=\frac{1}{\sqrt{2}}(\ket{00}+\ket{11})
\end{align}
then after measuring $Z_1$ on the first qubit, which is equivalent to measuring it in the $\{\ket{0},\ket{1}\}$ basis, the state would be
\begin{align}
\ket{\bar{\psi}^\prime} =
\begin{cases}
\ket{00} & 50\%\text{ of the time} \\
\ket{11} & 50\%\text{ of the time}
\end{cases},
\end{align}
which is, whatever the result, completely classical.  In order to detect errors without inducing collapse, we need to find a set of measurements that tells us nothing about the encoded state while still conveying all available error information.

We saw just such a measurement in Section~\ref{subsec:introduceParityCheck} where the syndrome obtained from the parity-check matrix was shown to contain error information exclusively.  Physically, each row of the parity-check matrix corresponds to an operator in which the parity of a group of bits are measured.  Classical bit measurements are equivalent to $Z$ measurements quantum mechanically, so, for the example of $\code{R}^X_2$, the measurement operator  corresponding to the parity-check matrix $\left[1\ 1\right]$ is $Z_1Z_2$.  Measuring $Z_1Z_2$ has no effect on the logical basis states since they are $+1$ eigenvectors of the operator,
\begin{align}
  Z_1Z_2 \ket{00} = \ket{00} && \text{and} && Z_1Z_2 \ket{11} = (-1)^2\ket{11} = \ket{11}.
\end{align}
Indeed, the check operator $Z_1Z_2$ completely defines our code, so that, for an initial state $\ket{\psi}$, the post-measurement state $\ket{\psi^\prime}$ is
\begin{align}
  \ket{\psi^\prime} =
  \begin{cases}
    \Pi_{\code{R}^X_2} \ket{\psi} & \text{with probability }\brakket{\psi}{\Pi_{\code{R}^X_2}}{\psi} \\
    \Pi_{\code{R}^{X\perp}_2} \ket{\psi} & \text{with probability }\brakket{\psi}{\Pi_{\code{R}^{X\perp}_2}}{\psi}
  \end{cases}
\end{align}
where the outcome is specified by the measurement and $\Pi_{\code{R}^X_2} = \ket{00}\bra{00} + \ket{11}\bra{11}$ and $\Pi_{\code{R}^{X\perp}_2}=I-\Pi_{\code{R}^X_2}$ are the projectors onto the codespace and the orthogonal space respectively.  This illustrates another important part of quantum error correction, which is how coherent errors get dealt with.  We might, after all, imagine errors that do not flip a bit entirely, but put it into a superposition of flipped and not flipped.  A state initially prepared in $\ket{\psi}=\ket{\bar{0}}=\ket{00}$ might, for instance, suffer a coherent bit error such that the new state is $\ket{\psi^\prime}=\sqrt{1-\alpha^2}\ket{00}+\alpha\ket{01}$.  Such errors look very worrisome because they appear to be continuous, and we are only extracting one bit of information.  But after the measurement of $Z_1Z_2$,
\begin{align}
  \ket{\psi^\prime} =
  \begin{cases}
    \Pi_{\code{R}^X_2} \ket{\psi} = \ket{00} & \text{with probability }\brakket{\psi}{\Pi_{\code{R}^X_2}}{\psi} = 1-\alpha^2 \\
    \Pi_{\code{R}^{X\perp}_2} \ket{\psi} = \ket{01} & \text{with probability }\brakket{\psi}{\Pi_{\code{R}^{X\perp}_2}}{\psi} = \alpha^2
  \end{cases},
\end{align}
showing that the process of measuring whether there is a bit flip error or not projects $\ket{\psi^\prime}$ into either a state where there is a bit flip error or one where there is not.

Along with superposition in quantum mechanics comes relative phase.  Indeed, superposition is meaningless in the absence of well-defined relative phases.  Averaging over all possible relative phases for a superposition state, say $\left(\ket{0}+\e^{\i\theta}\ket{1}\right)/\sqrt{2}$, yields
\begin{align}
  \begin{split}
    \rho_\textrm{AVE} =& \frac{1}{4\pi} \int_0^{2\pi} d\theta\left(\ket{0}+\e^{\i\theta}\ket{1}\right)\left(\bra{0}+\e^{-\i\theta}\bra{1}\right) \\
    =& \frac{1}{4\pi} \left[\ket{0}\bra{0} \int_0^{2\pi} d\theta +\ket{1}\bra{1} \int_0^{2\pi} d\theta \right.\\
    &\hspace{9em}\left.+ \ket{0}\bra{1} \int_0^{2\pi} d\theta \e^{-\i\theta} + \ket{1}\bra{0} \int_0^{2\pi} d\theta \e^{\i\theta} \right]\\
    =& \frac{1}{2}\ket{0}\bra{0}+\frac{1}{2}\ket{1}\bra{1}
  \end{split},
\end{align}
so a completely unknown relative phase looks identical to a classical mixture of the basis states.  If we're going to store quantum data, we need to be able to detect phase as well as bit errors.  As with bit errors, however, we don't need to worry about protecting against an arbitrary phase error; preventing sign-flip errors is sufficient.  Sign-flip errors are generated by the operator $Z$, whose effect is clarified by shifting to a superposition basis such as $\ket{\pm}$.  $Z\ket{\pm}=\ket{\mp}$, thus, $Z$ errors interchange the states $\ket{+}$ and $\ket{-}$ in just the same way that $X$ errors interchange the states $\ket{0}$ and $\ket{1}$.  This similarity suggests that phase errors might be detected using a code $\code{R}_2^Z$ which has basis vectors $\ket{\bar{\pm}}=\ket{\pm\pm}$ and a logical sign flip operator $Z_1Z_2$.  As expected, a single sign flip is detected by $X_1X_2$, the check operator for this code.

Having learned how to correct both bit and phase errors, it only remains to combine these two skills so that we have a fully functional quantum code.  Perhaps the most straightforward way to implement both bit and phase-error correction is to do them each separately.  The code $\code{R}_2^Z\circ\code{R}_2^X$ detects $X$ errors on the first level using $\code{R}_2^X$ and detects $Z$ errors one the second level (where the base components are $\code{R}_2^X$ codestates) using $\code{R}_2^Z$; its logical basis states are
\begin{align}
  \ket{\bar{\pm}} = \frac{1}{2} \left(\ket{00} \pm \ket{11}\right)^{\otimes 2}.
\end{align}
$\code{R}_2^Z\circ\code{R}_2^X$ detects a single bit error and/or a single phase error.
Quantum codes, such as this, that are created by concatenating a repetition code in the bases $X$ and $Z$ are called Shor codes after their inventor Peter Shor~\cite{Shor95}.

Shor's original code was actually a concatenation of the three-bit repetition code, not the two-bit repetition code.  The resulting 9-qubit code is capable of correcting one error and has the following logical basis states
\begin{align}
  \ket{\bar{0}} = \frac{1}{\sqrt{8}} \left(\ket{000} + \ket{111}\right)^{\otimes 3} && \text{and} &&
  \ket{\bar{1}} = \frac{1}{\sqrt{8}} \left(\ket{000} - \ket{111}\right)^{\otimes 3}.
\end{align}
Note that Shor's original $[[9,1,3]]$ code assigned the upper-level logical basis states differently than was done in the preceding paragraph.

Finally, as with classical codes, a notation has been adopted for referring to quantum codes by a triplet of important parameters.  The designation $[[n,k,d]]$ is applied to a quantum code that encodes $k$ qubits in $n$ qubits such that the minimum distance (for any combination of $X$, $Y$, and $Z$ errors) is $d$.


\subsection{Stabilizer Codes}
Though the previous section focused on states, quantum error correction, like the rest of quantum mechanics, can also be approached from the point of view of operators.  In particular, the stabilizer state formalism~\cite{Gottesman96,Calderbank97,Gottesman97} has proven to be a powerful tool for quantum coding.  Section~\ref{subsec:stabilizerStates} showed that a set of $n$ stabilizer generators specifies a unique quantum state on $\group{H}^n$.  Similarly, a set of $n-k$ stabilizers can be used to specify the codespace of a quantum code encoding $k$ logical qubits on $\group{H}^n$.  The logical states are specified implicitly by choosing an additional $2k$ Pauli operators that commute with the stabilizer of the code.  These additional operators represent logical $X$ and $Z$ for each of the $k$ encoded qubits and, as such, form a matched pair of stabilizers of the sort described in Section~\ref{subsec:CliffordGroup}.  The error correcting properties of the resultant code can be described completely in terms of the commutation properties of the stabilizer.

Consider a code $\code{C}$ whose logical states are simultaneous $+1$ eigenvectors of the set of $m=n-k$ stabilizer generators $\{A^i\}$.  Because the $A^i$ are Pauli operators they each either commute or anti-commute with other Pauli operators.  Consequently, a logical state $\ket{\bar{\psi}}$ modified by a Pauli error $E$ is, like $\ket{\bar{\psi}}$ itself, an eigenvector of the stabilizer corresponding to a well defined set of measurement outcomes for the stabilizer generators $A^i$,
\begin{align}
  \brakket{\bar{\psi}}{E^\dag A^i E}{\bar{\psi}} = \pm \brakket{\bar{\psi}}{|E|^2 A^i}{\bar{\psi}} = \pm \brakket{\bar{\psi}}{A^i}{\bar{\psi}} = \pm 1.
\end{align}
The error $E$ is detectable only if $\{E,A^i\}=0$ for some $A^i$.  Together with the stabilizer generators, the logical Pauli operators provide a basis for the group of undetectable errors.  Errors that are elements of the stabilizer group, $A^i$, are harmless since $A^i\ket{\bar{\psi}}=\ket{\bar{\psi}}$, but the errors corresponding to logical operations irreparably damage the encoded state.

As for classical error correction, a set of errors $\set{E}=\{E^i\}$ is distinguishable if no pair of them yield the same state, that is if ${E^{j}}^\dag E^i$ is detectable for all $E^i,E^j\in\set{E}$.  Most classical codes, however, fail to correct pairs of errors that they cannot distinguish since applying the wrong correction completes a logical operation changing the encoded state.  By contrast, quantum codes, like classical secret-sharing codes, utilize logical states that are unaffected (or stabilized) by certain non-trivial operations.  Application of a stabilizer has no effect on the logical state, so a set of errors $\set{E}=\{E^i\}$ is correctable if ${E^j}^\dag E^i$ is either a stabilizer or a detectable operation for all $E^i,E^j\in\set{E}$.

As can be seen most clearly by considering the binary representation, the stabilizer generator is closely related to the classical parity-check matrix. In addition to the independence requirements satisfied by the parity checks, however, the stabilizer generators must also satisfy the property of commutativity.  In binary notation, commutativity equates to orthogonality under the symplectic inner product defined in Section~\ref{subsubsec:binaryGeneratorMatrix}, and the symplectic inner product also replaces the standard one in determining the detectable error set. The analogy between parity-check matrices and stabilizer generators is most apt when applied to the operators that generate errors and logical gates rather than to states, reflecting the stronger distinction between states and operators in quantum coding.

The quantum bit-flip (sign-flip) codes discussed earlier are stabilizer codes where the stabilizer generators are obtained from the rows of the parity-check matrix by placing an $I$ in the generator at locations where there are $0$'s in the corresponding row of the parity check matrix and a $Z$ ($X$) where there are $1$'s.  The stabilizer generators of a concatenated code $\code{C}_B\circ\code{C}_A$ follow simply from replacing the Pauli operators in the stabilizer generator of $\code{C}_B$ by the corresponding logical Pauli operators of $\code{C}_A$ and adding the stabilizer generators of $\code{C}_A$ on the first level groups of qubits.  The need for this second step is perhaps most easily understood by thinking of the stabilizer generator as the generator of the identity elements of a code, in which case, the inclusion of the stabilizer generators of $\code{C}_A$ in the stabilizer generators of $\code{C}_B\circ\code{C}_A$ amounts to the inclusion of the logical identity operations for the code $\code{C}_A$.

Quantum codes based on classical binary constructions do not make full use of the stabilizer formalism.  A quantum code need not correct $X$ and $Z$ errors separately.  In fact, the smallest quantum error-correcting code, a $[[5,1,3]]$ code known as the $5$-qubit code, does not have this property.  Figure~\ref{fig:exStabilizerGenerators} shows the stabilizer generators for the $5$-qubit code $\code{Q}_5$ and the codes $\code{R}_2^Z$, $\code{R}_2^X$, and $\code{R}_2^Z\circ\code{R}_2^X$ covered in the last section.

\begin{figure}
\capstart
  \begin{tabular}{l@{\hspace{2.5em}}l@{\hspace{2.5em}}l@{\hspace{2.5em}}l}
    (a) & (b) & (c) & (d) \\
    \hspace{1em}
    $\left[
    \begin{array}{@{}c@{\hspace{.4em}}c@{}}
X & X \\
    \end{array}
    \right]$
  &
    \hspace{1em}
    $\left[
    \begin{array}{@{}c@{\hspace{.4em}}c@{}}
Z & Z \\
    \end{array}
    \right]$
  &
    \hspace{1em}
    $\left[
    \begin{array}{@{}c@{\hspace{.4em}}c@{\hspace{.4em}}c@{\hspace{.4em}}c@{}}
X & X & X & X \\
Z & Z & I & I \\
I & I & Z & Z
    \end{array}
    \right]$
  &
    \hspace{1em}
    $\left[
    \begin{array}{@{}c@{\hspace{.4em}}c@{\hspace{.4em}}c@{\hspace{.4em}}c@{\hspace{.4em}}c@{\hspace{.4em}}c@{\hspace{.4em}}c@{\hspace{.4em}}c@{}}
X & Z & Z & X & I \\
I & X & Z & Z & X \\
X & I & X & Z & Z \\
Z & X & I & X & Z
    \end{array}
    \right]$
\end{tabular}
\caption[Example stabilizer generators]{Stabilizer generators for a) $\code{R}_2^Z$ b) $\code{R}_2^X$ c) $\code{R}_2^Z\circ\code{R}_2^X$ and d) $\code{Q}_5$.  $\code{R}_2^Z$ and $\code{R}_2^X$ are simply the classical two-bit repetition code in the $X$ and the $Z$ bases.  The stabilizers generators for $\code{R}_2^Z\circ\code{R}_2^X$ are constructed by replacing the $X$ operators in the stabilizer generators of $\code{R}_2^Z$ by $X\otimes X$, the logical $X$ operation for $\code{R}_2^X$, and adding the stabilizer generators of $\code{R}_2^X$ on the first level pairs of qubits.  The $5$-qubit code $\code{Q}_5$ does not have such a simple relation to classical binary codes.\label{fig:exStabilizerGenerators}}
\end{figure}

\subsection{CSS Codes}

The Shor codes, introduced in Section~\ref{subsec:quantumCodes}, are constructed by concatenation of a quantum bit-flip code whose stabilizers contain only $Z$'s and the quantum sign-flip code obtained by exchanging $Z$ for $X$ in the stabilizers of the bit-flip code.  Another class of quantum codes that obeys a kind of $X$-$Z$ symmetry is the Calderbank-Shor-Steane (CSS) codes~\cite{Calderbank96,Steane96b}.  CSS codes are defined as quantum codes for which there exists a set of stabilizer generators consisting of two sub-generators, one containing only $X$-type Pauli operators and the other obtained from the first by exchanging $X$ for $Z$.  Using binary notation, this is the statement that the stabilizer generator matrix can be written in the form
\begin{align}
  \group{S}_\set{G} =
    \left[
      \begin{array}{c|c}
\mat{H} & \mat{0} \\
\mat{0} & \mat{H}
      \end{array}
    \right]
\end{align}
where $\mat{H}\cdot\mat{H}^\trans = 0$, ensuring commutativity.  The class of CSS codes can be generalized by dropping the symmetry requirement while retaining the segregation of the generators.  Generalized CSS codes permit independent detection of $X$ and $Z$ errors and have stabilizers whose generator matrices can be written in the form
\begin{align}
  \group{S}_\set{G} =
    \left[
      \begin{array}{c|c}
\mat{H} & \mat{0} \\
\mat{0} & \mat{F}
      \end{array}
    \right]
\end{align}
where $\mat{H}\cdot\mat{F}^\trans = 0$.  Of the quantum codes we have encountered thus far, all but the $5$-qubit code are generalized CSS codes.

The canonical, and original, example of a CSS code is the Steane code, a $[[7,1,3]]$ code named in honor of its inventor, Andrew Steane~\cite{Steane96}.  As shown in Figure~\ref{fig:SteaneCode}, the binary matrix $\mat{H}$ for the Steane code is the parity check matrix for the classical Hamming code $\code{H}_7$.  The logical $X$ and $Z$ operators of the code are chosen to be
\begin{align}
  \bar{X} = X_1X_2X_3X_4X_5X_6X_7 && \text{and} && \bar{Z} = Z_1Z_2Z_3Z_4Z_5Z_6Z_7,
\end{align}
making the code fully symmetric under the interchange of $X$ and $Z$.  The Steane code is the smallest of the error-correcting CSS codes and probably the most widely used of all quantum codes.  Its popularity derives both from its size and the convenient properties of CSS codes for fault-tolerant quantum computing, a topic covered in Section~\ref{sec:faultTolerance}.

\begin{figure}
\capstart
  \begin{tabular}{l@{\hspace{3em}}l}
    (a) & (b) \\
    \hspace{1em}
    $\left[
    \begin{array}{@{}c@{\hspace{.4em}}c@{\hspace{.4em}}c@{\hspace{.4em}}c@{\hspace{.4em}}c@{\hspace{.4em}}c@{\hspace{.4em}}c@{\hspace{.4em}}c@{\hspace{.4em}}c@{\hspace{.4em}}c@{}}
X & I & X & I & X & I & X \\
I & X & X & I & I & X & X \\
I & I & I & X & X & X & X \\
Z & I & Z & I & Z & I & Z \\
I & Z & Z & I & I & Z & Z \\
I & I & I & Z & Z & Z & Z
    \end{array}
    \right]$
    &
    \hspace{1em}
    $\left[
      \begin{array}{@{\hspace{.2em}}c@{\hspace{.8em}}c@{\hspace{.8em}}c@{\hspace{.8em}}c@{\hspace{.8em}}c@{\hspace{.8em}}c@{\hspace{.8em}}c@{\hspace{.8em}}|c@{\hspace{.8em}}c@{\hspace{.8em}}c@{\hspace{.8em}}c@{\hspace{.8em}}c@{\hspace{.8em}}c@{\hspace{.8em}}c@{\hspace{.2em}}}
1 & 0 & 1 & 0 & 1 & 0 & 1 & 0 & 0 & 0 & 0 & 0 & 0 & 0 \\
0 & 1 & 1 & 0 & 0 & 1 & 1 & 0 & 0 & 0 & 0 & 0 & 0 & 0 \\
0 & 0 & 0 & 1 & 1 & 1 & 1 & 0 & 0 & 0 & 0 & 0 & 0 & 0 \\
0 & 0 & 0 & 0 & 0 & 0 & 0 & 1 & 0 & 1 & 0 & 1 & 0 & 1 \\
0 & 0 & 0 & 0 & 0 & 0 & 0 & 0 & 1 & 1 & 0 & 0 & 1 & 1 \\
0 & 0 & 0 & 0 & 0 & 0 & 0 & 0 & 0 & 0 & 1 & 1 & 1 & 1
      \end{array}
    \right]$
  \end{tabular}
\caption[The Steane code]{A stabilizer generator for the Steane code in a) standard and b) binary form.\label{fig:SteaneCode}}
\end{figure}

\subsection{Non-Pauli Errors\label{subsec:nonPauliErrors}}
Up till now I have considered only a very limited set of errors, but the variety of things that might go wrong with a quantum system is great; an error might take the form of any quantum operation.  As discussed Section~\ref{sec:quantumGates}, an arbitrary trace-preserving\footnote{Trace-decreasing errors that correspond to loss of the physical system, e.g. an ion falling out of a trap or decaying into an inaccessible state, are not treated here.  Interested readers are referred to References~\cite{Aliferis07,Knill05,Mochon04,Preskill98,Gottesman97}.}
 quantum operation $\op{E}$ on a state $\rho$ can be written as
\begin{align}
  \op{E}(\rho) = \sum_a E^a\rho {E^a}^\dag
\end{align}
where the error operators $E^a$ need only satisfy $\sum_a {E^a}^\dag E^a = I$.

In the face of such a general error model, it would seem unlikely that correcting Pauli errors should be sufficient to guard quantum data.  Recall from Section~\ref{subsec:PauliGroup}, however, that an arbitrary operator on $\group{H}^n\times\group{H}^n$ can be expanded in terms of elements of the Hermitian $n$-qubit Pauli operators $\hat{\set{P}}^n$.  Consequently, each error operator can be written in terms of $\hat{\set{P}}^n=\{A^i\}$ as
\begin{align}
  E^a = \sum_i \alpha_{ai} A^i
\end{align}
where the $\alpha_{ai}$ are complex coefficients.  The action of $\op{E}$ can thus be expressed as
\begin{align}
  \op{E}(\rho) = \sum_{aij} \alpha_{ai}\alpha_{aj}^* A^i\rho A^j. \label{eq:PauliDecompositionError}
\end{align}

For the special case $\sum_a \alpha_{ai}\alpha_{aj}^*=|\beta_i|^2 \delta_{ij}$, Equation~(\ref{eq:PauliDecompositionError}) reduces to
\begin{align}
  \op{E}(\rho) = \sum_{i} |\beta_i|^2 A^i\rho A^i. \label{eq:probabilisticPauliError}
\end{align}
Equation~(\ref{eq:probabilisticPauliError}) defines a multi-qubit stochastic Pauli channel, that is, a channel that can be interpreted as applying the Pauli operator $A^i$ to $\rho$ with probability $|\beta_i|^2$.  Stochastic Pauli channels differ from the more general case in that there are no coherences between Pauli errors.  Either error $A^i$ happens or it doesn't; the qubits never suffer from a superposition of errors $A^i$ and $A^j$.  Consequently, everything we have learned about error correction up till now applies.  For $\rho$ initially encoded in an $[[n,k,d]]$ code $\code{C}_n$, each $A^i$ with weight less than $d$ corresponds to a detectable error and each $A^i$ with weight of at most $t=\left\lfloor(d-1)/2\right\rfloor$ corresponds to a correctable error.

The error operation given by Equation~(\ref{eq:PauliDecompositionError}) does entail coherences between Pauli errors, but these, like any other kind of coherence, are susceptible to destruction by measurement.  Moreover, projective measurements of the stabilizer generators are well suited to orchestrating such a collapse since the measurements are designed to distinguish between different kinds of Pauli errors.  Given a set of stabilizer generators $\group{S}_\set{G}=\{C^i\}$ for the code $\code{C}_n$, the projectors corresponding to each $C^i$ are $(I\pm C^i)/2$.  Thus, having projectively measured each $C^i$ and obtained the values $c_i$, the pre-measurement state $\rho^\prime=\op{E}(\rho)$ is transformed to
\begin{align}
  \begin{split}
    \rho^{\prime\prime} &\propto \left(\prod_k \frac{I+c_kC^k}{2}\right) \sum_{aij} \alpha_{ai} \alpha_{aj}^* A^i \rho A^j \left(\prod_l \frac{I+c_l C^l}{2}\right) \\
    &= \sum_{aij} \alpha_{ai} \alpha_{aj}^* A^i \left(\prod_k \frac{I+\chi_{ki} c_kC^k}{2}\right) \rho \left(\prod_l \frac{I+\chi_{lj}c_l C^l}{2}\right) A^j
  \end{split}\label{eq:postErrorDetectionState}
\end{align}
where $C^kA^i = \chi_{ki} A^iC^k$.  Since $\rho$ is in the codespace stabilized by $\{C^k\}$ the projectors $(I-C^k)/2$ annihilate $\rho$ while the projectors $(I+C^k)/2$ have no effect,
\begin{align}
  \left(\frac{I+C^k}{2}\right)\rho = \rho = \rho \left(\frac{I+C^k}{2}\right) && \text{and} && \left(\frac{I-C^k}{2}\right)\rho = 0 = \rho \left(\frac{I-C^k}{2}\right).
\end{align}
Thus, the only terms in Equation~(\ref{eq:postErrorDetectionState}) that survive are those with $i$ and $j$ such that $\chi_{hi}=\chi_{hj}=c_h$ for all $h$, that is, those for which $A^i$ and $A^j$ have the measured syndrome $\{c_h\}$.  The post-measurement state is therefore
\begin{align}
  \rho^{\prime\prime} \propto \sum_{a} \sum_{i,j\in\set{F}} \alpha_{ai}\alpha_{aj}^* A^i\rho A^j \label{eq:postSyndromeMeasurementState}
\end{align}
where $\set{F}=\{i|A^iC^k = c_k C^kA^i\ \forall k\}$.

Depending on the error operation, measuring the stabilizer generators can sometimes be sufficient to remove all harmful coherences.  Let $w$ be the weight such that $w\geq \max(\wt(E^a))$ for all error operators $E^a$ of the quantum operation $\op{E}$, where $\wt(E^a)$ refers to largest weight of any Pauli operator in the decomposition of $E^a$.

Error detection works flawlessly when $w<d$, since, by definition, all $A^i$ such that $\wt(A^i)<d$ and $[A^i,C^k]=0$ for all $k$ are in $\group{S}_\set{G}$.  In other words, any Pauli operator that commutes with all of the stabilizers and satisfies the weight restriction must be a stabilizer itself, and therefore harmless to the encoded state.

Likewise, when $w\leq t=\lfloor(d-1)/2\rfloor$ then error correction succeeds since $A^j A^i \in \group{S}_\set{G}$ for all $A^i$ and $A^j$ with the same syndrome and satisfying $\wt(A^i),\wt(A^j)\leq t$.  In words, any Pauli errors with the same syndrome and satisfying the weight restrictions differ by at most an element of the stabilizer and are therefore equivalent for the purposes of error diagnosis.

More typically, $w$ is greater than the maximum error weight that the code can tolerate, so error management may not succeed perfectly.  In this case, the probability of success can be quantified by the probability that a projective measurement on the actual state should find the ideal state.  This metric is often called the fidelity (though so is its square root), and is defined as
\begin{align}
  \fun{F}\left(\ket{\theta},\sigma\right) = \brakket{\theta}{\sigma}{\theta}.
\end{align}
for a state $\ket{\theta}$ and density matrix $\sigma$.  Taking my initial state to be pure, $\rho=\ket{\phi_1}\bra{\phi_1}$, and letting $\rho^{\prime\prime\prime}$ denote the state of $\rho^{\prime\prime}$ after error correction (if we're correcting), the probability of failing to obtain $\rho^{\prime\prime\prime}=\ket{\phi_1}\bra{\phi_1}$ is given by $1-\fun{F}\left(\ket{\phi_1},\rho^{\prime\prime\prime}\right)$.  Since $\rho^{\prime\prime\prime}$ is normalized, however,
\begin{align}
  1 = \tr(\rho^{\prime\prime\prime}) = \sum_h \brakket{\phi_h}{\rho^{\prime\prime\prime}}{\phi_h} =
  \fun{F}(\ket{\phi_1},\rho^{\prime\prime\prime}) + \sum_{h\neq1} \brakket{\phi_h}{\rho^{\prime\prime\prime}}{\phi_h}\label{eq:normalizationAndFidelity}
\end{align}
where $\{\ket{\phi_h}\}$ is an orthonormal basis.  Applying Equation~(\ref{eq:normalizationAndFidelity}), the failure probability is given by
\begin{align}
  \fun{P}(\textit{failure}) = 1-\fun{F}\left(\ket{\phi_1},\rho^{\prime\prime\prime}\right) = \sum_{h\neq1} \brakket{\phi_h}{\rho^{\prime\prime\prime}}{\phi_h}.\label{eq:generalFailureProbability}
\end{align}

Now let $v$ be the maximum error weight that the code can tolerate, either $d-1$ or $t$ depending on whether error correction or detection is being performed.  The terms of $\rho^{\prime\prime}$ that wind up contributing to the failure probability in Equation~(\ref{eq:generalFailureProbability}) are those for which both $A^i$ and $A^j$ have weight greater than $v$.  This is because, for all other terms of $\rho^{\prime\prime}$, the corrective action taken is appropriate for one of the two Pauli errors, thus yielding a term in $\rho^{\prime\prime\prime}$ where one of the state vectors is $\ket{\phi_{1}}$ and therefore orthogonal to all $\ket{\phi_{h\neq1}}$.

For independent, local errors with amplitude $\sqrt{p}$, each non-identity element in a Pauli error brings an additional factor of $\sqrt{p}$.  Consequently, terms in $\rho^{\prime\prime}$ with $\wt(A^i),\wt(A^j)\geq v$ occur with probability of at most order $p^v$, showing that the failure probability for independent, local errors goes like, at worst, $p^v$.

\section{Fault Tolerance\label{sec:faultTolerance}}

Our discussion of error correction was founded upon the assumption that errors on different components are independent, and, thus, that the most likely errors are those of low weight, that is, those affecting few qubits.  Errors are never perfectly isolated, however, so, even should the independence assumption hold for the creation of errors, there is nothing to guarantee that errors remain independent as a computation proceeds.  Indeed, quantum computers are particularly prone to spreading errors.  Contrary to the case of classical computing, where two-bit gates can be made essentially unidirectional, the unitary two-qubit gates employed for quantum computing are inherently bidirectional, that is, they transfer information, and therefore errors, between the qubits acted upon in both directions.  However errors are generated, if they are permitted to spread indiscriminately the final error distribution will violate the independence assumption necessary for making minimum distance error correction work.

Fault-tolerant design~\cite{DiVincenzo96,Shor97,Preskill98,Aharonov97} is an approach to quantum computing that seeks to minimize the spread of errors.  Formally we can define fault tolerance as
\begin{description}
  \item[Fault tolerance] The property that errors affecting $r<w$ components involved in an operation cannot result in more than $r$ errors on a single encoded block.
\end{description}
where $w$ is some weight of interest, typically, the number of correctable errors.
The focus on individual encoded blocks reflects the fact that encoded blocks are corrected independently.  Spreading errors between blocks increases the frequency of errors on individual qubits of a block, but spreading errors within a block generates correlated block errors from uncorrelated ones.

To clarify the distinction, consider the case of error detection using the two-qubit quantum bit-flip code $\code{R}_2^X$.  Suppose that we have a logical qubit prepared in the state $\ket{\bar{0}}=\ket{00}$ that has suffered, with probability $p$, an $X$ error on the first encoding qubit; the corresponding density operator is
\begin{align}
  \rho_{\bar{1}} = (1-p)\ket{00}\bra{00} + p \ket{10}\bra{10}.
\end{align}
At worst, $\rho_{\bar{1}}$ has a single, detectable error; this occurs with probability $p$.
Given a second copy of $\ket{\bar{0}}$ whose second encoding qubit is similarly uncertain, the joint state of the four qubits can be written
\begin{align}
  \rho =& ((1-p)\ket{00}\bra{00} + p \ket{10}\bra{10})\otimes((1-p)\ket{00}\bra{00} + p \ket{01}\bra{01}) \nonumber \\
    =& (1-p)^2(\ket{00}\otimes\ket{00})(\bra{00}\otimes\bra{00}) + p(1-p) (\ket{00}\otimes\ket{01})(\bra{00}\otimes\bra{01}) \nonumber \\
    &+ p(1-p) (\ket{10}\otimes\ket{00})(\bra{10}\otimes\bra{00}) + p^2 (\ket{10}\otimes\ket{01})(\bra{10}\otimes\bra{01}).
\end{align}

Now let a $\CX$ gate be applied from the second qubit of the second code block to the second qubit of the first code block.  The resulting state is
\begin{align}
  \rho^\prime =& \CX_{\bar{2}2,\bar{1}2}\rho\CX_{\bar{2}2,\bar{1}2}^\dag \nonumber \\
    =& (1-p)^2(\ket{00}\otimes\ket{00})(\bra{00}\otimes\bra{00}) + p(1-p) (\ket{01}\otimes\ket{01})(\bra{01}\otimes\bra{01}) \nonumber \\
    &+ p(1-p) (\ket{10}\otimes\ket{00})(\bra{10}\otimes\bra{00}) + p^2 (\ket{11}\otimes\ket{01})(\bra{11}\otimes\bra{01}).
\end{align}
Tracing over the second logical qubit yields
\begin{align}
  \begin{split}
    \rho^\prime_{\bar{1}} &= \tr_{\bar{2}}(\rho^\prime) \\
    &= (1-p)^2\ket{00}\bra{00} + p(1-p)\ket{01}\bra{01} + p(1-p)\ket{10}\bra{10} + p^2 \ket{11}\bra{11},
  \end{split}
\end{align}
showing that, while the total probability of an $X$ error on qubit $2$ has increased to roughly $p$, errors on the two qubits can still be considered independent since the probability of an $X$ error on both qubits is $p^2$.

By contrast, suppose that we had applied a $\CX$ gate between the first and second qubits of $\rho_{\bar{1}}$.
The final state is then
\begin{align}
  \rho^\prime_{\bar{1}} =& \CX_{1,2}\rho\CX_{1,2}^\dag = (1-p)\ket{00}\bra{00} + p \ket{11}\bra{11},
\end{align}
which suffers from two $X$ errors with probability $p$.
By applying the $\CX$ gate between qubits in the same encoded block, we have converted a single, detectable bit error into a pair of bit errors that occur with the same probability and are undetectable.

In addition to demonstrating the importance of fault tolerance, the preceding example suggests the most straightforward method of achieving it, which is through the avoidance of gates that couple qubits within the same encoded block.  For some codes a surprising number of encoded operations can be applied in this manner.  In this dissertation, I refer to all such operations as transversal, a designation differing from some of the
literature, where the term transversal is assumed additionally to imply that the same component operation is applied to each qubit of a block.  I use the term homogeneous is used to specify operations satisfying both properties.  Subsection~\ref{subsec:transversalGates} describes the encoded operations that can be implemented transversally for CSS codes.

Performing encoded gates fault-tolerantly does not allow us to avoid error correction; even errors that build nicely must be corrected periodically lest they become overwhelming.  As explained in Section~\ref{subsec:quantumCodes}, however, quantum error correction requires the measurement of non-local check operators, the results of which are then used to infer the location of errors on the data.  This introduces two possible paths by which errors might spread.  The first is that the process of determining the value of the check operators, which necessarily involves combining information from multiple qubits of the same encoded block, might allow the same error to leak into the data more than once.  The second issue is that a single mistaken check operator can sometimes lead us to misinterpret the syndrome, thereby yielding many errors due to our ``correction'' of the data.  The second of these problems is dealt with by simple repetition of the check operator measurements.  The first can be avoided in a variety of clever ways explained in Subsection~\ref{subsec:errorCorrection}, all of which rely, at heart, on coupling the data qubits to an ancillary state with nice error properties.

Ancillary states turn out to be quite useful for fault-tolerant quantum computing.  Aside from their employment in error correction, Subsection~\ref{subsec:nontransversalGates} shows how, in conjunction with teleportation, ancillae can be used to fault-tolerantly implement encoded gates that cannot be performed transversally.

Ancillary states are not free, of course, so a full fault-tolerant scheme must also include a method of constructing the necessary ancillae.  Typically, ancillae are constructed in ways that are not at all fault-tolerant but are subsequently verified to ensure that the states are of adequate quality.  Subsection~\ref{subsec:ancillaPreparation} covers the standard approach to ancilla construction while Chapter~\ref{chap:ancillaConstruction} presents a method of my own.

\subsection{Pauli-error Propagation\label{subsec:propagation}}

\begin{figure}
\capstart
\centerline{
\Qcircuit @R=.5em @C=.8em {
& \gate{X} & \gate{H} & \gate{P} & \ctrl{1} & \qw & & & \gate{H} & \gate{P} & \ctrl{1} & \gate{Z} & \qw \\
& \gate{Y} & \ctrl{1} & \gate{X} & \targ & \qw & \push{\rule{1em}{0em}} & & \ctrl{1} & \gate{X} & \targ & \gate{X} & \qw\\
& \gate{Z} & \targ & \targ & \gate{P} & \qw & \raisebox{1.5em}{=} & & \targ & \targ & \gate{P} & \gate{X} & \qw \\
& \qw & \gate{Y} & \ctrl{-1} & \gate{H} & \qw & & & \gate{Y} & \ctrl{-1} & \gate{H} & \gate{X} & \qw
}
}
\caption[Example of Pauli propagation]{An example of Pauli propagation.  Equality is up to an overall phase. \label{fig:eprop}}
\end{figure}
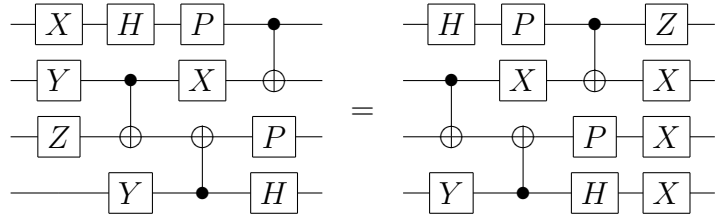

Before delving into the intricacies of fault tolerance, it is probably useful to introduce the notion of error propagation.
Pauli-error propagation relies on the fact, mentioned in Section~\ref{subsec:PauliGroup}, that the Pauli group is
invariant under conjugation by Clifford gates.  This implies that any
string of Pauli gates followed by a Clifford gate is equivalent to the
same Clifford gate followed by some (possibly different) string of Pauli
gates.  Consequently, it is possible to shuffle Pauli errors to the end of
a Clifford circuit, thus yielding a perfect circuit modified by the
resultant terminal Pauli operators (see Figure~\ref{fig:eprop}).  Moreover, for a circuit built up from gates in $\group{C}_\set{G}=\{H,P,\CX\}$, it is possible to perform this error propagation in a computationally efficient way by sequential application of a few of the circuit identities defined in Figure~\ref{fig:simpleCircuitIdentities}.

Error propagation is more generally applicable than one might imagine.  As subsequent sections show, fault-tolerant procedures based on CSS codes need never apply non-Clifford gates to the data qubits, so error propagation can typically be used to determine the impact of a Pauli error at any later point in a fault-tolerant computation.  Moreover, as discussed in Section~\ref{subsec:propagatingMoreGenErr}, Pauli-errors are frequently an acceptable substitute for more general errors and, even when this is not the case, they still provide a basis for any possible error (see Section~\ref{subsec:PauliGroup}) whose impact can thus be gauged by considering the evolution of this basis.

Knill~\cite{Knill05b,Knill05} has pointed out an additional benefit of applying only Clifford gates the data, which is that Pauli gates need never be applied.  Rather than applying Pauli gates, it is sufficient to determine by propagation the effect that they would have on subsequent measurements and to adjust the measurement outcomes accordingly.  This property is put to good use in Chapter~\ref{chap:thresholdsForHomogeneousAncillae}.

\subsection{Transversal Gates\label{subsec:transversalGates}}

The standard method of showing that a transversal operation implements some encoded gate is to demonstrate that it has the desired effect on an arbitrary logical state.  This might be called the Schr\"{o}dinger method of gate verification, since it concerns itself with transformations of the state.  Instead, I take here an operator-oriented approach to gate verification.  The two are equivalent since the requirement that encoded and unencoded measurement statistics match (which is all we really care about in the end),
\begin{align}
  \brakket{\psi}{O^\dag MO}{\psi} = \brakket{\bar{\psi}}{\bar{O}^\dag\bar{M}\bar{O}}{\bar{\psi}},
\end{align}
can be verified either by showing that encoded $O\ket{\psi}$ is $\bar{O}\ket{\bar{\psi}}$ or by showing that encoded $O^\dag MO$ is $\bar{O}^\dag\bar{M}\bar{O}$.  This second problem is made tractable by the fact, discussed in Subsection~\ref{subsec:PauliGroup}, that the Pauli operators are a basis for all Hermitian operators.  For encoded unitaries, since $\bar{Y}=-i\bar{Z}\bar{X}$ and $\bar{O}\bar{Z}\bar{X}\bar{O}^\dag=\bar{O}\bar{Z}\bar{O}^\dag\bar{O}\bar{X}\bar{O}^\dag$, it is sufficient to check the transformations of $\bar{M}=\bar{I},\bar{X},\text{ and } \bar{Z}$.  The operator $\bar{I}$ is retained here because I use it to represent not simply the unencoded identity but the entire stabilizer group and any convex combination of the elements thereof.

CSS codes are particularly well suited to transversal encoded gate constructions.  The following paragraphs cover, in quick succession, the implementation of various transversal gates on $\code{C}_n$, an $[[n,1,d]]$ CSS code for $n\in\text{Odd}$ where the logical $X$ and $Z$ operators have been chosen such that $\bar{X}=X^{\otimes n}$ and $\bar{Z}=Z^{\otimes n}$.  For the purposes of my analysis, it is convenient to choose a set of generators $\group{S}_\set{G}$ of the stabilizer of $\code{C}_n$ such that $\group{S}_\set{G}$ partitions into two subsets $\group{S}_\set{G}^X$ and $\group{S}_\set{G}^Z$, where $\group{S}_\set{G}^X$ contains only $X$-type Pauli operators and $\group{S}_\set{G}^Z$ becomes $\group{S}_\set{G}^X$ under exchange of $X$ and $Z$.

Destructive measurement need not really be transversal, but the destructive measurement of any transversal operator can, of course, be performed transversally by measuring each of the component operators and taking the product of the measurement outcomes.

Let $\bar{H}=H^{\otimes n}$.
\begin{align}
  \begin{split}
    \bar{H}\bar{X}\bar{H}&=H^{\otimes n}X^{\otimes n}H^{\otimes n} = Z^{\otimes n} = \bar{Z}\\
    \bar{H}\bar{Z}\bar{H}&=H^{\otimes n}Z^{\otimes n}H^{\otimes n} = X^{\otimes n} = \bar{X}\\
    \bar{H}\group{S}_\set{G}^{X i}\bar{H}&= H^{\otimes n}\group{S}_\set{G}^{X i}H^{\otimes n} = \group{S}_\set{G}^{Z i} \\
    \bar{H}\group{S}_\set{G}^{Z i}\bar{H}&= H^{\otimes n}\group{S}_\set{G}^{Z i}H^{\otimes n} = \group{S}_\set{G}^{X i},
  \end{split}
\end{align}
which verifies this choice of encoded $H$.
This form of $\bar{H}$ follows simply from the $X$-$Z$ symmetry of CSS codes and, consequently, applies to any fully $X$-$Z$ symmetric code.

Let $\overline{\CX}= \left(\CX\right)^{\otimes n}$, and thus, $\overline{\CX}_{jk}=\bigotimes_{h=1}^n\CX_{\bar{j}h,\bar{k}h}$.
\begin{align}
    \overline{\CX}_{jk}\bar{X}_j\overline{\CX}_{jk}&=\bar{X}_j\bar{X}_k &
    \overline{\CX}_{jk}\group{S}_{\set{G}\bar{j}}^{X i}\overline{\CX}_{jk}&=\group{S}_{\set{G}\bar{j}}^{X i} \group{S}_{\set{G}\bar{k}}^{X i} \nonumber \\
    \overline{\CX}_{jk}\bar{X}_k\overline{\CX}_{jk}&=\bar{X}_k &
    \overline{\CX}_{jk}\group{S}_{\set{G}\bar{k}}^{X i}\overline{\CX}_{jk}&=\group{S}_{\set{G}\bar{k}}^{X i} \nonumber \\
    \overline{\CX}_{jk}\bar{Z}_j\overline{\CX}_{jk}&=\bar{Z}_j &
    \overline{\CX}_{jk}\group{S}_{\set{G}\bar{j}}^{Z i}\overline{\CX}_{jk}&=\group{S}_{\set{G}\bar{j}}^{Z i} \label{eq:showTransversalCX}\\
    \overline{\CX}_{jk}\bar{Z}_k\overline{\CX}_{jk}&=\bar{Z}_j\bar{Z}_k &
    \overline{\CX}_{jk}\group{S}_{\set{G}\bar{k}}^{Z i}\overline{\CX}_{jk}&=\group{S}_{\set{G}\bar{j}}^{Z i} \group{S}_{\set{G}\bar{k}}^{Z i}, \nonumber
\end{align}
which verifies this choice of encoded $\CX$.
Since Equation~(\ref{eq:showTransversalCX}) utilizes only the segregation of $X$ and $Z$ operators, this form of $\overline{\CX}$ actually applies to general CSS codes as well.

Not all CSS codes possess a transversal implementation of the phase gate, $P$.  There exists a subset of CSS codes, however, those codes with $n=8l-1$ where $l\in\mathbb{Z}^+$ constructed from doubly-even dual-containing classical codes\footnote{A code is called doubly even if the weight of each of its codewords is a multiple of $4$.}, that satisfy the property that the number of $X$ and $Y$ operators in any stabilizer is a multiple of $4$ and likewise for $Z$ and $Y$ operators.  For such codes, the logical phase gate is given by $\bar{P}=(P^{\dagger})^{\otimes n}$, as can be seen below.
\begin{align}
  \bar{P}\bar{X}\bar{P}^\dag&=(P^\dag)^{\otimes n}X^{\otimes n}P^{\otimes n} = (i Z X)^{\otimes n} = -i\bar{Z}\bar{X} \nonumber \\
  \bar{P}\bar{Z}\bar{P}^\dag&=(P^\dag)^{\otimes n}Z^{\otimes n}P^{\otimes n} = Z^{\otimes n} = \bar{Z} \nonumber \\
  \bar{P} \group{S}_\set{G}^{X i}\bar{P}^\dag&=(P^\dag)^{\otimes n}\group{S}_\set{G}^{X i}P^{\otimes n} = i^{4g} \group{S}_\set{G}^{Z i}\group{S}_\set{G}^{X i} = \group{S}_\set{G}^{Z i}\group{S}_\set{G}^{X i} \\
  \bar{P} \group{S}_\set{G}^{Z i}\bar{P}^\dag&=(P^\dag)^{\otimes n}\group{S}_\set{G}^{Z i}P^{\otimes n} = \group{S}_\set{G}^{Z i} \nonumber
\end{align}
where $g\in\mathbb{N}$. 
The addition of a transversal implementation of logical $P$ to that for logical $H$ and logical $\CX$ completes a transversal basis for encoded Clifford operations.  It is notable that the Steane code permits such an implementation of $P$.

Given a transversal basis for encoded Clifford operations, the addition of any transversal non-Clifford encoded gate would complete a universal set of transversal encoded operations on $\code{C}_n$.  It is tempting, therefore, to seek such an implementation of, for example, the $\pi/4$-rotation, $T$.
Because $T$ is not a Clifford gate, however, it is ill suited to my Heisenberg analysis; conjugating a Pauli operator by $T^{\otimes n}$ generates sums of Pauli operators.  If, instead, we were to fall back on the Schr\"{o}dinger approach to encoded gate verification, we would find that a simple, transversal implementation of logical $T$ existed for CSS codes constructed from quadruply-even dual-containing classical codes.  Unfortunately, quadruply-even dual-containing classical codes do not exist (see Chapter 19 of Reference~\cite{MacWilliamsSloane}), and there is reason, having to do with the interaction between error correction and logical operators, to believe that transversal, universal encoded gate sets do not exist for non-trivial quantum codes.  To complete an encoded gate set, it is necessary to introduce auxiliary systems prepared in logical states that cannot be constructed using transversal encoded gates.

\subsection{Non-transversal Gates\label{subsec:nontransversalGates}}

Transversal encoded operations plus the logical state $\ket{\bar{0}}$ are not sufficient to implement an arbitrary encoded operation.  Universality is obtained, while satisfying the constraints of fault tolerance, by introducing the encoded form of some state such as
\begin{align}
  \ket{\e^{\i\pi/4}} = \frac{1}{\sqrt{2}}\left(\ket{0}+\e^{\i\pi/4}\ket{1}\right).
\end{align}
The usefulness of this state derives from its ability to serve as a proxy for the $T$ gate, as is shown below.

The circuit
\begin{align}
\begin{array}{l}
  \Qcircuit @R=.5em @C=1em {
& & \lstick{\ket{\psi}} & \targ & \meter \cwx[1]\\
& & \lstick{\ket{+}} & \ctrl{-1} & \gate{X} & \qw & \qw
  }
\end{array}
\end{align}
transfers the state $\ket{\psi}=\alpha\ket{0}+\beta\ket{1}$ from the first to the second qubit.  This can be seen from the action of the $\CX$ gate
\begin{align}
  \begin{split}
    \CX_{21} \ket{\psi}\ket{+} &= \CX_{21} (\alpha\ket{0}+\beta\ket{1})\frac{1}{\sqrt{2}}(\ket{0}+\ket{1}) \\
      &= \frac{1}{\sqrt{2}}((\alpha\ket{0}+\beta\ket{1})\ket{0}+(\alpha\ket{1}+\beta\ket{0})\ket{1}) \\
      &= \frac{1}{\sqrt{2}}(\ket{0}(\alpha\ket{0}+\beta\ket{1})+\ket{1}(\alpha\ket{1}+\beta\ket{0})) \\
      &= \frac{1}{\sqrt{2}}\left(\ket{0}\ket{\psi}+\ket{1}X_2\ket{\psi}\right)
  \end{split};\label{eq:transferState}
\end{align}
after measurement and the conditional correction, the state of the second qubit is $\ket{\psi}$.  We might, therefore, apply the $\pi/4$ rotation $T$ to the state $\ket{\psi}$ by applying the circuit in Equation~(\ref{eq:transferState}) and subsequently applying $T$ to the output.  This roundabout application is interesting because of the way that it can be transformed by circuit identities,
\begin{multline}
  \begin{array}{c}
  \Qcircuit @R=.5em @C=1em {
& & \lstick{\ket{\psi}} & \targ & \meter \cwx[1]\\
& & \lstick{\ket{+}} & \ctrl{-1} & \gate{X} & \gate{T} & \qw
  }
  \raisebox{-1em}{\hspace{.9em}$=$\hspace{1em}}
  \Qcircuit @R=.5em @C=1em {
& & \lstick{\ket{\psi}} & \targ & \qw & \meter \cwx[1]\\
& & \lstick{\ket{+}} & \ctrl{-1} & \gate{T} & \gate{T X T^\dag} & \qw
  }
  \end{array}
  \\
  \begin{array}{c}
  \raisebox{-1em}{$=$\hspace{1em}}
  \Qcircuit @R=.5em @C=1em {
& & \lstick{\ket{\psi}} & \qw & \targ & \meter \cwx[1]\\
& & \lstick{\ket{+}} & \gate{T} & \ctrl{-1} & \gate{\e^{-\i\pi/4} P X} & \qw
  }
  \raisebox{-1em}{\hspace{.9em}$\propto$\hspace{1em}}
  \Qcircuit @R=.5em @C=1em {
& & & \lstick{\ket{\psi}} & \targ & \meter \cwx[1] & \control \cw \cwx[1] \\
& & & \lstick{\ket{\e^{\i\pi/4}}} & \ctrl{-1} & \gate{X} & \gate{P} & \qw
  }
  \end{array}.\label{eq:equivalentTImplementationCircuits}
\end{multline}
The final circuit of Equation~(\ref{eq:equivalentTImplementationCircuits}) is composed entirely of Clifford gates.  Thus, since $T$ together with the Clifford gates constitutes a universal gate set, the ability to apply Clifford gates and to make the state $\ket{\e^{\i\pi/4}}$ is sufficient for universal quantum computation.  In terms of fault tolerance, this allows us to transmute the problem of non-transversal gates into a new problem, that of preparing encoded ancillae~\cite{Gottesman99}.  The details of ancilla preparation are discussed in Section~\ref{subsec:ancillaPreparation}.

\subsection{Error Correction\label{subsec:errorCorrection}}

Error correction entails projectively measuring non-local operators on the data qubits.  There are a variety of way to do this.  For any Hermitian operator $M$ with eigenvalues $\pm1$, the circuit
\begin{align}
  \Qcircuit @R=.4em @C=.8em {
& & \lstick{\ket{0}} & \gate{H} & \ctrl{1} & \gate{H} & \meter \\
& & \lstick{\ket{\psi}} & \qw & \gate{M} & \qw & \qw & \qw
  }
\end{align}
performs a projective measurement of $M$ on the state $\ket{\psi}$.  This follows from the action of the unitaries
\begin{align}
  H_1 \C{M}_{12} H_1 \ket{0} \ket{\psi} &= H_1 \C{M}_{12} \frac{1}{\sqrt{2}}(\ket{0}+\ket{1}) \ket{\psi} \nonumber \\
  &= H_1 \frac{1}{\sqrt{2}}(\ket{0} \ket{\psi}+\ket{1}  M_2\ket{\psi}) \nonumber \\
  &= \frac{1}{2} ((\ket{0}+\ket{1}) \ket{\psi}+(\ket{0}-\ket{1})  M_2\ket{\psi}) \label{eq:projectiveMeasurementCircuit}\\
  &= \ket{0} \frac{(I_2+M_2)}{2}\ket{\psi}+\ket{1} \frac{(I_2-M_2)}{2}\ket{\psi} \nonumber
\end{align}
and the fact that the projectors onto the $\pm1$ eigenspaces are $(I\pm M)/2$ for an operator $M$ with only $\pm1$ eigenvalues.
The same circuit works for an $w$-qubit measurement $\bar{M}$ with eigenvalues $\pm1$, in which case the control gate becomes $\C{\bar{M}}$, a multi-qubit operation controlled by a single qubit.  When $M$ decomposes into a tensor product of single qubit operators with $w$ non-trivial elements,
\begin{align}
M = \prod_{i=1}^w M^i_i, \label{eq:transversalMeasurement}
\end{align}
$\C{\bar{M}}$ decomposes into $w$ two-qubit controlled-gates
\begin{align}
  \C{\bar{M}}_{j\bar{k}} = \prod_{i=1}^w \C{(M^i)}_{j,\bar{k}i}, \label{eq:badTransversalMeasurementImplementation}
\end{align}
and, for $M\in\group{P}^n$, $\C{\bar{M}}$ can be implemented using Clifford gates exclusively.

\begin{figure}
\capstart
  \begin{tabular}{l}
    \Qcircuit @R=.2em @C=.52em {
& & \text{\small$\ket{0}$\hspace{1.3em}} & \gate{H} & \ctrl{7} & \ctrl{5} & \gate{X} & \ctrl{3} & \ctrl{1} & \gate{H} & \meter \\
\push{\rule[-.29em]{0em}{1.34em}}& \qw & \qw & \qw & \qw & \qw & \qw & \qw & \targ & \qw & \qw & \qw \\
& \qw & \qw & \qw & \qw & \qw & \qw & \qw & \qw & \qw & \qw & \qw \\
\push{\rule[-.29em]{0em}{1.34em}}& \qw & \qw & \qw & \qw & \qw & \qw & \targ & \qw & \qw & \qw & \qw \\
& \qw & \qw & \qw & \qw & \qw & \qw & \qw & \qw & \qw & \qw & \qw \\
\push{\rule[-.29em]{0em}{1.34em}}& \qw & \qw & \qw & \qw & \targ & \qw & \qw & \qw & \qw & \qw & \qw \\
& \qw & \qw & \qw & \qw & \qw & \qw & \qw & \qw & \qw & \qw & \qw \\
\push{\rule[-.29em]{0em}{1.34em}}& \qw & \qw & \qw & \targ & \qw & \qw & \qw & \qw & \qw & \qw & \qw
    }
    \raisebox{-3.5em}{$\hspace{1em}=\hspace{.7em}$}
    \Qcircuit @R=.2em @C=.52em {
& & \text{\small$\ket{0}$\hspace{1.3em}} & \gate{H} & \gate{X} & \ctrl{7} & \ctrl{5} & \ctrl{3} & \ctrl{1} & \gate{H} & \meter \\
\push{\rule[-.29em]{0em}{1.34em}}& \qw & \qw & \qw & \qw & \qw & \qw & \qw & \targ & \qw & \qw & \qw \\
& \qw & \qw & \qw & \qw & \qw & \qw & \qw & \qw & \qw & \qw & \qw \\
\push{\rule[-.29em]{0em}{1.34em}}& \qw & \qw & \qw & \qw & \qw & \qw & \targ & \qw & \qw & \qw & \qw \\
& \qw & \qw & \qw & \qw & \qw & \qw & \qw & \qw & \qw & \qw & \qw \\
\push{\rule[-.29em]{0em}{1.34em}}& \qw & \qw & \qw & \gate{X} & \qw & \targ & \qw & \qw & \qw & \qw & \qw \\
& \qw & \qw & \qw & \qw & \qw & \qw & \qw & \qw & \qw & \qw & \qw \\
\push{\rule[-.29em]{0em}{1.34em}}& \qw & \qw & \qw & \gate{X} & \targ & \qw & \qw & \qw & \qw & \qw & \qw
    }
    \\\\
    \raisebox{-3.5em}{$\hspace{2em}=\hspace{.7em}$}
    \Qcircuit @R=.2em @C=.52em {
& & \text{\small$\ket{0}$\hspace{1.3em}} & \gate{Z} & \gate{H} & \ctrl{7} & \ctrl{5} & \ctrl{3} & \ctrl{1} & \gate{H} & \meter \\
\push{\rule[-.29em]{0em}{1.34em}}& \qw & \qw & \qw & \qw & \qw & \qw & \qw & \targ & \qw & \qw & \qw \\
& \qw & \qw & \qw & \qw & \qw & \qw & \qw & \qw & \qw & \qw & \qw \\
\push{\rule[-.29em]{0em}{1.34em}}& \qw & \qw & \qw & \qw & \qw & \qw & \targ & \qw & \qw & \qw & \qw \\
& \qw & \qw & \qw & \qw & \qw & \qw & \qw & \qw & \qw & \qw & \qw \\
\push{\rule[-.29em]{0em}{1.34em}}& \qw & \qw & \qw & \gate{X} & \qw & \targ & \qw & \qw & \qw & \qw & \qw \\
& \qw & \qw & \qw & \qw & \qw & \qw & \qw & \qw & \qw & \qw & \qw \\
\push{\rule[-.29em]{0em}{1.34em}}& \qw & \qw & \qw & \gate{X} & \targ & \qw & \qw & \qw & \qw & \qw & \qw
    }
    \raisebox{-3.5em}{$\hspace{1em}=\hspace{.7em}$}
    \Qcircuit @R=.2em @C=.52em {
& & \text{\small$\ket{0}$\hspace{1.3em}} & \gate{H} & \ctrl{7} & \ctrl{5} & \ctrl{3} & \ctrl{1} & \gate{H} & \meter \\
\push{\rule[-.29em]{0em}{1.34em}}& \qw & \qw & \qw & \qw & \qw & \qw & \targ & \qw & \qw & \qw \\
& \qw & \qw & \qw & \qw & \qw & \qw & \qw & \qw & \qw & \qw \\
\push{\rule[-.29em]{0em}{1.34em}}& \qw & \qw & \qw & \qw & \qw & \targ & \qw & \qw & \qw & \qw \\
& \qw & \qw & \qw & \qw & \qw & \qw & \qw & \qw & \qw & \qw \\
\push{\rule[-.29em]{0em}{1.34em}}& \qw & \qw & \qw & \qw & \targ & \qw & \qw & \qw & \gate{X} & \qw \\
& \qw & \qw & \qw & \qw & \qw & \qw & \qw & \qw & \qw & \qw \\
\push{\rule[-.29em]{0em}{1.34em}}& \qw & \qw & \qw & \targ & \qw & \qw & \qw & \qw & \gate{X} & \qw
    }
  \end{tabular}
\caption[Error propagation in a non-fault-tolerant measurement circuit]{Error propagation in a non-fault-tolerant circuit for measuring $X_1X_3X_5X_7$.  Circuit identities show that a single $X$ error occurring on the ancillary qubit halfway through the measurement generates two $X$ errors on the data.  \label{fig:faultyMeasurement}}
\end{figure}
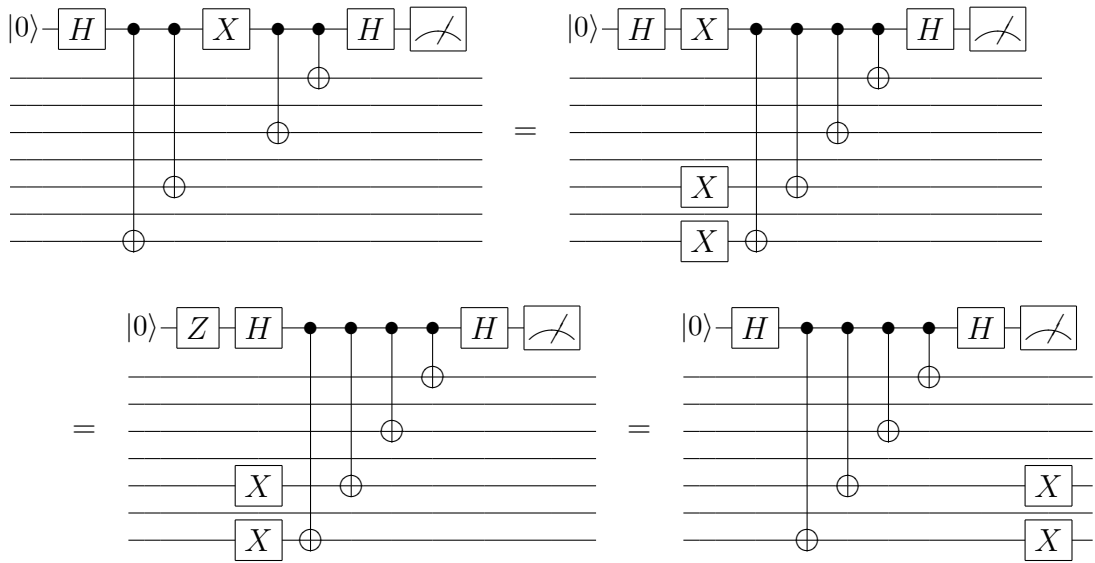

The problem with this form of measurement is that it is not fault tolerant.  Imagine, for instance, measuring $X\otimes X\otimes X\otimes X$ on the four-qubit error-detection code $\code{R}_2^Z\circ\code{R}_2^X$.  Figure~\ref{fig:faultyMeasurement} shows that a single $Z$ error on the ancilla qubit halfway through the measurement results in two errors on the data qubits.  This is possible because the measurement circuit couples the same ancillary qubit to multiple data qubits.  The problem can be avoided in at least three different ways through the use of more complex ancillae.

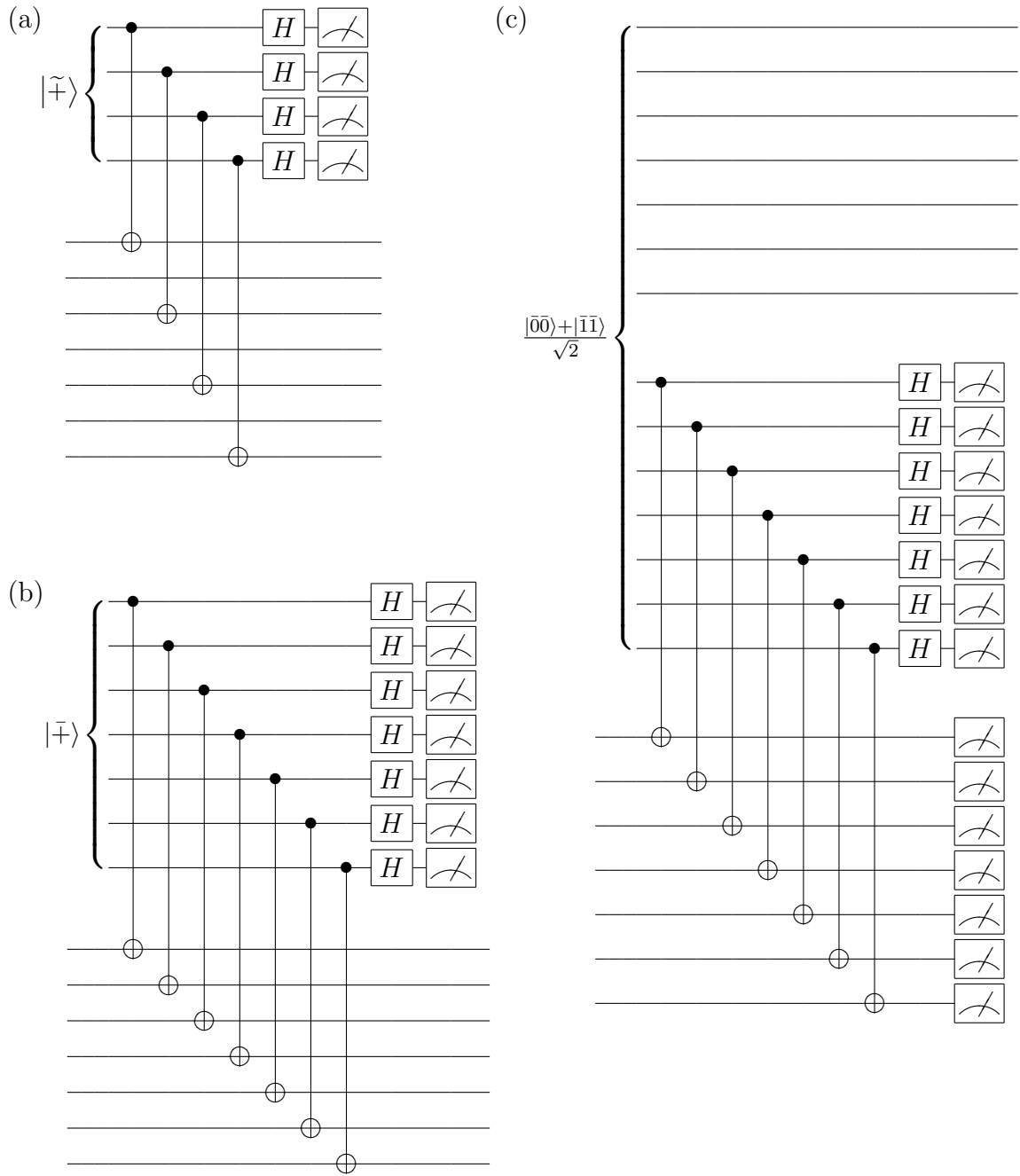
\begin{figure}
\capstart
\hspace{-1.3em}
  \begin{tabular}{l@{\hspace{0em}}l}
    \begin{tabular}{l}
  (a)
  \Qcircuit @R=.2em @C=.5em {
& & & & & \ctrl{5} & \qw & \qw & \qw & \gate{H} & \meter \\
& & & \lstick{\raisebox{-2.25em}{\small$\ket{\widetilde{+}}$}} & & \qw & \ctrl{6} & \qw & \qw & \gate{H} & \meter \\
& & & & & \qw & \qw & \ctrl{7} & \qw & \gate{H} & \meter \\
& & & & & \qw & \qw & \qw & \ctrl{8} & \gate{H} & \meter \\
\push{\rule[-.28em]{0em}{1.5em}} \\
& & \qw & \qw & \qw & \targ & \qw & \qw & \qw & \qw & \qw & \qw \\
\push{\rule[-.28em]{0em}{1.5em}} & & \qw & \qw & \qw & \qw & \qw & \qw & \qw & \qw & \qw & \qw \\
& & \qw & \qw & \qw & \qw & \targ & \qw & \qw & \qw & \qw & \qw \\
\push{\rule[-.28em]{0em}{1.5em}} & & \qw & \qw & \qw & \qw & \qw & \qw & \qw & \qw & \qw & \qw \\
& & \qw & \qw & \qw & \qw & \qw & \targ & \qw & \qw & \qw & \qw \\
\push{\rule[-.28em]{0em}{1.5em}} & & \qw & \qw & \qw & \qw & \qw & \qw & \qw & \qw & \qw & \qw \\
& & \qw & \qw & \qw & \qw & \qw & \qw & \targ & \qw & \qw & \qw
\gategroup{1}{4}{4}{5}{0em}{\{}
  }
  \vspace{4em}
  \\
  (b)
  \Qcircuit @R=.2em @C=.5em {
& & & & & \ctrl{8} & \qw & \qw & \qw & \qw & \qw & \qw & \gate{H} & \meter \\
& & & & & \qw & \ctrl{8} & \qw & \qw & \qw & \qw & \qw & \gate{H} & \meter \\
& & & & & \qw & \qw & \ctrl{8} & \qw & \qw & \qw & \qw & \gate{H} & \meter \\
& & & \lstick{\text{\small$\ket{\bar{+}}$}} & & \qw & \qw & \qw & \ctrl{8} & \qw & \qw & \qw & \gate{H} & \meter \\
& & & & & \qw & \qw & \qw & \qw & \ctrl{8} & \qw & \qw & \gate{H} & \meter \\
& & & & & \qw & \qw & \qw & \qw & \qw & \ctrl{8} & \qw & \gate{H} & \meter \\
& & & & & \qw & \qw & \qw & \qw & \qw & \qw & \ctrl{8} & \gate{H} & \meter \\
\push{\rule[-.28em]{0em}{1.5em}} \\
& & \qw & \qw & \qw & \targ & \qw & \qw & \qw & \qw & \qw & \qw & \qw & \qw & \qw \\
\push{\rule[-.28em]{0em}{1.5em}} & & \qw & \qw & \qw & \qw & \targ & \qw & \qw & \qw & \qw & \qw & \qw & \qw & \qw \\
& & \qw & \qw & \qw & \qw & \qw & \targ & \qw & \qw & \qw & \qw & \qw & \qw & \qw \\
\push{\rule[-.28em]{0em}{1.5em}} & & \qw & \qw & \qw & \qw & \qw & \qw & \targ & \qw & \qw & \qw & \qw & \qw & \qw \\
& & \qw & \qw & \qw & \qw & \qw & \qw & \qw & \targ & \qw & \qw & \qw & \qw & \qw \\
\push{\rule[-.28em]{0em}{1.5em}} & & \qw & \qw & \qw & \qw & \qw & \qw & \qw & \qw & \targ & \qw & \qw & \qw & \qw \\
& & \qw & \qw & \qw & \qw & \qw & \qw & \qw & \qw & \qw & \targ & \qw & \qw & \qw
\gategroup{1}{4}{7}{5}{0em}{\{}
  }
  \end{tabular}
  &
  \hspace{-1em}
  \raisebox{20.5em}{
  (c)\hspace{1.6em}
  \Qcircuit @R=.2em @C=.5em @!R {
& & & & & \qw & \qw & \qw & \qw & \qw & \qw & \qw & \qw & \qw & \qw \\
& & & & & \qw & \qw & \qw & \qw & \qw & \qw & \qw & \qw & \qw & \qw \\
& & & & & \qw & \qw & \qw & \qw & \qw & \qw & \qw & \qw & \qw & \qw \\
& & & & & \qw & \qw & \qw & \qw & \qw & \qw & \qw & \qw & \qw & \qw \\
& & & & & \qw & \qw & \qw & \qw & \qw & \qw & \qw & \qw & \qw & \qw \\
& & & & & \qw & \qw & \qw & \qw & \qw & \qw & \qw & \qw & \qw & \qw \\
& & & & & \qw & \qw & \qw & \qw & \qw & \qw & \qw & \qw & \qw & \qw \\
& & & \lstick{\text{\small$\frac{\ket{\bar{0}\bar{0}}+\ket{\bar{1}\bar{1}}}{\sqrt{2}}$}} \\
& & & & & \ctrl{8} & \qw & \qw & \qw & \qw & \qw & \qw & \gate{H} & \meter \\
& & & & & \qw & \ctrl{8} & \qw & \qw & \qw & \qw & \qw & \gate{H} & \meter \\
& & & & & \qw & \qw & \ctrl{8} & \qw & \qw & \qw & \qw & \gate{H} & \meter \\
& & & & & \qw & \qw & \qw & \ctrl{8} & \qw & \qw & \qw & \gate{H} & \meter \\
& & & & & \qw & \qw & \qw & \qw & \ctrl{8} & \qw & \qw & \gate{H} & \meter \\
& & & & & \qw & \qw & \qw & \qw & \qw & \ctrl{8} & \qw & \gate{H} & \meter \\
& & & & & \qw & \qw & \qw & \qw & \qw & \qw & \ctrl{8} & \gate{H} & \meter \\
\\
& & \qw & \qw & \qw & \targ & \qw & \qw & \qw & \qw & \qw & \qw & \qw & \meter \\
& & \qw & \qw & \qw & \qw & \targ & \qw & \qw & \qw & \qw & \qw & \qw & \meter \\
& & \qw & \qw & \qw & \qw & \qw & \targ & \qw & \qw & \qw & \qw & \qw & \meter \\
& & \qw & \qw & \qw & \qw & \qw & \qw & \targ & \qw & \qw & \qw & \qw & \meter \\
& & \qw & \qw & \qw & \qw & \qw & \qw & \qw & \targ & \qw & \qw & \qw & \meter \\
& & \qw & \qw & \qw & \qw & \qw & \qw & \qw & \qw & \targ & \qw & \qw & \meter \\
& & \qw & \qw & \qw & \qw & \qw & \qw & \qw & \qw & \qw & \targ & \qw & \meter
\gategroup{1}{4}{15}{5}{0em}{\{}
  }
  }
  \end{tabular}
\caption[Examples of fault-tolerant syndrome extraction]{Measurement of the Steane-code stabilizer $X_1X_3X_5X_7$ using a) Shor's, b) Steane's, and c) Knill's method of syndrome extraction.  Shor's method is the most widely applicable and necessitates the smallest ancillae.  Steane's method applies to all CSS codes and extracts all syndromes corresponding to a single kind of stabilizer ($X$ or $Z$) at once.  Knill's method requires the largest ancillae, but it extracts all $X$ and $Z$ syndromes while minimizing the number of gates applied to the data.   Given ancillae with uncorrelated errors, each of these methods is fault-tolerant.\label{fig:faultTolerantSyndromeExtractionExamples}}
\end{figure}

The first, and most broadly applicable, fault-tolerant method of performing syndrome extraction was developed by Peter Shor~\cite{Shor97}.
Figure~\ref{fig:faultTolerantSyndromeExtractionExamples}a demonstrates Shor-style measurement of $Z_1Z_3Z_5Z_7$ for the Steane code.  Compared to other fault-tolerant methods of syndrome extraction, Shor's method requires the smallest ancillae.  No assumptions about the nature of the code are required, though the measurement operator must be transversal and have eigenvalues $\pm1$.

Andrew Steane developed~\cite{Steane98} a method of syndrome extraction requiring fewer ancilla qubits and fewer gates to be applied to the data, but at the cost of requiring larger, more complex ancillae.
The process of $Z$ error extraction ($X$ generator measurement) using Steane's approach is illustrated for the Steane code in Figure~\ref{fig:faultTolerantSyndromeExtractionExamples}b.  Steane's method requires that the data be encoded using a generalized CSS code.

Knill's method of syndrome extraction~\cite{Knill05} is really less about extracting the syndrome than about determining the encoded operation that must be applied to complete teleportation of the data from one location to another.  Figure~\ref{fig:faultTolerantSyndromeExtractionExamples}c shows Knill's method for the Steane code.  It requires larger ancillae than either of the other two methods, and the ancillae must not have correlated errors of any type, but the number of gates applied to the data is reduced to a bare minimum, as are the effects of gate errors.  No additional assumptions about the nature of the code are necessary over those of Steane's method.

The next three sections provide detailed information on each of these methods of syndrome extraction.

\subsubsection{Shor's Method\label{subsubsec:ShorsMethod}}

Shor's method~\cite{Shor97} replaces the control qubit of the projective measurement circuit with a block of qubits prepared in the logical state
\begin{align}
  \ket{\widetilde{+}} = \frac{1}{\sqrt{2}}(\ket{0}^{\otimes w}+\ket{1}^{\otimes w})
\end{align}
of the $w$-qubit bit-flip code $\code{R}_w^X$, a state better known as the $w$-qubit cat state.  For measurement operators of the form given in Equation~(\ref{eq:transversalMeasurement}), this permits the transversal application of $\overline{\C{M}}$ since
\begin{align}
  \begin{split}
    E^\dag_{\bar{1}} \prod_{i=1}^w \C{\left(M^i\right)}_{\bar{1}i,\bar{2}i} &E_{\bar{1}} \ket{0}^{\otimes (w-1)}\frac{1}{\sqrt{2}}(\ket{0}+\ket{1})\ket{\bar{\psi}}_{\bar{2}} \\
    &= E^\dag_{\bar{1}} \prod_{i=1}^w \C{\left(M^i\right)}_{\bar{1}i,\bar{2}i} \frac{1}{\sqrt{2}}\left(\ket{0}^{\otimes w}+\ket{1}^{\otimes w}\right)\ket{\bar{\psi}}_{\bar{2}} \\
    &= E^\dag_{\bar{1}} \frac{1}{\sqrt{2}}\left(\ket{0}^{\otimes w}\ket{\bar{\psi}}_{\bar{2}}+\ket{1}^{\otimes w} M_{\bar{2}}\ket{\bar{\psi}}_{\bar{2}}\right) \\
    &= \ket{0}^{\otimes (w-1)} \frac{1}{\sqrt{2}}\left(\ket{0}\ket{\bar{\psi}}_{\bar{2}}+\ket{1}M_{\bar{2}}\ket{\bar{\psi}}_{\bar{2}}\right)
  \end{split}
\end{align}
where $E$ is the encoding unitary that takes $\ket{0}^{\otimes (w-1)}\ket{\phi}$ to the state $\ket{\widetilde{\phi}}$ encoded in $\code{R}_w^X$.  The encoding unitary $E$ is not fault tolerant, so the cat state must by prepared separately and tested for $X$ errors ($Z$ errors only affect the measurement outcome, not the data) prior to coupling it to the data.  The decoding unitary typically does not appear in the check measurement circuit either since decoding the ancilla and then measuring the output is equivalent to measuring each qubit in the $\ket{\pm}$ basis and interpreting the parity of the measured states as the outcome of the check operator measurement.  This works because
\begin{align}
  H^{\otimes w}\frac{1}{\sqrt{2}}\left(\ket{0}^{\otimes w} + \ket{1}^{\otimes w}\right) &= 2^{-(w+1)/2} \sum_{a\in\text{Even}} \ket{a} && \text{and} \nonumber \\
  H^{\otimes w}\frac{1}{\sqrt{2}}\left(\ket{0}^{\otimes w} - \ket{1}^{\otimes w}\right) &= 2^{-(w+1)/2} \sum_{a\in\text{Odd}} \ket{a}.
\end{align}

While fault tolerant, however, the outcome of such a measurement is not particularly reliable; a single error at any of a variety of locations results in a mistaken syndrome bit.  Thus, before any errors can be corrected, it is necessary to repeat the check measurement until a consecutive sequence of $c$ agreeing results are obtained where $c$ is equal to the lesser of the number of correctable errors plus one and the maximum number of errors that might be generated by accepting an inaccurate measurement.

\subsubsection{\label{subsubsec:SteanesMethod}Steane's Method}

Steane's method~\cite{Steane98} takes a very different approach to extracting syndrome information.  Rather than measuring each check operator individually, the existence of a set of segregated generators (into $X$-type and $Z$-type) for CSS codes is exploited to divide measurement of the entire syndrome into two phases, one where all $Z$ error information is copied to an ancillary logical basis state, and one where all $X$ error information is copied.
The logical circuit for extracting $X$ error information from the state $\ket{\bar{\psi}}$ is simply
\begin{align}
  \begin{array}{ccc}
  \Qcircuit @R=1em @C=.8em {
& & & \lstick{\bar{\ket{0}}} & \ctrl{1} & \gate{H} & \meter \\
& & & \lstick{\ket{\bar{\psi}}} & \control \qw & \qw & \qw & \qw
  }
  &\raisebox{-.7em}{$=$}&
  \Qcircuit @R=1em @C=.8em {
& & & \lstick{\bar{\ket{+}}} & \targ & \meter \\
& & & \lstick{\ket{\bar{\psi}}} & \ctrl{-1} & \qw & \qw
  }
  \end{array}.\label{eq:SteaneXExtractionLogicalCircuit}
\end{align}
That this circuit has no effect on the logical state of the data follows from the first form of Equation~(\ref{eq:SteaneXExtractionLogicalCircuit}), in absence of error, the initial $\CZ$ does nothing since the upper logical qubit is in the state $\ket{\bar{0}}$.  To see why it extracts $X$ error information, consider instead the data state $A\ket{\bar{\psi}}$ where $A$ is an $X$-type Pauli error.  Using circuit identities,
\begin{align}
  \begin{array}{ccccc}
  \Qcircuit @R=1em @C=.8em {
& & & \lstick{\bar{\ket{+}}} & \qw & \targ & \meter \\
& & & \lstick{\ket{\bar{\psi}}} & \gate{A} & \ctrl{-1} & \qw & \qw
  }
  &\raisebox{-.7em}{$=$}&
  \Qcircuit @R=1em @C=.8em {
& & & \lstick{\bar{\ket{+}}} & \targ & \gate{A} & \meter \\
& & & \lstick{\ket{\bar{\psi}}} & \ctrl{-1} & \gate{A} & \qw & \qw
  }
  &\raisebox{-.7em}{$=$}&
  \Qcircuit @R=1em @C=.8em {
& & & \lstick{\bar{\ket{+}}} & \gate{A} & \meter \\
& & & \lstick{\ket{\bar{\psi}}} & \gate{A} & \qw & \qw
  }
  \end{array}.\label{eq:whySteaneXExtractionLogicalCircuit}
\end{align}
Thus, $X$ errors are copied to the ancillary state where they result in measurement errors.  The location of those errors can be identified using the classical code corresponding to the $Z$ check operators.  Just as $X$ errors in the data migrate to the ancilla, however, $Z$ errors may enter the data from the ancilla, so it is necessary to check the ancillary logical qubit very carefully for $Z$ errors prior to use.  Extraction of $Z$ errors from the data proceeds in an identical manner via the logical circuit
\begin{align}
  \begin{array}{c}
  \Qcircuit @R=1em @C=.8em {
& & & \lstick{\bar{\ket{0}}} & \ctrl{1} & \gate{H} & \meter \\
& & & \lstick{\ket{\bar{\psi}}} & \targ \qw & \qw & \qw & \qw
  }
  \end{array}.\label{eq:SteaneZExtractionLogicalCircuit}
\end{align}

Given ancillae without correlated errors, the demands of fault tolerance are satisfied by a single application of Steane's method of syndrome extraction.  So long as the number of errors in the circuit does not exceed $t$, the number of errors correctable, the measurement can be decoded correctly to reveal location of $X$ errors on the ancilla.  In such a situation, no error can reach more than one data qubit since data qubits are not coupled and mistaking the value of a single ancilla qubit only results in a mistaken recovery operator on the associated data qubit.

While unnecessary for fault tolerance, however, Chapter~\ref{chap:thresholdsForHomogeneousAncillae} shows that multiple syndrome extractions can improve performance even when no correlations exist between errors on component qubits of ancillae.  Moreover, for correlated errors of the type that do not propagate to the data (the other kind are unsalvageable) verification of syndrome information by multiple extractions is an absolute necessity.  A typical approach is to repeat the syndrome extraction until $t+1$ consecutive measurements agree.  For $[[n,k,2t+1]]$ codes with $n$ and $t$ large this becomes impractical since extractions will almost never agree.  The appropriate number of repetitions then depends on the details of the error model describing the ancillae.

\subsubsection{Knill's Method\label{subsubsec:KnillsMethod}}

Like Steane's method, Knill's method~\cite{Knill05} of fault-tolerant syndrome extraction relies on the ability to segregate the stabilizer generators and the availability of ancillary states encoded in the same quantum code as the data.  All $X$- and $Z$-error information is extracted at once as the side effect of teleporting the logical state to a new location.  The basic logical circuit for this method is
\begin{align}
  \begin{array}{ccc}
  \Qcircuit @R=1.4em @C=.8em {
\push{\rule[-.28em]{0em}{1.5em}}& & & & & & \qw & \qw & \qw & \qw \\
& & & & & \lstick{\raisebox{3.3em}{$\frac{\ket{\bar{0}\bar{0}}+\ket{\bar{1}\bar{1}}}{\sqrt{2}}$\hspace{.7em}}} & \targ & \qw & \meter \\
& & & \lstick{\ket{\bar{\psi}}} & \qw & \qw & \ctrl{-1} & \gate{H} & \meter
\gategroup{1}{5}{2}{5}{0em}{\{}
  }
  &\raisebox{-3em}{\hspace{.8em}$=$\hspace{.5em}}&
  \Qcircuit @R=1.4em @C=.8em {
\push{\rule[-.28em]{0em}{1.5em}}& & & & & & \qw & \qw & \qw & \qw \\
& & & & & \lstick{\raisebox{3.3em}{$\frac{\ket{\bar{0}\bar{0}}+\ket{\bar{1}\bar{1}}}{\sqrt{2}}$\hspace{.7em}}} & \ctrl{1} & \gate{H} & \meter \\
& & & \lstick{\ket{\bar{\psi}}} & \qw & \qw & \targ & \qw & \meter
\gategroup{1}{5}{2}{5}{0em}{\{}
  }
  \end{array}.
\end{align}
Examining the effect of the unencoded form of the first circuit for $\ket{\psi}=\alpha\ket{0}+\beta\ket{1}$
\begin{align}
  \begin{split}
    H_3 &\CX_{32} \frac{1}{\sqrt{2}}(\ket{00}+\ket{11})\ket{\psi} =
    H_3 \CX_{32} \frac{1}{\sqrt{2}}(\ket{00}+\ket{11})(\alpha\ket{0}+\beta\ket{1}) \\
    &= \frac{1}{\sqrt{2}} H_3 (\alpha(\ket{000}+\ket{110})+\beta(\ket{011}+\ket{101})) \\
    &= \frac{1}{2} (\alpha(\ket{000}+\ket{001}+\ket{110}+\ket{111}) \\
    &\hspace{8em}+\beta(\ket{010}-\ket{011}+\ket{100}-\ket{101})) \\
    &= (\alpha\ket{0}+\beta\ket{1})\ket{00} + (\alpha\ket{0}-\beta\ket{1})\ket{01} \\
    &\hspace{8em}+(\alpha\ket{1}+\beta\ket{0})\ket{10}+(\alpha\ket{1}-\beta\ket{0})\ket{11} \\
    &= \ket{\psi}\ket{00}+Z_1\ket{\psi}\ket{01}+X_1\ket{\psi}\ket{10}+X_1Z_1\ket{\psi}\ket{00}
  \end{split}
\end{align}
we see that, modulo a corrective Pauli operator that depends on the outcome of the measurements, $\ket{\psi}$ is teleported from the last to the first qubit.  The logical circuit has precisely the same effect on logical states, but, just as in Steane's method, some (in this case the $X$) error information is transferred by the encoded $\CX$ to another logical qubit.  The final measurements thus reveal both the $X$-error locations and the $Z$-error locations, both mixed with the errors from the logical ancilla qubit.  This information is useful, however, only in that it comes in the process of decoding the logical measurement value.  The logical state no longer needs to be corrected for the discovered errors; it has been teleported clear of them.  In effect, Knill's method exchanges the error distribution of an encoded data block for that of one half of an ancilla prepared in a logical Bell state.  Failure occurs only when too many errors are present in the measurement outcomes to correctly identify the encoded Pauli operator needed to complete the teleportation.

By its nature, Knill's method of syndrome extraction is not verifiable; the original logical qubit is obliterated by the process.  It is therefore fortunate that, like Steane's method, it is fault tolerant even without repetition.  As in Steane's method, however, this only holds for ancillae with uncorrelated errors; correlated ancilla errors are fatal.

A convenient feature of this form of syndrome extraction and correction is that encoded single-qubit gates can be accomplished by performing the teleportation using a logical Bell state\footnote{If desired, multi-qubit gates can be performed using larger and more complicated ancillae.} prepared with the desired gate already applied to one half.

\subsection{Ancilla Preparation\label{subsec:ancillaPreparation}}

By now it should be clear that ancilla preparation is a crucial part of fault tolerance.  Ancillae in non-trivial states are the means by which a variety of non-fault-tolerant operations are made fault tolerant.  Such ancillae are not part of our assumed resources, however, so, in order to be complete, a procedure for fault-tolerant quantum computing must also include a method of making the requisite ancillae.

Useful ancillary states tend to be complex, highly-entangled objects.  By definition, entanglement cannot be generated between two systems without allowing them to interact, so fault-tolerant ancilla construction cannot simply rely on avoidance of couplings within an encoded block.  Instead, ancillae are typically constructed in ways that are not particularly resistant to errors, and fault-tolerance is achieved through verification and the discard of dubious ancillae.  Full verification is possible since the exact target state is known, and both the ability to discard suspect states and the low error rates assumed for bare qubits improve the output of state construction.

In this section I discuss ways of constructing the ancillae that I have used up till now which are appropriate for codes on small numbers of qubits.

\subsubsection{Cat states\label{subsubsec:prepareCats}}
I begin with an explanation of cat state construction, both because it is the simplest of the construction routines, and because cat states are a kind of primordial ancilla employed in the construction of many other ancillary states.  In the absence of errors, cat states are easy to construct; the unitary
\begin{align}
  \prod_{i=w-2}^{1} \CX_{i,i+1} H_1, \label{eq:prepareCat}
\end{align}
for example, transforms the state $\ket{0}^{\otimes w}$ to an $w$-qubit cat state since
\begin{align}
  \begin{split}
  \prod_{i=w-2}^{1}& \CX_{i,i+1} H_1 \ket{0}^{\otimes w} = \prod_{i=w-2}^{1} \CX_{i,i+1} \frac{1}{\sqrt{2}} (\ket{0}+\ket{1}) \ket{0}^{\otimes (w-1)} \\
  =& \prod_{i=w-2}^{2} \CX_{i,i+1} \frac{1}{\sqrt{2}} (\ket{00}+\ket{11}) \ket{0}^{\otimes (w-2)} = \ldots = \frac{1}{\sqrt{2}} (\ket{0}^{\otimes w}+\ket{1}^{\otimes w})
  \end{split}.
\end{align}
If the $j$th $\CX$ should fail in such a way that an $X$ error is generated on the target, however, the state produced would instead be
\begin{align}
  \frac{1}{\sqrt{2}} (\ket{0}^{\otimes j}\ket{1}^{\otimes (w-j)}+\ket{1}^{\otimes j}\ket{0}^{\otimes (w-j)}).
\end{align}
Employing such a damaged ancillae in, for example, Shor-style syndrome extraction would result in $\min(j,w-j)$ $X$ errors on the data, all from a single failed gate.

To verify cat states against $X$ errors we measure random $Z$ stabilizers $Z_iZ_j$ discarding the state if an error is found~\cite{Shor97}.  Two-qubit stabilizer measurements are easily performed fault-tolerantly using the unitary given in Equation~(\ref{eq:badTransversalMeasurementImplementation}).  (This is what makes the Bacon-Shor codes~\cite{Bacon06} so interesting.)  Ignoring uninvolved qubits, the circuit for measuring the stabilizer $A\otimes B$ is
\begin{align}
  \begin{array}{c}
  \Qcircuit @R=.5em @C=.8em {
& & & & \lstick{\ket{+}} & \ctrl{1} & \ctrl{2} & \gate{H} & \meter \\
& & & & \qw & \gate{A} & \qw & \qw & \qw & \qw \\
& & & & \qw & \qw & \gate{B} & \qw & \qw & \qw
  }
  \end{array}.
\end{align}
This circuit is fault-tolerant since any single error affecting both data qubits must occur before the application of $\C{B}$ and therefore must be transmitted to the second data qubit by the $\C{B}$ gate, implying that the error has the form $E\otimes B$, but $A\otimes B$ is a stabilizer, so $E\otimes B$ is equivalent to $(A\otimes B)\cdot(E\otimes B)=(AE)\otimes I$, which is a single qubit error.

In general, cat state verification requires measuring many $Z$ stabilizers.  For the case of a $w$-qubit cat state prepared using the unitary in Equation~(\ref{eq:prepareCat}) and utilized with an $[[n,k,d=3]]$ code, however, it is sufficient to measure $Z_1Z_w$ since any single $X$ error that spreads will continue to spread all the way to the last qubit.

\subsubsection{Logical Clifford states}
After the cat states, the most commonly used ancillae are phaseless logical Clifford states such as $\ket{\bar{0}}$, $\ket{\bar{+}}$, and $(\ket{\bar{0}\bar{0}}+\ket{\bar{1}\bar{1}})/\sqrt{2}$.  Since the encoded $\CX$ and $H$ gates are transversal for CSS codes, the ability to prepare any of these states fault-tolerantly implies the ability to prepare the rest.   In this section I discuss the preparation of $\ket{\bar{+}}$.

Let $\mat{G}$ and $\mat{H}$ be the generator matrix and the parity check matrix of the classical code associated with a CSS code $\code{C}_n$.  If the logical $X$ operators of $\code{C}_n$ are all $X$-type Pauli operators then the stabilizer generator
\begin{align}
  \group{S}_\set{G} =
    \left[
      \begin{array}{c|c}
\mat{G} & \mat{0} \\
\mat{0} & \mat{H}
      \end{array}
    \right]\label{eq:plusStabilizer}
\end{align}
specifies the logical eigenstate $\ket{\bar{+}}$.  From Section~\ref{subsec:stabilizerStates} we know how to express a stabilizer state in terms of an equally weighted sum over all stabilizers applied to some state,
\begin{align}
  \ket{\bar{+}} \propto 2^{-n} \sum_{C\in\group{S}} C\ket{0}^{\otimes n} = 2^{-n} \prod_{D\in\group{S}_\set{G}} (I+D)\ket{0}^{\otimes n}.
\end{align}
For a stabilizer with segregated generators, like that in Equation~(\ref{eq:plusStabilizer}),
\begin{align}
  \begin{split}
  \ket{\bar{+}} \propto 2^{-n} \sum_{C\in\group{S}} C\ket{0}^{\otimes n} &= 2^{-n} \sum_{A\in\group{S}^X} \sum_{B\in\group{S}^Z} AB\ket{0}^{\otimes n} \\
  &= 2^{-n} \sum_{A\in\group{S}^X} \sum_{B\in\group{S}^Z} A\ket{0}^{\otimes n} = 2^{-(n+k)/2} \sum_{A\in\group{S}^X} A\ket{0}^{\otimes n}
  \end{split},
\end{align}
which shows that $\ket{\bar{+}}$ is just an equally weighted superposition over all the bit strings in $\mat{G}$.  To create and verify this superposition, it is helpful to switch to a generator matrix $\check{\mat{G}}$ that is row reduced.  For simplicity I assume that we can do this without swapping qubits.  Let $\set{\check{G}}^i$ be the set of qubits $\{j|j>i\text{ and }\check{\mat{G}}^{ij}=1\}$, the unitary
\begin{align}
  \prod_{i=1}^{(n+k)/2} \prod_{j\in\set{\check{G}}^i} \CX_{ij} H_i \label{eq:plusPreparationUnitary}
\end{align}
applied to $\ket{0}^{\otimes n}$ produces $\ket{\bar{+}}$.  To see why, consider the $i=1$ term of Equation~(\ref{eq:plusPreparationUnitary}).  The effect of this unitary on $\ket{0}^{\otimes n}$ is to produce
\begin{align}
  \prod_{j\in\set{\check{G}}^1} \CX_{1j} H_1 \ket{0}^{\otimes n}
  = \prod_{j\in\set{\check{G}}^1} \CX_{1j} \frac{1}{\sqrt{2}} (\ket{0}+\ket{1})\ket{0}^{\otimes (n-1)}
  = \frac{1}{\sqrt{2}} \left(\ket{0}^{\otimes n}+\ket{\check{\mat{G}}^1}\right),
\end{align}
that is, it applies the operator $(I+\check{\group{S}}_\set{G}^{X1})$ to the state $\ket{0}^{\otimes n}$, where $\check{\group{S}}_\set{G}^{X}$ is the $X$-stabilizer generator corresponding to $\check{\mat{G}}$.  Since $\check{\mat{G}}$ is row reduced, the control qubits for the other values of $i$ are untouched by this process, so everything works the same for $i=2$, etc., and the effect of the entire unitary on $\ket{0}^{\otimes n}$ is to produce
\begin{align}
  2^{-(n+k)/2} \prod_{D\in\check{\group{S}}_\set{G}^X} (I+D)\ket{0}^{\otimes n} = 2^{-(n+k)/2} \sum_{A\in\group{S}^X} A\ket{0}^{\otimes n}=\ket{\bar{+}}.
\end{align}

\begin{sidewaysfigure}
\renewcommand{\baselinestretch}{1}\selectfont
\begin{tabular}{cc@{\hspace{3em}}cc@{\hspace{3em}}cc}
  (a) &
  \raisebox{-3em}{
    $\left[
      \begin{array}{@{\hspace{.2em}}c@{\hspace{.8em}}c@{\hspace{.8em}}c@{\hspace{.8em}}c@{\hspace{.8em}}c@{\hspace{.8em}}c@{\hspace{.8em}}c@{\hspace{.2em}}}
1 & 0 & 0 & 0 & 0 & 1 & 1 \\
0 & 1 & 0 & 0 & 1 & 0 & 1 \\
0 & 0 & 1 & 0 & 1 & 1 & 0 \\
0 & 0 & 0 & 1 & 1 & 1 & 1
      \end{array}
    \right]$
  }
  & (b) &
  \Qcircuit @R=.05em @C=.5em @!R {
& & \lstick{\ket{0}} & \gate{H} & \ctrl{1} & \qw & \qw & \ctrl{3} & \qw & \qw \\
& & \lstick{\ket{0}} & \qw & \targ & \ctrl{1} & \qw & \qw & \qw & \qw \\
& & \lstick{\ket{0}} & \qw & \qw & \targ & \ctrl{1} & \qw & \qw & \qw \\
& & \lstick{\ket{0}} & \qw & \qw & \qw & \targ & \targ & \meter
  }
& (c) &
  \Qcircuit @R=.05em @C=.5em @!R {
& & \lstick{\ket{0}} & \gate{H} & \ctrl{1} & \qw & \qw & \qw & \ctrl{4} & \qw & \qw \\
& & \lstick{\ket{0}} & \qw & \targ & \ctrl{1} & \qw & \qw & \qw & \qw & \qw \\
& & \lstick{\ket{0}} & \qw & \qw & \targ & \ctrl{1} & \qw & \qw & \qw & \qw \\
& & \lstick{\ket{0}} & \qw & \qw & \qw & \targ & \ctrl{1} & \qw & \qw & \qw \\
& & \lstick{\ket{0}} & \qw & \qw & \qw & \qw & \targ & \targ & \meter
  }
  \\
  \\
\end{tabular}\\
\begin{tabular}{ll}
  (d) &
  \Qcircuit @R=.3em @C=.2em {
& & & & & & & & & & & & & \ctrl{5} & \qw & \qw & \gate{H} & \measureD{} & & & \ctrl{6} & \qw & \qw & \gate{H} & \measureD{} & & & \ctrl{7} & \qw & \qw & \gate{H} & \measureD{} & & & \ctrl{8} & \qw & \qw & \qw & \gate{H} & \measureD{} \\
& & & & & & & & & & & & \lstick{\text{\small$\ket{\widetilde{+}}\ $}} & \qw & \ctrl{9} & \qw & \gate{H} & \measureD{} & \push{\rule{3.5em}{0em}} & \lstick{\text{\small$\ket{\widetilde{+}}\ $}} & \qw & \ctrl{8} & \qw & \gate{H} & \measureD{} & \push{\rule{3.5em}{0em}} & \lstick{\text{\small$\ket{\widetilde{+}}\ $}} & \qw & \ctrl{8} & \qw & \gate{H} & \measureD{} & \push{\rule{3.5em}{0em}} & \lstick{\raisebox{-2.4em}{\small$\ket{\widetilde{+}}\ $}} & \qw & \ctrl{8} & \qw & \qw & \gate{H} & \measureD{} \\
& & & & & & & & & & & & & \qw & \qw & \ctrl{9} & \gate{H} & \measureD{} & & & \qw & \qw & \ctrl{9} & \gate{H} & \measureD{} & & & \qw & \qw & \ctrl{8} & \gate{H} & \measureD{} & & & \qw & \qw & \ctrl{8} & \qw & \gate{H} & \measureD{} \\
& & & & & & & & & & & & & & & & & & & & & & & & & & & & & & & & & & \qw & \qw & \qw & \ctrl{8} & \gate{H} & \measureD{} \\
& & & & & & & & & & & & & & & & & \push{\rule{1.2em}{0em}} & & & & & & & \push{\rule{1.2em}{0em}} & & & & & & & \push{\rule{1.2em}{0em}} & & & & & & & & \push{\rule{1.2em}{0em}} \\
\push{\rule[-.4em]{0em}{1.1em}} & & \lstick{\text{\small$\ket{+}$}} & \ctrl{5} & \ctrl{6} & \qw & \qw & \qw & \qw & \qw & \qw & \qw & \qw & \targ & \qw & \qw & \qw & \qw & \qw & \qw & \qw & \qw & \qw & \qw & \qw & \qw & \qw & \qw & \qw & \qw & \qw & \qw & \qw & \qw & \qw & \qw & \qw & \qw & \qw & \qw & \qw & \qw & \qw & \qw & \qw & \qw & \qw & \qw \\
\push{\rule[-.4em]{0em}{1.1em}} & & \lstick{\text{\small$\ket{+}$}} & \qw & \qw & \ctrl{3} & \ctrl{5} & \qw & \qw & \qw & \qw & \qw & \qw & \qw & \qw & \qw & \qw & \qw & \qw & \qw & \targ & \qw & \qw & \qw & \qw & \qw & \qw & \qw & \qw & \qw & \qw & \qw & \qw & \qw & \qw & \qw & \qw & \qw & \qw & \qw & \qw & \qw & \qw & \qw & \qw & \qw & \qw & \qw \\
\push{\rule[-.4em]{0em}{1.1em}} & & \lstick{\text{\small$\ket{+}$}} & \qw & \qw & \qw & \qw & \ctrl{2} & \ctrl{3} & \qw & \qw & \qw & \qw & \qw & \qw & \qw & \qw & \qw & \qw & \qw & \qw & \qw & \qw & \qw & \qw & \qw & \qw & \targ & \qw & \qw & \qw & \qw & \qw & \qw & \qw & \qw & \qw & \qw & \qw & \qw & \qw & \qw & \qw & \qw & \qw & \qw & \qw & \qw \\
\push{\rule[-.4em]{0em}{1.1em}} & & \lstick{\text{\small$\ket{+}$}} & \qw & \qw & \qw & \qw & \qw & \qw & \ctrl{1} & \ctrl{2} & \ctrl{3} & \qw & \qw & \qw & \qw & \qw & \qw & \qw & \qw & \qw & \qw & \qw & \qw & \qw & \qw & \qw & \qw & \qw & \qw & \qw & \qw & \qw & \qw & \targ & \qw & \qw & \qw & \qw & \qw & \qw & \qw & \qw & \qw & \qw & \qw & \qw & \qw \\
\push{\rule[-.4em]{0em}{1.1em}} & & \lstick{\text{\small$\ket{0}$}} & \qw & \qw & \targ & \qw & \targ & \qw & \targ & \qw & \qw & \qw & \qw & \qw & \qw & \qw & \qw & \qw & \qw & \qw & \targ & \qw & \qw & \qw & \qw & \qw & \qw & \targ & \qw & \qw & \qw & \qw & \qw & \qw & \targ & \qw & \qw & \qw & \qw & \qw & \qw & \qw & \qw & \qw & \qw & \qw & \qw \\
\push{\rule[-.4em]{0em}{1.1em}} & & \lstick{\text{\small$\ket{0}$}} & \targ & \qw & \qw & \qw & \qw & \targ & \qw & \targ & \qw & \qw & \qw & \targ & \qw & \qw & \qw & \qw & \qw & \qw & \qw & \qw & \qw & \qw & \qw & \qw & \qw & \qw & \targ & \qw & \qw & \qw & \qw & \qw & \qw & \targ & \qw & \qw & \qw & \qw & \qw & \qw & \qw & \qw & \qw & \qw & \qw \\
\push{\rule[-.4em]{0em}{1.1em}} & & \lstick{\text{\small$\ket{0}$}} & \qw & \targ & \qw & \targ & \qw & \qw & \qw & \qw & \targ & \qw & \qw & \qw & \targ & \qw & \qw & \qw & \qw & \qw & \qw & \targ & \qw & \qw & \qw & \qw & \qw & \qw & \qw & \qw & \qw & \qw & \qw & \qw & \qw & \qw & \targ & \qw & \qw & \qw & \qw & \qw & \qw & \qw & \qw & \qw & \qw
\gategroup{1}{13}{3}{13}{.5em}{\{}
\gategroup{1}{20}{3}{20}{.5em}{\{}
\gategroup{1}{27}{3}{27}{.5em}{\{}
\gategroup{1}{34}{4}{34}{.5em}{\{}
  }
\end{tabular}
\caption[Preparing $\ket{\bar{+}}$ for the Steane code]{Preparing $\ket{\bar{+}}$ for the Steane code.  Shown are a) $\check{G}$, the row-reduced binary matrix representing the $X$ stabilizers for $\ket{\bar{+}}$ of the Steane code and circuits for constructing the b) $3$-qubit cat state, c) $4$-qubit cat state, and d) logical $\ket{+}$ for the Steane code.  If the measurements in either b) or c) yield the outcome $-1$ the state is discarded.  In d) the state is retained only when the product of the measurements in each set are $+1$.\label{fig:plusSteaneAncillaConstruction}}
\end{sidewaysfigure}
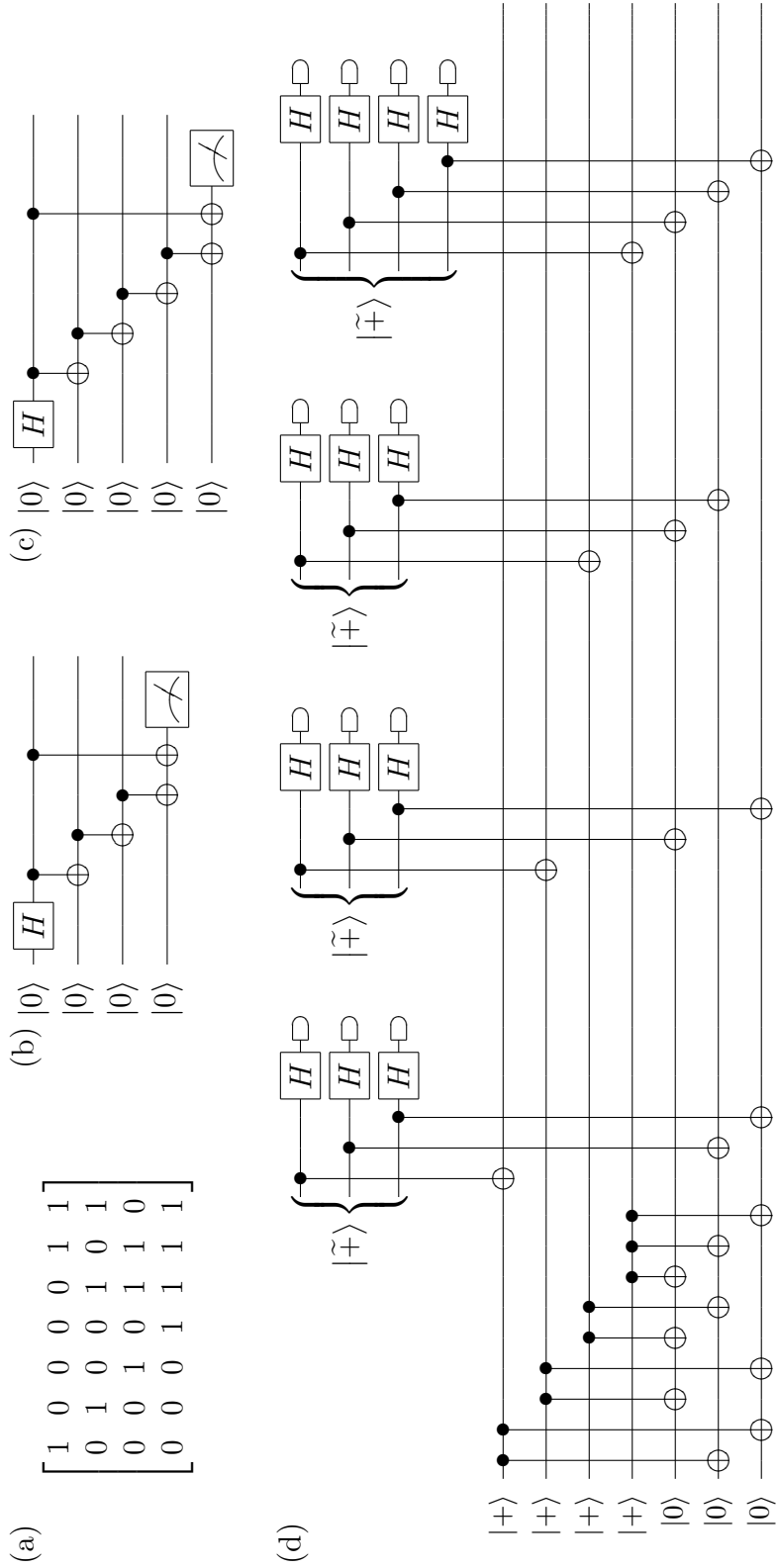

This preparation procedure for $\ket{\bar{+}}$ is not fault tolerant, however, so the state must be checked for errors.  To check for $Z$ errors the $X$-stabilizer generators are measured using Shor's syndrome extraction method.  Any state which passes all checks (failures are typically discarded) has been verified sufficiently for use in Steane's method of syndrome extraction.  For other uses, $X$ errors must also be tested for, either using Shor's method or by Steane-style syndrome extraction.  The logical circuit for a single round of verification of $\ket{\bar{+}}$ using Steane's method is
\begin{align}
    \begin{array}{c}
\Qcircuit @R=.8em @C=.8em {
& & \lstick{\ket{\bar{+}}} & \ctrl{1} & \qw & \qw & \qw \\
& & \lstick{\ket{\bar{+}}} & \targ & \gate{H} & \meter
}
    \end{array}.
\end{align}
Figure~\ref{fig:plusSteaneAncillaConstruction} shows explicitly the circuits involved in constructing $\ket{\bar{+}}$ for the Steane code and checking it for $Z$ errors.


\subsubsection{The $\pi/4$ state}

The state $\ket{\overline{\e^{\i\pi/4}}}$ was used in Section~\ref{subsec:nontransversalGates} to complete our encoded gate set, but none of the construction methods yet covered applies to this state.  In this section I remedy that oversight by presenting two different methods of constructing $\ket{\overline{\e^{\i\pi/4}}}$.

\subsubsubsection{Constructing the $\pi/4$ state through measurement}
One way that we have not yet used to prepare logical states is through measurement.  Projective measurement leaves behind the eigenstate corresponding to the measured eigenvalue, effectively extracting an eigenvector.  Thus, we can probabilistically prepare the eigenstates of any operator that we can measure, providing only that our initial state is not orthogonal to the desired state.  To use this fact to prepare $\ket{\e^{\i\pi/4}}$, we need to measure an operator with eigenvectors $\ket{\pm\e^{\i\pi/4}}$.  As shown below, $\e^{-\i\pi/4}P X$ has just these eigenvalues.
\begin{align}
  \begin{split}
    \e^{-\i\pi/4}P X \ket{\pm\e^{\i\pi/4}} &= \e^{-\i\pi/4}P X \frac{1}{\sqrt{2}}(\ket{0} \pm \e^{\i\pi/4}\ket{1}) \\
    &= \e^{-\i\pi/4}P \frac{1}{\sqrt{2}}(\ket{1} \pm \e^{\i\pi/4}\ket{0}) = \frac{\e^{-\i\pi/4}}{\sqrt{2}}(\i \ket{1} \pm \e^{\i\pi/4}\ket{0})\\
    &= \pm \frac{1}{\sqrt{2}}(\ket{0} \pm \e^{\i\pi/4}\ket{1}) = \pm\ket{\pm\e^{\i\pi/4}}
  \end{split}.
\end{align}
Moreover, the state $\ket{0}$ can be decomposed as
\begin{align}
  \ket{0}&=\frac{1}{2}(\ket{0}+\e^{\i\pi/4}\ket{1})+\frac{1}{2}(\ket{0}-\e^{\i\pi/4}\ket{1})
  =\frac{1}{\sqrt{2}}(\ket{\e^{\i\pi/4}}+\ket{-\e^{\i\pi/4}}),
\end{align}
so a projective measurement of $\e^{-\i\pi/4}P X$ on $\ket{0}$ yields the state $\ket{\e^{\i\pi/4}}$ and the state $\ket{-\e^{\i\pi/4}}$ with equal probability.

\begin{figure}
\capstart
  \begin{tabular}{l}
\Qcircuit @R=.4em @C=.8em {
& & \lstick{\ket{\widetilde{+}}} & \gate{T^\dag} & \ctrl{1} & \gate{H} & \meter \\
& & \lstick{\ket{\bar{0}}} & \qw & \gate{P X} & \qw & \qw & \qw
}
  \\ \\
\raisebox{-.5cm}{\hspace{3.5em}=\hspace{.7em}}
\Qcircuit @R=.4em @C=.8em {
& & & \lstick{\ket{\widetilde{+}}} & \gate{T_1^\dagger} & \ctrl{1} & \gate{H} & \meter \\
& & & \lstick{\ket{\bar{0}}} & \qw & \gate{(P^\dag X)^{\otimes n}} & \qw & \qw & \qw
}
  \\ \\
\raisebox{-.5cm}{\hspace{9.5em}=\hspace{.7em}}
\Qcircuit @R=.4em @C=.8em {
& & & \lstick{\ket{\widetilde{+}}} & \gate{T_1^\dagger} & \ctrl{1} & \gate{H} & \meter \\
& & & \lstick{\ket{\bar{0}}} & \gate{T^{\otimes n}} & \gate{(\e^{-\i\pi/4}X)^{\otimes n}} & \gate{T^{\dag \otimes n}} & \qw & \qw
}
  \\ \\
\raisebox{-.5cm}{\hspace{16em}=\hspace{.7em}}
\Qcircuit @R=.4em @C=.8em {
\rule{0em}{1.2em} & & & \lstick{\ket{\widetilde{+}}} & \gate{T_1^\dagger} & \gate{T^{\dag \otimes n}} & \ctrl{1} & \gate{H^{\otimes n}} & \meter & {\hspace{-.4em}\rule{0em}{1.4em}^{\otimes n}}\\
& & & \lstick{\ket{\bar{0}}} & \qw & \gate{T^{\otimes n}} & \targ & \gate{T^{\dag \otimes n}} & \qw & \qw
}
  \end{tabular}
\caption[Equivalent circuits for measuring $\overline{\e^{-\i\pi/4}P X}$]{A string of circuit identities decomposing a logical measurement circuit for $\overline{\e^{-\i\pi/4}P X}$ into one involving only single qubit gates.  These equivalences make use of the facts $\bar{P}={P^\dag}^{\otimes n}$, $\e^{\i\pi/4}TXT^\dag=PX$, and $\widetilde{T}=T_i$ and the equivalence for codestates of the repetition code (see Section~\ref{subsubsec:prepareCats}) of transversal Hadamard followed by transversal measurement to logical Hadamard followed by logical measurement.  As before, the tilde overtop states indicates they are encoded using the repetition code. \label{fig:encodedPiOver4MeasurementCircuits}}
\end{figure}
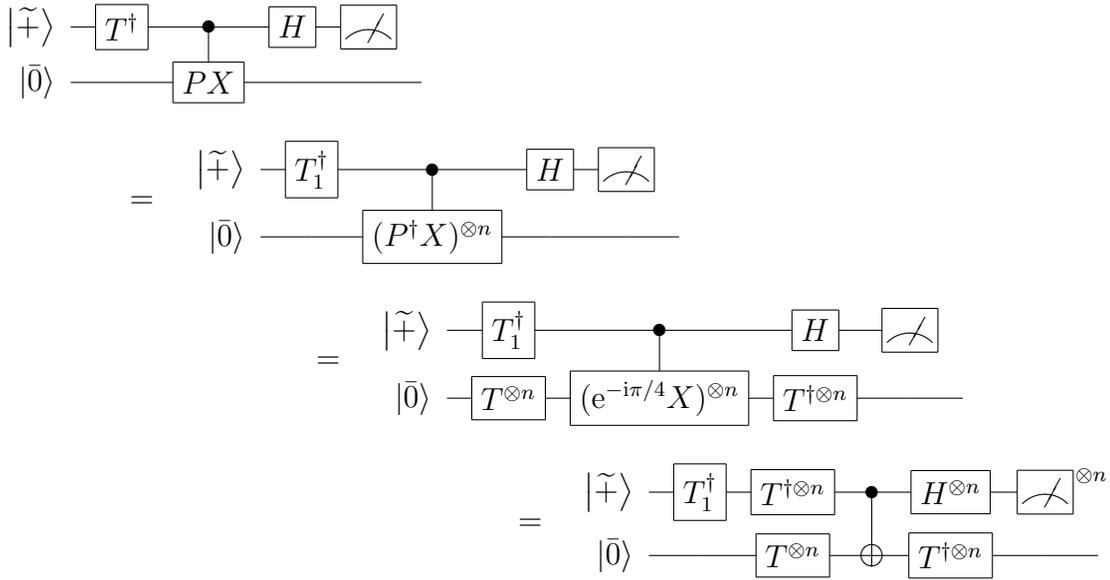

From the preceding paragraph it is clear that one method of preparing $\ket{\overline{\e^{\i\pi/4}}}$ is to measure the encoded operator $\overline{\e^{-\i\pi/4}P X}$ on the logical state $\ket{\bar{0}}$, either trying again or applying $\bar{Z}$ when the outcome is $-1$.
Implementing the encoded measurement circuit
\begin{multline}
    \begin{array}{c}
\Qcircuit @R=.4em @C=.8em {
& & \lstick{\ket{\bar{0}}} & \gate{H} & \ctrl{1} & \gate{H} & \meter \\
& & \lstick{\ket{\bar{0}}} & \qw & \gate{e^{-\i\pi/4}P X} & \qw & \qw & \qw
}
    \end{array}
    \\
    \begin{array}{c}
    \raisebox{-.7em}{\hspace{8em}$=$\hspace{1em}}
\Qcircuit @R=.4em @C=.8em {
& & \lstick{\ket{\bar{0}}} & \gate{H} & \gate{T^\dag} & \ctrl{1} & \gate{H} & \meter \\
& & \lstick{\ket{\bar{0}}} & \qw & \qw & \gate{P X} & \qw & \qw & \qw
}
    \end{array},
\end{multline}
however, requires the encoded $T^\dag$ gate, and the entire point of this exercise is to allow us to implement $\bar{T}$.  Taking our cue from Section~\ref{subsubsec:ShorsMethod} we replace the control qubit with a logical qubit prepared, not in the code of interest, but in the $n$-qubit repetition code.  For CSS codes such that $\bar{P}={P^\dag}^{\otimes n}$, the resulting encoded measurement circuit can be reexpressed in terms of unencoded gates exclusively, as shown in Figure~\ref{fig:encodedPiOver4MeasurementCircuits}.

As with other Shor-style measurements, the encoded measurement presented here is not robust, necessitating $t+1$ extractions for a $t$-error-correcting code.  The problem is more dire here, however, since a single error on the data can change every measurement result, causing us to mistakenly identify the prepared state.  To guard against this possibility, error detection must be performed between the measurements.

\subsubsubsection{Constructing the $\pi/4$ state through teleportation (State Injection)}
We saw in Section~\ref{subsec:errorCorrection} how an arbitrary state could be teleported using an ancilla prepared in the Bell state $\ket{\beta_{00}}$.  Teleportation is achieved by performing a measurement in the Bell basis on the qubit to be teleported and one of the qubits of the Bell state.  Modulo possible corrections indicated by the measurement results, this operation transfers the state of interest to the unmeasured qubit of the Bell pair.

For this section, the crucial fact to note about the teleportation process is that the two halves of the Bell pair do not need to be encoded using the same code.  Thus, the state $(\ket{0\bar{0}}+\ket{1\bar{1}})/\sqrt{2}$ can be used to teleport an arbitrary unencoded state to the encoded qubit of the Bell pair. The relevant circuit diagram is
\begin{align}
  \begin{array}{c}
    \Qcircuit @R=.5em @C=1em {
& & \lstick{\ket{\psi}} & \qw & \qw & \qw & \ctrl{1} & \gate{H} & \meter \cwx[2] \\
& & & & \lstick{\raisebox{-2.8em}{$\frac{\ket{\bar{0}\bar{0}}+\ket{\bar{1}\bar{1}}}{\sqrt{2}}$}} & \gate{\mathcal{D}} & \targ & \meter \cwx[1] \\
& & & & & \qw & \qw & \gate{X} & \gate{Z} & \qw & \rstick{\ket{\bar{\psi}}} \qw
\gategroup{2}{5}{3}{5}{.5em}{(}
    }
  \end{array},
\end{align}
where $\mathcal{D}$ is a logical decoder.
The teleported state must subsequently be verified since a single error either during the decoding process or on the unencoded state can result in a logical error on the encoded state, but verification is feasible since the correct state is produced with probability near $1$.  This method of logical state preparation, known as state injection~\cite{Knill05}, can be applied for any unencoded state, including the state $\ket{\e^{\i\pi/4}}$.

The techniques required for state injection have already been introduced.  Earlier in Section~\ref{subsec:ancillaPreparation}, a means of constructing encoded $\ket{+}$ was described.  As mentioned there, the transversal gate implementations of $\bar{H}$ and $\overline{\CX}$ given in Section~\ref{subsec:transversalGates} permit the fault-tolerant construction of encoded $(\ket{\bar{0}\bar{0}}+\ket{\bar{1}\bar{1}})/\sqrt{2}$ from the state $\ket{\bar{+}}$.  Decoding of the second logical qubit in the encoded Bell pair can be accomplished by reversing an encoding process similar (though not quite identical since care must be taken to design an encoder, minus verification procedures, that works for arbitrary input states) to that presented for $\ket{\bar{+}}$.  A single round of verification of $\ket{\overline{\e^{\i\pi/4}}}$ can be implemented using a second copy of $\ket{\overline{\e^{\i\pi/4}}}$ and a modified form of the logical measurement procedure (no cat states required) described for the previously presented method of preparing $\ket{\overline{\e^{\i\pi/4}}}$.

\section{Thresholds\label{sec:thresholds}}
No matter how skillfully designed the fault-tolerant procedure, there
remains for any finite code afflicted by independent errors a nonzero probability that too many
errors will occur in a computational step and our data will
become irreparably corrupted.  Consequently, as the length of the computation increases, the probability of failure
approaches one.  To ensure that
we can perform computations of arbitrary length, we need a way of making
the probability of an uncorrectable set of unencoded errors, i.e., an
encoded error, arbitrarily small.

Currently, the only known, viable method of achieving an arbitrarily low encoded error rate
is by concatenation of fault-tolerant procedures.  As explained in Section~\ref{sec:faultTolerance}, a fault-tolerant procedure replaces the states and gates in an unencoded circuit with encoded versions satisfying certain desirable restrictions with regard to errors.  The resulting circuit implements the unencoded circuit in an encoded subspace where qubits and operations display different error properties.  Concatenation of fault-tolerant procedures works in roughly the same way as code concatenation; the encoded states and gates of a fault-tolerant procedure are used as the ``unencoded'' states and gates for the subsequent level of encoding.  The basic idea behind this process is the following:
``If encoding qubits and gates reduces the effective error rate then encoding the
encoded qubits and gates should reduce the error rate even more.''  This provides us
with a plausible sounding way of achieving an arbitrarily small error
rate; we simply add layers of encoding until the error rate is acceptable.

But encoding will not always decrease the error rate.  It is possible for
hardware to be so error prone that the process of applying an encoded
gate and error correcting is less likely to succeed than simply applying
the unencoded gate.  From this observation arises the idea of a threshold
error probability $p_\text{th}$ for quantum computation.  The threshold is the
unencoded error probability below which we can achieve an arbitrarily low
encoded error probability using a number of qubits that scales
polynomially in the size of the problem.  Put another way, it is the error
probability below which we can compute indefinitely.

Determining the threshold exactly for a given set of assumptions has
proven to be a hard problem, but we can get some idea of its value through
bounds and estimates.

\subsection{Threshold Estimates\label{subsec:thresholdEstimates}}

Estimates of the threshold for quantum computation are generally made by
analyzing a particular fault-tolerant implementation using a specific
finite code under concatenation.  Fundamentally, these estimates derive
from the idea that encoding is undesirable if
\begin{equation}
\left\{
\parbox{5.3em}{\centering Encoded error rate}\rule{0em}{1.3em}
\right\}
>
\left\{
\parbox{5.5em}{\centering Unencoded error rate}\rule{0em}{1.3em}
\right\}. \label{eq:thresholdRestriction}
\end{equation}
Intuitively, this makes sense; we would not expect error correction to be
advantageous when an encoded gate or qubit is more likely to fail than an
unencoded one.  Equation~(\ref{eq:thresholdRestriction}) harbors some ambiguity since there are many sorts of
errors, and it might well be the case that the encoded error rate
increases for some of them, but not all.  That ambiguity can be resolved either by specifying that the inequality holds for the lowest encoded error rate and the highest unencoded error rate, or, for a finite set of possible errors, by specifying that it holds for each kind of error.

Putting aside the matter of diverse error species, however, consider the implications of Equation~(\ref{eq:thresholdRestriction}).
Showing that a fault-tolerant procedure satisfies Equation~(\ref{eq:thresholdRestriction}) does not provide an upper bound on the threshold for quantum computing since some other procedure might perform better given the same unencoded error model.  Nor does proving the opposite provide a lower bound.  While an increase in the error rate at the first level of concatenation implies that
subsequent layers of concatenation also increase the error rate, the
converse is not true.  To see why, imagine that we have some fault-tolerant
procedure for which the encoded failure rate is less than the unencoded
failure rate.  At the first level our code is constructed of unencoded
qubits that are either perfect or have failed.  At the second level of
encoding, however, our code is constructed of singly encoded qubits that
may be perfect, insufficiently corrupted to result in failure, or failed,
yet only the last case is considered an encoded error.  In some sense, the
qubits that have not failed are now of lower quality than they were at the
previous level.  Thus, the fact that encoding worked at the previous level
does not guarantee that it will work at the current one.

Given a precise mapping between encoded error rates and unencoded error rates, the threshold error probability is bounded below by the unencoded error rate such that, for a particular fault-tolerant procedure, the encoded error rate is the same.  In absence of a precise mapping, the aforementioned probability might be called an approximate lower bound on the threshold.  Generally, however, I think that the goal of researchers who perform such approximate calculations is not to bound the threshold so much as to estimate its value for a particular fault-tolerant procedure.  Consequently, I more often use the appellation ``threshold estimate'' than ``approximate lower bound on the threshold''.

Following Aharonov and Ben-Or~\cite{Aharonov97,Aharonov99}, we can estimate the threshold analytically by counting the number of unencoded gates $g$ involved in the most complex encoded gate of a particular procedure.  A fault-tolerant procedure does not spread errors, so $t+1$ errors must occur to generate $t+1$ errors on an encoded block.  The number of ways to choose $t+1$ errors on $g$ gates is $\binom{g}{t+1}$, so, if there are no memory errors or initial errors on the data blocks, the encoded failure probability, i.e., the probability of $t+1$ or more errors occurring, is bounded by
\begin{align}
\left\{
\parbox{5.3em}{\centering Encoded error rate}\rule{0em}{1.3em}
\right\}
\leq \binom{g}{t+1} p^{t+1}
\end{align}
where $t$ is the number of correctable errors and $p$ is the probability of an unencoded error (or the largest such probability if there are several). The corresponding threshold estimate is
\begin{align}
p_\text{th} = \binom{g}{t+1} p_\text{th}^{t+1} && \rightarrow && p_\text{th}= \binom{g}{t+1}^{-1/t}.
\end{align}
This is an estimate rather than a bound because the data blocks to which the encoded gate is applied are not necessarily free of errors initially.  An accurate accounting these errors is difficult to make since they ultimately depend on the errors, successes, and failures of preceding gates.

The standard numerical approach to estimating the threshold, introduced by Christof Zalka~\cite{Zalka96}, is to program a Monte-Carlo routine that propagates Pauli errors through a sequence of gates corresponding to an encoded gate while, with some pre-assigned probability, causing unencoded gates randomly to fail and generate additional Pauli errors.  Over the course of many runs, statistics on encoded gate failures are collected and used to approximate the encoded failure rate associated with the unencoded error probabilities utilized.  By repeating this process for a sequence of unencoded error probabilities enough data can be obtained to make a fit of the encoded error rate as a function of some small number of parameters in the unencoded error model.  The threshold is then taken to be the point (or surface) where the encoded error probability becomes less than the unencoded error probability.  A number is obtained for the threshold only when the encoded error probability is a function of a single parameter.  This is generally accomplished by setting the failure probability of all unencoded gates to be equal and assuming that gate failures obey the depolarizing error model, that is, that they produce all possible Pauli errors with equal probability.  Aside from the fact that it is approximate, the primary disadvantage of this method of threshold estimation is that it is computationally prohibitive to apply to large codes or error models with many degrees of freedom.

Until recently, most results regarding thresholds have been one sort of estimate or another.  Extraordinarily, these estimates have currently settled near the $1\%$ mark~\cite{Reichardt04,Knill05}.
For a more detailed discussion of the possible pitfalls of such calculations, the reader is referred to the work of Svore and others~\cite{Svore05,Svore06}.  Chapters~\ref{chap:channelDependencyOfTheThreshold} and~\ref{chap:thresholdsForHomogeneousAncillae} focus largely on producing further estimates of the threshold in various cases of interest.

\subsection{Upper Bounds on the Threshold\label{subsec:upperBoundThreshold}}

Several rigorous upper bounds on the threshold for fault-tolerant quantum computing have been produced by proving that quantum gates suffering from exceedingly probable Pauli errors can be simulated by a classical computer.  I do not discuss upper bounds on the threshold further, but it is interesting to note that such proofs have pushed the depolarizing threshold below
$50\%$~\cite{Buhrman06,Razborov03}.

\subsection{Lower Bounds on the Threshold\label{subsec:lowerBoundThreshold}}

The art of obtaining rigorous lower bounds on the threshold is experiencing a renaissance~\cite{Aliferis06,Reichardt05,Aliferis07,Aliferis07b}; recently lower bounds have reached $10^{-4}$.  In what follows, I briefly review a method of bounding the threshold developed by Aliferis, Gottesman, and Preskill.  A much more detailed treatment is contained in their very readable paper~\cite{Aliferis06}.

The analytical approach to calculating the threshold presented in Section~\ref{subsec:thresholdEstimates} could not be used to establish a lower bound because encoded and unencoded errors were incommensurate, effectively precluding an induction step that might prove an arbitrary reduction in the error rate were possible.  The primary origin of this difficultly lay in the dependence of encoded errors on failures outside of the scope of the encoded gate, though a secondary problem arose from the complexity available to a quantum error model.

Aliferis et al.\ do away with the complexity of quantum error models by adopting an adversarial local error model at each level of encoding.  As with other local error models, failures are assumed to strike individual components randomly and independently, but the errors induced are assumed to be the most destructive ones possible on the faulty component.  This error model is unphysical since the most destructive errors at one level of encoding are not necessarily compatible with the most destructive errors at the next level of encoding, but the threshold can only be reduced by considering such harsh errors, so the associated threshold probability can safely be interpreted as a lower bound.  Moreover, this choice eliminates the enormous burden of identifying the encoded failure that results from an excess of component failures.

Establishing the independence of successive levels of encoding is more complicated.  I begin by defining some terms for a $d=3$ error correcting code; in these definitions, $k$-Ga and $k$-EC stand for gates and error corrections at the $k$-th level of encoding.
\begin{description}
  \item[$k$-Rec] An encoded gate and the subsequent error corrections in a $k$th level circuit. \\
    \centerline{
          \Qcircuit @R=1em @C=1em {
& \gate{k\text{-Ga}} & \gate{k\text{-EC}} & \gate{k\text{-Ga}} & \gate{k\text{-EC}} & \qw
\gategroup{1}{2}{1}{3}{.5em}{.}
\gategroup{1}{4}{1}{5}{.5em}{--}
          }
    }
  \item[$k$-exRec] An encoded gate and the preceding and subsequent error corrections in a $k$th level circuit. \\
    \centerline{
      \Qcircuit @R=1em @C=1em {
& \gate{k\text{-EC}} & \gate{k\text{-Ga}} & \gate{k\text{-EC}} & \gate{k\text{-Ga}} & \gate{k\text{-EC}} & \qw
\gategroup{1}{2}{1}{4}{.5em}{.}
\gategroup{1}{4}{1}{6}{1em}{--}
      }
    }
  \item[Bad] A $k$-exRec is bad if it contains two independently bad $(k-1)$-exRecs.
  \item[Dependent] A pair of bad $k$-exRecs are dependent if they overlap and the first $k$-exRec is not bad when the overlapping $k$-EC is ignored.\\
    \centerline{
      \Qcircuit @R=0em @C=1em {
& \gate{k\text{-EC}} & \gate{k\text{-Ga}} & \gate{k\text{-EC}} & \gate{k\text{-Ga}} & \gate{k\text{-EC}} & \qw \\
& & \raisebox{2.1em}{\tiny\color{red}{X}} & \raisebox{.7em}{\tiny\color{red}{X}} & \raisebox{1.2em}{\tiny\color{red}{X}}
\gategroup{1}{2}{1}{4}{.5em}{.}
\gategroup{1}{4}{1}{6}{1em}{--}
      }
    }
  \item[Good] A $k$-exRec is good if it is not bad.
\end{description}

From these definitions it follows that a good $k$-exRec takes a valid input block, that is, a data block having at most one error at the beginning of the $k$-Rec, to a valid output block.  In order for there to be one error at the beginning of the $k$-Rec, at least one error must have occurred during the leading error correction, implying, since the $k$-exRec is good, that no additional errors occur and the output state is flawless.  By contrast, if there are no errors on the data at the location preceding the $k$-Rec then an error might happen during the $k$-Rec, but, since the procedure is fault-tolerant, this yields no more than a single error on the output.

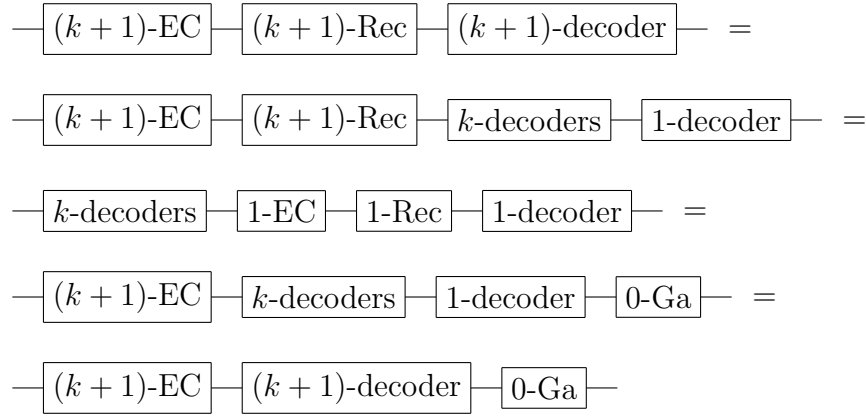
\begin{figure}
\capstart
\centerline{
\begin{tabular}{l}
\Qcircuit @R=.5em @C=1em {
& \gate{(k+1)\text{-EC}} & \gate{(k+1)\text{-Rec}} & \gate{(k+1)\text{-decoder}} & \qw
}
\raisebox{-.25em}{\ \ =}
\\\\
\Qcircuit @R=.5em @C=1em {
& \gate{(k+1)\text{-EC}} & \gate{(k+1)\text{-Rec}} & \gate{k\text{-decoders}} & \gate{1\text{-decoder}} & \qw
}
\raisebox{-.25em}{\ \ =}
\\\\
\Qcircuit @R=.5em @C=1em {
& \gate{k\text{-decoders}} & \gate{1\text{-EC}} & \gate{1\text{-Rec}} & \gate{1\text{-decoder}} & \qw
}
\raisebox{-.25em}{\ \ =}
\\\\
\Qcircuit @R=.5em @C=1em {
& \gate{(k+1)\text{-EC}} & \gate{k\text{-decoders}} & \gate{1\text{-decoder}} & \gate{0\text{-Ga}} & \qw
}
\raisebox{-.25em}{\ \ =}
\\\\
\Qcircuit @R=.5em @C=1em {
& \gate{(k+1)\text{-EC}} & \gate{(k+1)\text{-decoder}} & \gate{0\text{-Ga}} & \qw
}
\end{tabular}
}
\caption[The threshold dance]{The threshold dance, the inductive step of the proof that a $k$-Rec which is part of a good $k$-exRec is correct.  The dance is performed with ideal decoders.  The $k$-decoders pictured act on the bottom $k$ levels of encoding.  If the $k$-exRec is good, then the resulting $0$-Ga is perfect. \label{fig:thresholdDance}}
\end{figure}

Ultimately, of course, we care whether a circuit gives the correct answer, not how good it is.  A correct $k$-Rec should satisfy
\begin{align}
\Qcircuit @R=.5em @C=1em {
& \gate{\parbox{1.1cm}{\centering $k$-Rec}} & \gate{\parbox{1.8cm}{\centering ideal $k$-decoder}} & \qw & \push{\text{\large$=$}} & & \gate{\parbox{1.8cm}{\centering ideal $k$-decoder}} & \gate{\parbox{1.0cm}{\centering ideal $0$-Ga}} & \qw
}.
\end{align}
For a $1$-Rec contained in a good $1$-exRec, fault-tolerance implies that
\begin{align}
\Qcircuit @R=.5em @C=1em {
& \gate{\parbox{1.1cm}{\centering $1$-Rec}} & \gate{\parbox{1.8cm}{\centering ideal $1$-decoder}} & \qw & \push{\text{\large$=$}} & & \gate{\parbox{1.8cm}{\centering ideal $1$-decoder}} & \gate{\parbox{1.0cm}{\centering ideal $0$-Ga}} & \qw
}. \label{eq:thresholdBoundProofBaseCase}
\end{align}
Using Equation~(\ref{eq:thresholdBoundProofBaseCase}) as a base case, Figure~\ref{fig:thresholdDance} shows the inductive argument necessary to prove that a $k$-Rec that is part of a good $k$-exRec is correctly decoded by an ideal decoder.

From all of this we learn that the analytical approach of Section~\ref{subsec:thresholdEstimates} was basically right, but that, instead of counting the number of gates in an encoded gate (a.k.a. a $1$-Rec), we should have been counting the number of gates in an encoded gate plus the leading error correction (a.k.a. a $1$-exRec).

Thus, a number bounding the threshold for quantum computation can be obtained by, for instance, setting the failure probabilities of all gates to be equal (multiple free parameters yields a surface instead of a number), and counting the gates involved in a $1$-exRec.  If $g$ is the number of gates in a $1$-exRec and no memory errors occur then a loose bound on the threshold $p_\text{th}$ is
\begin{align}
  p_\text{th} > \binom{g}{2}^{-1}.
\end{align}
A tighter bound can be obtained by only counting malignant pairs of faults, that is, those pairs faults that might possibly cause an encoded failure.  For a particular error model, even higher values of the threshold result from factoring in the probability that a pair of failures generates a fatal error, but the probability thus obtained is not a bound since the encoded error model will not match the unencoded error model.

\subsection{Propagating More General Errors\label{subsec:propagatingMoreGenErr}}

The codes and procedures considered in this chapter are intended to combat arbitrary independent, local errors, but my analysis of them has been almost exclusively in terms of stochastic Pauli channels, that is, error models in which only Pauli errors occur.

In Section~\ref{subsec:nonPauliErrors} we saw that the measurement of check operators can be used to project arbitrary errors affecting at most $t=\lfloor(d-1)/2\rfloor$ qubits into the Pauli basis.  This observation does not, by itself, absolve us of the need to consider other error models, however, since many gates are typically performed between each measurement of a check operator.  To determine whether a stochastic Pauli channel is an suitable substitute for another error model, it is necessary to consider how each builds up over the course of a computation.  This question has been examined by several authors~\cite{Aharonov99,Knill98,Preskill98}; my own treatment is given below.

Begin by considering an arbitrary trace-preserving error operator $\mathcal{E}$.
The action, on a state $\rho$, of any such error operator can be written as
\begin{align}
\mathcal{E}(\rho) = \sum_j E^j\rho {E^j}^\dagger & & \mathrm{where} & & \sum_j {E^j}^\dagger E^j = I. \label{eq:QOpTrPr}
\end{align}
By interspersing errors of this form with perfect quantum gates it is possible to model any faulty quantum circuit that does not suffer from leakage.  When the error operators are local, it makes sense to approximate them by stochastic Pauli channels.  Given a local error operator satisfying Equation~(\ref{eq:QOpTrPr}), I define the associated stochastic Pauli channel to have error probabilities
\begin{align}
\begin{split}
\po{X} &= \sum_j \left\lvert\mathrm{tr}(E^jX)\right\rvert^2, \\
\po{Y} &= \sum_j \left\lvert\mathrm{tr}(E^jY)\right\rvert^2\textrm{, and} \\
\po{Z} &= \sum_j \left\lvert\mathrm{tr}(E^jZ)\right\rvert^2.
\end{split}\label{eq:apstoch}
\end{align}
By design, this channel correctly reproduces the probability of measuring that a given Pauli error occurred after a single application of the general error operator.  Its suitability in more varied circumstances is the subject of the remainder of this section.

In practice, many gates are required (and therefore many error operators
act) between each error correction, so it is important to know how
errors accumulate.  As for the case of stochastic errors, general
trace-preserving errors can be separated from the associated circuit
providing that it is composed exclusively of Clifford gates.  The
separation is accomplished by applying Pauli propagation to each term in
the Pauli-basis decomposition of the elements, e.g. $E^j$, of the error
operator.  If the Clifford circuit is fault tolerant, then error
propagation maps single-qubit errors to single-qubit errors (on a given
encoded block).  Each of the resultant error operators differs from the actual error
by a simple relabeling of the local Pauli basis.  Thus, the intervening circuit can be
disregarded; it is sufficient to consider how the transformed local errors accumulate.

A sequence of $s$ single-qubit trace-preserving errors acting on a state
$\rho$ can be written as
\begin{align}
\begin{split}
\op{E}^s\circ&\cdots\circ\op{E}^2\circ\op{E}^1(\rho) \\
&= \sum_{j_s}\cdots\sum_{j_2}\sum_{j_1} \left(\prod_{k=s}^{1} E^{j_k k} \right)\rho\left(\prod_{k=1}^{s} {E^{j_k k}}^\dag \right). \\
\end{split}
\end{align}
The probability of measuring, for example, an $X$ error on the resulting state is
\begin{align}
\po{\mathrm{E}X} &= \sum_{j_s}\cdots\sum_{j_2}\sum_{j_1} \left\lvert\mathrm{tr}\left(\prod_{k=s}^{1} E^{j_k k}X\right)\right\rvert^2. \label{eq:actXerr}
\end{align}
By contrast, replacing the error operators with their associated stochastic Pauli errors, as defined in Equation~(\ref{eq:apstoch}), yields
\begin{align}
\po{\mathrm{S}X} &= \sum_{k=1}^s \sum_{j_k} \left\lvert\mathrm{tr}(E^{j_k k}X)\right\rvert^2 + O(p^2). \label{eq:appXerr}
\end{align}
When $\po{\mathrm{E}X}$ and $\po{\mathrm{S}X}$~(and the equivalent probabilities for $Y$ and $Z$) agree, the associated stochastic Pauli channel is a good substitute for the actual error channel.  To the lowest nontrivial order in $p$, the condition for equality can be derived as follows.

Consider each error operator $\mathcal{E}$ as a function of the total
single application error probability $p$.  Taylor expanding the
elements of $\mathcal{E}$ in $\sqrt{p}$ yields
\begin{align}
\begin{split}
E^0 &= I + \sqrt{p}(\alpha_0 X + \beta_0 Y + \gamma_0 Z + \delta_0 I) + O(p) \\
E^{j\neq0} &= \sqrt{p}(\alpha_j X + \beta_j Y + \gamma_j Z + \delta_j I) + O(p)
\end{split}\label{eq:TaylorErrOp}
\end{align}
where the freedom in the $E^j$ has been used to assure that $E^0$ contains
the only term independent of $p$.

Inserting expanded error operators of the form given in
Equation~(\ref{eq:TaylorErrOp}) into Equation~(\ref{eq:actXerr}) and discarding
terms of order greater than $p$ yields
\begin{align}
\begin{split}
\po{\mathrm{E}X} &\approx \sum_{k=1}^s\sum_{j_k\neq0} \left\lvert\mathrm{tr}\left(E^{j_k k}X\right)\right\rvert^2 + \left\lvert\mathrm{tr}\left(\prod_{k=s}^{1} E^{0 k}X\right)\right\rvert^2 \\
&\approx \sum_{k=1}^s\sum_{j_k\neq0} \left\lvert\mathrm{tr}\left(E^{j_k k}X\right)\right\rvert^2 + \left\lvert\sum_{k=1}^s\mathrm{tr}\left(E^{0 k}X\right)\right\rvert^2.
\end{split}\label{eq:appExXerr}
\end{align}
Thus, to first order in $p$, the difference between $\po{\mathrm{S}X}$ and
$\po{\mathrm{E}X}$ is
\begin{align}
\begin{split}
\po{\mathrm{E}X}-\po{\mathrm{S}X} &\approx \left\lvert\sum_{k=1}^s\mathrm{tr}\left(E^{0 k}X\right)\right\rvert^2 -\sum_{k=1}^s \left\lvert\mathrm{tr}(E^{0 k}X)\right\rvert^2\\
&\approx p\left\lvert\sum_{k=1}^s\alpha_{0 k}\right\rvert^2 -p\sum_{k=1}^s \left\lvert\alpha_{0 k}\right\rvert^2\\
&= p \sum_{k=1}^s \sum_{l\neq k} \alpha_{0 k}\alpha_{0 l}^*.
\end{split}
\label{eq:appExXerrDiff}
\end{align}
As suggested by Preskill~\cite{Preskill98}, this expression has a simple
interpretation in terms of a $2$-D walk composed of $s$ steps of sizes
$\lvert\sqrt{p}\alpha_{0 k}\rvert$.  Equation~(\ref{eq:appExXerrDiff}) is equal to the difference
between the square of the displacement for such a walk and the expectation of the square of the displacement assuming that the walk is random, that is, that stepping forward and backward are equiprobable.  Thus, the expectation of Equation~(\ref{eq:appExXerrDiff}) vanishes if the sign of
$\alpha_{0 k}$ is random.  Taking a slightly different approach, we can treat the entire expression as the displacement of a $2$-D random
walk composed of $s(s-1)$ steps of sizes $\lvert p \alpha_{0 k}\alpha_{0 l\neq k}\rvert$.  The expectation is again seen to vanish when the sign of $\alpha_{0 k}$ is random, but now it becomes clear that the standard deviation will scale like $s$.  Since $\po{\mathrm{E}X}$ is also proportional to $s$, this implies that the associated stochastic Pauli channel is only a really exacting substitute when the number of qubits being considered is large.

An identical argument holds for $Y$ and $Z$ errors, showing that, for large $n$, error models for which the sign of the coherent error is random are well approximated by their associated stochastic Pauli channel.  Conveniently, this restriction is preserved under any local relabeling of the Pauli bases and therefore applies equally well to the original error operators.  Examples satisfying the restriction include all stochastic
errors, which have no coherent component, and unitary rotation errors
where under and over rotation are equally likely. Systematic errors, such
as amplitude damping or a bias towards over rotation, are not well
modeled, though, in practice, the local relabeling of the Pauli bases
imposed by gates will randomize these errors somewhat.

\chapter{Channel Dependency of the Threshold\label{chap:channelDependencyOfTheThreshold}}
The approach to fault tolerance presented in the background material was built around the idea that quantum gates fail independently, producing an arbitrary error on the involved qubits.  Consequently, it was necessary to construct circuits satisfying the strictures of fault tolerance for any conceivable Pauli error since a single gate failure might entail Pauli errors of any sort on the participating qubits.  To simplify the analysis, it was subsequently assumed for the analytical lower bound on the threshold that all gates failed with the same probability.  In numerical estimates of the threshold, it is generally additionally assumed that all Pauli errors that might result from a single gate failure are equally likely.  These are common, reasonable assumptions, but they do not reflect the diversity of error models that appear in actual physical systems.  We might therefore ask the question, ``What effect do these assumptions have?''

It is certainly possible to reduce the threshold for quantum computation by discarding some of the standard assumptions, e.g., independence.  Instead, in this chapter I investigate whether the threshold can be increased by modifying the standard assumptions about the error model.  The error model is a a promising candidate for modification since it is typically chosen for generality and convenience rather than performance.

An analytical threshold bound for an arbitrary error channel would be impractical due, if nothing else, to the complexity of the result.  Given a quantum code and fault-tolerant procedure, however, the effect of any particular stochastic Pauli channel on the threshold can be investigated numerically using a Monte-Carlo routine simply by varying the probabilities of different errors in the simulation.  This approach has been used previously to study the dependence of the threshold on the relative probabilities of various gate failures~\cite{Steane03}.  I chart a somewhat more involved course by first tailoring a fault-tolerant procedure to a particular error model and then investigating the concomitant threshold.  Knill~\cite{Knill05b} has produced impressive results using this kind of approach for error models with heralded errors.

\section{Threshold Estimation\label{sec:thresholdEstimation}}

The simulation that yields the bulk of the data for this chapter is a Monte-Carlo error-propagation routine of the general sort introduced in Section~\ref{subsec:thresholdEstimates}.  The program used here is specialized to the Steane code and includes functions implementing all of the operations necessary for applying Pauli error propagation on encoded Clifford gates laid out following Steane's method of fault-tolerant quantum computation.  The circuits corresponding to these functions (as well as a few additional ones) are given in Section~\ref{sec:faultTolerance}; in the unspecialized procedure, each syndrome is extracted twice.

\begin{figure}
\capstart
  \centerline{
    \begin{pgfpicture}{0cm}{0cm}{9.25cm}{6.65cm}
      \pgfputat{\pgfxy(.5,.25)}{\includegraphics[clip=true, trim=0cm 7cm 0cm 7cm, width=10cm]{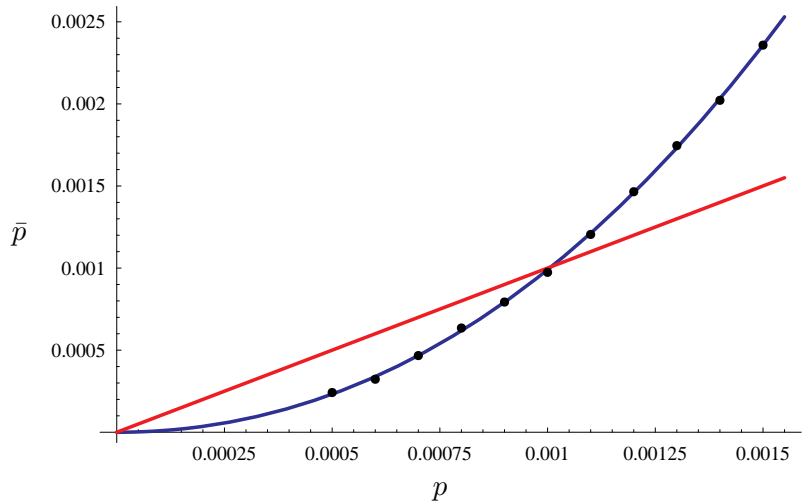}}
      \pgfputat{\pgfxy(5.7,.1)}{\pgfbox[center,center]{\footnotesize $p$}}
      \pgfputat{\pgfxy(.1,3.5)}{\pgfbox[center,center]{\footnotesize $\bar{p}$}}
    \end{pgfpicture}\hspace{2cm}
  }
  \caption[Estimating the depolarizing threshold]{The encoded $\CX$ error probability $\bar{p}$ versus the unencoded gate error probability $p$.  The unencoded error channel here is depolarizing, and the code and fault-tolerant method used are those of Steane.  For reference, a diagonal line demarcating the break-even point for encoding is drawn in red.  The intersection of this line with the blue curve fitting the data gives a depolarizing threshold estimate of $p_\text{Dth}=0.001$.  Error bars fit within the dots. \label{fig:SFTIDepData}}
\end{figure}

The threshold estimation portion of my code prepares $16$ logical qubits in the state $\ket{\bar{0}}$ and then enters a loop which randomly applies an encoded gate from the generating set $\{H,S,\CX\}$ of the Clifford group, checks for an encoded failure, prepares anew any failed logical qubits, and repeats.  The encoded failure rate for the gate of interest is estimated by counting the number of times the gate is applied between each time it fails.  Statistics are taken for this counting data, and the loop exits when the variance in the average reaches the target value.  The output of my simulation for the case of Steane's method and a depolarizing error channel is plotted in Figure~\ref{fig:SFTIDepData}.  As in all subsequent plots, only data for the encoded $\CX$ gate is shown since its encoded error rate is roughly a factor of two greater than either of the other two gates.

\section{Symmetric Two-qubit Error Channel}
In place of the depolarizing channel, I consider a symmetric two-qubit-gate error channel, that is, an error model such that errors are generated exclusively through the failure of two-qubit gates (which, in this chapter, means $\CX$ gates) where the generated error can be decomposed into the Pauli operators $X\otimes X$, $Y\otimes Y$, and $Z\otimes Z$.

This error model is not chosen for its physical plausibility.  Indeed, I consider it unlikely that the $\CX$ gate should be implemented in such a fashion that symmetric two-qubit errors dominate.  Rather, my choice of error model reflects an attempt to pick a non-trivial stochastic Pauli channel for which real gains might plausibly be expected in the threshold.

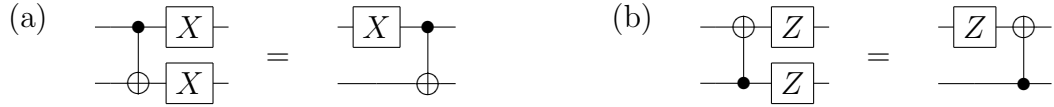
\begin{figure}
\capstart
  \begin{tabular}{cccc}
  (a) &
    \Qcircuit @R=.5em @C=.5em {
& & \qw & \ctrl{1} & \gate{X} & \qw \\
& & \qw & \targ & \gate{X} & \qw
    }
    \raisebox{-1.2em}{\hspace{1.2em}$=$\hspace{.7em}}
    \Qcircuit @R=.5em @C=.5em {
& & \gate{X} & \ctrl{1} & \qw \\
& & \qw & \targ & \push{\rule[-.3em]{0em}{1.32em}} \qw
    }
    \hspace{4em}
    &
  (b) &
    \Qcircuit @R=.5em @C=.5em {
& & \qw & \targ & \gate{Z} & \qw \\
& & \qw & \ctrl{-1} & \gate{Z} & \qw
    }
    \raisebox{-1.2em}{\hspace{1.2em}$=$\hspace{.7em}}
    \Qcircuit @R=.5em @C=.5em {
& & \gate{Z} & \targ & \qw \\
& & \qw & \ctrl{-1} & \push{\rule[-.3em]{0em}{1.32em}} \qw
    }
  \end{tabular}
\caption[Accurate reporting by $\protect\CX$ gates suffering symmetric errors]{The accurate reporting property of $\CX$ gates which generate exclusively symmetric errors.  The identities in a) and b) show that an $X\otimes X$ or $Z \otimes Z$ error after a $\CX$ gate is equivalent to a single error prior to it, implying that $\CX$ gates used in error extraction never misreport the state of the data. \label{fig:littleSymmetricRobustCircuits}}
\end{figure}

The symmetric $\CX$ error channel is a promising candidate for improving the threshold because the correlated errors generated by a gate failure are exactly those that would be produced by the propagation of single-qubit errors initially present on the input qubits.
Since $\CX$ converts single-qubit errors to symmetric two-qubit errors, these errors form a kind of default two-qubit-gate error set.  Moreover, symmetric errors interfere minimally with $\CX$ gates used for error extraction since, even in the event of a failure, the indicated error is actually on the data.  Figure~\ref{fig:littleSymmetricRobustCircuits} illustrates this useful property.

\section{Tailored Fault-tolerant Procedure\label{sec:symmetricFTP}}

Starting from Steane's method, this section lays out a modified fault-tolerant procedure tailored to combat symmetric two-qubit $\CX$ errors.

Given the fact, discussed in the previous section, that $\CX$ gates which fail symmetrically always accurately report data errors, it makes sense to try to reduce the number of times that ancillae are coupled to the data for the purpose of extracting error information.  In order to do so, however, the quality of the logical basis states used in Steane-style syndrome extraction must be improved; repeated syndrome extraction guards against massively faulty ancillae as much as against faulty coupling gates.  Ancillae free (to first order) of correlated errors might be produced using the method discussed in Section~\ref{subsec:ancillaPreparation}, but instead I describe how especially low-error ancillae can be produced by taking advantage of our knowledge of the error model.

\begin{figure}
\capstart
    \begin{tabular}{l@{\hspace{1.8em}}c}
    \hspace{-.84em}
    \begin{tabular}{lc}
  (a) &
    \Qcircuit @R=.5em @C=.5em {
& & & \text{\small$\ket{0}$\hspace{1.5em}} & \gate{H} & \ctrl{1} & \gate{Z^a} & \qw \\
& & & \text{\small$\ket{0}$\hspace{1.5em}} & \qw & \targ & \gate{X^b} & \qw
    }
  \\
  \\
  (b) &
    \Qcircuit @R=.2em @C=.5em {
& & & \text{\small$\ket{0}$\hspace{1.5em}} & \gate{H} & \ctrl{1} & \qw & \qw & \ctrl{4} & \qw & \qw & \qw & \qw \\
& & & \text{\small$\ket{0}$\hspace{1.5em}} & \qw & \targ & \ctrl{1} & \qw & \qw & \qw & \qw & \qw & \qw \\
& & & \text{\small$\ket{0}$\hspace{1.5em}} & \qw & \qw & \targ & \ctrl{1} & \qw & \push{\rule[-.3em]{0em}{1.32em}} \qw & \qw & \qw & \qw \\
& & & \text{\small$\ket{0}$\hspace{1.5em}} & \gate{H} & \ctrl{1} & \qw & \targ & \qw & \ctrl{1} & \gate{H} & \measureD{} \\
& & & \text{\small$\ket{0}$\hspace{1.5em}} & \qw & \targ & \qw & \qw & \targ & \targ & \qw & \measureD{}
    }
    \end{tabular}
      &
    \begin{tabular}{lc}
  (c) &
    \Qcircuit @R=.2em @C=.5em {
& & & \text{\small$\ket{0}$\hspace{1.5em}} & \gate{H} & \ctrl{1} & \qw & \qw & \qw & \ctrl{5} & \qw & \qw & \qw & \qw \\
& & & \text{\small$\ket{0}$\hspace{1.5em}} & \qw & \targ & \ctrl{1} & \qw & \qw & \qw & \qw & \qw & \qw & \qw \\
& & & \text{\small$\ket{0}$\hspace{1.5em}} & \qw & \qw & \targ & \ctrl{1} & \qw & \qw & \push{\rule[-.3em]{0em}{1.32em}} \qw & \qw & \qw & \qw \\
& & & \text{\small$\ket{0}$\hspace{1.5em}} & \qw & \qw & \qw & \targ & \ctrl{1} & \qw & \qw & \qw & \qw & \qw \\
& & & \text{\small$\ket{0}$\hspace{1.5em}} & \gate{H} & \ctrl{1} & \qw & \qw & \targ & \qw & \ctrl{1} & \gate{H} & \measureD{} \\
& & & \text{\small$\ket{0}$\hspace{1.5em}} & \qw & \targ & \qw & \qw & \qw & \targ & \targ & \qw & \measureD{}
    }
    \vspace{3.85em}
    \end{tabular}
    \end{tabular}
  \begin{tabular}{lc}
    \\
  (d) &
    \Qcircuit @R=.2em @C=.4em {
& & & & & & \lstick{\raisebox{-1.85em}{\small$\ket{\beta_{00}}$}} & \ctrl{3} & \ctrl{6} & \qw & \ctrl{1} & \gate{H} & \measureD{} & & & & & \lstick{\raisebox{-1.85em}{\small$\ket{\beta_{00}}$}} & \ctrl{4} & \ctrl{5} & \qw & \ctrl{1} & \gate{H} & \measureD{} \\
& & & & & & & \qw & \qw & \ctrl{7} & \targ & \qw & \measureD{} & & & & & & \qw & \qw & \ctrl{7} & \targ & \qw & \measureD{} \\
& & & & \qw & \qw & \qw & \qw & \qw & \qw & \qw & \qw & \targ & \qw & \qw & \qw & \qw & \qw & \qw & \qw & \qw & \qw & \qw & \qw & \qw & \qw & \qw & \qw & \qw & \qw \\
& & & \lstick{\text{\small$\ket{\widetilde{+}}$\ }} & \qw & \qw & \qw & \targ & \qw & \qw & \qw & \qw & \qw & \qw & \qw & \qw & \qw & \qw & \qw & \qw & \qw & \qw & \qw & \qw & \qw & \qw & \qw & \qw & \qw & \qw \\
& & & & \qw & \qw & \qw & \qw & \qw & \qw & \qw & \qw & \qw & \qw & \qw & \qw & \qw & \qw & \targ & \qw & \qw & \qw & \qw & \targ & \qw & \qw & \qw & \qw & \qw & \qw \\
& & & & \qw & \qw & \qw & \qw & \qw & \qw & \qw & \qw & \qw & \qw & \qw & \qw & \qw & \qw & \qw & \targ & \qw & \qw & \qw & \qw & \qw & \qw & \qw & \qw & \qw & \qw \\
& & & \lstick{\raisebox{-1.7em}{\small$\ket{\widetilde{+}}$\ }} & \qw & \qw & \qw & \qw & \targ & \qw & \qw & \qw & \qw & \qw & \qw & \qw & \qw & \qw & \qw & \qw & \qw & \qw & \qw & \qw & \targ & \qw & \qw & \qw & \qw & \qw \\
& & & & \qw & \qw & \qw & \qw & \qw & \qw & \qw & \qw & \qw & \targ & \qw & \qw & \qw & \qw & \qw & \qw & \qw & \qw & \qw & \qw & \qw & \targ & \qw & \qw & \qw & \qw \\
& & & & \qw & \qw & \qw & \qw & \qw & \targ & \qw & \qw & \qw & \qw & \targ & \qw & \qw & \qw & \qw & \qw & \targ & \qw & \qw & \qw & \qw & \qw & \qw & \qw & \qw & \qw \\
& & & & & & & & & & & \lstick{\raisebox{-1.85em}{\small$\ket{\beta_{00}}$}} & \ctrl{-7} & \ctrl{-2} & \qw & \ctrl{1} & \gate{H} & \measureD{} & & & & & \lstick{\raisebox{-1.85em}{\small$\ket{\beta_{00}}$}} & \ctrl{-5} & \ctrl{-3} & \qw & \ctrl{1} & \gate{H} & \measureD{} \\
& & & & & & & & & & & & \qw & \qw & \ctrl{-2} & \targ & \qw & \measureD{} & & & & & & \qw & \qw & \ctrl{-3} & \targ & \qw & \measureD{}
\gategroup{1}{7}{2}{7}{.4em}{(} \gategroup{1}{18}{2}{18}{.4em}{(} \gategroup{3}{4}{5}{4}{.4em}{(} \gategroup{6}{4}{9}{4}{.4em}{\{} \gategroup{10}{12}{11}{12}{.4em}{(} \gategroup{10}{23}{11}{23}{.4em}{(}
    }
  \end{tabular}
\caption[Ancilla construction circuits tolerant of individual two-qubit symmetric $\protect\CX$ errors]{Circuits for constructing a) $\ket{\beta_{ab}}$, b) the $3$-qubit cat state, c) the $4$-qubit cat state, and d) logical $\ket{+}$ for the Steane code, such that any single symmetric error on a $\CX$ is either harmless or detected.  As before, the overtop tilde is used to denote logical qubits encoded in the classical repetition code. \label{fig:symmetricFaultTolerantAncillaConstruction}}
\end{figure}
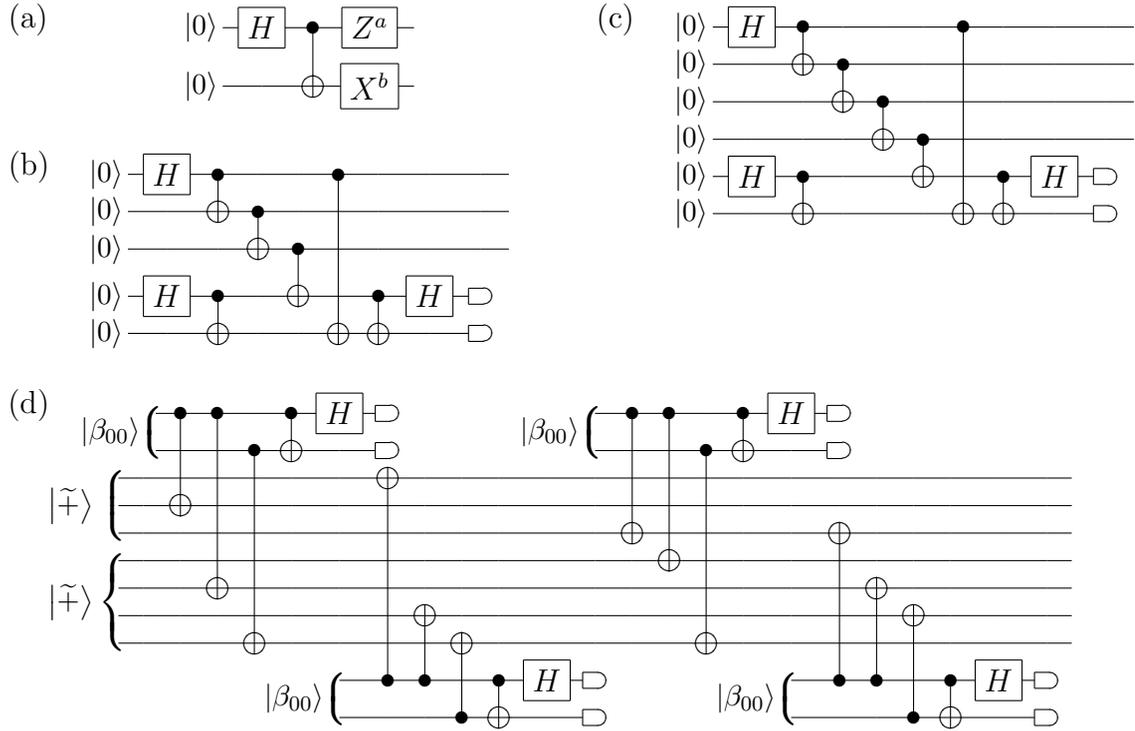

I employ two basic tricks for constructing high quality ancillae, each relying on the fact that only a subset of Pauli errors can result from most circuits.  The first is to design circuits such that some fraction of the possible Pauli errors are, in fact, stabilizers, a trick epitomized by the circuit for constructing Bell states shown in Figure~\ref{fig:symmetricFaultTolerantAncillaConstruction}a.  Since $X\otimes X$, $Y\otimes Y$, and $Z \otimes Z$ all stabilize $\ket{\beta_{00}}$, this circuit is completely unaffected by a symmetric error on the $\CX$.  The same trick also plays a role in the cat-state construction circuits of Figure~\ref{fig:symmetricFaultTolerantAncillaConstruction}b and Figure~\ref{fig:symmetricFaultTolerantAncillaConstruction}c where many symmetric $\CX$ errors are equivalent to some combination of the stabilizers $\{Z_iZ_j,X^{\otimes m}\}$, where $m$ is the size of the cat state and $1\leq i,j\leq m$ and $i\neq j$.  The design of these circuits is such that all other symmetric errors resulting from a single gate failure are detected by one or both of the measurements, so, conditional on positive measurement outcomes, the prepared cat states are, to first order, flawless.  This is sufficient to insure fault tolerance for the Steane code since it corrects only single errors.

The construction circuit that we need for Steane-style syndrome extraction, however, is the one for $\ket{\bar{+}}$ in part d) of Figure~\ref{fig:symmetricFaultTolerantAncillaConstruction}.  This circuit takes advantage of both the techniques and the states generated in the other examples.  That it generates the desired state, absent errors and conditional on positive measurement results, can be seen most easily by considering the evolution of the stabilizer.  Subsequent to preparing the two input cat states, a stabilizer generator for the state is
\begin{align}
  \{Z_1Z_2,Z_2Z_3,X_1X_2X_3,Z_4Z_5,Z_5Z_6,Z_6Z_7,X_4X_5X_6X_7\}
  \label{eq:stepOneLogicalPlusStabilizerGeneratorSymmetricFaultTolerance}
\end{align}
To account for the measurement of $+1$ for $X_2X_5X_7$, it is sufficient to take products of stabilizer generators until only a single generator anti-commutes with $X_2X_5X_7$ and substitute $X_2X_5X_7$ for it, yielding, for example,
\begin{align}
  \{Z_1Z_3,Z_1Z_2Z_4Z_5,Z_1Z_2Z_5Z_6,Z_1Z_2Z_6Z_7,X_1X_2X_3,X_4X_5X_6X_7,X_2X_5X_7\}.
\end{align}
Doing the same for $X_1X_6X_7$ gives
\begin{align}
  \{Z_2Z_3Z_4Z_5,Z_1Z_2Z_5Z_6,Z_2Z_3Z_6Z_7,X_1X_2X_3,X_4X_5X_6X_7,X_2X_5X_7,X_1X_6X_7\}
\end{align}
which is equivalent, under products of the stabilizer generators, to
\begin{align}
  \begin{split}
  \{&Z_1Z_3Z_5Z_7,Z_2Z_3Z_6Z_7,Z_4Z_5Z_6Z_7, \\
  &X_1X_3X_5X_7,X_2X_3X_6X_7,X_4X_5X_6X_7,X_1X_2X_3\}
  \end{split},
\end{align}
the standard generator for $\ket{\bar{+}}$ for the Steane code.  The remaining two measurements are needed only for error detection; being measurements of stabilizers of $\ket{\bar{+}}$ they do not, ideally, change the state.

Verifying that the circuit in Figure~\ref{fig:symmetricFaultTolerantAncillaConstruction}d is robust against any single symmetric $\CX$ failure is more complicated than for the Bell or cat states since the state $\ket{\bar{+}}$ is partially formed by the measurements.  Given perfect input ancillae, which, to first order, we have, it is clear that any single $X\otimes X$ or $Y\otimes Y$ error will be detected by one of the measurements on the lower qubits of the Bell states.  A $Z\otimes Z$ error on any of the first eight $\CX$ gates, however, will flip the result of one or both of the initial two stabilizer measurements, causing us to retain the wrong state.  The state that we actually produce can be written in terms of an undamaged copy of $\ket{\bar{+}}$, the residual error on $\ket{\bar{+}}$ from the gate failure, and the errors $Z_1Z_3$ and $Z_2Z_3$ which account for our mistaken selection.  The errors $Z_1Z_3$ and $Z_2Z_3$ are an appropriate choice for representing the result of retaining the wrong state, since they both commute with all of the generators in Equation~(\ref{eq:stepOneLogicalPlusStabilizerGeneratorSymmetricFaultTolerance}) or, in other words, with the stabilizers of the input state, but each anti-commutes with one of the two measurements.  Thus,
\begin{align}
  \begin{split}
    \brakket{\bar{+}}{Z_1Z_3X_1X_6X_7Z_1Z_3}{\bar{+}} &= -\brakket{\bar{+}}{X_1X_6X_7}{\bar{+}} = -1 \\
    \brakket{\bar{+}}{Z_1Z_3X_2X_5X_7Z_1Z_3}{\bar{+}} &= \brakket{\bar{+}}{X_2X_5X_7}{\bar{+}} = 1 \\
    \brakket{\bar{+}}{Z_2Z_3X_1X_6X_7Z_2Z_3}{\bar{+}} &= \brakket{\bar{+}}{X_1X_6X_7}{\bar{+}} =1 \\
    \brakket{\bar{+}}{Z_2Z_3X_2X_5X_7Z_2Z_3}{\bar{+}} &= -\brakket{\bar{+}}{X_2X_5X_7}{\bar{+}} = -1
  \end{split}.
\end{align}
A $Z\otimes Z$ error on the first $\CX$ in Figure~\ref{fig:symmetricFaultTolerantAncillaConstruction}d, for example, would yield the state
\begin{align}
  Z_2\ Z_2Z_3\ \ket{\bar{+}} = Z_3\ket{\bar{+}}
\end{align}
half-way through the circuit, but a single $Z$ error on the third qubit would be detected by the error checking portion of the circuit.  In a similar way, it can be shown that a $Z\otimes Z$ error on any single $\CX$ gate in the first half of the circuit is detected.  A $Z \otimes Z$ error on a single $\CX$ gate in the second half of the circuit flips some certain measurement outcome and is thus easily detectable.

\begin{figure}
\capstart
\begin{tabular}{l@{\hspace{6em}}l}
\Qcircuit @R=.3em @C=.4em @!R {
& \text{(a)} & & & & & & & & & & {\ F_X} \\
& & & & & & & & {\ket{\bar{+}}} & & & \targ & \meter \\
& & & & & \qw & \push{\rule[-.4em]{0em}{.4em}} \qw & \qw & \qw & \qw & \qw & \ctrl{-1} & \qw & \qw & \qw
\gategroup{3}{7}{2}{13}{.6em}{--}
}
&
\Qcircuit @R=.3em @C=.4em @!R {
& \text{(b)} & & & & & & & & & & & {\!\!\!\!F_Z} \\
& & & & & & & & {\ket{\bar{0}}} & & & \ctrl{1} & \gate{H} & \meter \\
& & & & & \qw & \push{\rule[-.6em]{0em}{.8em}} \qw & \qw & \qw & \qw & \qw & \targ & \qw & \qw & \qw & \qw
\gategroup{3}{7}{2}{14}{.6em}{--}
}
\end{tabular}
\caption[Fault-tolerant syndrome extraction for symmetric errors]{Logical syndrome extraction circuits for a) $X$ and b) $Z$ errors.  Extraction follows Steane's method, but no repetition is necessary due to the quality of the ancillae employed. \label{fig:symmetricFaultTolerantSyndromeExtraction}}
\end{figure}
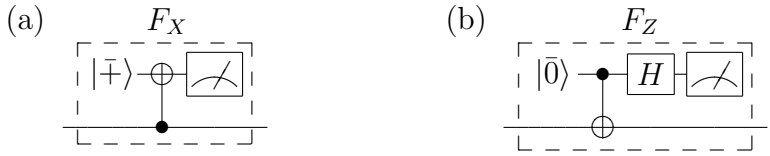

The upshot of the preceding paragraph is that $\ket{\bar{+}}$ states constructed using the circuit in Figure~\ref{fig:symmetricFaultTolerantAncillaConstruction}d are not only free of correlated errors to first order, which was our primary goal, but are, in fact, free of any errors whatsoever.  The availability of a transversal encoded Hadamard gate means that the same is true for $\ket{\bar{0}}$.  Combining such ancillae with the accurate copying property ensured by our error channel, it makes sense to reduce the number of syndrome extractions in Steane's fault-tolerant procedure to one per kind of error.  In this modified procedure, the logical circuits for performing $X$ and $Z$ syndrome extraction are simply those given in Figure~\ref{fig:symmetricFaultTolerantSyndromeExtraction}.

\section{Bounding the Threshold for Symmetric Errors}

The method of analytically bounding the encoded error rates introduced in Section~\ref{sec:thresholds} applies equally well to the error model and fault-tolerant procedure just described.  The only modification necessary is to restrict the space of possible errors to the symmetric $\CX$ errors.

The question of a bound on the threshold is somewhat more subtle, however, since the error model describing the encoded gates will not be that of symmetric $\CX$ errors.  In lieu of the more standard restriction, that the encoded error probability be less than the unencoded error probability, I require that the encoded error probability for this procedure and error model be no greater than the threshold for more general kinds or errors.  Thus, $p_\text{Sth}$, the threshold for symmetric errors, is bounded below by the symmetric error probability $p_\text{S}$ such that
\begin{align}
  G p_\text{S}^2 = p_\text{Ath}
\end{align}
where $G$ is the number of malignant pairs in the my most heinous exRec and $p_\text{Ath}$ is the threshold for adversarial errors in a more general fault-tolerant procedure.  In what follows, I actually use a slight modification of Aliferis's counting method suited to the case that $X \otimes X$, $Y \otimes Y$, and $Z\otimes Z$ are all equally likely; I replace $G p_\text{S}^2$ with $K p_\text{S}^2/9$ where $K$ counts different kinds of malignant errors at a given pair of locations as different malignant pairs.

The problem of counting malignant pairs is greatly simplified by the fact that my ancilla construction routines all output perfect states when success is indicated and fewer than two gate failures have occurred.  A single failure occurring during the construction of a Bell, cat, or $\ket{\bar{+}}$ state is of no consequence whatsoever.  Thus, any malignant pair of errors involving ancilla construction must both appear in the construction procedure.

There are no such pairs in the Bell state circuit.  For the $n$-qubit cat state circuit, however, there are
\begin{align}
  4 \binom{n-2}{2} + 6 (n-2) + 9
\end{align}
pairs that result in some kind of error on the state; I assume pessimistically that all of these are fatal.  Given that there are no errors on the input states, the circuit for constructing $\ket{\bar{+}}$ includes $68$ or fewer malignant pairs, so, including cat state construction, the total number of malignant pairs in the construction circuit for $\ket{\bar{+}}$ is $108$.  The probability of occurrence for these malignant pairs is actually bigger than $108p_\text{S}^2/9$ since we are discarding events in which detectable errors are produced, but for $p_\text{S}\ll 0.01$ the difference here is slight.

\begin{figure}
\capstart
\centerline{
\Qcircuit @R=.4em @C=.8em {
& \gate{F_{X}} &  \gate{F_{Z}} &  \ctrl{1} &  \gate{F_{X}} &  \gate{F_{Z}} & \qw \\
& \gate{F_{X}} &  \gate{F_{Z}} &  \targ &  \gate{F_{X}} &  \gate{F_{Z}} & \qw
}
}
\caption[The exRec for $\protect\CX$]{The exRec for $\protect\CX$ in my fault-tolerant procedure.\label{fig:CXExRec}}
\end{figure}
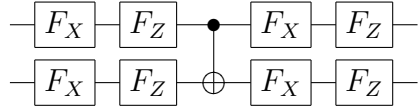

For this fault-tolerant procedure, the exRec with the largest number of malignant pairs is the one for the encoded $\CX$ gate.
Ignoring ancilla errors, this exRec, shown in Figure~\ref{fig:CXExRec}, has a total of
\begin{align}
  9\times\binom{7\times9}{2} = 17577
\end{align}
possible distinct error pairs.  Counting the number of malignant ones with any kind of accuracy is not an easy mental exercise, but a short \command{Python} program hacked together for the purpose yields up the answer $3927$.  Thus, including ancilla construction, the total number of malignant pairs in the exRec for $\CX$ is $4791$, and the corresponding bound on the threshold is given by
\begin{align}
  532.3 p_\text{Sth}^2 \geq p_\text{Ath}. \label{eq:symmetricThresholdBound}
\end{align}

The information content of the bound in Equation~(\ref{eq:symmetricThresholdBound}) depends strongly on the value of $p_\text{Ath}$.  For $p_\text{Ath}=10^{-5}$, it tells us that the threshold for symmetric $\CX$ errors is more than an order of magnitude larger than the threshold for general adversarial errors.  For $p_\text{Ath}>1/532.3\approx 0.19\%$, it tells us nothing at all.  An approximate number for this bound can be obtained by replacing $p_\text{Ath}$ with $p_\text{Dth}=0.001$, the numerical estimate of the depolarizing threshold obtained in Section~\ref{sec:thresholdEstimation}.  Doing so yields
\begin{align}
  p_\text{Sth} \geq \sqrt{p_\text{Dth}/523.3} > 0.0014 = 1.4 p_\text{Dth},
\end{align}
a rather small improvement.  Even this estimate of the bound is overly optimistic, however, analytical threshold bounds in the literature range as high as $1.94\times10^{-4}$~\cite{Aliferis07b} and estimates placing the threshold above $0.2\%$ are already common~\cite{Reichardt04,Knill05}.

These analytical results do not bode well for my program of tailoring fault-tolerant procedures, but the lower bound given here is not tight, due both to my approximate treatment of ancilla errors and to the nature of the exRec method.  For further evidence, I turn in the following section to numerical techniques.

\section{Estimating the Threshold for Symmetric Errors Numerically}

As for the analytical estimate, a numerical estimate of the threshold for symmetric errors requires that the encoded error rate for the tailored procedure be referenced to a more general threshold result, the implied structure being that of a tailored fault-tolerant procedure implemented at the first level of encoding combined with a more general procedure at higher levels.

I determined the error rates for the initial level of encoding numerically using the simulation described in Section~\ref{sec:thresholdEstimation}.  To adapt my code for depolarizing threshold estimation to the case of threshold estimation for symmetric $\CX$ errors using a tailored fault-tolerant procedure, I simply changed the input error model and defined new functions for ancilla construction and error correction based on the circuits of Section~\ref{sec:symmetricFTP}.

\begin{figure}
\capstart
  \centerline{
    \begin{pgfpicture}{0cm}{0cm}{9.25cm}{6.65cm}
      \pgfputat{\pgfxy(.5,.25)}{\includegraphics[clip=true, trim=0cm 7cm 0cm 7cm, width=10cm]{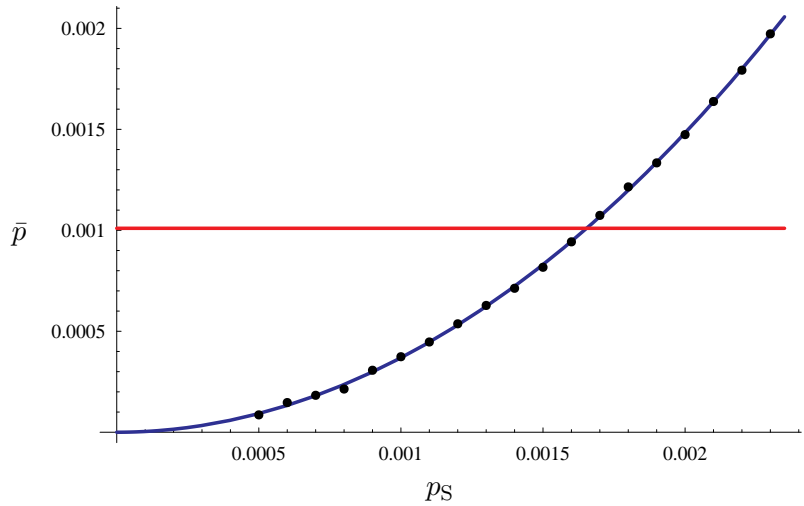}}
      \pgfputat{\pgfxy(5.7,.1)}{\pgfbox[center,center]{\footnotesize $p_\text{S}$}}
      \pgfputat{\pgfxy(.1,3.5)}{\pgfbox[center,center]{\footnotesize $\bar{p}$}}
    \end{pgfpicture}\hspace{2cm}
  }
  \caption[Estimating the threshold for the symmetric $\protect\CX$ error model]{The encoded $\CX$ error probability $\bar{p}$ versus the unencoded $\CX$ error probability $p_\text{S}$.  The unencoded error channel here is that of symmetric $\CX$ errors, and the fault-tolerant procedure used is that described in Section~\ref{sec:symmetricFTP}.  For reference, a line marking my earlier estimate of the depolarizing threshold is drawn in red.  The intersection of this line with the blue curve fitting the data gives an estimate of the threshold for symmetric errors of $p_\text{Sth}\approx.0017$.  Error bars fit within the dots. \label{fig:MyFTISymData}}
\end{figure}

Figure~\ref{fig:MyFTISymData} shows the encoded failure probability of the $\CX$ gate as a function of $p_{S}$, the probability of a symmetric $\CX$ error.  The horizontal line on the graph marks the value of the depolarizing threshold calculated in Section~\ref{sec:thresholdEstimation}.  In order for subsequent levels of encoding to succeed, the encoded error probability after the initial level of encoding must be less than the threshold for the new encoded error model.  Taking the new error model to be depolarizing\footnote{Even taking the encoded error model to also be symmetric yields only a total factor of $2.6$ increase in the threshold.}, this restriction yields $p_\text{Sth} \approx 0.0017$
as an estimate of the threshold for symmetric errors on my modified fault-tolerant procedure.  This estimate agrees reasonably well with the conclusions of the previous section.  Only a small improvement, less than a factor of two, is achieved by my tailored fault-tolerant procedure.

Increasing the threshold by a factor of two is not without merit, but my investigations up till now have indicated this improvement only for the case of an error model with exclusively symmetric $\CX$ errors.  That a physical implementation of a quantum computer should suffer primarily from symmetric $\CX$ errors is perhaps unlikely, but that such a device should suffer from them exclusively is unthinkable.  To be useful at all, my tailored procedure must be robust against small perturbations in the error model.

\begin{figure}
\capstart
  \centerline{
    \begin{pgfpicture}{0cm}{0cm}{10.83cm}{10cm}
      \pgfputat{\pgfxy(10.4,5.2)}{\pgfbox[center,center]{\footnotesize $\bar{p}$}}
      \pgfputat{\pgfxy(0,0)}{\includegraphics[width=10cm]{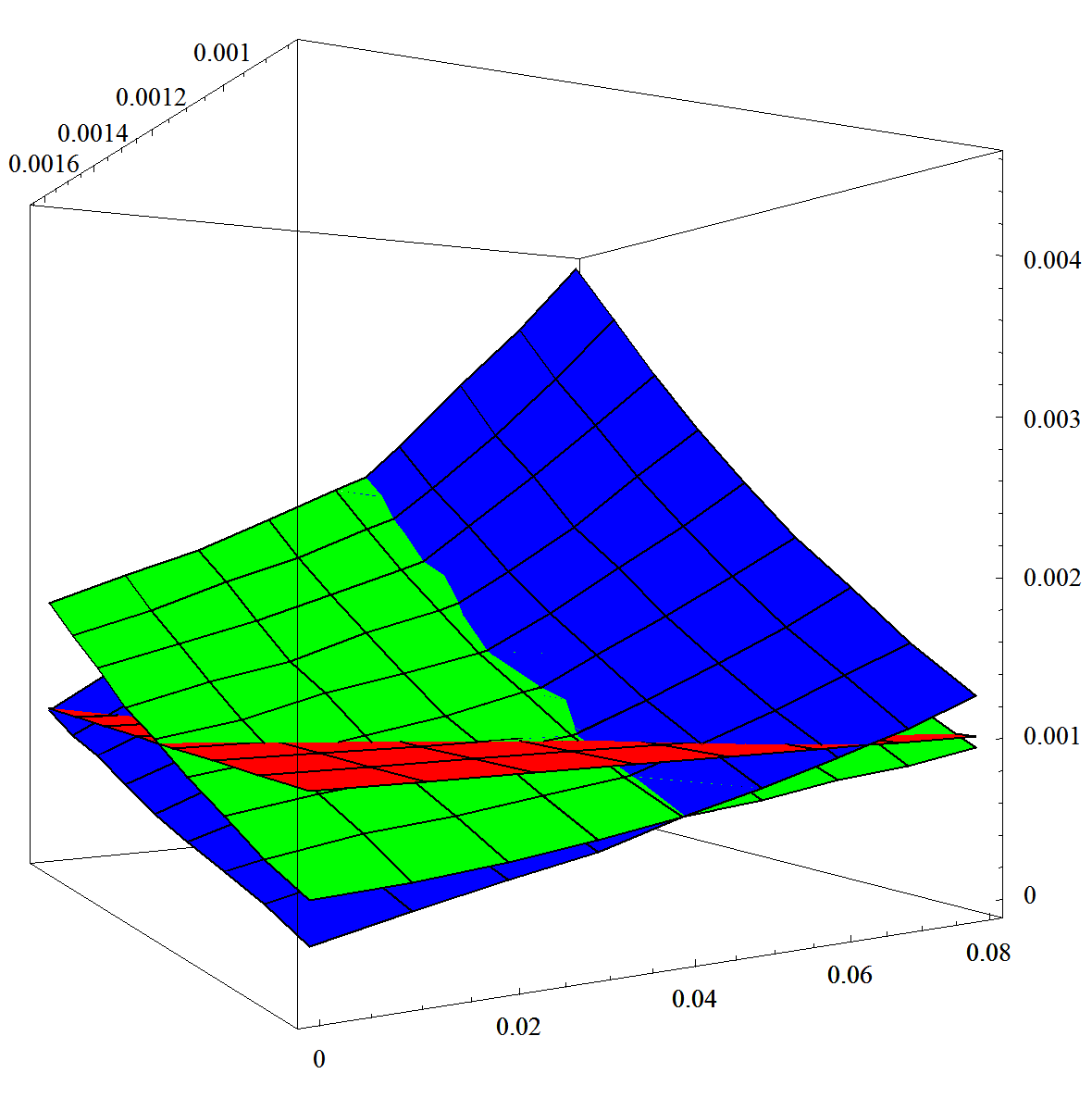}}
      \pgfputat{\pgfxy(.7,9.6)}{\pgfbox[center,center]{\footnotesize $p_\text{S}$}}
      \pgfputat{\pgfxy(6.6,.5)}{\pgfbox[center,center]{\footnotesize $\Delta$}}
      \pgfputat{\pgfxy(6.95,6.6)}{\pgfbox[center,center]{\footnotesize Tailored}}
      \pgfputat{\pgfxy(1.4,5.4)}{\pgfbox[center,center]{\footnotesize Steane}}
    \end{pgfpicture}
  }
  \caption[Relative performance of my tailored fault-tolerant procedure]{Plots of the encoded $\CX$ error rate $\bar{p}$ for my tailored procedure (in blue) and the standard Steane-style procedure (in green) for various values of the symmetric error rate $p_\text{S}$ and the portion of the depolarizing channel added $\Delta$.  The red horizontal plane marks the estimate of the depolarizing threshold obtained earlier, so gates with error rates above that are unlikely to benefit from further encoding.\label{fig:CombinedPartialDepData}}
\end{figure}

\begin{figure}
\capstart
  \centerline{
    \begin{pgfpicture}{0cm}{0cm}{10.7cm}{10cm}
      \pgfputat{\pgfxy(.7,.17)}{\includegraphics[width=10cm]{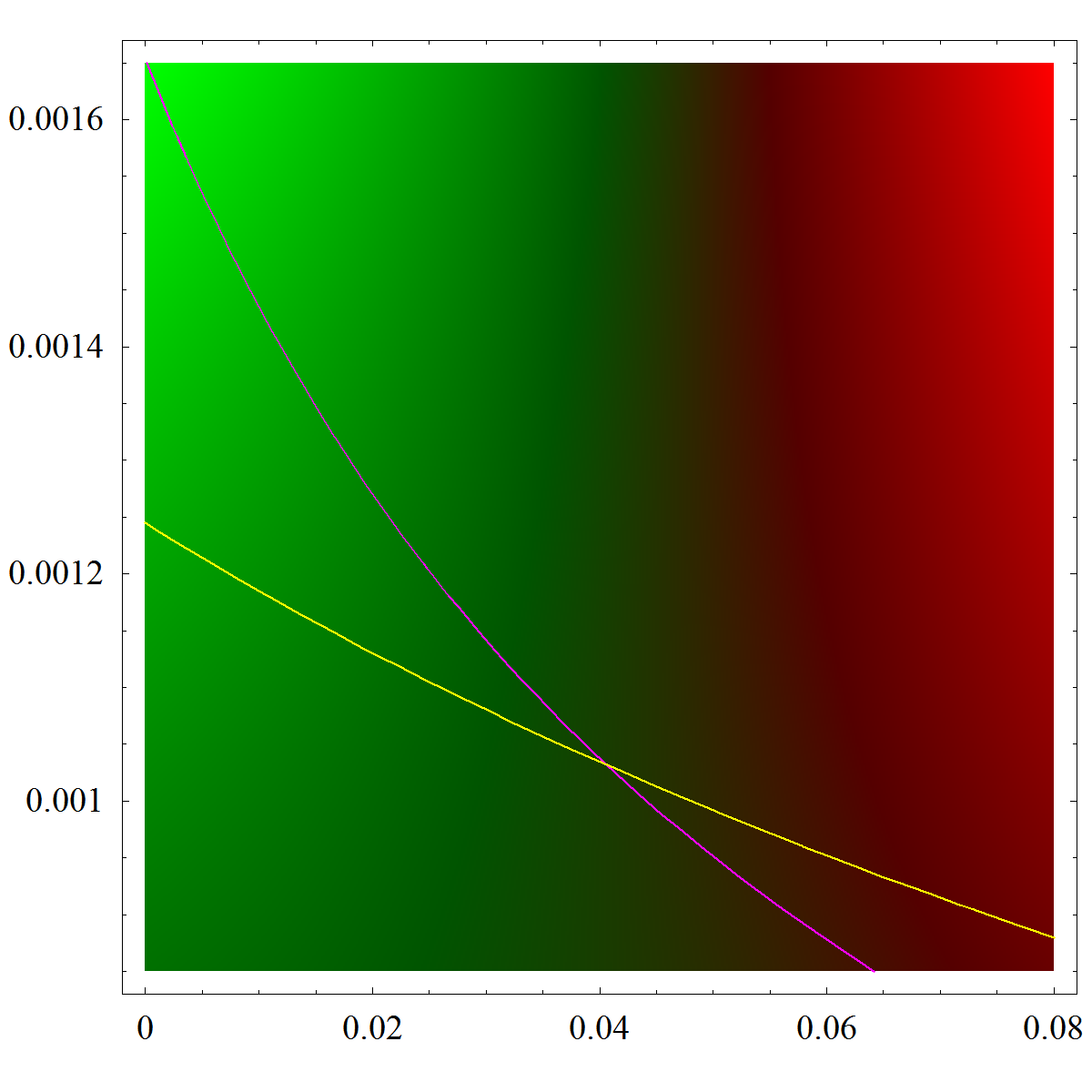}}
      \pgfputat{\pgfxy(6.16,.17)}{\pgfbox[center,center]{\footnotesize $\Delta$}}
      \pgfputat{\pgfxy(.17,5.6)}{\pgfbox[center,center]{\footnotesize $p_\text{S}$}}
    \end{pgfpicture}
  \hspace{1cm}
  }
  \caption[Differential performance of my tailored fault-tolerant procedure]{Fits of the data in Figure~\ref{fig:CombinedPartialDepData} displayed as a density plot of the difference between the encoded $\CX$ error rate for my tailored procedure and that of Steane.  Green indicates that my procedure has a smaller error rate, while red favors Steane.  The magenta line marks the fit of the threshold for my procedure, and the yellow line likewise marks the fit for Steane's. \label{fig:CombinedPartialDepFitDiffDen}}
\end{figure}

To investigate the stability of my procedure, I added a variable depolarizing component to the symmetric $\CX$ error model in my code.  The amount of depolarizing error was quantified by the parameter $\Delta$ where $0\leq \Delta\leq 1$.  $\Delta=0$ gives the symmetric $\CX$ error model and $\Delta=1$ gives the standard depolarizing model.  The encoded $\CX$ error rates for values of $\Delta$ and $p_\text{S}$ such that $0\leq \Delta \leq 0.08$ and $0.0085\leq p_\text{S} \leq 0.00165$ are plotted in Figure~\ref{fig:CombinedPartialDepData} for both my tailored and the standard fault-tolerant procedures.  From Figure~\ref{fig:CombinedPartialDepData} it can be seen that, for values of $p_\text{S}$ greater than $0.00085$, the standard fault-tolerant implementation quickly overtakes my tailored procedure as the amount of depolarizing error is increased.  An advantage is obtained only when the strength of the depolarizing errors is less than roughly $4\%$ of what it would be for a depolarizing channel given the size of the symmetric $\CX$ errors.  From Figure~\ref{fig:CombinedPartialDepFitDiffDen}, which displays a fit of the data using a density plot, we see that, compared to the standard procedure, the largest increase in the threshold over this range was less than $35\%$, showing that much of the observed improvement for symmetric $\CX$ errors was not due to changing the fault-tolerant procedure at all.

\section{Analysis}

The results of my effort to tailor Steane's fault-tolerant method to a particular error model were largely negative.  In ideal circumstances a small increase in the threshold can be achieved, but perturbations of the error model on the order of only $5\%$ of the symmetric $\CX$ error probability are sufficient not only to wipe out the improvement but to reduce the effectiveness of the procedure below that of the standard approach.

In retrospect, the narrowness of the window for improvement is unsurprising since the construction routine for $\ket{\bar{+}}$ is not fault tolerant against general errors.  In order for the procedure to function as designed, non-symmetric first-order errors must be negligible, which is to say that they should occur with probability of at most $p_\text{S}^2$ or so.  For very small values of $p_\text{S}$ this is a physically unreasonable asymmetry to expect in the probabilities of various $\CX$ errors, and for much larger values of $p_\text{S}$ there is, as we have seen, little advantage to be had from my procedure.

More interesting is the reason for the rather small increase of the threshold with the introduction of improved ancillae.  Naively, it would seem that utilizing ancillae that are, to first order, perfect would be advantageous, but the analysis of this chapter appears to suggest otherwise.

\begin{figure}
\capstart
  \centerline{
  \begin{pgfpicture}{0cm}{0cm}{9.25cm}{6.65cm}
    \pgfputat{\pgfxy(.5,.25)}{\includegraphics[clip=true, trim=0cm 7cm 0cm 7cm, width=10cm]{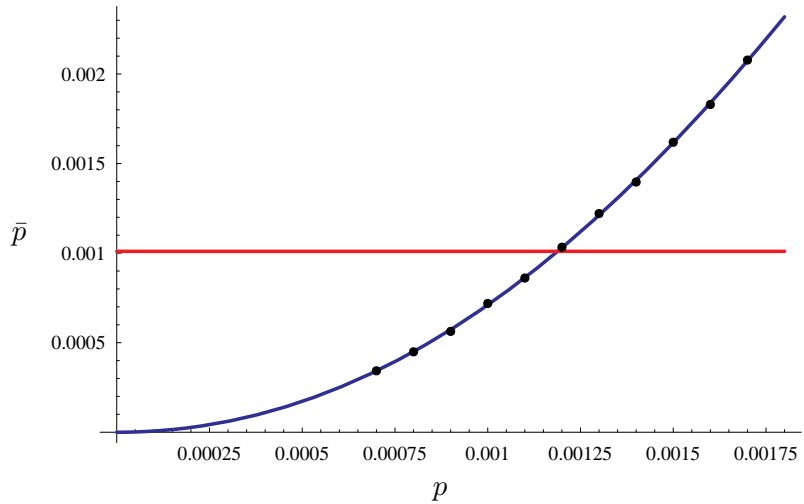}}
    \pgfputat{\pgfxy(5.7,.1)}{\pgfbox[center,center]{\footnotesize $p$}}
    \pgfputat{\pgfxy(.1,3.5)}{\pgfbox[center,center]{\footnotesize $\bar{p}$}}
  \end{pgfpicture}\hspace{2cm}
  }
  \caption[Estimating the depolarizing threshold for perfect ancillae]{The encoded $\CX$ $\bar{p}$ error probability versus the unencoded $\CX$ error probability $p$.  The error channel here is depolarizing, but the ancillae are assumed to be flawless.  The code and fault-tolerant method used are those of Steane.  For reference, a line marking my earlier estimate of the depolarizing threshold is drawn in red.  The intersection of this line with the blue curve fitting the data gives an estimate of the threshold for depolarizing errors and perfect ancillae of $p_\text{Sth}\approx 0.0012$.  Error bars fit within the dots. \label{fig:SFTIDepPerfectAncillaeData}}
\end{figure}

To settle this point I performed one final simulation to determine the encoded error rate for $\CX$ using Steane's fault-tolerant procedure and the full depolarizing error model but with the ancilla error probabilities all set to zero.  The output of this simulation is displayed in Figure~\ref{fig:SFTIDepPerfectAncillaeData} along with a line marking the threshold for subsequent encoding.  The estimate for the depolarizing threshold associated with this case, where ancillae are absolutely perfect, is $0.0012$, only $20\%$ bigger than the depolarizing threshold I obtained for ancillae constructed using the prepare and discard method.

This fact suggests an interesting idea.  The best threshold estimates employ construction routines that discard large numbers of qubits during ancilla construction~\cite{Knill05,Reichardt04}, so the verified ancillae are quite good.  Consequently, ancillary errors are likely to play a less dramatic part in threshold estimates than one might expect.  Taking this idea to the extreme, one might contemplate new methods of estimating the threshold in which ancillae play only a minor role.  Indeed, the topic of the next chapter is exactly that.

\chapter{Thresholds for Homogeneous Ancillae\label{chap:thresholdsForHomogeneousAncillae}}

Based on the observations of Chapter~\ref{chap:channelDependencyOfTheThreshold}, it is uncertain how much impact ancilla construction has on the threshold for fault-tolerant quantum computation, but it is clear that improved ancillae have very little effect on the method of procedure specific threshold estimation employed there.  This is an intriguing finding (though not entirely unheard of~\cite{Reichardt04}) given the amount of effort that is expended on designing construction procedures for various kinds of ancillary states, since it implies that threshold estimation might be accomplished without any reference to ancilla construction whatsoever.  Ancilla construction is a messy affair to treat analytically, so the option to basically ignore ancillae also provides hope that approximate values of the threshold for selected procedures might be obtained analytically.  In this chapter I describe such an analytical method and use it to investigate the relative merits of a selection of error models and fault-tolerant procedures.

\section{Assumptions\label{sec:assumptions}}
Some large part of the variance in fault-tolerant threshold calculations is
due to the variety of assumptions employed by various authors.  In an
effort both to combat confusion and to facilitate comparisons, my assumptions for this chapter are listed below in roughly
the order of decreasing novelty.
\begin{enumerate}
\item All ancillary qubits have independent, identical error distributions.\label{ass:ident}
\item There are no memory errors. \label{ass:nomemory}
\item $\CX$ is the only two-qubit gate. \label{ass:CXonly}
\item Any pair of qubits can interact via a two-qubit gate. \label{ass:nonlocal}
\item Error operators are trace preserving and lack systematic coherent terms. \label{ass:errtype}
\item Gate failures are uncorrelated. \label{ass:uncorrelated}
\item Classical computation is freely available. \label{ass:cheapclassical}
\end{enumerate}
Assumptions~\ref{ass:ident}, \ref{ass:nonlocal}, \ref{ass:errtype}, \ref{ass:uncorrelated}, and \ref{ass:cheapclassical} are necessary for my analysis; the others are convenient but optional.  Assumption~\ref{ass:nomemory} obviates the need to consider questions of parallelism, gate timing, and the speed of classical computation, while Assumption~\ref{ass:CXonly} reduces the number of cases that must be considered.

\section{Thresholds for Homogeneous Methods\label{sec:homogeneous}}
Two important observations from Chapter~\ref{chap:background} provide the
foundation for my method of threshold calculation.  The first is that fault-tolerant procedures for CSS codes are, to a large degree, transversal.
The second is that, for the kind of CSS codes typically employed, these
transversal operations can be implemented by applying the same gate to
every qubit in a block.  The sum of these observations is that most
operations performed in a fault-tolerant procedure consist of doing the
same thing to each of the qubits in a block.  If one could arrange for all
operations to have this property, which I refer to henceforth as
homogeneity, analyzing the behavior of fault-tolerant circuits would be
greatly simplified.

Three components of the typical fault-tolerant method stand in the way of full
homogeneity: ancilla production, syndrome extraction, and recovery.  Starting from
the perspective of threshold estimation, this section addresses each of these aspects, partly by keeping in mind that the
eventual goal is to model errors, not computation.  Ultimately, the method derived is meaningful both as a form of threshold estimation and
as a threshold bound for idealized resources.

\subsection{Ancillae}
The ancillae used in CSS-code fault-tolerant procedures are typically
prepared in highly entangled states, i.e. in logical basis states.  By
definition, entangled states cannot be constructed without the interaction
of the constituent parts, so there is no a priori reason to think that the
qubits composing an ancilla will have either independent or identical
error distributions.  For fault-tolerant procedures, however, the
production of an entangled ancilla is usually followed by a homogeneous
verification circuit, so one might expect that
most of the residual error probability (of the kind tested for) arises
during this verification step.  With this is mind, I approximate ancillae
as having uniform error distributions.

It is important to realize that this assumption is far less innocuous than
it sounds.  Implicitly, I am assuming that ancillae of the desired size
can be constructed for use in a verification circuit, but in subsequent
sections I take the limit $n\rightarrow\infty$.  Thresholds given in this
limit are only practically achievable if an efficient procedure exists to
prepare logical ancillae.  To be efficiently scalable, however, a
construction routine must have nonvanishing probability of generating an
ancilla that has good fidelity with the desired state.  Fault-tolerant
schemes using concatenated codes provide a method of achieving this for
arbitrarily large ancillae, but as a side effect of universal quantum
computation.  At present, there is no known method of preparing a logical
qubit encoded using a CSS code of arbitrary size that does not depend on
the ability to perform universal quantum computation.  Thus, absent an
explicit recipe for ancilla preparation, the algorithm presented in this
chapter does not constitute a constructive procedure for achieving any
threshold.

\subsection{Error Location}
Current techniques for locating errors require performing a complicated
and distinctly non-homogeneous function on the output of ancilla
measurement.  But while this classical processing requires knowledge of
all the measurements, its effect, assuming that no more than the
correctable number of errors has occurred, can be described in terms
of the individual qubits.  So long as the total number of errors present
on a measured ancilla is less than half the minimum distance\footnote{Some
higher weight errors will also be correctable, but I only lower my
threshold by ignoring them.}, the effect of the classical
processing is to determine a subset of the measured bits that can be
flipped to yield an undamaged codeword.  For the purposes of
error correction, knowing this string is equivalent to knowing the
location of all the errors. While the second kind of information is not
directly available to a quantum computer, it is quite accessible to a
theorist treating errors probabilistically.  I can therefore model the
effect of classical processing in two steps.  First, I determine whether
too many errors have occurred on a block to permit proper decoding, and,
if this is not the case, I treat the location
of bit flips on the measured qubits as revealed.

I have reduced the process of error location to a non-homogeneous failure
check and an arguably homogeneous revelation step.  For a Monte-Carlo
simulation, the failure check would consist of polling all of the other
qubits and counting up the number of errors that have occurred to see
whether they exceeded half the minimum distance of the code.  If instead
we performed a probability flow analysis, the expected probability of
passing the check would be simply
\begin{equation}
E_\mathrm{p}(p_\loc{L}) = \sum_{i=0}^t\binom{n}{i}p_\loc{L}^i(1-p_\loc{L})^{n-i}\label{eq:ppass}
\end{equation}
where $n$ is the number of qubits in the block, $t$ is the maximum number
of errors that can be corrected with certainty by the code, and
$p_\loc{L}$ is the probability that a particular qubit has an $X$ error at
some location (step) $\loc{L}$, here chosen to be just after the time of
measurement.

Equation~(\ref{eq:ppass}) suggests a way of recovering full homogeneity.
Letting $\tau=t/n$, in the limit of large $n$, Equation~(\ref{eq:ppass}) becomes
\begin{equation}
E_\mathrm{p}(p_\loc{L}) =
\begin{cases}
0 & \text{if $p_\loc{L}<\tau$} \\
1 & \text{if $p_\loc{L}>\tau$} \\
\end{cases}
\end{equation}
which is again homogeneous from the perspective of a simulation.  As an
added benefit, it is no longer necessary to concatenate many layers of
coding to achieve a rigorous threshold; instead a vanishing error
probability is achieved as the limit of a very large code.  This
alternative to concatenation is known as large block coding or, simply,
block coding.

The preceding paragraphs demonstrate that, for homogeneous~(independent,
identically distributed) errors, whether or not an encoded state on a
large number of qubits fails is determined by the error probability of an
individual qubit.  For this result to be useful, an infinite family of CSS
codes with nonvanishing fractional minimum distance must exist.
Fortunately, it has been shown~\cite{Steane96b,Calderbank96} that CSS
codes exist such that $\tau\approx 5.5\%$ for asymptotic values of $n$.
It is not known whether a similar claim can be made for CSS codes in which
the encoded phase gate can be implemented transversally, but this
convenience is not necessary for my construction.

The analysis of this section assumes minimum distance error correction,
but an identical result applies to error correction up to the channel
capacity. Gottesman and Preskill~\cite{Gottesman01} have shown that
families of general CSS codes exist that are asymptotically capable of
correcting errors up to $\tau\approx11\%$.  Hamada~\cite{Hamada04} has
shown (my own version of this proof can be found in Appendix~\ref{app:asymCor}) that this result applies to CSS codes with $X$-$Z$ exchange symmetry
as well.

The appropriate choice for $\tau$ depends on the purpose of the calculation.
When the goal is to estimate the threshold that would be obtained by
running a Monte-Carlo simulation of a minimum distance decoder, $\tau$
should be chosen to be $5.5\%$, or, perhaps better still, the limit of the correctable error fraction as the number of concatenations of the code in question goes to infinity.  To obtain the largest bound on the threshold
for homogeneous ancillae or for comparison to threshold estimates that use
the channel capacity, it is best to choose $\tau=11\%$. In other cases it
may be desirable to choose a value of $\tau$ specific to a family of
quantum error correcting codes with special properties, such as ease of
syndrome decoding or the possession of low weight stabilizer operators.

\subsection{Recovery}
Having diagnosed the location of our errors, the obvious way of dealing
with them is to apply to each qubit the gate which reverses its error.
Such a recovery operation is inherently inhomogeneous since not all qubits
will be in error, and thus not all qubits will have recovery gates applied
to them.  There are a number of ways to deal with this problem, but I
take the approach, described in Section~\ref{subsec:propagation}, of dispensing with recovery altogether.
As explained there, the effect of Pauli gates on Clifford circuits can be efficiently dealt with
through post-processing.

\begin{figure}
\capstart
\centerline{
\xy \Qcircuit @R=.2em @C=.4em {
& & & & \ctrl{8} & \qw & \qw & \qw & \qw & \qw & \qw & \gate{H} & \meter \\
& & & & \qw & \ctrl{8} & \qw & \qw & \qw & \qw & \qw & \gate{H} & \meter \\
& & & & \qw & \qw & \ctrl{8} & \qw & \qw & \qw & \qw & \gate{H} & \meter \\
& \push{\ket{\bar{0}}\ } & & & \qw & \qw & \qw & \ctrl{8} & \qw & \qw & \qw & \gate{H} & \meter \\
& & & & \qw & \qw & \qw & \qw & \ctrl{8} & \qw & \qw & \gate{H} & \meter \\
& & & & \qw & \qw & \qw & \qw & \qw & \ctrl{8} & \qw & \gate{H} & \meter \\
& & & & \qw & \qw & \qw & \qw & \qw & \qw & \ctrl{8} & \gate{H} & \meter \\
\\
& & \qw & \qw & \targ & \qw & \qw & \qw & \qw & \qw & \qw & \qw & \qw & \qw & \qw & \qw & \qw \\
& & \qw & \qw & \qw & \targ & \qw & \qw & \qw & \qw & \qw & \qw & \qw & \qw & \qw & \qw & \qw \\
& & \qw & \qw & \qw & \qw & \targ & \qw & \qw & \qw & \qw & \qw & \qw & \qw & \qw & \qw & \qw \\
& & \qw & \qw & \qw & \qw & \qw & \targ & \qw & \qw & \qw & \qw & \qw & \qw & \qw & \qw & \qw \\
& & \qw & \qw & \qw & \qw & \qw & \qw & \targ & \qw & \qw & \qw & \qw & \qw & \qw & \qw & \qw \\
& & \qw & \qw & \qw & \qw & \qw & \qw & \qw & \targ & \qw & \qw & \qw & \qw & \qw & \qw & \qw \\
& & \qw & \qw & \qw & \qw & \qw & \qw & \qw & \qw & \targ & \qw & \qw & \qw & \qw & \qw & \qw
\gategroup{1}{3}{7}{3}{.1em}{\{}
}
\POS
,<21em,-4.7em>*\xycircle<5em,3em>{}="c"
;<15.3em,-6.4em>*{} ** @{-}
,"c";<17em,-14.5em>*{} ** @{-}
,<17.1em,-4em>
\Qcircuit @R=.5em @C=.4em {
& & & & \ctrl{1} & \gate{H} & \meter \\
& \qw & \qw & \qw & \targ & \qw & \qw & \qw & \qw & \qw
}
\endxy
}
\caption[Illustration of a strand]{A homogeneous sequence of operations on the Steane code and a single strand thereof.  The inset on the right shows a single strand which has been extracted from the coding blocks.  Gates between pairs of qubits
bind then into the same strand.  Modulo a label indicating the ancilla's starting state, the strand is identical
to the encoded circuit. \label{fig:strand} }
\end{figure}
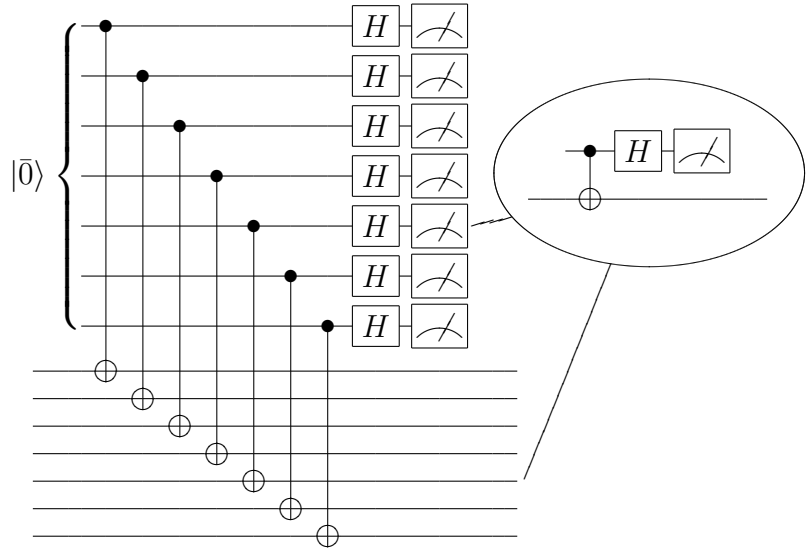

\section{Error Counting\label{sec:counting}}
In the previous section we saw how to modify fault-tolerant procedures based on CSS
codes so that they are fully homogeneous.  The advantage of doing this is that
the error probabilities of qubits within an encoded block then become independent and identical.
Blocks can therefore be separated into strands (see Figure~\ref{fig:strand}), where each strand is made up of a single qubit of a block and all qubits that directly or
indirectly couple to it.
Since each strand is functionally identical, it suffices to determine the
error spectrum for one of them; the probability of an encoded failure
at any point can be predicted from the error probabilities of an individual strand
of the transversal procedure.  As the number of encoding qubits becomes large, the fraction of
qubits with a particular error approaches the expectation for that error.
In the limit that $n\rightarrow\infty$, we can say for certain whether our
procedure fails on any given step since, in that limit, the
probability of an encoded failure becomes a step function.  Thus, the threshold
is completely determined by the probability of an encoded failure, and the
probability of an encoded failure is completely determined by the error
probability of a single strand of the blocks. Therefore, in order to calculate
the threshold I need only determine the error probability on a single
strand at every point in the fault-tolerant circuit.  This can be
accomplished through a combination of error propagation and exhaustive
bookkeeping which I describe in the following sections.

\subsection{Error Bookkeeping}
Given a gate, say the Hadamard, and a set of probabilities describing the
likelihood of various Pauli errors, say $\po{X}$, $\po{Y}$, and $\po{Z}$
for the errors $X$, $Y$, and $Z$, the post gate state can be written as a
probabilistically selected pure state, such that
\begin{equation}
\ket{\Psi^\prime}=
\begin{cases}
X H \ket{\Psi} & \text{with probability } \po{X}\\
Y H \ket{\Psi} & \text{with probability } \po{Y}\\
Z H \ket{\Psi} & \text{with probability } \po{Z}\\
H \ket{\Psi} & \text{otherwise.}
\end{cases}
\end{equation}
The effect of applying further Clifford gates is to change, via error propagation, which Pauli error corresponds to each probability, and then to add additional layers of probabilistic errors.  If, for example, we were to apply another Hadamard gate our state would become
\begin{equation}
\ket{\Psi^{\prime\prime}}=
\begin{cases}
X \ket{\Psi} & \text{with probability } \po{Z}+\po{X}+\po{Y} \po{Z}+\po{X}^2\\
Y \ket{\Psi} & \text{with probability } 2 \po{Y}+2 \po{X} \po{Z}\\
Z \ket{\Psi} & \text{with probability } \po{X}+\po{Z}+\po{Y} \po{X}+ \po{Z}^2\\
\ket{\Psi} & \text{otherwise.}
\end{cases}
\end{equation}
By repeated application of this process it is possible to determine the
probability of various kinds of errors at any point in a circuit composed
of Clifford gates.  Armed with this knowledge we can
determine\footnote{For non-transversal circuits the number of terms in our
bookkeeping rapidly becomes unmanageable as the size of the code
increases.} the likelihood of an encoded failure or, in the infinite
limit, whether an encoded failure will happen or not.

\section{Practicalities\label{sec:practicalities}}
While the previous section presented the basic algorithm for determining
whether an encoded failure occurs, this section deals with details of the error model
and the implementation that must be considered in any actual application
of the method.

\subsection{Applicable Error Models}

Like much of the work on fault-tolerance, the analysis throughout this chapter employs stochastic Pauli errors.
Its applicability, however, is not limited to that case.  We found in Section~\ref{subsec:propagatingMoreGenErr} that, on average, coherent errors with random phases add like stochastic errors.  The variance in that average turned out to be large, but the variance is suppressed as the number of samples increases, and, in this chapter, the number of samples is the number of qubits, which is taken to be infinite.  Consequently, the method described here applies to any unbiased error model.

\subsection{Implementation\label{sec:implementation}}

In the examples that follow, single-strand error rates were determined for three fault-tolerant procedures.  For each procedure, error rates were calculated for a
encoded gate set, $H$, $\CX$, $P$, and $T$ gates, as well as for an idle step
that accounted for the possibility of changing the order of $X$ and $Z$
error correction.  No checks were made on the encoded $T$ gate following
its first error correction, since the remainder of the gate consists of
applying the encoded $P$ gate.  The encoded $P$ gate was assumed
pessimistically to be implemented via a teleportation process akin to that
used for $T$.  For each encoded gate, the maximum was taken over the strand error probabilities at all measurement steps
since in the limit that $n\rightarrow\infty$ encoded failures are caused exclusively by the largest relevant error probability at a measurement
location.

The error probabilities for unencoded gates, measurements, and ancillae were left as
free parameters.  $\po{\Gamma}$, $\pt{\Lambda}{\Xi}$, $\po{M}$,
$\pt{A}{\Gamma}$, and $\pt{B}{\Gamma}$ are used to denote one-qubit,
two-qubit, measurement, $A$-type-ancilla, and $B$-type-ancilla error
probabilities where $\Gamma$ ranges over the single-qubit Pauli errors and
$\Lambda\Xi$ ranges over the two-qubit Pauli errors.  Note that ancillae
are labeled irrespective of what they encode.  $A$-type ancillae are used
in locations where $Z$ errors are more disruptive than $X$ errors, and
contrariwise for $B$-type ancillae.  In the absence of better information,
I assume that $A$-type ancillae are tested using a homogeneous coupling,
with discard on failure, for first $X$ and then $Z$ errors; the opposite
order is used for $B$-type ancillae.  In this case I approximate the
ancilla error distributions as
\begin{equation}
\begin{split}
\pt{A}{X} &= \pt{X}{Z} + \pt{X}{I} + \pt{I}{X} + \pt{X}{X}\\
\pt{A}{Y} &= \pt{I}{Y} + \pt{X}{Y} \\
\pt{A}{Z} &= \pt{I}{Z} + \pt{X}{Z} \\
\pt{B}{X} &= \pt{X}{I} + \pt{X}{Z} \\
\pt{B}{Y} &= \pt{Y}{I} + \pt{Y}{Z} \\
\pt{B}{Z} &= \pt{I}{Z} + \pt{X}{Z} + \pt{Z}{I} + \pt{Z}{Z}\text{,}
\end{split}\label{eq:ancillaErrors}
\end{equation}
which is (to first order) what one would expect if the only errors on a
verified ancilla were due to undetectable errors on the $\CX$ gates used
to check it.

It should be emphasized that the ancilla error probabilities given by
Equation~(\ref{eq:ancillaErrors}) are not the only possible choice.  They were
chosen as a good approximation to the residual error following a
verification procedure that discards the state whenever a problem is
indicated. Depending on the purpose of the calculation, it will sometimes
be more appropriate to assign, for example, higher error probabilities
associated with less resource intensive verification or different
probabilities for different kinds of ancillae.

\begin{table}
\capstart
\centerline{
\begin{tabular}{|c|c|}
\hline
\parbox{3.1em}{\centering \rule[.9em]{0em}{0em}Error Model\rule[-.2em]{0em}{0em}} & \parbox{22.5em}{\centering Nonzero Error Probabilities} \\
\hline
\#1 &
\parbox{22.5em}{\centering
$\po{\Gamma}=\frac{p}{4}$\rule{0em}{1.1em},\; $\pt{\Lambda}{\Xi}=\frac{p}{16}$,\; $\po{M}=\frac{p}{2}$,\\
$\pt{A}{X}=\pt{B}{Z}=\frac{p}{4}$,\; $\pt{A}{Y}=\pt{A}{Z}=\pt{B}{X}=\pt{B}{Y}=\frac{p}{8}$\rule[-.5em]{0em}{1.5em} }
\\
\hline
\#2 &
\parbox{22.5em}{\centering
$\po{\Gamma}=\frac{4p}{15}$\rule{0em}{1.1em},\; $\pt{\Lambda}{\Xi}=\frac{p}{15}$,\; $\po{M}=4p$,\\
$\pt{A}{X}=\pt{B}{Z}=\frac{4p}{15}$,\; $\pt{A}{Y}=\pt{A}{Z}=\pt{B}{X}=\pt{B}{Y}=\frac{2p}{15}$\rule[-.5em]{0em}{1.5em} }
\\
\hline
\#3 &
\parbox{22.5em}{\centering
$\pt{\Lambda}{\Xi}=\frac{p}{15}$\rule{0em}{1.1em},\; $\pt{A}{X}=\pt{B}{Z}=\frac{4p}{15}$,\\
$\pt{A}{Y}=\pt{A}{Z}=\pt{B}{X}=\pt{B}{Y}=\frac{2p}{15}$\rule[-.5em]{0em}{1.5em} }
\\
\hline
\#4 &
\parbox{22.5em}{\centering
$\pt{I}{X}=\pt{X}{I}=\pt{I}{Z}=\pt{Z}{I}=\frac{p}{4}$\rule{0em}{1.1em},\\
$\pt{A}{X}=\pt{B}{Z}=\frac{p}{2}$,\; $\pt{A}{Z}=\pt{B}{X}=\frac{p}{4}$\rule[-.5em]{0em}{1.5em} }
\\
\hline
\end{tabular}
}
\caption[Reduced error models]{Four reduced error models considered in the text and in Table \ref{tab:thresholdCoefficients}.  $\po{\Gamma}$, $\pt{\Lambda}{\Xi}$, $\po{M}$, $\pt{A}{\Gamma}$, and $\pt{B}{\Gamma}$ represent various one-qubit, two-qubit, measurement, $A$-type-ancilla, and $B$-type-ancilla error probabilities where $\Gamma\in\{X,Y,Z\}$ and $\Lambda\Xi\in\{I,X,Y,Z\}^{\otimes2}/\{II\}$.  Unspecified probabilities are zero.
\label{tab:errorModels}}
\end{table}

The Mathematica program that I use to calculate encoded error rates retains terms
up to second order in the
base error probabilities, but the results given in the following sections
include only first-order terms.  Second-order terms were found to be
negligible for any plausible choice of error model.  To understand why,
consider a simplified error model in which gates can fail in only a single
way.  Let $p$ be the probability of an individual gate failing,
and let $g_1$ be the number of gates on which a single failure results in
an error at location $\loc{L}$.  Further, let $g_2$ be the number of gates
that might participate in some pair of failures to yield an error at
location $\loc{L}$.  The expected error at location $\loc{L}$ is then
bounded by
\begin{align}
E_{\loc{L}} <& \binom{g_1}{1} p (1-p)^{g_2-1} \nonumber \\
&+ \binom{g_2}{2} p^2 (1-p)^{g_2-2} + \mathcal{O}(p^3) \\
=& g_1 p -g_1(g_2-1)p^2 + \frac{g_2}{2}(g_2-1) p^2 + \mathcal{O}(p^3). \nonumber
\end{align}
The inequality arises from the fact that not all pairs of failures will
necessarily produce an error at $\loc{L}$.  Even ignoring that, however,
the second-order terms will be negative unless $g_2>2g_1$; negative terms
may safely be neglected since their omission only lowers the threshold.
Among the examples of the following section, the double-coupling Steane
procedure, when applied to error model \#1, has relatively large second-order terms.  Yet the worst location in that procedure corresponds,
roughly, to a single-error situation where $g_1=13$, $g_2=27$, and
$p<.018$, for which the ratio of second to first-order terms is
less than $.02$.  Again, this does not even take into consideration the
fact that many second-order errors will be harmless.

\section{Special Cases\label{sec:examples}}

Having described the operation of my algorithm for calculating
thresholds, I now apply it to three cases of interest.  Two of these are
variants on the fault-tolerant method of Steane,
while the third case is a Knill-style fault-tolerant telecorrection procedure.  The mechanics of syndrome extraction for both Steane- and Knill-style procedures was described in Section~\ref{subsec:errorCorrection}.

\begin{sidewaystable}
\renewcommand{\baselinestretch}{1}\selectfont
\capstart
\begin{tabular}{|c|c|c|}
\hline
\raisebox{-.4em}{\rule{0em}{1.6em}} & Gate & Maximal Single-strand Error Probability\\
\hline
& \text{None} & \raisebox{-1.5em}{\rule{0em}{3.5em}}\parbox{18.4cm}{$2 \pt{A}{Y} + 2 \pt{B}{Y} + \pt{I}{Y} + 2 \po{M} + \pt{X}{Y} + \pt{X}{Z} + \pt{Y}{I} + 3 \pt{Y}{X} + 3 \pt{Y}{Y} + \pt{Y}{Z} + 4 \pt{Z}{X} + 3 \pt{Z}{Y} + \max(2 \pt{A}{Z} + 2 \pt{B}{Z} + \pt{I}{Z} + 2 \po{X} + 2 \po{Y} + 3 \pt{Y}{I} + \pt{Y}{X} + 2 \pt{Y}{Z} + 4 \pt{Z}{I} + 3 \pt{Z}{Z},2 \pt{A}{Y} + 2 \pt{B}{Y} + \pt{I}{Y} + 2 \po{M} + \pt{X}{Y} + \pt{X}{Z} + \pt{Y}{I} + 3 \pt{Y}{X} + 3 \pt{Y}{Y} + \pt{Y}{Z} + 4 \pt{Z}{X} + 3 \pt{Z}{Y})$} \\
\cline{2-3}
& $H$ & \raisebox{-1.5em}{\rule{0em}{3.5em}}\parbox{18.4cm}{$\pt{A}{X} + 2 \pt{A}{Y} + \pt{A}{Z} + \pt{B}{Y} + \pt{B}{Z} + \pt{I}{X} + \pt{I}{Y} + 2 \po{M} + \po{X} + \pt{X}{Z} + 2 \po{Y} + 2 \pt{Y}{I} + 2 \pt{Y}{X} + 2 \pt{Y}{Y} + \pt{Y}{Z} + 2 \pt{Z}{I} + 3 \pt{Z}{X} + 2 \pt{Z}{Y} + \pt{Z}{Z} + \max(\pt{X}{I} + \pt{Y}{I} + 2 \pt{Y}{Z} + \po{Z} + \pt{Z}{Y} + \pt{Z}{Z},\pt{I}{Y} + \pt{I}{Z} + \po{X} + \pt{X}{X} + 2 \pt{X}{Y} + \pt{Y}{X})$} \\
\cline{2-3}
& $\CX$ & \raisebox{-1.5em}{\rule{0em}{3.5em}}\parbox{18.4cm}{$\pt{A}{Y} + \pt{B}{Y} + 2 \pt{I}{Y} + 3 \po{M} + 2 \pt{X}{Y} + 2 \pt{X}{Z} + 2 \pt{Y}{I} + 3 \pt{Y}{X} + 3 \pt{Y}{Y} + 2 \pt{Y}{Z} + 5 \pt{Z}{X} + 3 \pt{Z}{Y} + \max(\pt{A}{Z} + 2 \pt{B}{Y} + 3 \pt{B}{Z} + 2 \pt{I}{Z} + 3 \po{X} + 3 \po{Y} + 3 \pt{Y}{I} + 2 \pt{Y}{X} + \pt{Y}{Z} + 5 \pt{Z}{I}, 3 \pt{A}{X} + 2 \pt{A}{Z} + \pt{B}{X} + 5 \pt{I}{X} + 3 \pt{I}{Y} + 2 \pt{X}{I} + 3 \pt{X}{X} + \pt{X}{Y} + 2\pt{Z}{Y})$} \\
\cline{2-3}
\rule{.9em}{0em}\turnbox{90}{Single-coupling Steane}\rule{.3em}{0em} & $T$, $P$ & \raisebox{-1.5em}{\rule{0em}{3.5em}}\parbox{18.4cm}{$\pt{B}{Y} + \po{M} + \pt{Y}{I} + \pt{Y}{X} + \pt{Y}{Y} + \pt{Y}{Z} + \pt{Z}{X} + \pt{Z}{Y} + \max(\pt{A}{X} + \pt{A}{Y} + \pt{B}{X} + 2 \pt{I}{X} + 2 \pt{I}{Y} + \po{M} + \pt{X}{I} + \pt{X}{X} + \pt{X}{Y} + \pt{X}{Z} + \pt{Z}{X} + \pt{Z}{Y},\pt{B}{Y} + 2 \pt{B}{Z} + \po{X} + \po{Y} + \pt{Z}{I} + \pt{Z}{Z})$} \\
\hline
& \text{None} & \raisebox{-1.5em}{\rule{0em}{3.5em}}\parbox{18.4cm}{$\pt{A}{Y} + \pt{B}{Y} + 3 \pt{I}{Y} + \po{M} + 3 \pt{X}{Y} + 3 \pt{X}{Z} + 3 \pt{Y}{I} + 5 \pt{Y}{X} + 7 \pt{Y}{Y} + 3 \pt{Y}{Z} + 5 \pt{Z}{X} + 5 \pt{Z}{Y} + \max(3 \pt{A}{Y} + 4 \pt{A}{Z} + \pt{B}{Z} + 3 \pt{I}{Z} + \po{X} + \po{Y} + 2 \pt{Y}{I} + 4 \pt{Y}{Z} + 5 \pt{Z}{I} + 2 \pt{Z}{Y} + 7 \pt{Z}{Z},\pt{A}{X} + 4 \pt{B}{X} + 3 \pt{B}{Y} + 5 \pt{I}{X} + 2 \pt{I}{Y} + 3 \pt{X}{I} + 7 \pt{X}{X} + 4 \pt{X}{Y} + 2 \pt{Y}{X})$} \\
\cline{2-3}
& $H$ & \raisebox{-1.5em}{\rule{0em}{3.5em}}\parbox{18.4cm}{$2 \pt{A}{Y} + 2 \pt{A}{Z} + \pt{I}{Y} + \pt{I}{Z} + \po{M} + \po{X} + \pt{X}{I} + \pt{X}{X} + 2 \pt{X}{Y} + 3 \pt{X}{Z} + \po{Y} + 3 \pt{Y}{I} + 4 \pt{Y}{X} + 5 \pt{Y}{Y} + 5 \pt{Y}{Z} + 2 \pt{Z}{I} + 3 \pt{Z}{X} + 4 \pt{Z}{Y} + 3 \pt{Z}{Z} + \max(\pt{A}{X} + \pt{A}{Y} + \pt{I}{X} + 2 \pt{I}{Y} + \pt{I}{Z} + \pt{X}{X} + 2 \pt{X}{Y},\pt{B}{Y} + \pt{B}{Z} + \pt{X}{I} + \po{Y} + 2 \pt{Y}{I} + 2 \pt{Y}{Z} + \po{Z} + \pt{Z}{I} + \pt{Z}{Z})$} \\
\cline{2-3}
& $\CX$ & \raisebox{-1.5em}{\rule{0em}{3.5em}}\parbox{18.4cm}{$\pt{A}{Y} + \pt{B}{Y} + 4 \pt{I}{Y} + \po{M} + 5 \pt{X}{Y} + 5 \pt{X}{Z} + 4 \pt{Y}{I} + 4 \pt{Y}{X} + 7 \pt{Y}{Y} + 5 \pt{Y}{Z} + 4 \pt{Z}{X} + 4 \pt{Z}{Y} + \max(\pt{A}{Y} + 2 \pt{A}{Z} + \pt{B}{Z} + \pt{I}{Y} + 5 \pt{I}{Z} + \po{X} + \po{Y} + 2 \pt{Y}{Z} + 4 \pt{Z}{I} + 3 \pt{Z}{Y} + 7 \pt{Z}{Z}, \pt{A}{X} + 2 \pt{B}{X} + \pt{B}{Y} + 4 \pt{I}{X} + 5 \pt{X}{I} + 7 \pt{X}{X} + 2 \pt{X}{Y} + \pt{Y}{I} + 3 \pt{Y}{X})$} \\
\cline{2-3}
\rule{.9em}{0em}\turnbox{90}{Double-coupling Steane}\rule{.3em}{0em} & $T$, $P$ & \raisebox{-2em}{\rule{0em}{4.5em}}\parbox{18.4cm}{$2 \pt{B}{Y} + \pt{I}{Y} + \po{M} + \pt{X}{Y} + \pt{Y}{X} + 2 \pt{Y}{Y} + \pt{Z}{X} + 2 \pt{Z}{Y} + \max(2 \pt{B}{X} + 2 \pt{I}{X} + \pt{I}{Y} + 2 \pt{X}{I} + 3 \pt{X}{X} + 2 \pt{X}{Y} + 2 \pt{X}{Z} + 2 \pt{Y}{I} + 2 \pt{Y}{X} + \pt{Y}{Y} + 2 \pt{Y}{Z} + \pt{Z}{X},2 \pt{B}{Z} + \pt{I}{Z} + \po{X} + \pt{X}{Z} + \po{Y} + \pt{Y}{I} + 2 \pt{Y}{Z} + \pt{Z}{I} + 2 \pt{Z}{Z},\pt{A}{X} + \pt{A}{Y} + 3 \pt{B}{X} + \pt{B}{Y} + 3 \pt{I}{X} + 2 \pt{I}{Y} + 3 \pt{X}{X} + 2 \pt{X}{Y} + 2 \pt{Y}{X} + \pt{Y}{Y} + 2 \pt{Z}{X} + \pt{Z}{Y})$} \\
\hline
\end{tabular}
\end{sidewaystable}
\addtocounter{table}{-1}

\begin{sidewaystable}
\renewcommand{\baselinestretch}{1}\selectfont
\capstart
\begin{tabular}{|c|c|c|}
\hline
\raisebox{-.4em}{\rule{0em}{1.6em}} & Gate & Maximal Single-strand Error Probability (Continued)\\
\hline
& \parbox{2.65em}{None, $T$, $P$} & \raisebox{-1.5em}{\rule{0em}{3.5em}}\parbox{18.4cm}{$\pt{A}{Y} + \pt{B}{Y} + \po{M} + \pt{Y}{X} + \pt{Y}{Y} + \pt{Z}{X} + \pt{Z}{Y} + \max(\pt{A}{Z} + \pt{B}{Z} + \po{X} + \po{Y} + \pt{Y}{I} + \pt{Y}{Z} + \pt{Z}{I} + \pt{Z}{Z},\pt{A}{X} + \pt{B}{X} + \pt{I}{X} + \pt{I}{Y} + \pt{X}{X} + \pt{X}{Y})$} \\
\cline{2-3}
& $H$ & \raisebox{-1em}{\rule{0em}{2.5em}}\parbox{18.4cm}{$\pt{B}{X} + 2 \pt{B}{Y} + \pt{B}{Z} + \po{M} + \po{X} + \po{Y} + \pt{Y}{X} + \pt{Y}{Y} + \pt{Z}{X} + \pt{Z}{Y} + \min(\pt{I}{X} + \pt{I}{Y} + \pt{X}{X} + \pt{X}{Y},\po{Y} + \pt{Y}{I} + \pt{Y}{Z} + \po{Z} + \pt{Z}{I} + \pt{Z}{Z})$} \\
\cline{2-3}
\rule{.9em}{0em}\turnbox{90}{\hspace{2em}Knill}\rule{.3em}{0em} & $\CX$ & \raisebox{-2em}{\rule{0em}{4.5em}}\parbox{18.4cm}{$\pt{A}{Y} + \pt{B}{Y} + \po{M} + \pt{Y}{X} + 2 \pt{Y}{Y} + \pt{Z}{X} + \pt{Z}{Y} + \max(\pt{A}{Y} + 2 \pt{A}{Z} + \pt{B}{Z} + \po{X} + \po{Y} + 2 \pt{Y}{I} + \pt{Y}{X} + 2 \pt{Y}{Z} + 2 \pt{Z}{I} + \pt{Z}{X} + \pt{Z}{Y} + 2 \pt{Z}{Z},\pt{A}{X} + \pt{B}{X} + \pt{I}{X} + \pt{I}{Y} + \pt{X}{I} + 2 \pt{X}{X} + 2 \pt{X}{Y} + \pt{X}{Z} + \pt{Y}{I} + \pt{Y}{X} + \pt{Y}{Z},\pt{A}{X} + 2 \pt{B}{X} + \pt{B}{Y} + 2 \pt{I}{X} + 2 \pt{I}{Y} + 2 \pt{X}{X} + 2 \pt{X}{Y} + \pt{Y}{X} + \pt{Z}{X} + \pt{Z}{Y},\pt{A}{Z} + \pt{B}{Z} + \pt{I}{Y} + \pt{I}{Z} + \po{X} + \pt{X}{Y} + \pt{X}{Z} + \po{Y} + \pt{Y}{I} + 2 \pt{Y}{Z} + \pt{Z}{I} + \pt{Z}{Y} + 2 \pt{Z}{Z})$} \\
\hline
\end{tabular}
\caption[Maximum error probabilities for a single strand of various encoded gates for a selection of fault-tolerant procedures]{Maximum error probabilities for a single strand of various encoded gates for a selection of fault-tolerant procedures.  The subscripted $p$'s refer to probabilities of error of various unencoded operations with $X$, $Y$, $Z$ denoting Pauli errors, $M$ measurement errors, and $A$- and $B$-ancilla errors.  Thus $\po{X}$, $\pt{I}{Z}$, $\pt{Z}{I}$, $\po{M}$, and $\pt{A}{Y}$ denote the probabilities of a single-qubit-gate $X$ error, a $Z$ error on the target of a $\protect \CX$, a $Z$ error on the control of a $\protect \CX$, a measurement error, and a $Y$ error on an $A$-type ancilla.  A procedure is below threshold when the maximum error probability for all gates is less than $\tau$, the fraction of errors corrected asymptotically. \label{tab:encodedErrorRates}}
\end{sidewaystable}

The unrefined output of this endeavor is the set of maximum
strand error probabilities listed in Table~\ref{tab:encodedErrorRates}.
This table specifies a kind of high-dimensional threshold surface in the space of generic
stochastic error models.  An error channel is below the threshold for a particular procedure
whenever the maximal strand error probabilities for that procedure are lower than the fraction
of errors that are correctable asymptotically.

For the purpose of illustration, however, it is more useful to consider
less complicated error models.  Table~\ref{tab:errorModels} defines four
reduced error models in terms of the generic stochastic error model.  Since these
reduced error models have only a single free parameter, their threshold surfaces are
simply numbers.  Table~\ref{tab:thresholdCoefficients} lists thresholds for three
procedures and four reduced error models in terms of $\tau$, the asymptotic
correctable error fraction.

The following sections provide supplemental information specific to
each procedure, including qualitative reviews of the procedures, circuit
diagrams for encoded gates, and commentary on the thresholds given in
Table~\ref{tab:thresholdCoefficients}.

\begin{table}
\capstart
\centerline{
\begin{tabular}{|c|r@{.}l|r@{.}l|r@{.}l|r@{.}l|}
\hline
\multicolumn{9}{|c|}{Thresholds for Homogeneous Ancillae ($\tau$)\rule[-.8em]{0em}{2.2em}} \\
\hline
\setlength{\unitlength}{1em}
\begin{picture}(12,1.79)
\put(-.53,1.78){\line(6,-1){12.71}}
\put(-.1,.02){Procedure}
\put(6.4,.82){Error Model}
\end{picture}
& \multicolumn{2}{|c|}{\raisebox{.4em}{\#1}} & \multicolumn{2}{|c|}{\raisebox{.4em}{\#2}} & \multicolumn{2}{|c|}{\raisebox{.4em}{\#3}} & \multicolumn{2}{|c|}{\raisebox{.4em}{\#4}} \\
\hline
Single-coupling Steane\rule[-.5em]{0em}{1.6em} & 0&15 & 0&06 & 0&24 & 0&29 \\
\hline
Double-coupling Steane\rule[-.5em]{0em}{1.6em} & 0&16 & 0&10 & 0&18 & 0&29 \\
\hline
Knill\rule[-.5em]{0em}{1.6em} & \:0&35\: & \:0&15\: & \:0&50\: & \:0&67\: \\
\hline
\end{tabular}
}
\caption[Thresholds for ancillae with homogeneous errors]{Thresholds for ancillae with homogeneous errors given in units of $\tau$, the correctable error fraction, for the three procedures and four error models I consider as examples.  These thresholds were obtained by substituting the parameters given in Table~\ref{tab:errorModels} into Table~\ref{tab:encodedErrorRates} and requiring that single-strand error probabilities not exceed $\tau$. \label{tab:thresholdCoefficients} }
\end{table}

\subsection{Steane's Method}

Typical instantiations of Steane's method employ multiple extractions to guard against errors made during the coupling and measurement process.
Often \cite{Steane03,Reichardt04,Zalka96} the number of extractions performed is conditional on their output.  I deviate from this rule by demanding a fixed number of couplings.  This a sensible choice for my analysis since, up to rearrangement of qubits, the output of an extraction becomes deterministic as the number of qubits in an encoding approaches infinity.  Moreover, if $p$ is the probability of an error occurring, requiring the sequential agreement of $j$ extractions reduces the probability of misdiagnosing an error on a particular line to order $p^j$.  Since I ultimately retain only first-order terms I need only consider single and double-coupling Steane procedures.

\subsubsection{Single-coupling Steane Procedure}

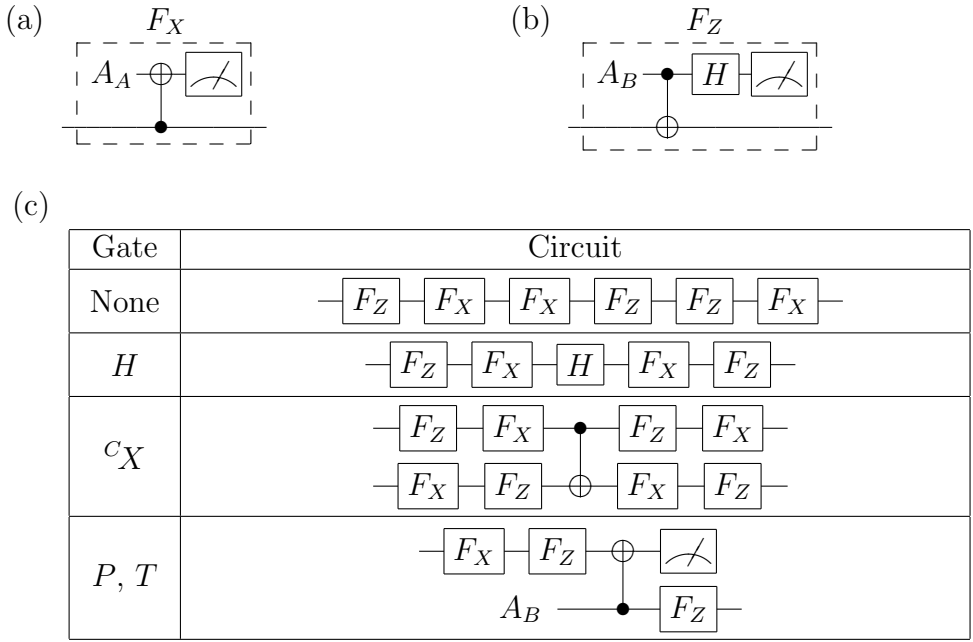
\begin{figure}[tbp]
\capstart
\begin{tabular*}{34em}{@{\extracolsep{\fill}}lll}
\Qcircuit @R=.3em @C=.4em @!R {
& \text{(a)} & & & & & & & & & & {\ F_X} \\
& & & & & & & & {A_A} & & & \targ & \meter \\
& & & & & \qw & \push{\rule[-.4em]{0em}{.4em}} \qw & \qw & \qw & \qw & \qw & \ctrl{-1} & \qw & \qw & \qw
\gategroup{3}{7}{2}{13}{.6em}{--}
}
&
\Qcircuit @R=.3em @C=.4em @!R {
& \text{(b)} & & & & & & & & & & & {\!\!\!\!F_Z} \\
& & & & & & & & {A_B} & & & \ctrl{1} & \gate{H} & \meter \\
& & & & & \qw & \push{\rule[-.6em]{0em}{.8em}} \qw & \qw & \qw & \qw & \qw & \targ & \qw & \qw & \qw & \qw
\gategroup{3}{7}{2}{14}{.6em}{--}
}
\\
\\
(c)
\end{tabular*}
\\
\rule{1.5em}{0em}
\begin{tabular*}{29em}{@{\extracolsep{\fill}}|c|c|}
\hline
\:Gate\: & Circuit \\
\hline
\raisebox{.1em}{None} &
\rule{3.25em}{0em}
\raisebox{.45em}{\rule[-1em]{0em}{2em}
\Qcircuit @R=.4em @C=.8em {
& \gate{F_{Z}} &  \gate{F_{X}} &  \gate{F_{X}} &  \gate{F_{Z}} & \gate{F_{Z}} &  \gate{F_{X}} & \qw
}
}
\rule{3.25em}{0em}
\\
\hline
\raisebox{.1em}{$H$} &
\raisebox{.45em}{\rule[-1em]{0em}{2em}
\Qcircuit @R=.4em @C=.8em {
& \gate{F_{Z}} &  \gate{F_{X}} &  \gate{H} &  \gate{F_{X}} &  \gate{F_{Z}} & \qw
}
}
\\
\hline
\raisebox{-.4em}{$\CX$} &
\raisebox{.9em}{\rule[-2.8em]{0em}{3.8em}
\Qcircuit @R=.4em @C=.8em {
& \gate{F_{Z}} &  \gate{F_{X}} &  \ctrl{1} &  \gate{F_{Z}} &  \gate{F_{X}} & \qw \\
& \gate{F_{X}} &  \gate{F_{Z}} &  \targ &  \gate{F_{X}} &  \gate{F_{Z}} & \qw
}
}
\\
\hline
\raisebox{-.2em}{$P$, $T$} &
\raisebox{.9em}{\rule[-2.8em]{0em}{3.9em}
\Qcircuit @R=.4em @C=.8em {
& \gate{F_{X}} &  \gate{F_{Z}} &  \targ &  \meter \\
& & \lstick{A_B} & \ctrl{-1} & \gate{F_Z} & \qw
}
}
\\
\hline
\end{tabular*}
\caption[Encoded circuits for the single-coupling Steane procedure]{Encoded circuits for the single-coupling Steane procedure.  Handling of the data is minimized at the cost of syndrome verification.  Parts a and b display the circuits for finding $X$ and $Z$ errors, respectively.  Part c) lists the circuits analyzed to determine the encoded error rates for this example.\label{fig:singleSteane}}
\end{figure}

Two-qubit gates require a controlled interaction between two otherwise
isolated quantum systems.  Consequently, they are often the most error
prone gates in a universal set.  In such cases, the factor limiting the
probability of successful error correction may be the number of times that
two-qubit gates must be applied to the data in order to reliably diagnose
errors.  For Steane's method, this interaction is minimized by coupling to
the data once per $X$ correction and once per $Z$ correction, as shown in
Figure~\ref{fig:singleSteane}.

Table~\ref{tab:thresholdCoefficients} shows that, relative to the other
procedures considered, the single-coupling Steane procedure performs most
strongly for error model \#3.  This is in line with our expectations since
model \#3 includes only two-qubit gate errors and the resultant ancilla
errors.  Surprisingly, it also does rather well overall, suffering in
comparison to the double-coupling Steane procedure only for error model
\#2 where measurement errors dominate.  The single-coupling Steane
procedure lacks a means of syndrome verification, so any errors in
syndrome measurement are transferred directly to the data. Nevertheless,
my results demonstrate that moderate single-qubit and measurement error
probabilities can be tolerated when high quality ancillae are available.

\subsubsection{Double-coupling Steane Procedure}

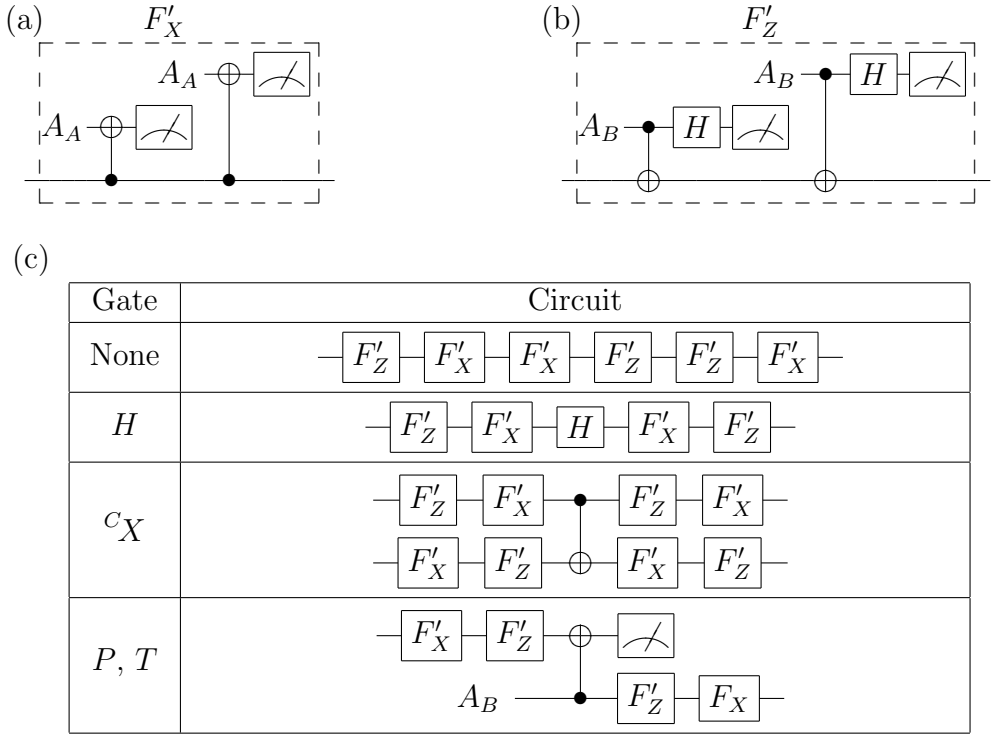
\begin{figure}[btp]
\capstart
\begin{tabular*}{32em}{@{\extracolsep{\fill}}ll}
\Qcircuit @R=.3em @C=.4em @!R {
& \text{(a)} & & & & & & & {F^\prime_X} \\
& & & & & & & & & {A_A\;\;\;\;\;\;} & \targ &  \meter \\
& & & & {A_A} & & & \targ & \meter \\
& & \qw & \push{\rule[-.6em]{0em}{.8em}} \qw & \qw & \qw & \qw & \ctrl{-1} & \qw & \qw & \ctrl{-2} & \qw & \qw & \qw
\gategroup{4}{4}{2}{12}{.6em}{--}
}
&
\Qcircuit @R=.3em @C=.4em @!R {
& \text{(b)} & & & & & & & & {F^\prime_Z} \\
& & & & & & & & & & {A_B\;\;\;\;\;\;} & \ctrl{2} & \gate{H} &  \meter \\
& & & & {A_B} & & & \ctrl{1} & \gate{H} & \meter \\
& & \qw & \push{\rule[-.6em]{0em}{.8em}} \qw & \qw & \qw & \qw & \targ & \qw & \qw & \qw & \targ & \qw & \qw & \qw & \qw
\gategroup{4}{4}{2}{14}{.6em}{--}
}
\\
\\
(c)
\end{tabular*}
\\
\rule{1.5em}{0em}
\begin{tabular*}{29em}{@{\extracolsep{\fill}}|c|c|}
\hline
\:Gate\: & Circuit \\
\hline
\raisebox{.2em}{None} &
\rule{3.25em}{0em}
\raisebox{.45em}{\rule[-1.1em]{0em}{2.2em}
\Qcircuit @R=.4em @C=.8em {
& \gate{F^\prime_Z} &  \gate{F^\prime_X} &  \gate{F^\prime_X} &  \gate{F^\prime_Z} & \gate{F^\prime_Z} &  \gate{F^\prime_X} & \qw
}
}
\rule{3.25em}{0em}
\\
\hline
\raisebox{.2em}{$H$} &
\raisebox{.45em}{\rule[-1.1em]{0em}{2.2em}
\Qcircuit @R=.4em @C=.8em {
& \gate{F^\prime_Z} &  \gate{F^\prime_X} &  \gate{H} &  \gate{F^\prime_X} &  \gate{F^\prime_Z} & \qw
}
}
\\
\hline
\raisebox{-.4em}{$\CX$} &
\raisebox{.9em}{\rule[-3.1em]{0em}{4.3em}
\Qcircuit @R=.4em @C=.8em {
& \gate{F^\prime_Z} &  \gate{F^\prime_X} &  \ctrl{1} &  \gate{F^\prime_Z} &  \gate{F^\prime_X} & \qw \\
& \gate{F^\prime_X} &  \gate{F^\prime_Z} &  \targ &  \gate{F^\prime_X} &  \gate{F^\prime_Z} & \qw
}
}
\\
\hline
\raisebox{-.2em}{$P$, $T$} &
\raisebox{.9em}{\rule[-3.1em]{0em}{4.3em}
\Qcircuit @R=.4em @C=.8em {
& \gate{F^\prime_X} &  \gate{F^\prime_Z} &  \targ &  \meter \\
& & \lstick{A_B} & \ctrl{-1} & \gate{F^\prime_Z} & \gate{F_X} & \qw
}
}
\\
\hline
\end{tabular*}
\caption[Encoded circuits for the double-coupling Steane procedure]{Encoded circuits for the double-coupling Steane procedure.  Syndrome information is extracted twice, and qubits implicated both times are presumed to be in error.  Parts a and b display the circuits for finding $X$ and $Z$ errors, respectively.  Part c) lists the circuits analyzed to determine the encoded error rates for this example.\label{fig:doubleSteane}}
\end{figure}

When two-qubit gates are relatively reliable, the damage done during the
extraction of error information can be limited by preparing ancillae such
that they include few errors capable of propagating to the data.   Under
these circumstances, it is often advantageous to verify error diagnoses by
coupling to the data more than once, as shown in
Figure~\ref{fig:doubleSteane}.

For the double-coupling Steane procedure, error model \#1 is especially
interesting because it was chosen in imitation of the error model used by
Reichardt~\cite{Reichardt04} in his numerical estimation of the threshold
for a Steane style procedure on a 49 qubit code.  My threshold of $0.90\%$
for asymptotic minimum distance decoding is quite close to his value of
roughly $0.88\%$.  The extraordinary agreement of these two estimates is a
coincidence, as can be seen from my discussion of finite codes in
Section~\ref{sec:finiteCodes}, but their rough equivalence illustrates the value
of my idealized algorithm for approximating the encoded error rates used in
threshold estimation.  As the number of qubits grows, the threshold estimate obtained by Monte-Carlo
simulation approaches that predicted my analytical method.
In fact, Section~\ref{sec:finiteCodes} shows that reasonably good bounds can be
placed on outcome of threshold estimates even for very small numbers of qubits
provided that the fault-tolerant procedure is exactly implementable in my form.

As expected, relative to the other two procedures, the double-coupling
Steane procedure performs most favorably for error model \#2.  Somewhat
surprisingly, however, it still underperforms the Knill procedure.  The
reason for this is most easily understood by considering the limiting case
in which only measurement errors occur.  In the absence of any other
source of error, measurement errors have no effect on either the double-coupling Steane or the Knill procedure until their probability exceeds
$\tau$; beyond that point both procedures fail with certainty.  Thus, the
two procedures cope with measurement errors equally well, but the Knill
procedure handles other kinds of gate errors more effectively.

Error models \#3 and \#4 demonstrate small gains in the threshold that can
result when two-qubit gate errors have some underlying structure.  Model
\#3 is a pure two-qubit-gate depolarizing error model (which includes the
associated ancilla errors) while \#4 is a model in which two-qubit gates
malfunction by producing either an $X$ or a $Z$ error on either the
control or the target.  Given the highly restricted form of error model
\#4 it is discouraging that the threshold increases by less than a factor
of two over that of error model \#3.  Though, in light of the already high
value for the threshold in error model \#3, it is perhaps unsurprising.

\subsection{Knill's Method}

For my implementation of Knill's method, error model \#4 achieves the highest threshold, though physical systems displaying this sort of error seem unlikely.  Error model \#1 provides another check of my algorithm, since its parameters are also roughly those used by Knill \cite{Knill05} in a paper on telecorrection.  Setting $\tau$ to $11\%$ for the channel capacity for CSS codes, I find that the threshold for this model is $3.9\%$ compared to Knill's estimate of $3\%$ and his extrapolation of up to $5\%$.  The approximate agreement between these values is satisfying, though an exact match is not expected since Knill assumes that errors on up to $19\%$ of the qubits can be corrected, an assumption that derives from bounds on the channel capacity for general quantum codes~\cite{DiVincenzo98}.

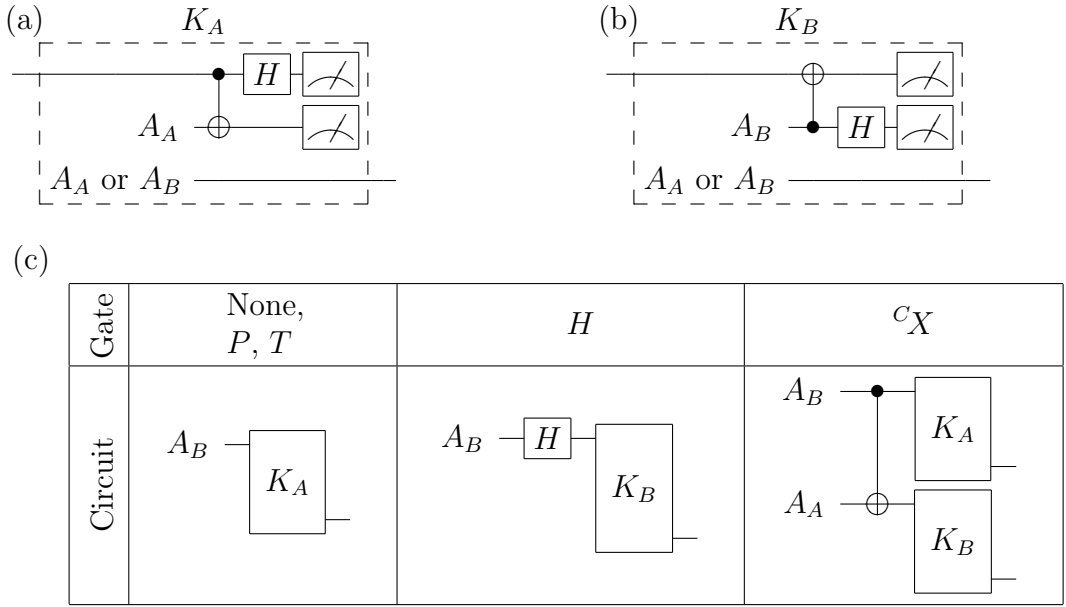
\begin{figure}[tbp]
\capstart
\begin{tabular*}{32em}{@{\extracolsep{\fill}}ll}
\Qcircuit @R=.3em @C=.4em @!R {
& \text{(a)} & & & & {\!\!\!\!\!\!K_A}\\
& \qw & \qw & \qw & \qw & \ctrl{1} & \gate{H} & \meter \\
& & & & \lstick{A_A} & \targ & \qw & \meter \\
& & & \push{A_A\text{ or } A_B} & & \qw & \qw & \qw & \qw & \qw & \qw
\gategroup{4}{4}{2}{8}{.6em}{--}
}
&
\Qcircuit @R=.3em @C=.4em @!R {
& \text{(b)} & & & & {\!\!\!\!\!\!K_B}\\
& \qw & \qw & \qw & \qw & \targ & \qw & \meter \\
& & & & \lstick{A_B} & \ctrl{-1} & \gate{H} & \meter \\
& & & \push{A_A\text{ or } A_B} & & \qw & \qw & \qw & \qw & \qw & \qw
\gategroup{4}{4}{2}{8}{.6em}{--}
}
\\
\\
(c)
\end{tabular*}
\\
\rule{1.5em}{0em}
\begin{tabular*}{32em}{@{\extracolsep{\fill}}|c|c|c|c|}
\hline
\raisebox{-.8em}{\rule{.9em}{0em}\turnbox{90}{Gate}\rule[-.2em]{0em}{2.6em}} & \parbox{3em}{None, $P$, $T$} & $H$ & $\CX$
\\
\hline
\raisebox{-3em}{\rule{.9em}{0em}\turnbox{90}{Circuit}} &
\raisebox{-.2em}{
\Qcircuit @R=.3em @C=.8em @!R {
& & \lstick{A_B} & \multigate{2}{K_A}\\
& & & \pureghost{K_A}\\
& & & \pureghost{K_A} & \qw
}
}\rule{1em}{0em}
&
\rule{.2em}{0em}
\Qcircuit @R=.3em @C=.8em @!R {
& & \lstick{A_B} & \gate{H} & \multigate{2}{K_B}\\
& & & & \pureghost{K_B}\\
& & & & \pureghost{K_B} & \qw
}\rule{1em}{0em}
&
\raisebox{1.5em}{
\Qcircuit @R=.3em @C=.8em @!R {
& & \lstick{A_B} & \ctrl{3} & \multigate{2}{K_A}\\
& & & & \pureghost{K_A}\\
& & & & \pureghost{K_A} & \qw \\
& & \lstick{A_A} & \targ & \multigate{2}{K_B}\\
& & & & \pureghost{K_B}\\
& & & & \pureghost{K_B} & \qw
}
}\rule[-5.3em]{0em}{7.6em}\rule{1em}{0em}
\\
\hline
\end{tabular*}
\caption[Encoded circuits for the Knill procedure]{Encoded circuits for the Knill procedure.  Error correction is performed by teleporting the data, minus the errors, using an entangled two-logical-qubit ancilla; the precise location of errors is unimportant so long as the result of the logical-qubit measurement is correctly decoded.  Parts a and b display the circuits for correcting errors when the output of the previous step was dominated by $Z$ and $X$ errors respectively.  Part c) lists the circuits analyzed to determine the encoded error rates for this example.  Ancilla error distributions are used as the input to these circuits since that is the only remaining source of error after a successful teleportation correction.\label{fig:Knill}}
\end{figure}

Of course the most striking aspect of
Table~\ref{tab:thresholdCoefficients} is that the Knill procedure yields a
higher threshold for every error model.  As with the single-coupling
Steane case, this derives partly from my assumptions regarding ancillae.
In particular, Steane's method was designed to utilize ancillae for
which either correlated $X$ or correlated $Z$ errors could be minimized,
but not both, a situation certain to favor his approach.  A second but
lesser objection can be made that I set the ancilla error probabilities
equal for all gates and all methods, ignoring the fact that some methods,
such as Knill's, and some gates, such as the $T$ gate, will require more
complex ancillae which may in turn be more error prone.  Substantially
more detailed ancilla information would be needed to evaluate the
importance of this effect, but the overall character of my results is
unlikely to change since that would entail in excess of a two-fold
increase in the error probabilities for logical two-qubit ancillae over
those for ancillae prepared in a single-qubit logical state.  Thus, so
long as resource considerations do not limit our ability to discard
suspect ancillae, and therefore to make very high quality ancillae,
Knill's method will provide the highest thresholds.

\section{Finite Codes\label{sec:finiteCodes}}
Prior to taking the limit $n\rightarrow\infty$, the expression for the
probability of an encoded error at a location $\loc{L}$ was
\begin{align}
E_{\mathrm{f}}(p_\loc{L}) = \sum_{i=t+1}^n\binom{n}{i}p_\loc{L}^i(1-p_\loc{L})^{n-i},\label{eq:pfail}
\end{align}
where $t$ is the number of correctable errors and $p_\loc{L}$ is the
probability of a relevant error on a single qubit at the location in
question.  Using this expression, the programme of Section~\ref{sec:homogeneous}
can be implemented for finite $n$.  In doing so, however, the simplicity
of the algorithm suffers somewhat, and its interpretation as an idealized
threshold bound is completely lost.  Fundamentally, the complications that
arise are all due to the fact that the success or failure of various
portions of an encoded gate are no longer deterministic.  This section
explains how to deal with the associated difficulties and concludes with a
brief demonstration of the algorithm for a $[[49,1,9]]$ and $[[7,1,3]]$ code.

In the examples of Section~\ref{sec:examples}, I establish a background error
rate by performing an initial error correction, but for finite $n$ this
initialization is not guaranteed to succeed.  Though the failure of the
initial error correction is properly assigned to the previous encoded
gate, the residual errors will differ dramatically depending on whether it
occurred.  This presents no problem when only a single level of encoding
is employed since any encoded failure is considered a failure of the
computation.  In concatenated coding schemes, however, failed encoded
qubits are corrected at higher levels of encoding.  Their continued use is
problematic since an encoded gate failure may be correlated with
subsequent encoded failures.  Nevertheless, I recommend calculating the
encoded error rate for finite codes using the assumption that the
initialization did not fail, a choice that requires no modification to the
case for large $n$.

Likewise, calculation of the single line error rate $p_\loc{L}$ proceeds
without modification.  For finite codes, however, the maximum tolerable
single line error rate becomes a nontrivial function of the encoded error
rate that we wish to achieve.  The probability of an unrecoverable error
never goes to zero, so it is necessary to perform the summation in
Equation~(\ref{eq:pfail}) to determine the portion of the encoded error rate
due to any particular location.

The possibility of failure must be considered at many points in the
circuit since statistical fluctuations will produce unrecoverable errors
at a variety of locations.  Typically, encoded failure probabilities at
various locations will be strongly correlated, but the exact nature of
these correlations is difficult to predict.  Thus, the best I can do
is to bound the encoded failure probability,
\begin{align}
\max_{\loc{L}\in\loc{S}} E_{\mathrm{f}}(p_\loc{L}) \leq \left\{
\parbox{5.3em}{\centering Encoded error rate}\rule{0em}{1.3em}
\right\} \leq \sum_{\loc{L}\in\loc{S}} E_{\mathrm{f}}(p_\loc{L})\label{eq:encodedRange}
\end{align}
where $\loc{S}$ ranges over the locations of every post-initialization output, that is, syndrome measurements and the final state of the data with regard to both $X$ and $Z$ errors.

To clarify the changes outlined above, consider the example of the double-coupling Steane procedure implemented using a $[[49,1,9]]$ quantum code and subject to the error channel defined by error model \#1.  For the encoded $\CX$ gate, the set of single line error probabilities corresponding to $X$ errors at the eight locations of post-initialization syndrome measurement and $X$ and $Z$ errors at the two output locations of the data is
\begin{align}
\begin{split}
\{p_\loc{S}\}=\left\{\frac{47p}{8},\frac{43p}{8},\frac{43p}{8},\right.&\frac{41p}{8},\frac{39p}{8},\frac{37p}{8},\frac{37p}{8},\\
&\left.\frac{33p}{8},\frac{9p}{4},\frac{9p}{4},\frac{3p}{4},\frac{3p}{4}\right\}.
\end{split}
\end{align}
Solving Equation~(\ref{eq:encodedRange}) subject to the restriction that the encoded error rate is exactly $p$ yields solutions in the range
\begin{align}
0.0036 \geq p \geq 0.0023.
\end{align}
Repeating this process for each of the other encoded gates and taking the minimum over the upper and lower bounds produces a threshold of
\begin{align}
0.0034 \geq p_\text{th} \geq 0.0025
\end{align}
where, of course, the caveats discussed in Section~\ref{subsec:thresholdEstimates} regarding concatenated threshold estimates all apply.  This example provides a particularly apt comparison to Reichardt's threshold estimate for the $[[49,1,9]]$ code~\cite{Reichardt04}.  The threshold calculated here is roughly a third of that estimated by Reichardt.  The difference presumably springs from the superiority of his rule for syndrome extraction when $n=49$.

\begin{figure}
  \capstart
  \centerline{
    \begin{pgfpicture}{0cm}{0cm}{9.25cm}{6.65cm}
      \pgfputat{\pgfxy(.5,.25)}{\includegraphics[clip=true, trim=0cm 7cm 0cm 7cm, width=10cm]{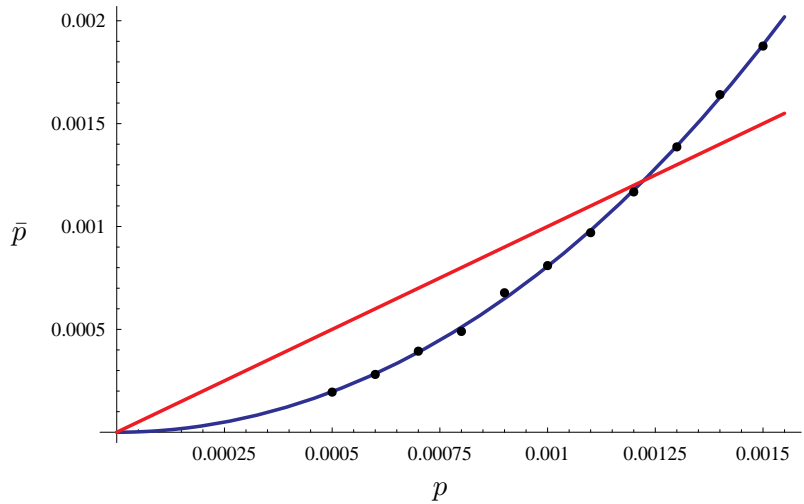}}
      \pgfputat{\pgfxy(5.7,.1)}{\pgfbox[center,center]{\footnotesize $p$}}
      \pgfputat{\pgfxy(.1,3.5)}{\pgfbox[center,center]{\footnotesize $\bar{p}$}}
    \end{pgfpicture}\hspace{2cm}
  }
  \caption[Estimating the depolarizing threshold]{The encoded $\CX$ error probability $\bar{p}$ versus the unencoded gate error probability $p$.  The data was taken for Steane's code and the double-coupling Steane procedure using error model \#1, a rescaled depolarizing channel.  For reference, a diagonal line demarcating the break-even point for encoding is drawn in red.  The intersection of this line with the blue curve fitting the data gives a depolarizing threshold estimate of $p_\text{Dth}=0.0012$.  Error bars fit within the dots. \label{fig:SFTIReichDepData}}
\end{figure}

Readers puzzled by the discrepancy between the threshold estimates obtained in this chapter and those calculated in Chapter~\ref{chap:channelDependencyOfTheThreshold} should be relieved to know that the equivalent bounds for Steane's $[[7,1,3]]$ code are
\begin{align}
0.0013 \geq p_\text{th} \geq 0.0003 \label{eq:thresholdBoundRangeSteane}
\end{align}
showing once again that, in addition to being approximate, the values of the threshold estimated by such methods are highly procedure dependent.  As a consistency check, encoded $\CX$ error rates obtained for exactly the same error model (excepting ancilla errors which are determined by the ancilla construction) and fault-tolerant procedure using the Monte-Carlo algorithm from Chapter~\ref{chap:channelDependencyOfTheThreshold} are plotted in Figure~\ref{fig:SFTIReichDepData}.  From the figure, it can be seen that the threshold estimate falls within the range specified by Equation~(\ref{eq:thresholdBoundRangeSteane}).  The notable difference between the estimate of the depolarizing threshold here and in the last chapter arises because the error model used to represent the depolarizing channel here, chosen for comparison with Reichardt's work, actually produces fewer Pauli errors for a given value of $p$ than that in the previous chapter.

\section{Analysis}

The algorithm that I have described for generating thresholds can be viewed in two possible lights.  First, it might thought of as a way of establishing rigorous bounds on the threshold for fault-tolerant quantum computation given the, admittedly elusive, resource of ancillae with independent, identically distributed errors.  Second, it can be considered a fast, flexible method for establishing threshold estimates, yielding the estimate that would ultimately be obtained for a method given a large enough quantum code and sufficient computer time.

My approach applies to most fault-tolerant procedures employing CSS codes.  It relies on the fact that nearly all elements of such a procedure are homogeneous, that is, transversal with identical components.  Inhomogeneous elements are either eliminated, as for classical syndrome processing and the application of recovery unitaries, or, in the case of ancillae, replaced with homogeneous equivalents.  This allows me to calculate the probability of failure for encoded gates in terms of the error probabilities associated with a single strand of the encoded blocks.  In the limit that the number of encoding qubits approaches infinity, a criterion for success becomes simply that the probability of finding an error never exceed the fraction of the encoded qubits on which said error can be corrected.  When this is satisfied, it is possible, in the limit of infinite block size, to compute indefinitely, and our base error rates are, by definition, below threshold.

The value of considering thresholds for homogeneous ancillae is that they can easily be calculated for a variety of fault-tolerant procedures and error models, thereby providing a relatively simple metric for comparison.  Section~\ref{sec:examples} includes thresholds for computation for three fault-tolerant procedures and four error models.  One of the procedures considered is based on a method of telecorrection used by Knill, while the other two are variations, in that the number of syndrome extractions is fixed, on Steane's approach to achieving fault tolerance.  The error models considered are a full depolarizing error model, a depolarizing error model with increased measurement errors, a depolarizing error model for two-qubit gates exclusively, and a restricted two-qubit-gate error model.  Holding the total probability of an error constant, small improvements are observed in the threshold for certain choices of the two-qubit-gate error model.  For stochastic errors and CSS codes, improvements of a grander scale are unlikely because the threshold coefficient for the depolarizing channel using Knill's procedure is already $1/3$.  For errors that actually reach the data, the largest threshold coefficient one would expect is $1$ (though $2$ might be achieved using multiple versions of gates with different highly unusual errors), so at best we might look for a factor of $3$.  Interestingly, I find by inspection that the threshold coefficient for measurement errors exclusively, which need not reach the data, is $1$ for the double-coupling Steane and Knill procedures.
With regard to comparisons between procedures, the single-coupling Steane procedure is shown to outperform the double-coupling procedure when two-qubit depolarizing errors dominate, but the double-coupling Steane procedure does notably better when measurement errors are likely.  I also find that Knill's approach outperforms that of Steane for all error models considered, a conclusion that is likely to hold so long as correlated ancillary errors are rare and the ancillae needed for Knill's method are not appreciably more error prone than those employed by Steane.

Idealized thresholds aside, my algorithm is useful as a means of approximately computing the logical error rate for a single level of encoding, which is an established method of estimating the threshold for quantum computation.  The two treatments yield similar outcomes because numerical estimates of the encoded error rate typically prepare ancillae in a way that maximizes their quality at the cost of additional resource overhead.  Ancillae prepared in this manner have error distributions approximating my ideal of independent, identically distributed errors.  The basic algorithm uses the infinite limit to obtain simple analytic results, but an alternative (and less rigorous) algorithm for finite codes is described in Section~\ref{sec:finiteCodes}.  Both methods were shown to yield results in rough accordance with the depolarizing threshold determined by Reichardt~\cite{Reichardt04} for the $[[49,1,9]]$ code, and the finite version was found to yield reasonable bounds even for the case of Steane's $[[7,1,3]]$ code.  For telecorrection, in the limit $n\rightarrow\infty$, my estimate of the depolarizing threshold was consistent with the range of values determined by Knill~\cite{Knill05}.

Much further work remains to be done on this subject.  One topic of interest is the rate of convergence of my threshold estimate ranges for finite codes with the estimates obtained in the large $n$ limit.  A second possibility is the extension of my analysis to include memory errors, which promises to be a straightforward, if unbeautiful, endeavor.  The most valuable addition, however, would be to explicitly define methods of ancilla construction and determine the degree to which they differ from my ideal.  Constructing ancillae to my specifications is an extremely difficult problem, but one whose solution would have a strong impact on the theory of quantum computing in general and this work in particular.  A scalable method for producing ancillae with independent, identically distributed errors would enable the algorithm presented here to be employed for the calculation of rigorous lower bounds on the threshold without any caveats about idealized resources.  Which is why I take up that problem in the following chapter.

\chapter{Ancilla Construction\label{chap:ancillaConstruction}}
In Chapter~\ref{chap:channelDependencyOfTheThreshold} I found little advantage to having a detailed knowledge of the Pauli error channel afflicting a system.  This was due to a pair of facts.  First, the error overhead due to the application of transversal gates is only marginally affected by the particular errors that happen to be generated by the gates.  Second, ancillae prepared using a very liberal discard policy already have such low error rates that they have little effect on the threshold at all, so improving them is of no consequence.  The second observation led me to investigate the threshold separately from ancilla construction in Chapter~\ref{chap:thresholdsForHomogeneousAncillae} where I found that large ancillae with homogeneous error distributions were indeed a valuable resource, sufficient to permit quantum computation at error rates on the order of a few percent.  The construction of large, high-quality ancillae, however, is a difficult matter.  One that I now turn to.

\section{Graph States}
A graph state~\cite{VandenNest04} is a stabilizer state for which there exists a set of stabilizer generators, $\set{S}_g=\{A^j\}$, such that
\begin{equation}
  A^j = X_j \prod_{k \in \mathcal{N}(j)} Z_k
  \label{eq:graphgenerators}
\end{equation}
where $\set{N}(j)$ denotes the neighbors of node $j$ on some graph.  Using the binary representation of the Pauli group, such a set of generators has the form
\begin{equation}
  \mathcal{S}_g = \left[\begin{array}{c|c} \mat{I}  & \mat{B} \end{array}\right]
  \label{eq:canonicalgraph}
\end{equation}
where $\mat{B}$ is the adjacency matrix of the graph and, thus, a symmetric matrix with $0$'s on the diagonal.

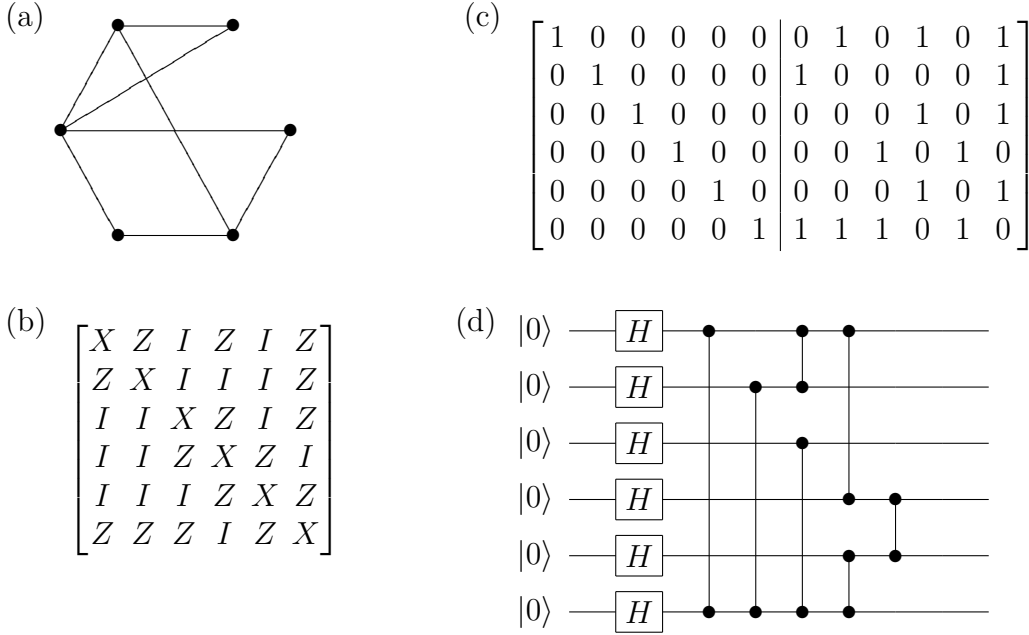
\begin{figure}
  \capstart
  (a)
  \Qcircuit[.35em] @C=1.5em @R=3em {
& \sink \link{0}{2} \link{2}{2} & & \sink \\
\sink \link{-1}{1} \link{-1}{3} \link{1}{1} \link{0}{4} & & & & \sink \\
& \sink \link{0}{2} & & \sink \link{-1}{1}
}
  \hspace{5.4em}(c)
  \raisebox{-3.75em}{
    $\left[
    \begin{array}{@{\hspace{.2em}}c@{\hspace{.8em}}c@{\hspace{.8em}}c@{\hspace{.8em}}c@{\hspace{.8em}}c@{\hspace{.8em}}c|c@{\hspace{.8em}}c@{\hspace{.8em}}c@{\hspace{.8em}}c@{\hspace{.8em}}c@{\hspace{.8em}}c@{\hspace{.2em}}}
1 & 0 & 0 & 0 & 0 & 0 & 0 & 1 & 0 & 1 & 0 & 1 \\
0 & 1 & 0 & 0 & 0 & 0 & 1 & 0 & 0 & 0 & 0 & 1 \\
0 & 0 & 1 & 0 & 0 & 0 & 0 & 0 & 0 & 1 & 0 & 1 \\
0 & 0 & 0 & 1 & 0 & 0 & 0 & 0 & 1 & 0 & 1 & 0 \\
0 & 0 & 0 & 0 & 1 & 0 & 0 & 0 & 0 & 1 & 0 & 1 \\
0 & 0 & 0 & 0 & 0 & 1 & 1 & 1 & 1 & 0 & 1 & 0
    \end{array}
    \right]$
  }
  \\
  \\
  \\
  (b)
  \raisebox{-3.75em}{
    $\left[
    \begin{array}{@{}c@{\hspace{.4em}}c@{\hspace{.4em}}c@{\hspace{.4em}}c@{\hspace{.4em}}c@{\hspace{.4em}}c@{}}
X & Z & I & Z & I & Z \\
Z & X & I & I & I & Z \\
I & I & X & Z & I & Z \\
I & I & Z & X & Z & I \\
I & I & I & Z & X & Z \\
Z & Z & Z & I & Z & X
    \end{array}
    \right]$
  }
  \hspace{3em}(d)\hspace{.5em}
  \Qcircuit @R=.5em @C=1.5em {
& \lstick{\ket{0}} & \gate{H} & \ctrl{5} & \qw & \ctrl{1} & \ctrl{3} & \qw & \qw & \qw \\
& \lstick{\ket{0}} & \gate{H} & \qw & \ctrl{4} & \control \qw & \qw & \qw & \qw & \qw \\
& \lstick{\ket{0}} & \gate{H} & \qw & \qw & \ctrl{3} & \qw & \qw & \qw & \qw \\
& \lstick{\ket{0}} & \gate{H} & \qw & \qw & \qw & \control \qw & \ctrl{1} & \qw & \qw \\
& \lstick{\ket{0}} & \gate{H} & \qw & \qw & \qw & \ctrl{1} & \control \qw & \qw & \qw \\
& \lstick{\ket{0}} & \gate{H} & \control \qw & \control \qw & \control \qw & \control \qw & \qw & \qw & \qw
  }
 \caption[The 6-qubit graph state]{A a) graph, b) stabilizer generator, c) binary stabilizer generator, and d) preparation circuit for an example 6-qubit graph state.  Nodes
 in a) are numbered sequentially starting from the upper left and moving clockwise.\label{fig:graphStateEx}}
\end{figure}

From this choice of stabilizer generators derives a particularly elegant preparation procedure for the associated state.  The quantum circuit
\begin{align}
  \prod_{j<k|j\in\set{N}(k)} \CZ_{jk} H^{\otimes n} \ket{0}^{\otimes n} \label{eq:standardGraphStateConstructionCircuit}
\end{align}
prepares the $n$-qubit graph state associated with the graph described by $\set{N}$.  The origin of this circuit is most easily understood by considering the effect that the unitaries have on $\{Z_h\}_{h=1}^n$, the stabilizer generator of $\ket{0}^{\otimes n}$.  Employing a few of the identities from Section~\ref{sec:quantumCircuitDiagrams}, each stabilizer $Z_h$ can be shown to transform to a new stabilizer
\begin{align}
  \begin{split}
    \prod_{j<k|j\in\set{N}(k)} \CZ_{jk} H^{\otimes n} &Z_h H^{\otimes n} \prod_{l<m|l\in\set{N}(m)} \CZ_{lm} \\
    &= \prod_{j<k|j\in\set{N}(k)} \CZ_{jk} X_h \prod_{l<m|l\in\set{N}(m)} \CZ_{lm} = X_h \prod_{k \in \mathcal{N}(h)} Z_k
  \end{split}
\end{align}
which corresponds exactly to a stabilizer for the desired graph state, hence, the unitary prepares the graph state.  Figure~\ref{fig:graphStateEx} provides an example of each of these representations of a graph state.

Much of the interest with regard to graph states in the literature has been focused on their suitability as a substrate for measurement based quantum computation~\cite{Raussendorf01}.  It has been shown, for a variety of classes of graphs, that single qubit measurements are sufficient to implement an arbitrary quantum computation.  For the purposes of this chapter, however, I am more interested in the fact, proven by Van den Nest~\cite{VandenNest04}, that, up to the application of local Clifford gates, the class of stabilizer states is equivalent to the class of graph states.  I include my own proof of this below.

\begin{defi}
  For any stabilizer generator $\group{S}_\set{G}=\{A^i\}$, let the term gap form refer to any column $j$ such that $A^{ij}=I$ for all $i\geq j$.
\end{defi}
\begin{defi}
  For any stabilizer generator $\group{S}_\set{G}=\{A^i\}$, let the term pivot form refer to any column $j$ such that $A^{jj}\in \{X,Y\}$ and $A^{ij}\in\{I,Z\}$ for all $i\neq j$.
\end{defi}
\begin{defi}
  For any stabilizer generator $\group{S}_\set{G}=\{A^i\}$, let the term echelon form refer to any column $j$ in either gap or pivot form.
\end{defi}
\begin{lem}
Given an $m$-element stabilizer generator $\group{S}_\set{G}=\{A^i\}$ such that the first $k-1<m$ columns are in echelon form, we can construct a stabilizer generator $\group{S}_\set{G}^{\prime\prime}$ such that the first $k$ columns of $\group{S}_\set{G}^{\prime\prime}$ have echelon form and $\group{S}$ and $\group{S}^{\prime\prime}$ are equivalent up to conjugation by local Clifford gates.\label{lem:nearGraphFormInductiveStep}
\end{lem}
\prove{Either the $k$th column of $\group{S}$ satisfies $A^{ik}=I$ for all $i\geq k$, showing that it is already in echelon form, or there exists an $i\geq k$ such that $A^{ik}\neq I$.  In the second case, by exchanging this row $i$ with row $k$ and conjugating by $H_k$ as necessary, we can transform to a stabilizer generator $\group{S}_\set{G}^{\prime}=\{B^h\}$ such that $B^{kk}\in\{X,Y\}$.  Multiplying any other row (generator) of $\group{S}^{\prime}$ by $B^k$ preserves echelon form on the first $k-1$ columns since, for all $j<k$, either $B^{kj}=I$ or $B^{kj}=Z$ and column $j$ is in pivot form, which is impervious to $Z$ Pauli operators.  Consequently, multiplying by $B^k$ each other row $B^i$ such that $B^{ik}\in\{X,Y\}$ yields a new stabilizer $\group{S}_\set{G}^{\prime\prime}$ whose first $k$ columns are in echelon form.
}
\begin{thm}
Given an $n$-qubit, $n$-element stabilizer generator $\group{S}_\set{G}=\{A^i\}$, we can construct a graph-form stabilizer generator $\group{S}_\set{G}^{\prime\prime}$ such that $\group{S}$ and $\group{S}^{\prime\prime}$ are equivalent up to conjugation by local Clifford gates.
\end{thm}
\prove{Applying Lemma~\ref{lem:nearGraphFormInductiveStep} $n$ times yields a new stabilizer in echelon form $\group{S}_\set{G}^\prime=\{B^i\}$.  Let $j$ be the index of the last gap column of $\group{S}_\set{G}^\prime$, by definition $B^{ji}=I$ for any $i$ indexing a gap column.  By assumption, row $j$ is independent, so there exists an $i$ such that $B^{ji}=Z$, but this implies that row $j$ and row $i$ anti-commute since row $i$ has an $X$ or $Y$ at position $i$ and $Z$'s every else that row $j$ might have $Z$'s.  Thus, by contradiction, we know that there are no gap columns and so every column in $\group{S}_\set{G}^\prime$ has pivot form.  Similarly, commutativity requires that the $Z$'s be symmetrically distributed since if $B^{ij}\neq B^{ji}$ then rows $i$ and $j$ anti-commute.  Consequently, $\group{S}_\set{G}^\prime$ differs from graph form only in the possible presence of $Y$'s on the diagonal or signs on generators.  Conjugating by $P$ and $Z$ appropriately remove these offending traits yielding a graph-form stabilizer $\group{S}_\set{G}^{\prime\prime}$.}

\begin{figure}
  \capstart
  \raisebox{1.75em}{
  (a)
  }
    $\left[
    \begin{array}{@{}c@{}c@{\hspace{.4em}}c@{\hspace{.4em}}c@{\hspace{.4em}}c@{}}
& Z & I & I & I \\
& Z & Z & X & Z \\
& Z & X & Z & Z \\
& Z & Y & I & X
    \end{array}
    \right]
    \xrightarrow{H_1}
    \left[
    \begin{array}{@{}c@{}c@{\hspace{.4em}}c@{\hspace{.4em}}c@{\hspace{.4em}}c@{}}
& X & I & I & I \\
& I & Z & X & Z \\
& I & X & Z & Z \\
& I & Y & I & X
    \end{array}
    \right]
    =
    \left[
    \begin{array}{@{}c@{}c@{\hspace{.4em}}c@{\hspace{.4em}}c@{\hspace{.4em}}c@{}}
& X & I & I & I \\
& I & X & Z & Z \\
& I & Z & X & Z \\
-& I & Z & Z & Y
    \end{array}
    \right]
    \xrightarrow{Z_4 P_4}
    \left[
    \begin{array}{@{}c@{}c@{\hspace{.4em}}c@{\hspace{.4em}}c@{\hspace{.4em}}c@{}}
& X & I & I & I \\
& I & X & Z & Z \\
& I & Z & X & Z \\
& I & Z & Z & X
    \end{array}
    \right]$
    \ \vspace{1em} \\
  \raisebox{1.75em}{
  (b)
  }
    $\left[
    \begin{array}{@{}c@{}c@{\hspace{.4em}}c@{\hspace{.4em}}c@{\hspace{.4em}}c@{}}
& I & I & Z & X \\
& Y & Y & I & X \\
& X & Y & X & Y \\
& I & I & X & Z
    \end{array}
    \right]
    =
    \left[
    \begin{array}{@{}c@{}c@{\hspace{.4em}}c@{\hspace{.4em}}c@{\hspace{.4em}}c@{}}
& Y & Y & I & X \\
& I & I & Z & X \\
& Z & I & X & Z \\
& I & I & X & Z
    \end{array}
    \right]
    =
    \left[
    \begin{array}{@{}c@{}c@{\hspace{.4em}}c@{\hspace{.4em}}c@{\hspace{.4em}}c@{}}
& Y & Y & I & X \\
& I & I & Z & X \\
& Z & I & X & Z \\
& Z & I & I & I
    \end{array}
    \right]$
 \caption[Illustrations of the proof of stabilizer and graph state equivalence]{Illustrations of the transformations involved in the constructive proof of the local-Clifford equivalence of stabilizer states and graph states.  The examples given are for a) a full-rank 5-qubit stabilizer and b) a non-commuting set of Pauli operators.  In each sequence, the third step and beyond are in echelon form.}
\end{figure}

\section{Compressed Graph-state Construction\label{sec:compressedGraphstateConstruction}}

Due to the phenomenon of memory errors, it is usually a good idea to design circuits such that they take as few time steps as possible.  While I neglect such errors in this work, a more compact form of the circuit for graph-state construction is also desirable for the development of a fault-tolerant version of the process.

Using the circuit from Equation~\ref{eq:standardGraphStateConstructionCircuit}, the construction time for large graph states is almost completely determined by $r$, the number of time steps spent applying $\CZ$ gates.  Consequently, compression of the circuit can be approached as a matter of rearranging $\CZ$ gates to achieve the maximum amount of parallelism, a task made possible by the fact that $\CZ$ gates commute with each other.  Since qubits cannot (I assume) participate in more than one gate at a time, $r$ is bounded below by $2e/n$ where $e$ is the number of $\CZ$ gates in the construction circuit or, equivalently, the number of edges in the corresponding graph.  Additionally, $r\leq e$ since one cannot possibly do worse than applying one gate per time step.

Finding the ideal configuration for an arbitrary selection of $\CZ$ gates is an excessively ambitious problem, but we can get an idea of the number of time steps required by bounding $r$ in the worst-case scenario.  For the problem of compressing graph-state circuits, the worst case scenario is that of the complete graph, where each node is connected to every other node.  The construction of any other graph state can be performed in equal or fewer time steps by omitting the appropriate $\CZ$ gates from the construction circuit for the graph state corresponding to the complete graph.  The number of edges in the complete graph is $\binom{n}{2}$, so we begin knowing that $n-1 \leq r \leq n(n-1)/2$.

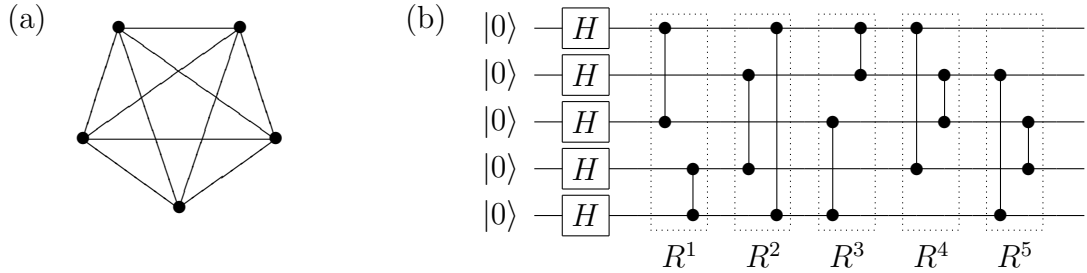
\begin{figure}
  \capstart
\begin{tabular}{llll}
(a) &
\Qcircuit[.3em] @R=.7em @C=.15em {
& \sink \link{0}{2} \link{2}{3} \link{2}{-1} \link{4}{1} & \push{\rule{1.9em}{0em}\rule[-3.4em]{0em}{0em}}& \sink \link{2}{1} \link{4}{-1} \link{2}{-3} \\
\\
\sink \link{2}{2} \link{0}{4} & & & & \sink \link{2}{-2} \\
\push{\rule[-.5em]{0em}{0em}}\\
& \push{\rule{1.7em}{0em}} & \sink & \push{\rule{1.7em}{0em}}
}
\hspace{3em} &
(b) &
\Qcircuit @R=.2em @C=.9em {
& & \lstick{\ket{0}} & \gate{H} & \qw & \ctrl{2} & \qw & \qw & \qw & \control \qw & \qw & \qw & \ctrl{1} & \qw & \control \qw & \qw & \qw & \qw & \qw & \qw & \qw \\
& & \lstick{\ket{0}} & \gate{H} & \qw & \qw & \qw & \qw & \ctrl{2} & \qw & \qw & \qw & \control \qw & \qw & \qw & \ctrl{1} & \qw & \control \qw & \qw & \qw & \qw \\
& & \lstick{\ket{0}} & \gate{H} & \qw & \control \qw & \qw & \qw & \qw & \qw & \qw & \ctrl{2} & \qw & \qw & \qw & \control \qw & \qw & \qw & \ctrl{1} & \qw & \qw \\
& & \lstick{\ket{0}} & \gate{H} & \qw & \qw & \ctrl{1} & \qw & \control \qw & \qw & \qw & \qw & \qw & \qw & \ctrl{-3} & \qw & \qw & \qw & \control \qw & \qw & \qw \\
& & \lstick{\ket{0}} & \gate{H} & \qw & \qw & \control \qw & \qw & \qw & \ctrl{-4} & \qw & \control \qw & \qw & \qw & \qw & \qw & \qw & \ctrl{-3} & \qw & \qw & \qw \\
\push{\rule{0em}{.8em}} & & & & & & {\!\!\!\!\!R^1} & & & {\!\!\!\!\!R^2} & & & {\!\!\!\!\!R^3} & & & {\!\!\!\!\!R^4} & & & {\!\!\!\!\!R^5}
\gategroup{1}{6}{5}{7}{.9em}{..}
\gategroup{1}{9}{5}{10}{.9em}{..}
\gategroup{1}{12}{5}{13}{.9em}{..}
\gategroup{1}{15}{5}{16}{.9em}{..}
\gategroup{1}{18}{5}{19}{.9em}{..}
}
\end{tabular}
\caption[An example of compact graph state construction]{a) The complete graph on $5$ qubits and b) a compact construction circuit of the kind described in the text for the associated state.\label{fig:compactGraphStateConstructionExample}}
\end{figure}

In order to construct the graph state corresponding to the complete graph, a $\CZ$ gate must be applied between every pair of qubits.  If we think of the qubits as being ordered on a ring, this can be accomplished utilizing any pattern of simultaneously applicable $\CZ$ gates such that for any distance $d\leq\lfloor n/2\rfloor$ (distances $d$ and $n-d$ are the same for a ring) exactly one $\CZ$ connects a pair of qubits separated by distance $d$.  Rotating the pattern through one cycle produces all possible pairings exactly once.  For $n=2a+1$, a workable gate pattern is
\begin{align}
  \prod_{i=1}^{\lfloor(a+1)/2\rfloor} \CZ_{i,a+2-i} \prod_{j=1}^{\lfloor(a-1)/2\rfloor} \CZ_{a+1+j,2a+2-j}.
\end{align}
For $n=2a+2$ this pattern does not quite work since it excludes $\CZ$ gates connecting qubits at distance $a+1$ from one another.  In order to establish these connections, the gate pattern
\begin{align}
  \prod_{i=1}^{a+1} \CZ_{i,a+1+i}
\end{align}
must be applied using one additional round.  A compact construction circuit for the complete graph on $5$ qubits is given in Figure~\ref{fig:compactGraphStateConstructionExample}.

The circuits just described provide a method of constructing any graph state such that all $\CZ$ gates are applied in a number of rounds $r\leq 2\lfloor n/2\rfloor+1$.  For the complete graph, these circuits are nearly optimal, and, for typical graphs, which have half as many connections, they are roughly within a factor of two of the lower bound on the number of required time steps.  For certain interesting classes of graph states, however, the constructive procedure given requires a number of time steps far in excess of the lower bound.  One such case is that of graph states with small, fixed generator weight, the graph state equivalent of low-density-parity-check codes~\cite{MacKay95}.

For a graph state with generators of weight $w$, the total number of edges in the corresponding graph is $e=(w-1)n/2$, yielding a lower bound on the number of required rounds of $r\geq w-1$.  For $w\ll n$, a significant gap exists between this bound and the number of time steps required by my explicit circuit.  The number of rounds required is actually at most $2w-3$, as can be seen from the following argument.

Imagine designing a circuit for constructing a graph state with weight $w$ generators by sequentially inserting the necessary $\CZ$ gates into one of $2(w-2)+1$ rounds of $\CZ$ application.  It might be impossible to insert a gate of the required set $\CZ_{ij}$ into any particular round due to qubit $i$ and/or $j$ participating in some other (previously inserted) $\CZ$ gate.  Qubits $i$ and $j$ participate in $w-1$ $\CZ$ gates apiece, however, so they may be previously engaged for at most $2(w-2)$ rounds.  Thus, it is always possible to insert the additional $\CZ$ gate into one of the $2(w-2)+1$ rounds, showing that $r\leq 2(w-2)+1=2w-3$.

\section{Fault-tolerant graph state construction}

The goal of this chapter is to develop a method of constructing large ancillae that does not depend on concatenated coding and does not require discarding large numbers of attempts.  This is a significant impediment to ancilla construction since, as discussed in Section~\ref{subsec:ancillaPreparation}, the primary advantages of constructing ancillae over applying operations directly to the data are low starting error rates, exact knowledge of what the state ought to be, and the ability discard failed construction attempts.  My decision to design routines for graph states was based on the hope that their simple structure would allow something to be made of the second feature.

\subsection{Tracking Errors}
\begin{figure}
  \capstart
  \centerline{
\Qcircuit @R=.4em @C=.4em {
\push{\rule{0em}{1.5em}\rule{1.5em}{0em}} & \lstick{\text{\small$\ket{\bar{0}}$}} & \gate{H} & \ctrl{1} & \ctrl{2} & \ctrl{3} & \ctrl{4} & \ctrl{5} & \ctrl{6} & \gate{R^1} & \ctrl{1} & \gate{R^2} & \ctrl{2} & \gate{R^3} & \ctrl{3} & \gate{R^4} & \ctrl{4} & \gate{R^5} & \ctrl{5} & \ctrl{6} & \qw & \qw \\
& \lstick{\text{\small$\ket{\bar{0}}$}} & \qw & \targ & \qw & \qw & \qw & \qw & \qw & \qw & \targ & \qw & \qw & \qw & \qw & \qw & \qw & \qw & \qw & \qw & \measureD{\rule{0em}{.35em}} \\
& \lstick{\text{\small$\ket{\bar{0}}$}} & \qw & \qw & \targ & \qw & \qw & \qw & \qw & \qw & \qw & \qw & \targ & \qw & \qw & \qw & \qw & \qw & \qw & \qw & \measureD{\rule{0em}{.35em}} \\
& \lstick{\text{\small$\ket{\bar{0}}$}} & \qw & \qw & \qw & \targ & \qw & \qw & \qw & \qw & \qw & \qw & \qw & \qw & \targ & \qw & \qw & \qw & \qw & \qw & \measureD{\rule{0em}{.35em}} \\
& \lstick{\text{\small$\ket{\bar{0}}$}} & \qw & \qw & \qw & \qw & \targ & \qw & \qw & \qw & \qw & \qw & \qw & \qw & \qw & \qw & \targ & \qw & \qw & \qw & \measureD{\rule{0em}{.35em}} \\
& \lstick{\text{\small$\ket{\bar{0}}$}} & \qw & \qw & \qw & \qw & \qw & \targ & \qw & \qw & \qw & \qw & \qw & \qw & \qw & \qw & \qw & \qw & \targ & \qw & \measureD{\rule{0em}{.35em}} \\
\push{\rule{0em}{1.3em}}& \lstick{\text{\small$\ket{\bar{0}}$}} & \qw & \qw & \qw & \qw & \qw & \qw & \targ & \qw & \qw & \qw & \qw & \qw & \qw & \qw & \qw & \qw & \qw & \targ & \measureD{\rule{0em}{.35em}} \\
\push{\rule{0em}{1.2em}} & & & & {\textrm{CATS}}
\gategroup{1}{1}{7}{9}{.2em}{.}
}
  }
\caption[Error tracking graph-state preparation example]{An example of the collective circuit for graph state preparation with error tracking.  The circuit shown prepares a graph state on $5$-qubits.  In order to track errors properly, the portion labeled \textrm{CATS} must be prepared separately in a way such that correlated errors are rare.\label{fig:errorTrackingGraphStatePreparationExample}}
\end{figure}
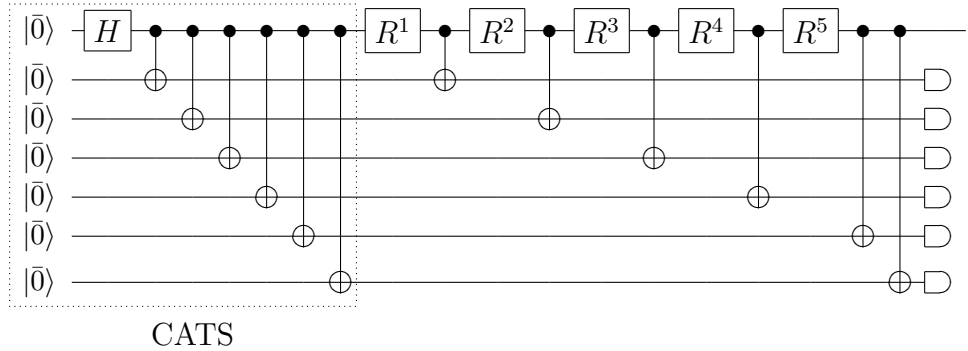

Consider the compact method of $n$-qubit graph-state construction described in the previous section.  Let $R^i$ label the pattern of $\CZ$ gates applied in round $i$ of the $r$ rounds of $\CZ$ gate application.  The circuit for graph state construction can then be written as
\begin{align}
  \prod_{i=r}^1 R^i H^{\otimes n} \ket{0}^{\otimes n} = \prod_{i=r}^1 R^i \bar{H} \ket{\bar{0}}, \label{eq:graphStatePreparation}
\end{align}
where I am abusing my notation somewhat by allowing the bar overtop to indicate states and gates for a collection of unencoded qubits, or, if you prefer, a very boring $[[n,n,1]]$ code.  Since $\CZ$ gates commute with the control of any controlled-operation, this circuit is equivalent to
\begin{align}
  \overline{\CX}_{1,r+2} \prod_{i=r}^1 \overline{\CX}_{1,i+1} R^i \prod_{j=r+2}^{2} \overline{\CX}_{1,j} \bar{H}_{1} \ket{\bar{0}}^{\otimes (r+2)},\label{eq:errorTrackingGraphStatePreparation}
\end{align}
where I have supplemented the primary ancilla qubits needed for graph-state construction by a host of secondary ancilla qubits.
An example of this circuit for $n=5$ is given in Figure~\ref{fig:errorTrackingGraphStatePreparationExample}.

The purpose behind adding all of those gates is that while Equation~(\ref{eq:errorTrackingGraphStatePreparation}) and Equation~(\ref{eq:graphStatePreparation}) produce the same output in the absence of errors, their action in the presence of errors is very different.  The extra gates in Equation~(\ref{eq:errorTrackingGraphStatePreparation}) copy out $X$ errors at various points in time to the auxiliary ancilla qubits.  If the initial portion of the circuit, corresponding to the production of $n$ ($r+2$)-qubit cat states, is implemented without error, then the latter portion serves to track the location of $X$ errors on the primary ancilla qubits during the construction.  Both the spatial and temporal localization of $X$ errors is important since $\CZ$ gates convert $X$ errors into $X\otimes Z$ errors. Contrariwise, $Z$ errors are not tested for at all, reflecting the fact that $Z$ errors do not spread.

There is no hope of making this circuit fault-tolerant in the strict sense defined in Section~\ref{sec:faultTolerance} since a single failure on any $\CZ$ gate can generate a pair of $Z$ errors on the output state.  At best, I can achieve a weaker form of fault-tolerance where rather than preventing the spread of errors I limit it to some fixed amount, optimally, to one additional location.  To achieve this, I need two things: cat states without correlated errors and a method of predicting the location of $X$ and $Z$ errors on the primary ancilla qubits given the tracking information obtained by measuring the secondary ancilla qubits.  I do not discuss a means of preparing suitable cat states since I have not yet managed to design one.  In what follows, I introduce a method of interpreting the measurement record and discuss its properties.

\subsection{Interpreting Error Tracks}

\begin{figure}
  \capstart
  \begin{tabular}{cccc}
    \raisebox{11em}{(a)} &
    \pgfimage[height=4.5cm]{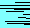} \hspace{3em} &
    \raisebox{11em}{(b)} &
    \pgfimage[height=4.5cm]{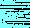}
  \end{tabular}
  \caption[Clean and noisy error track examples]{Examples of a) a perfect error track and b) the sort of error track that is actually observed.  The examples are for $30$ primary qubits where the qubits are arranged vertically and time runs from left to right.  Black indicates an $X$ error.  Noisy tracks can be obtained by layering from left to right the results of measuring the secondary ancilla qubits coupled to after each round of graph-state construction.\label{fig:cleanAndNoisyErrorTracks}}
\end{figure}

A typical $X$ error during graph-state construction presents many opportunities for its detection.  $X$ errors persist unless canceled by other $X$ errors, so a clairvoyant view of $X$ errors on the primary ancilla qubits at a sequence of times would tend to reveal a collection of streaks, like those illustrated in Figure~\ref{fig:cleanAndNoisyErrorTracks}a.  A noisy version of this insight, as in Figure~\ref{fig:cleanAndNoisyErrorTracks}b can be obtained by measuring the tracking qubits and ordering the results by the primary qubit coupled to and the time of coupling.

Observed error tracks include noise deriving from errors in both cat-state preparation and the measurement process which should be ignored when determining the location of errors on the primary ancilla qubits.  These inconsequential errors are generally recognizable in the measurement record by their failure to persist from one measurement to another, lending the noise in error tracks an appearance reminiscent of static on an analog television.  To accurately infer the locations of errors on the constructed graph-state, this static should be filtered from the error tracks.
Below I develop filtering rules based on the order of various events, that is, the number of errors required to make them happen.  Whenever two possible events have different orders I assume that the one requiring fewer errors actually occurred.

\begin{figure}
  \capstart
    \begin{tabular}{@{}cccccc}
      (a) & \pgfimage[height=.25cm]{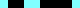} & $\rightarrow$ & \pgfimage[height=.25cm]{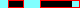} & $\rightarrow$ & \pgfimage[height=.25cm]{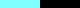} \\
      (b) & \pgfimage[height=.25cm]{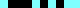} & $\rightarrow$ & \pgfimage[height=.25cm]{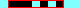} & $\rightarrow$ & \pgfimage[height=.25cm]{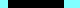} \\
      (c) & \pgfimage[height=.25cm]{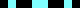} & $\rightarrow$ & \pgfimage[height=.25cm]{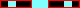} & $\rightarrow$ & \pgfimage[height=.25cm]{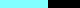}
    \end{tabular}
  \caption[Error track noise filtering examples]{Three examples of my standard noise filtering routine for error tracks.  Each shows the error track initially obtained, the segments considered, and the post-filtration track.  The filter breaks the track corresponding to each primary qubit into segments demarcated by sequential pairs of measurements indicating no error.  The segments are then judged error-free if $l-w+a>w$ and entirely in-error otherwise; $l$, $w$, and $a$ are the length, weight, and number of endpoints (discounting the end of the track) of the segment in question.\label{fig:standardFilteringExample}}
\end{figure}

In brief, the basic algorithm I employ to filter the noise from a set of error tracks is as follows.  The error track for each primary qubit is filtered separately. The track is divided up into segments separated by two or more measurements which indicate that no error occurred.  Each segment is then taken to be entirely in error if $l-w+a<w$ where $l$ and $w$ are the length and Hamming weight of the segment and $a$ is $1$ if the segment includes the end of the error track and $2$ otherwise.  This algorithm, which I refer to henceforth as the liberal filter, was used to filter the error track in Figure~\ref{fig:cleanAndNoisyErrorTracks}b to obtain Figure~\ref{fig:cleanAndNoisyErrorTracks}a.  Figure~\ref{fig:standardFilteringExample} illustrates its application for some informative examples.

To explain the origin of the liberal filtering routine, it is easiest to adopt a simplified terminology.  In what follows, ``track'' always refers to the error track (corresponding to a single primary qubit) being filtered, ``black'' and ``white'' describe the state of the primary qubit or the outcome of a measurement with black corresponding to an $X$ error and white to no $X$ error, and the term ``interim'' is used to denote errors that happen between measurements.

Schematically, the filtering problem is this: An initially white qubit undergoes a sequence of $r$ color measurements.  An error during a measurement can cause the wrong result to be reported and/or invert the color, but an error during the interim between measurements can only change the color.  There is a minimum number of errors that must occur for the observed measurement results to be possible.  Our task is to find any sequence of colors that can be generated and made consistent with the measurement results using the minimal number of errors.

At its core, my filtering algorithm relies on an observation about the reliability of sequential pairs of measurements.
A second order event is necessary for two agreeing sequential measurements to be wrong, but at most two interim errors are required to produce the same measurement results without any errors in measurement having occurred.  In the worst case, the pair of measurements disagree with the known color on either side, but this can result from two interim errors, namely, inverting the color before and after the measurements.  Consequently, it is always reasonable to assume that a sequence of agreeing measurement results are accurate.  My filtering algorithm takes advantage of this fact by partitioning each track into segments where the boundaries of segments are delineated by pairs of white measurements.

Partitioning the track simplifies the problem of filtering in two ways.  First, we may assume that the incoming color is white and, unless the segment includes the end of the track, likewise for the outgoing color.  Second, the minimal-error scenarios consistent with the observed measurements include one of the two following cases: the entire segment is black and all the white measurements were wrong or the entire segment is white and all the black measurements were wrong.  To see why this is so, imagine that some portion of the segment is black.  The nearest white measurement (in the segment) may be interpreted either as signaling that an error has occurred bringing the black region to end or, since every white measurement is bracketed by a pair of black measurements, that the white measurement was wrong.  Thus, any scenario involving a portion of the segment being black requires at least as many errors as the scenario in which the entire segment is black.  The only other scenario that must be considered, then, is the one in which the entire segment is white.

It is straightforward to compare the number of errors required for an entire segment to be black versus white.  If the entire segment is white, then all of the black measurements must have been wrong, so the number of errors is $w$ where, confusingly, $w$ is the weight of the segment or the number of black errors.  If the entire segment is black then all $l-w$ white measurements must have been wrong where $l$ is the length of the segment.  Additionally, if the entire segment is black then $1$ error must have occurred to change the color to black at the beginning of the segment and, unless the segment contains the end of the track, $1$ error must have occurred to change the color back to white at the end of the segment.  Consequently, I assume the entire segment is black when $l-w+a<w$, where $a$ is $1$ if the segment includes the end of the track and $2$ otherwise.

After having filtered the tracking data, I translate the filtered error tracks into an expected error distribution on the prepared graph state using a simple set of rules.  First, streaks that include the final round are taken to indicate the presence of an $X$ error on the final state.  Second, a $\CZ$ gate is assumed to have spread a $Z$ error if it was applied to a qubit at a time spanned by one of the qubit's streaks.  Third $\CZ$ gates applied just before a streak are assumed to have spread $Z$ errors, but $\CZ$ gates applied just after one are assumed not to have.  Since $Z$ errors do not themselves spread, these rules suffice to predict the final error distribution.

\subsection{Error spread\label{subsec:errorSpread}}

The liberal filtering algorithm was designed to ensure that one of the most probable error scenarios consistent with the observed measurements was adopted.  This is no guarantee, however, that the presumed error actually occurred.  To understand the fault-tolerance properties of the liberal filter, it is necessary to determine how different the error distribution of the output state might be for other error scenarios capable of generating the observed measurement results.

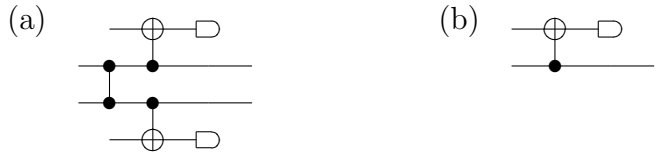
\begin{figure}
  \capstart
\begin{tabular}{ll@{\hspace{6em}}ll}
(a) &
  \Qcircuit @R=.5em @C=1em @!R {
& & \targ & \measureD{} \\
& \ctrl{1} & \ctrl{-1} & \qw & \qw \\
& \control \qw & \ctrl{1} & \qw & \qw \\
& & \targ & \measureD{}
  }
& (b) &
  \Qcircuit @R=.5em @C=1em @!R {
& \targ & \measureD{} \\
& \ctrl{-1} & \qw & \qw
  }
\end{tabular}
\caption[Circuit fragments for analyzing the fault-tolerance of my graph-state construction procedure when no track filter is used]{Circuit fragments for analyzing the fault-tolerance of my graph-state construction procedure when no track filter is used.  Modulo errors, each measurement yields the value $1$ since the qubits coupled by $\CX$ gates are initially prepared in cat states.  The fragment b) applies only to the last round of error tracking where the measurement is used to determine whether an $X$ error is on the primary qubit.  All other rounds are described by a) and use the measurements to determine whether $Z$ errors were spread in either direction. \label{fig:foolConstructionCircuitFragments}}
\end{figure}

I begin my analysis of error spread by considering the result of interpreting error tracks using no filter at all, a case I refer to henceforth as the fool's filter.  The tremendous advantage of this filter is that it is very easy to analyze its fault-tolerance properties.  Each measurement outcome is utilized individually and applied to the locality where it was obtained, so it is only necessary to consider small pieces of the construction circuit.  In fact, it is sufficient to analyze the two circuits in Figure~\ref{fig:foolConstructionCircuitFragments}.

The circuit fragment in Figure~\ref{fig:foolConstructionCircuitFragments}a depicts the operations applied in every round of error tracking but the last, which is shown in Figure~\ref{fig:foolConstructionCircuitFragments}b.  For each fragment, in the absence of errors, measurements yield the outcome $1$, while a single $X$ error on a primary ancilla qubit propagates to a measurement and is detected as an outcome of $-1$.  Whenever a measurement outcome of $-1$ is obtained in the circuit in Figure~\ref{fig:foolConstructionCircuitFragments}a a $Z$ error is assumed to have been spread to the opposing primary qubit.  A $Z$ error on either of the secondary ancilla qubits is spread to one of the primary ancilla qubits.  An $X$ error on either of the secondary ancilla qubits results in a mistaken inference (an effective $Z$ error) regarding the opposite primary qubit.

Independent of whether there was initially an $X$ error on either of the primary qubits, the circuit in Figure~\ref{fig:foolConstructionCircuitFragments}a correctly determines which primary locations $Z$ errors were spread to so long as neither the secondary ancilla qubits nor the gates introduced additional errors.  When one or more other errors occurs, the errors, together with diagnosis, may sometimes result in a $Z$ error on each primary qubit in the circuit fragment.  $X$ errors are irrelevant since they do not impede the remainder of the construction circuit and are tested for separately in the final round.  Thus, the maximum scale-up per error for this circuit fragment is $2$.  The circuit in Figure~\ref{fig:foolConstructionCircuitFragments}b detects an $X$ error on the primary ancilla qubit so long as neither the secondary ancilla qubit nor the gates introduced additional errors.  When one or more other errors occurs, a $Y$ error may result on the primary qubit.  Counting $X$ and $Z$ errors individually, the maximum scale-up per error is again 2.

If the error spread of the liberal filtering routine is to be greater than $2$, it must be for sequences of measurements where it acts differently than the fool's.  Thus, we need only concern ourselves with segments delineated by pairs of measurements indicating no error, excluding those segments of length greater than $2$ where all measurements are in agreement.  Recall that the restriction, using the liberal filter, for deciding whether a segment is in error is $(l-w)+2<w$.  Suppose that the liberal filter completely misidentifies a collection of measurement errors as representing an error of length $l$ on the primary qubits.  At most this may result in $l+1$ $Z$ errors ($X$ errors are irrelevant for segments that do not include the end of a track), but, in order for the algorithm to have reached this conclusion, it must be the case that $w>(l+2)/2$, implying that $2w>l+1$, i.e., twice the number of errors is greater than the number of errors generated.  Now suppose that the liberal filter completely misidentifies an error of length $l$ and a collection of measurement errors as simply the result of measurement errors.  At most, this may result in $l+2$ $Z$ errors, but, in order for the algorithm to have reached this conclusion, it must be the case that $l-w+2\geq w$, implying that $2(l-w+2)=2(l+2)-2w\geq 2(l+2) - (l+2)=l+2$, i.e., twice the number of errors is greater than or equal to the number of errors generated.  For the case of a segment including the end of the track, the relevant inequality, number of errors generated on the primary ancilla qubits, and number of errors that occurred are $(l-w)+1<w$, $l+1$, and $w$ for the first case and $(l-w)+1\geq w$, $l+1$, and $l-w+1$ for the second.  The results are the same.

In conclusion, neither filtering algorithm yields a fault-tolerant construction routine since they both permit the generation of correlated errors.  The preceding paragraphs show, however, that they do limit the spread of errors to $2$ per original failure.

\section{Numerical Investigations}

\subsection{Filtering with the Viterbi Algorithm}

The liberal filter interprets the measurement results from a single track as indicating an error of the lowest order possible, but, among errors of that order, the particular one chosen is rather arbitrary.  It is natural to wonder how this filter compares to an idealized filter that always yields the most probable error scenario consistent with the observed measurements for the entire set of error tracks.  Unfortunately, implementing an ideal filter as a brute-force maximum-likelihood decoder is impractical because the number of possible error scenarios grows exponentially in both the number of qubits and the number of rounds.  It is possible, however, to efficiently find the most probable sequence of $X$ error states corresponding to the measurements from a single track.  The exponential scaling in the number of rounds can be avoided by using the Viterbi algorithm to determine the most probable sequence of states.  The Viterbi algorithm, which is explained in detail in Appendix~\ref{chap:viterbiAlgorithm}, is a method for efficiently finding the maximum likelihood path for problems on directed graphs.  In this section I apply it to the problem of filtering error tracks.

Adapting a problem to the Viterbi algorithm is entirely a matter of writing down the appropriate graph.  The graph must be directed and acyclic, the transition and starting probabilities must be known, and, practically speaking, the number of simultaneously relevant states must be manageable.  Graphs describing time-ordered sequences are easily made to satisfy the first criterion by requiring distinct nodes for each state at each time.  The states that we care about for the filtering problem are the presence or absence, at each time, of an $X$ error on the primary qubit corresponding to the track being filtered.  The transition probabilities from one state to another follow very simply from the error propagation rules for $X$ and the probabilities of various gate errors, but only if we additionally include the $X$-error state of each qubit that might transmit an $X$ error to the qubit of interest.  It is possible to do this without violating the injunction against unmanageable numbers of states since the $X$ state of at most one additional qubit is relevant to the evolution of the primary qubit at any particular time and each additional qubit interacts only with the primary qubit.

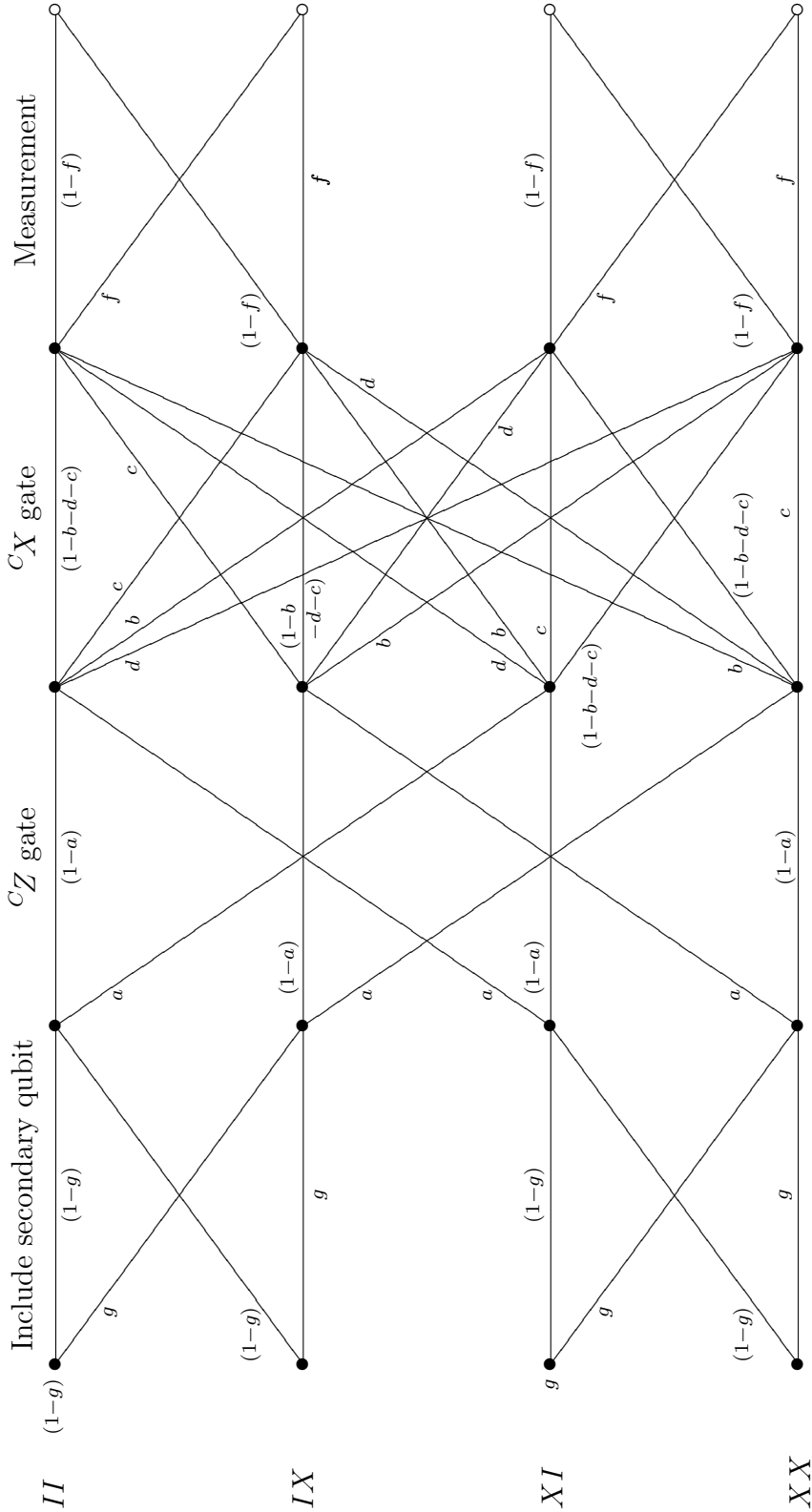
\begin{sidewaysfigure}
\renewcommand{\baselinestretch}{1}\selectfont
  \begin{pgfpicture}{0cm}{0cm}{21cm}{11.3cm}
    \pgfputat{\pgfxy(1.8,.15)}{\pgfbox[left,bottom]{
        \Qcircuit[.35em] @C=11em @R=8em {
\sink \link{0}{1} \link{1}{1} & \sink \link{0}{1} \link{2}{1} & \sink \link{0}{1} \link{1}{1} \link{2}{1} \link{3}{1} & \sink \link{0}{1} \link{1}{1} & \source \\
\sink \link{0}{1} \link{-1}{1} & \sink \link{0}{1} \link{2}{1} & \sink \link{0}{1} \link{1}{1} \link{2}{1} \link{-1}{1} & \sink \link{0}{1} \link{-1}{1} & \source \\
\sink \link{0}{1} \link{1}{1} & \sink \link{0}{1} \link{-2}{1} & \sink \link{0}{1} \link{1}{1} \link{-2}{1} \link{-1}{1} & \sink \link{0}{1} \link{1}{1} & \source \\
\sink \link{0}{1} \link{-1}{1} & \sink \link{0}{1} \link{-2}{1} & \sink \link{0}{1} \link{-3}{1} \link{-2}{1} \link{-1}{1} & \sink \link{0}{1} \link{-1}{1} & \source
        }
    }}
    \pgfputat{\pgfxy(0,10.6)}{\pgfbox[left,center]{$II$}}
    \pgfputat{\pgfxy(0,7.125)}{\pgfbox[left,center]{$IX$}}
    \pgfputat{\pgfxy(0,3.65)}{\pgfbox[left,center]{$XI$}}
    \pgfputat{\pgfxy(0,.2)}{\pgfbox[left,center]{$XX$}}
    \pgfputat{\pgfxy(1.8,10.6)}{\pgfbox[right,center]{$\scriptstyle (1-g)$}}
    \pgfputat{\pgfxy(1.8,3.65)}{\pgfbox[right,center]{$\scriptstyle g$}}
    \pgfputat{\pgfxy(4.35,11)}{\pgfbox[center,center]{Include secondary qubit}}
    \pgfputat{\pgfxy(9.03,11)}{\pgfbox[center,center]{$\CZ$ gate}}
    \pgfputat{\pgfxy(13.72,11)}{\pgfbox[center,center]{$\CX$ gate}}
    \pgfputat{\pgfxy(18.4,11)}{\pgfbox[center,center]{Measurement}}
    \pgfputat{\pgfxy(4.35,10.35)}{\pgfbox[center,center]{$\scriptstyle (1-g)$}}
    \pgfputat{\pgfxy(4.35,6.875)}{\pgfbox[center,center]{$\scriptstyle g$}}
    \pgfputat{\pgfxy(4.35,3.9)}{\pgfbox[center,center]{$\scriptstyle (1-g)$}}
    \pgfputat{\pgfxy(4.35,.4)}{\pgfbox[center,center]{$\scriptstyle g$}}
    \pgfputat{\pgfxy(2.8,9.8)}{\pgfbox[right,center]{$\scriptstyle g$}}
    \pgfputat{\pgfxy(2.8,7.85)}{\pgfbox[right,center]{$\scriptstyle (1-g)$}}
    \pgfputat{\pgfxy(2.8,2.9)}{\pgfbox[right,center]{$\scriptstyle g$}}
    \pgfputat{\pgfxy(2.8,.99)}{\pgfbox[right,center]{$\scriptstyle (1-g)$}}
    \pgfputat{\pgfxy(9.03,10.35)}{\pgfbox[center,center]{$\scriptstyle (1-a)$}}
    \pgfputat{\pgfxy(9.03,.4)}{\pgfbox[center,center]{$\scriptstyle (1-a)$}}
    \pgfputat{\pgfxy(7.5,7.32)}{\pgfbox[center,center]{$\scriptstyle (1-a)$}}
    \pgfputat{\pgfxy(7.5,3.9)}{\pgfbox[center,center]{$\scriptstyle (1-a)$}}
    \pgfputat{\pgfxy(7.2,9.7)}{\pgfbox[right,center]{$\scriptstyle a$}}
    \pgfputat{\pgfxy(7.2,6.225)}{\pgfbox[right,center]{$\scriptstyle a$}}
    \pgfputat{\pgfxy(7.2,4.55)}{\pgfbox[right,center]{$\scriptstyle a$}}
    \pgfputat{\pgfxy(7.2,1.1)}{\pgfbox[right,center]{$\scriptstyle a$}}
    \pgfputat{\pgfxy(13.72,10.35)}{\pgfbox[center,center]{$\scriptstyle (1-b-d-c)$}}
    \pgfputat{\pgfxy(11.9,7.3)}{\pgfbox[left,center]{$\scriptstyle (1-b$}}
    \pgfputat{\pgfxy(12.9,6.95)}{\pgfbox[right,center]{$\scriptstyle -d-c)$}}
    \pgfputat{\pgfxy(12,3.1)}{\pgfbox[right,center]{$\scriptstyle (1-b-d-c)$}}
    \pgfputat{\pgfxy(12.6,.99)}{\pgfbox[left,center]{$\scriptstyle (1-b-d-c)$}}
    \pgfputat{\pgfxy(12.37,9.5)}{\pgfbox[right,center]{$\scriptstyle b$}}
    \pgfputat{\pgfxy(12.05,6)}{\pgfbox[right,center]{$\scriptstyle b$}}
    \pgfputat{\pgfxy(12.2,4.4)}{\pgfbox[right,center]{$\scriptstyle b$}}
    \pgfputat{\pgfxy(11.7,1.1)}{\pgfbox[right,center]{$\scriptstyle b$}}
    \pgfputat{\pgfxy(12.7,9.7)}{\pgfbox[left,center]{$\scriptstyle c$}}
    \pgfputat{\pgfxy(14.44,9.5)}{\pgfbox[right,center]{$\scriptstyle c$}}
    \pgfputat{\pgfxy(12.1,3.8)}{\pgfbox[left,center]{$\scriptstyle c$}}
    \pgfputat{\pgfxy(13.72,.4)}{\pgfbox[left,center]{$\scriptstyle c$}}
    \pgfputat{\pgfxy(11.75,9.5)}{\pgfbox[right,center]{$\scriptstyle d$}}
    \pgfputat{\pgfxy(15.5,6.225)}{\pgfbox[left,center]{$\scriptstyle d$}}
    \pgfputat{\pgfxy(11.75,4.4)}{\pgfbox[right,center]{$\scriptstyle d$}}
    \pgfputat{\pgfxy(15,4.3)}{\pgfbox[right,center]{$\scriptstyle d$}}
    \pgfputat{\pgfxy(18.4,6.875)}{\pgfbox[center,center]{$\scriptstyle f$}}
    \pgfputat{\pgfxy(18.4,10.35)}{\pgfbox[center,center]{$\scriptstyle (1-f)$}}
    \pgfputat{\pgfxy(18.4,6.875)}{\pgfbox[center,center]{$\scriptstyle f$}}
    \pgfputat{\pgfxy(18.4,3.9)}{\pgfbox[center,center]{$\scriptstyle (1-f)$}}
    \pgfputat{\pgfxy(18.4,.4)}{\pgfbox[center,center]{$\scriptstyle f$}}
    \pgfputat{\pgfxy(16.85,9.8)}{\pgfbox[right,center]{$\scriptstyle f$}}
    \pgfputat{\pgfxy(16.85,7.85)}{\pgfbox[right,center]{$\scriptstyle (1-f)$}}
    \pgfputat{\pgfxy(16.85,2.9)}{\pgfbox[right,center]{$\scriptstyle f$}}
    \pgfputat{\pgfxy(16.85,.99)}{\pgfbox[right,center]{$\scriptstyle (1-f)$}}
  \end{pgfpicture}
\caption[State graph for tracking graph-state preparation]{The state graph for one primary qubit and the secondary qubit it interacts with during a single round of graph-state preparation.  Error states are labeled at the far left with the order being primary, secondary.  The labels at top indicate the gate corresponding to each set of transitions; time runs to the right.  Selection on the rightmost nodes ensures agreement with the observed measurement.  Transitions are labeled by the transition probability, and the starting probabilities for the first round in the preparation are given at left.  Subsequent rounds are appended at right and obey the same graph except that they inherit starting probabilities from the previous round. \label{fig:viterbiTrackingGraph}}
\end{sidewaysfigure}

A segment of the resulting directed (time runs to the right) graph is shown in Figure~\ref{fig:viterbiTrackingGraph}.  The graph in Figure~\ref{fig:viterbiTrackingGraph} depicts the allowed state transitions for any but the final round of tracking graph-state preparation; in the final round, the $\CZ$ transitions are omitted.  The graph for multiple rounds can be constructed by appending graph segments to each other, merging the hollow nodes on the right of each segment with the leftmost solid nodes of the next.  The secondary ancilla qubit referred to by the second state label changes from one segment to the another, but the transitions labeled ``Include secondary qubit'' erase any record of the previous secondary qubit.

Starting probabilities for the first round of graph-state construction are given at the left in Figure~\ref{fig:viterbiTrackingGraph}.  At the end of each round, paths terminating on a measurement outcome different from that which was observed are deleted, and the remaining unnormalized probabilities are fed into the subsequent round.  The transition probabilities $h$, $a$, $b$, $c$, $d$, and $f$ denote the probability of generating an $X$ error during cat state preparation, on the relevant end of the $\CZ$ gate, on the control end of the $\CX$ gate, on the target end of the $\CX$ gate, on both ends of the $\CX$ gate, and during measurement.  Thus,
\begin{align}
  \begin{split}
    h =& \po{PX}+\po{PY} \\
    a =& \qt{X}{X}+\qt{Y}{Y}+\frac{1}{2}(\qt{X}{I}+\qt{I}{X}+\qt{Y}{I}+\qt{I}{Y}+\qt{X}{Z}+\qt{Z}{X}+\qt{Y}{Z}+\qt{Z}{Y})\\
    &+\qt{X}{Y}+\qt{Y}{X} \\
    b =& \pt{X}{I}+\pt{Y}{I}+\pt{X}{Z}+\pt{Y}{Z} \\
    c =& \pt{I}{X}+\pt{I}{Y}+\pt{Z}{X}+\pt{Z}{Y} \\
    d =& \pt{X}{X}+\pt{X}{Y}+\pt{Y}{X}+\pt{Y}{Y} \\
    f =& \po{M}
  \end{split}
\end{align}
where, as in Chapter~\ref{chap:thresholdsForHomogeneousAncillae}, $\pt{P}{\Gamma}$, $\pt{\Lambda}{\Xi}$, $\qt{\Lambda}{\Xi}$, and $\po{M}$ denote the probabilities of preparation, $\CX$ gate, $\CZ$ gate, and measurement errors of the kinds indicated by $\Gamma$ and $\Lambda\Xi$ where $\Gamma$ ranges over the single-qubit Pauli errors and $\Lambda\Xi$ ranges over the two-qubit Pauli errors.

The output of running the Viterbi algorithm on the graph just described is the most probable sequence of $X$ error states for the primary qubit given the observed data.  The locations of $Z$ errors on the prepared graph state are then inferred using error propagation and, for indicted $\CZ$ gates, by assuming the most probable failure mode consistent with an $X$ error being generated.

\subsection{Code and Results}

To check my analytical results and to collect more detailed error information I wrote yet another simulation.  The code implements the error filters described\footnote{The code also implements a conservative filter that I do not discuss because it is only marginally different from the liberal filter.} as well as a Monte-Carlo error generation and propagation code complete with functions for performing tracking graph-state construction.  The cat states necessary for the simulation are assumed to be prepared to specifications elsewhere.

\begin{figure}
  \capstart
  (a)\\
  \centerline{
    \begin{pgfpicture}{0cm}{0cm}{13.41cm}{9.64cm}
      \pgfputat{\pgfxy(.24,.36)}{\includegraphics[clip=true, trim=0cm 7cm 0cm 7cm, width=14.5cm]{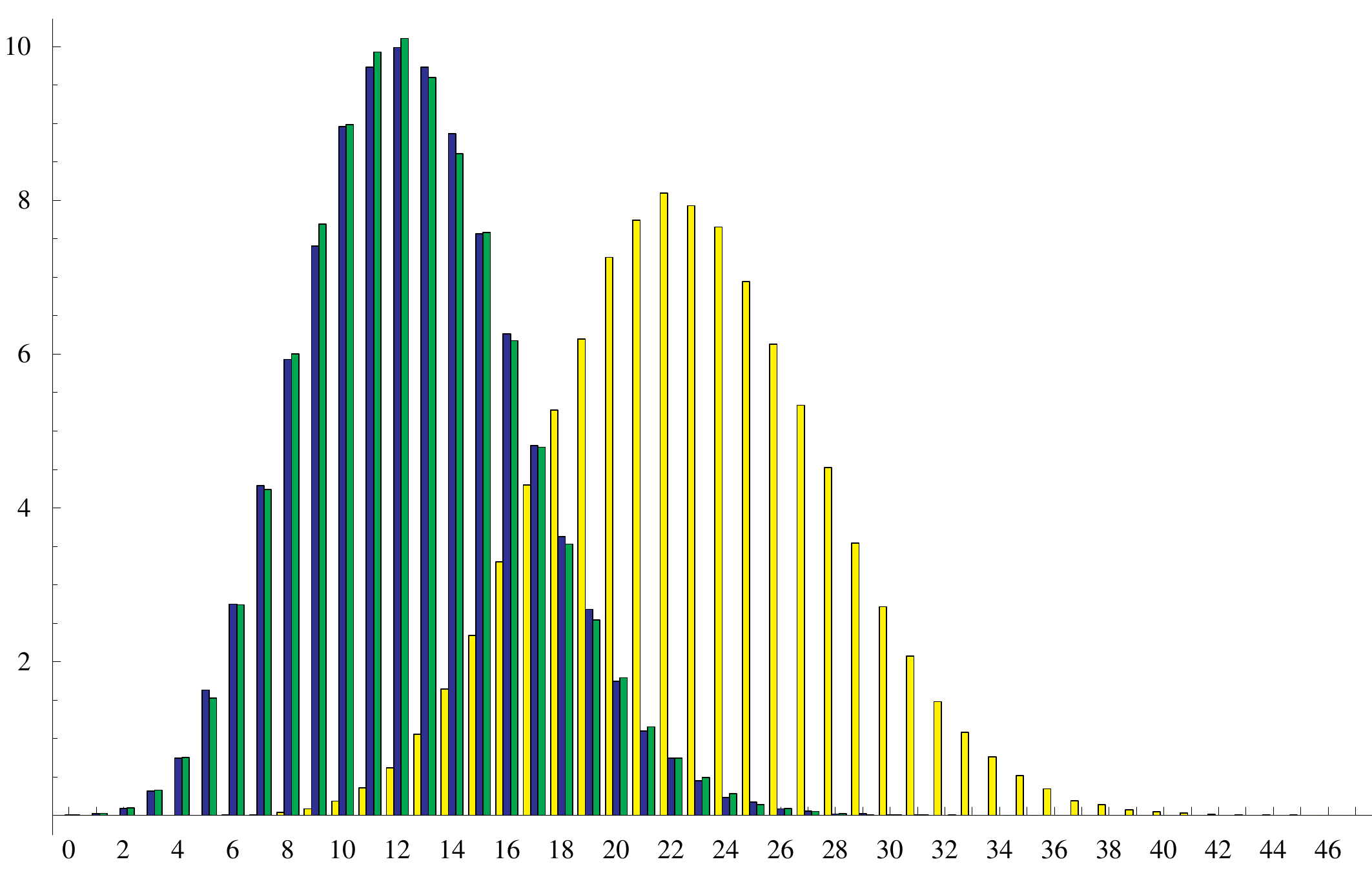}}
      \pgfputat{\pgfxy(7.25,.15)}{\pgfbox[center,center]{\footnotesize \# of $Z$ errors}}
      \pgfputat{\pgfxy(.15,5.08)}{\pgfbox[center,center]{\footnotesize \%}}
      \pgfputat{\pgfxy(10.55,8.85)}{\pgfbox[left,center]{\footnotesize Fool's filter}}
      \pgfputat{\pgfxy(10.55,8.25)}{\pgfbox[left,center]{\footnotesize Liberal filter}}
      \pgfputat{\pgfxy(10.55,7.65)}{\pgfbox[left,center]{\footnotesize Viterbi filter}}
      \color{yellow}
      \pgfrect[fill]{\pgfxy(10.2,8.7)}{\pgfxy(.3,.3)}
      \color{uglyBlue}
      \pgfrect[fill]{\pgfxy(10.2,8.1)}{\pgfxy(.3,.3)}
      \color{uglyGreen}
      \pgfrect[fill]{\pgfxy(10.2,7.5)}{\pgfxy(.3,.3)}
    \end{pgfpicture}\hspace{1cm}
  }
  \vspace{.8em}\\(b)\\
  \centerline{
    \begin{pgfpicture}{0cm}{0cm}{9.45cm}{6.75cm}
      \pgfputat{\pgfxy(.45,.35)}{\includegraphics[clip=true, trim=0cm 7cm 0cm 7cm, width=10cm]{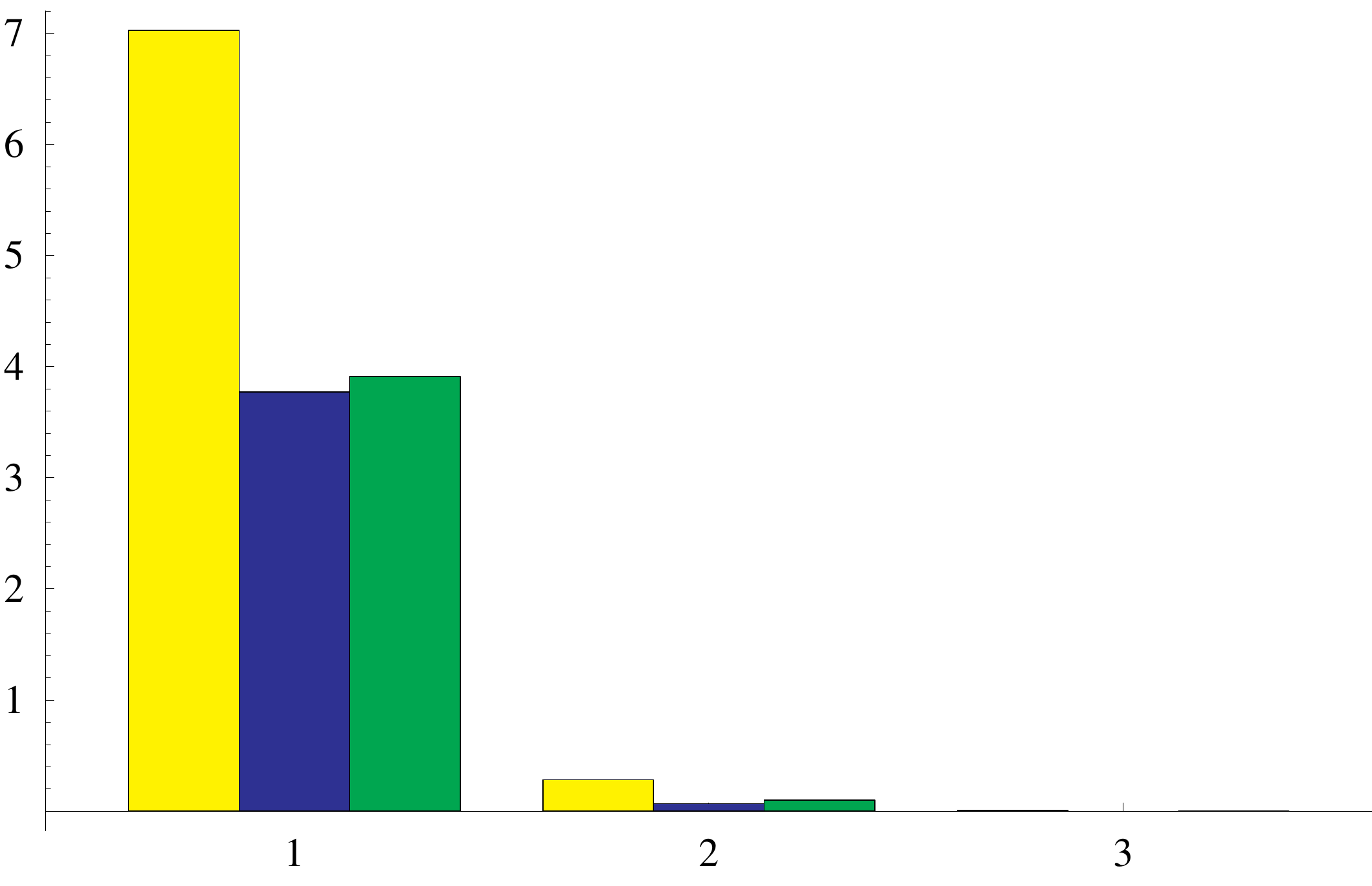}}
      \pgfputat{\pgfxy(5.5,.1)}{\pgfbox[center,center]{\footnotesize \# of $X$ errors}}
      \pgfputat{\pgfxy(.1,3.75)}{\pgfbox[center,center]{\footnotesize \%}}
      \pgfputat{\pgfxy(7.05,6.05)}{\pgfbox[left,center]{\footnotesize Fool's filter}}
      \pgfputat{\pgfxy(7.05,5.45)}{\pgfbox[left,center]{\footnotesize Liberal filter}}
      \pgfputat{\pgfxy(7.05,4.85)}{\pgfbox[left,center]{\footnotesize Viterbi filter}}
      \color{yellow}
      \pgfrect[fill]{\pgfxy(6.7,5.9)}{\pgfxy(.3,.3)}
      \color{uglyBlue}
      \pgfrect[fill]{\pgfxy(6.7,5.3)}{\pgfxy(.3,.3)}
      \color{uglyGreen}
      \pgfrect[fill]{\pgfxy(6.7,4.7)}{\pgfxy(.3,.3)}
    \end{pgfpicture}\hspace{1cm}
  }
  \caption[Post-construction $Z$- and $X$-error histograms]{Histograms showing the percent probability of various numbers of undetected a) $Z$ and b) $X$ errors surviving the tracking graph-state construction process for $n=200$ and $p_{dep}=1/4000$.  Yellow, blue, and green bars identify data for the fool's, liberal, and Viterbi filters respectively.  On average roughly $25$ failures occur during tracking construction, so the error scale-up is typically quite modest. \label{fig:errorHistograms}}
\end{figure}

Using this code, I collected data on the error composition resulting from tracking preparation of the graph-state corresponding to the complete graph on $200$ nodes.  The total failure probability for each operation was set to $1/4000$, and a depolarizing error model was employed for all operations except measurements, which produced only bit errors.  Data was collected for $100000$ runs using each of the three filters.  A normalized histogram of the number of $Z$ errors remaining after preparation is shown in Figure~\ref{fig:errorHistograms}a, and a normalized histogram of the number of $X$ errors remaining after preparation is shown (with the no-error column omitted) in Figure~\ref{fig:errorHistograms}b.  Some additional statistics are displayed in Table~\ref{tab:trackingConstructionStatistics}.

\begin{table}
  \capstart
\centerline{
  \begin{tabular}{c@{\hspace{1em}}|@{\hspace{1em}}c@{\hspace{2em}}c@{\hspace{2em}}c}
    \hspace{4em}Filter & Fool's & Liberal & Viterbi \\
    \hline
    Average scale-up & 0.912 & 0.499 & 0.498 \\
    Max scale-up & 1.5 & 1.25 & 1.21 \\
  \end{tabular}
}
\caption[Tracking construction error statistics]{The average scale-up and the maximum observed scale-up for the fool's, liberal, and Viterbi filters.  In each case the average number of failures per construction was approximately 25.\label{tab:trackingConstructionStatistics}}
\end{table}

All filtering algorithms perform substantially better than the maximum possible scale-up of $2$ predicted for the fool's and liberal filter in Section~\ref{subsec:errorSpread}.  As interpreted using the fool's filter, graph states prepared with tracking have, on average, a number of errors equal to the number of failures that occurred during the preparation.  The liberal and Viterbi filters both yield graph states with about $50\%$ as many errors as actually occurred during construction.  These two filters are identical to within the margin of error, indicating that the liberal filter is well suited to filtering depolarizing errors.

\section{Scaling}

The purpose of developing a new method of ancilla construction was to permit the production of large ancillae with improved overhead costs in terms of qubits, so it is important to consider the scaling properties my design.

First, it must be noted that the approach presented in this chapter is not applicable to graph states of arbitrarily large size.  The direct approach to ancilla construction tends to fail for two reasons.  Typically, the limiting factor is the propagation of errors between qubits during the construction process, a pitfall which I have made every effort to avoid.  As the number of applied gates grows, however, sheer accumulation of uncorrected, independent errors will eventually cause the construction process to fail.  I have made no attempt at correcting uncorrelated errors (though incidentally I have done so for $X$ errors), so this effect bounds the size of the ancillae which I might produce.

In each round of tracking graph-state construction, there are four sources from which $Z$ errors might be injected into each primary qubit.  When preparing the graph state corresponding to the complete graph, the total opportunities for each primary qubit to suffer a $Z$ error is thus roughly $4n$. If the construction is to succeed, however, it must be the case that the expected number of $Z$ errors per qubit is much less than $1$.  Taking all gates and ancilla qubits to fail with probability $p$ yields the bound
\begin{align}
  4np\ll 1 && \rightarrow && n\ll \frac{1}{4p}
\end{align}
on the size of the complete graph.

As mentioned in Section~\ref{sec:compressedGraphstateConstruction}, however, graph states corresponding to graphs with $(w-1)$ connections per node can be constructed using only $2w-3$ rounds.  In such a construction, the number of opportunities for a primary qubit to suffer a $Z$ error is only about $8w$, yielding the bound
\begin{align}
  8wp\ll 1 && \rightarrow && w\ll \frac{1}{8p}
\end{align}
on the weight of the stabilizer generators of the graph state.

In terms of qubit resources, tracking construction of the complete graph requires $n$ roughly $n$-qubit cat states.  By comparison, measuring a single weight $n$ generator using Shor's method of syndrome extraction also requires an $n$-qubit cat state, but the measurement must be repeated $t+1$ times where $t$ is the number of errors that we wish to be able to tolerate without failure.  Thus, verifying the complete graph state by Shor's method after it has been constructed requires roughly $nt$ $n$-qubit cat states.  Similarly, for graph states with weight $w$ generators, tracking construction requires $n$ roughly $2w$-qubit cat states while verification via Shor's method requires roughly $\text{Min}(nt,nw/2)$ $w$-qubit cat states.  Generally, therefore, the total number of qubits prepared in various cat states will be smaller (often much smaller) for my procedure.  Shor's method, however, is capable of tolerating cat states with much higher frequencies of $Z$ errors, so a fair comparison would require that I include the number of qubits needed to make each kind of cat state, a quantity which I do not presently know.

\section{Analysis\label{sec:analysisAncillaConstruction}}

Much work remains to be done on this topic.  The construction procedure I have developed displays a number of interesting properties: it requires relatively few cat states for its implementation, responds very differently to different two-qubit error models, constructs and verifies graph states without ever having measured any of their stabilizers, and generates states with an enormous asymmetry in the number of $Z$ and $X$ errors.  On the other hand, it requires higher quality cat states than are typically necessary and generates states with weight $2$ correlated errors.

The asymmetry between $X$ and $Z$ errors is a especially intriguing property.  Steane's method of syndrome extraction makes good use of ancillae with exactly this sort of asymmetry.  Non-trivial graph states cannot also be CSS codes, so this particular avenue is closed to me, but a variety of purification strategies are promising.  The graph state corresponding to the complete graph on an even number of nodes, for example, is invariant under the transversal application of $H$, and can thus be prepared with either minimal $X$ or minimal $Z$ errors.  Such states bring to mind the work of Glancy, Knill, and Vasconcelos~\cite{Glancy06} who have identified a $[[6,2,2]]$ code that can correct any single $X$ or $Z$ error so long as it is known which half of the code it occurred in.

The presence of weight $2$ correlated errors on the constructed ancillae is a definite drawback.  Such errors effectively reduce the order of the worst-case correctable error on an encoded state by half.  Ideally, then, ancillae constructed by the method described in this chapter would either be further purified or used in some specialized task like Steane-style syndrome extraction.

In one very pertinent special case, however, further verification might be unnecessary.  The graphs associated with CSS code states are bipartite, meaning that the graph can be divided into two sets of nodes such that no nodes in the same set are connected.  To recover a CSS code state from the graph state it is only necessary to apply $H$ to the qubits corresponding to all of the nodes in one set.  The correlated errors left behind by the fool's filter, however, only occur on qubits connected by a $\CZ$ gate, and those are only (effectively) $Z\otimes Z$ errors.  Thus, applying $H$ to all of the qubits corresponding to one set of the bipartite graph yields correlated errors only of the form $Z\otimes X$.  So long as the phase gate is applied by teleportation, this separation is maintained throughout encoded Clifford operations.  In such a case, the correlations can be ignored since they have no effect on CSS code error correction, which separately corrects $X$ and $Z$ errors.

Finally, the construction of adequate cat states has turned out to be an unexpectedly troublesome problem.  For large cat states, the standard approach involves making many pairwise parity ($Z\otimes Z$) measurements, during which the $Z$ error probability only builds.  For my procedure, $n$-qubit cat states thus produced must be corrected for $Z$ errors when $n p\not \ll 1$, but effective correction of $Z$ errors in an $n$-qubit cat state is impossible unless $np\ll 1$.  Without a novel technique for creating cat states, it is difficult to see how this problem might be resolved, and, without a solution, the window of probabilities and qubit numbers for my procedure is small.

\chapter{Conclusion\label{sec:conclusion}}

The primary conclusions of this dissertation are twofold.  First, I have found that a detailed knowledge of the kinds of errors produced by gate failures is, with one possible exception, not particularly useful.  Second, I have shown that large ancillae prepared in logical basis states are a sufficient resource to permit computation at quite high rates of error.  Also of interest are the following observations: (i) some restricted error models yield small gains in the threshold without any modifications to the standard approach to fault tolerance, (ii) improvements in ancilla preparation are of little consequence to threshold estimates, (iii) Knill's fault-tolerant procedure outperforms that of Steane given ancillae with uncorrelated errors, and (iv) $n$-qubit graph-state construction can be compressed to roughly $n$ time steps.  In addition to these findings, I develop a general tool for understanding thresholds and a novel technique for generating ancillae, an important resource for quantum computation.

My conclusions regarding unusual error models are based on the results of Chapters~\ref{chap:channelDependencyOfTheThreshold} and~\ref{chap:thresholdsForHomogeneousAncillae}.  In Chapter~\ref{chap:channelDependencyOfTheThreshold}, I investigate the impact of knowledge about the error model by tailoring a fault-tolerant procedure to a highly structured stochastic error channel, namely, symmetric $\CX$ errors.  Through bounds and estimates I then examine the threshold for quantum computation using this procedure.  Comparing my results with threshold estimates for the depolarizing channel and threshold bounds for adversarial errors, I find only a small increase in the threshold for my tailored procedure and error model of choice.  Moreover, I show numerically that the window of error models for which the procedure yields an advantage is quite small; adding a depolarizing channel at $1/10$ the strength of the symmetric $\CX$ errors completely disrupts my procedure's function.  Oddly, such small gains come in spite of the fact that my tailored procedure dramatically reduces the frequency of errors on constructed ancillae.  I resolve this mystery by estimating the threshold given perfect ancillae.  For the Steane code, these turn out not to give much higher thresholds than ancillae constructed using the standard prepare and discard approach.  Thus, in line with predictions by Reichardt~\cite{Reichardt04}, I find that improvements in the construction of small ancillae are largely irrelevant to the threshold.  The increase in the threshold that is observed is thus primarily due to the error properties of the gates applied to the data.  Consequently, the gain would be expected, and is observed, to apply to Steane's fault-tolerant method as well.  I show in Chapter~\ref{chap:thresholdsForHomogeneousAncillae}, however, that, effectively ignoring ancilla construction, the increase in the threshold due to a quite restricted error model is less than a factor of $2$.  On the other hand, given that I determine the threshold coefficient (the ratio of the threshold to the correctable error rate of the code) for the depolarizing error channel using Knill's procedure to be $.35$, it is unreasonable to hope for more than a factor of $3$.  This is because the threshold coefficient corresponding to a single error on the data with probability $p$ is $1$; a higher coefficient would require that the probability of an error on the data be less than $p$, the probability of a gate error.  Thus, for the cases of small ancilla preparation and of data gates, I have basically ruled out major gains in the threshold due to expanded knowledge of the form taken by gate errors.  The remaining potential for improvements in the threshold due to knowledge of the error model therefore lies either in the construction of large ancillae or in error models for which more is known than the kind of Pauli errors produced.  The work of Knill regarding heralded errors~\cite{Knill05b} is an example of the latter.

The utility of large ancillae is demonstrated in Chapter~\ref{chap:thresholdsForHomogeneousAncillae}.  There I bound the threshold given the availability of ancillae whose component qubits sport identical, uncorrelated error distributions.  For ancillary qubit errors that occur with probability on the order of the gate error probability and in the limit that the size of the code goes to infinity, I find that these resources permit computation at error rates in excess of $1\%$.  While these threshold bounds are only rigorous given the necessary ancillae, I observe fair agreement between my thresholds and recent estimates in the literature.  In addition, I develop a finite version of the algorithm for threshold estimates using small codes that yields predictions for the Steane code in accordance with threshold estimates derived from the simulation used in  Chapter~\ref{chap:channelDependencyOfTheThreshold}.  The success of my algorithm at threshold estimation depends on the feature of threshold estimates that dampened the results of Chapter~\ref{chap:channelDependencyOfTheThreshold}, ancillae have a relatively minor role to play in threshold estimation. In addition, to the comparisons between error models discussed in the previous paragraph, I also compare the threshold for Knill's fault-tolerant procedure and two procedures based on Steane's method.  Knill's telecorrection procedure is found always to have a higher threshold, a result that is likely to hold so long as ancillae with uncorrelated errors are available.

In the absence of ancillae with identically distributed, uncorrelated errors, however, my results from Chapter~\ref{chap:thresholdsForHomogeneousAncillae} do not establish rigorous bounds on the threshold, and the construction of sufficiently large ancillae is a non-trivial problem.  In an effort to address this problem, I develop a novel method of ancilla construction in Chapter~\ref{chap:ancillaConstruction}.  My method employs a compressed form of the standard circuit for constructing graph states, but the $\CZ$ gates corresponding to edges are interspersed with $\CX$ gates intended to extract $X$-error information during the process of construction.  Through post-processing of the collected information, locations of both $X$ errors and propagated $Z$ errors are inferred.  The process is not fault-tolerant, but I prove that each gate failure leads to at most $2$ errors on the ancilla, and, numerically, I find that the typical error scale-up is small and that ancilla errors are limited almost exclusively to $Z$ errors. Given these facts, the prospects for further verification, or even direct use in special situations, seem promising.  In fact, in Section~\ref{sec:analysisAncillaConstruction} I suggest a possible avenue by which verification might be avoided altogether for CSS codestates.  Ironically, the most difficult part of my ancilla construction procedure may prove to be the construction of the cat states necessary to perform it.  Otherwise it compares favorably to other means of preparing ancillae.

Even should direct ancilla construction prove impossible, however, the method developed in Chapter~\ref{chap:thresholdsForHomogeneousAncillae} provides a new tool for studying thresholds.  I have found it useful for comparative studies because it provides a quick and simple means of predicting the outcome of Monte-Carlo threshold estimates on large codes.  In addition, by simplifying the complexity associated with estimating thresholds, I believe it helps to provide insight into the factors that shape and limit them.

\chapter*{Appendices}

\appendix
\chapter[The asymptotic correctable error fraction for CSS codes]{Asymptotic correctable error fraction for CSS
codes\label{app:asymCor}}
In reference~\cite{Gottesman01} Gottesman and Preskill find that the asymptotic correctable error fraction for general CSS codes approaches $11\%$.  Their result follows from two separate applications of Shannon's noisy channel coding theorem.  Since they apply the random coding argument to the $X$ stabilizers and the $Z$ stabilizers separately, quantum mechanics plays a role only by restricting the number total number of stabilizers to be less than or equal to the number of qubits, n.  As a consequence, the value they obtain for the asymptotic correctable error fraction is exactly the maximum error rate $\tau$ for which a classical code with data rate $\frac{1}{2}$ exists, that is, $\tau$ such that $.5=H_2(\tau)=-\tau\log_2\tau-(1-\tau)\log_2(1-\tau)$, $\tau\approx.11$.

I would like to apply the same argument in this appendix, but the CSS codes considered here have the additional property that the $X$ stabilizers can be obtained from the $Z$ stabilizers simply by replacing each $Z$ with an $X$.  Since the $X$ and $Z$ stabilizers must commute, this restriction corresponds to requiring that the binary matrix representing the $X$ (or $Z$) stabilizer generators, known as the parity check matrix, be dual-contained.  To apply Shannon's noisy channel coding theorem\footnote{For a clear, detailed exposition of Shannon's noisy coding theorem for random linear codes see Section~14.2 of Reference~\cite{MacKay}.}, I must show that any $2w_m$ columns of the parity check matrix, where $w_m$ is largest weight of any error being corrected, can be treated as though the entries were randomly and equiprobably assigned values of $0$ or $1$.  In this case, $w_m\approx.11n$; the remainder of this appendix is devoted to showing that $.5n$ columns can be randomly assigned.

Consider the following non-standard way to construct an $m \times n$ dual-contained parity check matrix, $\mat{H}$, where $m=\frac{n}{2}-n\epsilon$ and $\epsilon\ll1$.  Divide the matrix horizontally into two matrices of width $\frac{n}{2}$ and denote them $\mat{L}$ and $\mat{R}$.  Now randomly assign the entries in $\mat{L}$ to be $0$ or $1$ with equal probability.  The probability is $2^{i-1}/2^\frac{n}{2}$ that the $i$th row of $\mat{L}$ is dependent given that the previous $i-1$ rows are independent.  The total probability that the rows of $\mat{L}$ are dependent, $\fun{P}_{\textrm{D}\mat{L}}$, is bounded by the sum of these terms,
\begin{align}
\fun{P}_{\textrm{D}\mat{L}} <2^{-\frac{n}{2}} \sum_{i=1}^m 2^{i-1} = 2^{-\frac{n}{2}} \frac{1-2^m}{1-2} < 2^{-n\epsilon} .
\label{eq:problefterror}
\end{align}
As $n$ becomes large, for any fixed $\epsilon$, $\fun{P}_{\textrm{D}\mat{L}}$ rapidly goes to zero and independent matrices come to dominate the output.

Now I move to the problem of assigning $\mat{R}$.  I require that $\mat{H}$ be dual-contained, that is, that the rows be orthogonal to themselves and each other.  Given that the rows of $\mat{L}$ consist of $\frac{n}{2}-n\epsilon$ independent, randomly chosen vectors, the restriction on $\mat{H}$ can be restated as the requirement that every row of $\mat{R}$ satisfy a different randomly chosen constraint with every other row in $\mat{R}$ and with itself (or equivalently, the vector of all $1$s).  In addition to being random, the constraints are uncorrelated because the rows of $\mat{L}$ are independent.  Constructing $\mat{R}$ one row at a time, the number of vectors that satisfy the constraints on the $i$th row is $2^{\frac{n}{2}-i}$ assuming that the set of all previous rows and the all $1$s vector are independent.  The probability of a binary string satisfying a binary condition with a randomly chosen constraint is $\frac{1}{2}$ since every binary string either satisfies a constraint or satisfies its negation. This means that the probability of any particular string satisfying $i$ such constraints is $2^{-i}$.  Consequently, the probability of picking a dependent vector for the $i$th row of $\mat{R}$ given that none of the previous $i-1$ vectors were dependent is $2^i2^{-i}/2^{\frac{n}{2}-i} = 2^{i-\frac{n}{2}}$.  As before, this yields a bound on the total probability of the rows of $\mat{R}$ being dependent,
\begin{align}
\fun{P}_{\textrm{D}\mat{R}}<2^{-\frac{n}{2}} \sum_{i=1}^m 2^{i} = 2\cdot 2^{-\frac{n}{2}} \sum_{i=1}^m 2^{i-1} < 2^{-n\epsilon+1},
\end{align}
which goes to zero as $n$ goes to infinity.

The probability of my matrix construction procedure halting due to the generation of dependent rows goes to zero, but that does not necessarily imply that it generates all dual-contained parity check matrices.  It is conceivable that the cases where a dependent vector is chosen, though rare, correspond to many more possible $\mat{H}$ matrices than the cases where an independent vector is chosen.  To verify that this is not the case, it is sufficient to count the number of matrices generated by my procedure and to compare it with the total number of dual-contained $m \times n$ matrices.

The number of possible $\mat{L}$ matrices generated by my procedure approaches $N_\mat{L}=2^{\frac{nm}{2}}$, and the number of possible $\mat{R}$ matrices approaches
\begin{align}
N_\mat{R}=\prod_{i=1}^m 2^{\frac{n}{2}-i} = 2^{\frac{nm}{2}-\sum_{i=1}^m i} = 2^{\frac{nm}{2} - \frac{m^2}{2} - \frac{m}{2}} .
\end{align}
By comparison, the total number of dual-contained parity check matrices of size $m\times n$ where all rows are linearly independent is
\begin{align}
N_\mat{H} = \prod_{i=1}^{m} (2^{n-i}-2^i) = 2^{nm - \sum_{i=1}^m i} \prod_{i=1}^{m} (1-2^{2i-n}) = 2^{nm - \frac{m^2}{2} - \frac{m}{2}} \prod_{i=1}^{m} (1-2^{2i-n})
\end{align}
which, of course, approaches $2^{nm - \frac{m^2}{2} - \frac{m}{2}}$ as $n$ becomes large.

Having found that $N_\mat{H}=N_LN_\mat{R}$ in the limit that $n$ goes to infinity, I am now free to treat a random $\mat{H}$ as though as many as half of the columns are filled with randomly generated binary digits.  This means that the probability that a randomly chosen $\mat{H}$ satisfies $\mat{H}\cdot(\vec{x}+\vec{y})=0$ is $2^{-m}$ for any two error vectors $\vec{x}$ and $\vec{y}$ such that $\vec{x}\neq\vec{y}$ and the weight of $\vec{x}+\vec{y}$ is less than $\frac{n}{2}$.  Given that, Shannon's noisy coding theorem proceeds exactly as it did in reference~\cite{Gottesman01}.  There exist classical dual-contained codes that, with probability approaching $1$ as $n\rightarrow\infty$, correct errors on up to $11\%$ of the bits.  Consequently, there exist CSS codes capable of correcting $.11n$ $X$ errors and a like number of $Z$ errors with arbitrarily high probability.

\chapter{Code}
The code used in this dissertation is available at \href{http://info.phys.unm.edu}{http://info.phys.unm.edu}.  The function of each file is explained briefly in the remainder of this appendix.

\section{Monte-Carlo Threshold Estimation Code \label{sec:thresholdEstimationCode}}

The backbone of my Monte-Carlo threshold estimator is composed of the \command{C} files \command{mt19937ar-cok.c}, \command{7QCode.h}, \command{7QCode.cpp}, \command{Threshold.h}, and \command{Threshold.cpp}.  The file \command{mt19937ar-cok.c} was coded by Takuji Nishimura and Makoto Matsumoto and implements a Mersenne Twister pseudorandom number generator.  Basic gates and functions are defined in \command{7QCode.cpp} for propagating errors using an array of length $2$ arrays of type \command{char} where each $2$-element array represents the $X$ and $Z$ errors on a set of $8$ qubits; the necessary declarations are given in \command{7QCode.h}.  \command{Threshold.cpp} contains the code for initializing and managing the simulation and taking data and statistics, while \command{Threshold.h} declares the functions for implementing encoded gates that are obtained from \textbf{either} \command{StandardFTI.cpp} or \command{MyFTI.cpp}.

\command{StandardFTI.cpp} and \command{MyFTI.cpp} implement encoded gates for the fault-tolerant method of Steane and for my own tailored method, respectively.  Only one or the other can be included on compilation, otherwise the compiler will crash.

\section{Homogeneous Ancillae Threshold Code \label{sec:homogeneousAncillaThresholdCode}}

The \command{Mathematica} notebooks I use for calculating thresholds for homogeneous ancillae are \command{Infinite CSS code CX counter.nb} and \command{Finite CSS Code Bounder.nb}.  The first file implements error generation and propagation routines for encoded and unencoded gates for the single-coupling Steane, double-coupling Steane, and Knill procedures defined in Section~\ref{sec:examples}.  All possible Pauli errors requiring two or fewer failures are stored along with an algebraic representation of their associated probabilities (also up to second order).  Pauli errors are stored as arrays of integers using the mapping $\{I,X,Y,Z\}=\{0,1,2,3\}$.  The second file contains the code used to determine the range of possible threshold estimates for quantum codes with finite (small) numbers of qubits.

\section{Monte-Carlo Ancilla Construction Code \label{sec:ancillaConstructionCode}}

\command{GraphStateConstruction.py} implements a Monte-Carlo routine for estimating the encoded failure probability of my method of graph-state construction using the language \command{Python}.  That file contains all of the necessary error propagation functions as well as functions for performing tracking graph-state construction and interpreting the error trace.  As in my \command{Mathematica} code, Pauli errors are stored as integers.  The associated file \command{GraphStatePreparationTraceViterbi.py} contains code for interpreting error tracks using the Viterbi algorithm.

\chapter{The Viterbi Algorithm\label{chap:viterbiAlgorithm}}
The Viterbi algorithm is a method of determining the most probable sequence of hidden states given limited observational data.  The algorithm employs a kind of message passing routine to efficiently find the most probable sequence.  This appendix explains the mechanics of the Viterbi algorithm and presents both a worked example and functional code.  For a general treatment of message passing, the reader is referred to \textit{Information Theory, Inference, and Learning Algorithms}~\cite{MacKay}.  Other informative and entertaining introductions to the Viterbi algorithm can be found online.

\section{Explanation}

Life is full of situations where it's important to figure out the most likely sequence of events based on limited observational data.
Given a set of observations $\{o_i\}$ on a system occupying an unknown sequence of states, the most probable state sequence, or path, is that which maximizes the conditional probability $P\left(\bigwedge_j S_j|\bigwedge_i O_i=o_i\right)$ where $S_j$ and $O_i$ are random variables labeling elements of the sequence of states and the set of observations respectively.  Using Bayes' rule this probability can be written in terms of more accessible quantities as
\begin{align}
  P\left(\bigwedge_j S_j\left|\bigwedge_i O_i=o_i\right.\right) = \frac{P\left(\bigwedge_j O_i=o_i|\bigwedge_j S_j\right)P\left(\bigwedge_j S_j\right)}{P\left(\bigwedge_i O_i=o_i\right)}.
\end{align}
$P\left(\bigwedge_i O_i=o_i\right)$ does not vary during the maximization and can thus be discarded, thereby reducing the problem of finding the most likely path to that of maximizing $P\left(\bigwedge_i O_i=o_i|\bigwedge_j S_j\right)P\left(\bigwedge_j S_j\right)$.

While conceptually simple, this maximization is frequently computationally infeasible because the number of possibilities that must be considered grows exponentially in the length of the sequence.  In certain cases, however, there exist more efficient methods of solution than exhaustively searching all possibilities.

One such case is that of a Markov process, that is, a process in which the state of the system at any time $t$ depends on the previous states only in that it depends on the state of the system at time $t-1$.  In terms of conditional probabilities this is the statement that
\begin{align}
  P\left(S_{t}\left|\bigwedge_j^{t-1} S_{j}\right.\right) = P(S_{t}|S_{t-1}).
\end{align}
Using this fact it is possible to expand $P\left(\bigwedge_j S_j\right)$ as
\begin{align}
  P\left(\bigwedge_j S_j\right) = P\left(S_{t}\left|\bigwedge_j^{t-1} S_{j}\right.\right)P\left(\bigwedge_j^{t-1}S_j\right) = P(S_t|S_{t-1})P\left(\bigwedge_j^{t-1}S_j\right). \label{eq:viterbiPS}
\end{align}
If each observation likewise depends only on the state of the system at a single time then $P\left(\left.\bigwedge_i O_i=o_i\right|\bigwedge_j S_j\right)$ can be expanded as
\begin{align}
  P\left(\left.\bigwedge_i O_i=o_i\right|\bigwedge_j S_j\right) = \prod_i P(O_i=o_i|S_i) \label{eq:viterbiPO}
\end{align}
where, of course, $P(O_i=o_i|S_i)=1$ if no observation occurs during time step $i$.
Applying the identities in Eqs.~\ref{eq:viterbiPS} and \ref{eq:viterbiPO} to $P\left(\bigwedge_i O_i=o_i|\bigwedge_j S_j\right)P\left(\bigwedge_j S_j\right)$, the probability that we wish to maximize over, yields
\begin{align}
  \begin{split}
    P&\left(\bigwedge_i O_i=o_i\left|\bigwedge_j S_j\right.\right)P\left(\bigwedge_j S_j\right) \\
    &= P(O_t=o_t|S_t) P(S_t|S_{t-1}) P\left(\bigwedge_i^{t-1}S_i\right) \prod_i^{t-1} P(O_i=o_i|S_i)
  \end{split} \label{eq:viterbiExpandedProbOfInterest}
\end{align}
which can, as we shall see, be maximized in an incremental fashion.

Suppose $\{u_i\}_{i=1}^t$ is the most probable sequence of states leading to $S_t=u_t$.  From equation~\ref{eq:viterbiExpandedProbOfInterest} we know that
\begin{align}
  P\left(\bigwedge_i^t S_i=u_i\right) = P\left(S_t=u_t|S_{t-1}=u_{t-1}\right)P\left(\bigwedge_i^{t-1} S_i=u_i\right),
\end{align}
implying that the sequence $\{u_i\}_{i=1}^{t-1}$ must be the most probable sequence of states leading to $S_{t-1}=u_{t-1}$.  Were it not, there would exist a different sequence $\{v_i\}_{i=1}^{t}$ such that $v_{t}=u_{t}$, $v_{t-1}=u_{t-1}$ and
\begin{align}
  P\left(\bigwedge_i^{t-1} S_i=v_i\right)>P\left(\bigwedge_i^{t-1} S_i=u_i\right).
\end{align}
But this would imply that
\begin{align}
  \begin{split}
    P&\left(\bigwedge_i^t S_i=v_i\right) = P\left(S_t=u_t|S_{t-1}=u_{t-1}\right)P\left(\bigwedge_i^{t-1} S_i=v_i\right) \\
    & > P\left(S_t=u_t|S_{t-1}=u_{t-1}\right)P\left(\bigwedge_i^{t-1} S_i=v_i\right) = P\left(\bigwedge_i^t S_i=u_i\right)
  \end{split}
\end{align}
contradicting our assumption that the most probable sequence of states leading to $S_t=u_t$ is $\{u_i\}_{i=1}^t$.

The preceding paragraph shows that, for any Markov process, the most probable sequence of states concomitant with a particular set of observations can be calculated in a step-wise fashion.  The procedure for doing so is known as the Viterbi algorithm and determines the most probable path to each state at each time step by starting from the most probable path to each state of the previous time step (and the associated probabilities) and calculating which of these paths leads most probably to a given state in the current time step.

\section{Example}

Imagine that you are a professor lecturing a class of overworked and drowsy students.  Long experience experience has taught you that students typically occupy one of two states, `learning' or `sleeping', and that students have a memory of about 15 minutes.  For the purpose of assigning participation points, you keep track of which students are both in attendance and conscious.  Unfortunately, one of your students has taken to wearing mirror shades.  You can test whether he is awake by asking him a question, but asking the same student questions throughout the period would be disruptive.  Instead, you decide to make a few observations and determine from those his most probable sequence of states.

Over the course of 15 minutes, students who are learning have a $30\%$ chance of going to sleep while students who are sleeping have a $20\%$ chance of waking up.  Additionally, $90\%$ of students are awake (learning) when class starts.  Thus, for a 45 minute lecture, students are modeled succinctly by the graph
\[
\Qcircuit[.35em] @C=1em @R=1em {
{\text{S}\ \ .1} & \sink \link{0}{5} \link{5}{5} & & & & & \sink \link{0}{5} \link{5}{5} & & & & & \sink \link{0}{5} \link{5}{5} & & & & & \sink \\
& & & {\raisebox{1.5em}{\ \ \ .8}} & & & & & {\raisebox{1.5em}{\ \ \ .8}} & & & & & {\raisebox{1.5em}{\ \ \ .8}} \\
& & {\raisebox{.7em}{.2}} & & & & & {\raisebox{.7em}{.2}} & & & & & {\raisebox{.7em}{.2}} \\
& & {\raisebox{-1.3em}{.3}} & & & & & {\raisebox{-1.3em}{.3}} & & & & & {\raisebox{-1.3em}{.3}} \\
& & & {\raisebox{-2em}{\ \ \ .7}} & & & & & {\raisebox{-2em}{\ \ \ .7}} & & & & & {\raisebox{-2em}{\ \ \ .7}} \\
{\text{L}\ \ .9} & \sink \link{0}{5} \link{-5}{5} & & & & & \sink \link{0}{5} \link{-5}{5} & & & & & \sink \link{0}{5} \link{-5}{5} & & & & & \sink \\
& \text{0min} & & & & & \text{15min} & & & & & \text{30min} & & & & & \text{45min} \\
}
\]
where the rows labeled S and L represent the states `sleeping' and `learning' respectively and the columns correspond to the labeled times.

At the 15 minute mark, you ask your blinkered student a question and receive no reply, a response that conscious students offer only $30\%$ of the time.  At the end of class, however, he promptly stands up and walks out, indicating that he was awake.  What was the student doing during your lecture?

\begin{table}
  \capstart
\centerline{$
\begin{array}{l@{\hspace{4em}}l}
P(\{\text{S},\text{S},\text{S},\text{S}\},\text{O}) = 0 & P(\{\text{S},\text{S},\text{S},\text{L}\},\text{O}) = 0.0128 \\
P(\{\text{S},\text{S},\text{L},\text{S}\},\text{O}) = 0 & P(\{\text{S},\text{S},\text{L},\text{L}\},\text{O}) = 0.0112 \\
P(\{\text{S},\text{L},\text{S},\text{S}\},\text{O}) = 0 & P(\{\text{S},\text{L},\text{S},\text{L}\},\text{O}) = 0.00036\\
P(\{\text{S},\text{L},\text{L},\text{S}\},\text{O}) = 0 & P(\{\text{S},\text{L},\text{L},\text{L}\},\text{O}) = 0.00294\\
P(\{\text{L},\text{S},\text{S},\text{S}\},\text{O}) = 0 & P(\{\text{L},\text{S},\text{S},\text{L}\},\text{O}) = 0.0432\\
P(\{\text{L},\text{S},\text{L},\text{S}\},\text{O}) = 0 & P(\{\text{L},\text{S},\text{L},\text{L}\},\text{O}) = 0.0378\\
P(\{\text{L},\text{L},\text{S},\text{S}\},\text{O}) = 0 & P(\{\text{L},\text{L},\text{S},\text{L}\},\text{O}) = 0.01134\\
P(\{\text{L},\text{L},\text{L},\text{S}\},\text{O}) = 0 & P(\{\text{L},\text{L},\text{L},\text{L}\},\text{O}) = 0.09261\\
\end{array}
$}
\caption[Brute force approach to finding the most probable sequence of states]{Brute force approach to finding the student's most probable sequence of states.  Each probability is found by multiplying the probability of the state sequence sequence in question by the probability of the observed data given that sequence.  For example, $P$($\{$L,L,L,L$\}$,O)$\ =(.9)(.7)^3 \times .3 = .09261$.\label{tab:bruteForceMostProbPathEx}}
\end{table}
\begin{table}
  \capstart
\centerline{
\begin{tabular}{ll@{\hspace{3em}}c@{\hspace{3em}}c@{\hspace{3em}}c@{\hspace{3em}}c}
\multicolumn{6}{c}{Most-likely Path v.s. Time} \\
\hline
\hline
& & \text{0min} & \text{15min} & \text{30min} & \text{45min} \\
\hline
\text{S} & \text{path} & \text{S} & \text{L},\text{S} & \text{L},\text{S},\text{S} & ? \\
& \text{probability} & 0.1 & 0.27 & 0.216 & 0\\
\hline
\text{L} & \text{path} & \text{L} & \text{L},\text{L} & \text{L},\text{L},\text{L} & \text{L},\text{L},\text{L},\text{L}\\
& \text{probability} & 0.9 & 0.189 & 0.1323 & 0.09261\\
\end{tabular}
}
\caption[Finding the most probable path using the Viterbi algorithm]{Finding the most probable path using the Viterbi algorithm.  Given a set of observations, the Viterbi algorithm calculates the most probable path to each state at each time starting from the most probable paths to the states of the previous time; all probabilities account for the observations made thus far.  At the 15 minute mark, for example, the most probable path to `learning' is \{L,L\} since the probability of being in state L initially and transitioning to state L is greater than the probability of being in state S initially and transitioning to state L; the corresponding probability is simply the product of the probability of being in state L, the probability of transitioning from there to state L, and the probability of L given the measurement, $.9\times.7\times.3=0.189$.\label{tab:viterbiMostProbPathEx}}
\end{table}
On the basis of the information given you should conclude that \textit{he was most likely awake for the entire lecture}.  One method of reaching this conclusion is simply to enumerate every possible sequence of states and their corresponding probabilities of occurring in conjunction with your observations.  This is done in table~\ref{tab:bruteForceMostProbPathEx}.  A more elegant, and generally more practical, approach is to apply the Viterbi algorithm as illustrated in table~\ref{tab:viterbiMostProbPathEx}.  Rather than calculate every probability, the Viterbi algorithm calculates for each time step the probabilities associated with an extension of the most probable paths from the previous time step.  As a consequence, the number of values that must be calculated by the Viterbi algorithm scales only linearly with the length of the sequence, while the brute force approach requires a number of calculations that is exponential in the length.

Finally, it should be noted that graphs can be constructed that include the measurement outcome explicitly as a state.  This is particularly useful when, as in chapter~\ref{chap:ancillaConstruction}, the measurement can change the state of the system.  If, in our example, students sometimes awoke due to being asked a question, then it would be necessary to include the result of the question in the graph.

\section{Code \label{sec:viterbiCode}}
The \command{Python} code for implementing a single step of the Viterbi algorithm is given below.

{
\small
\begin{verbatim}
def viterbiStep(lViterbi,multiplier):
    """Implements a single step of the Viterbi algorithm."""
    # lViterbi contains the previous Viterbi probabilities and paths
    # multiplier contains the transition probabilities
    nViterbi = [] # new set of Viterbi probabilities and paths
    for i in xrange(len(multiplier[0])): # loop over destination states
        pathMax = None
        probMax = 0
        for j in xrange(len(multiplier)): # loop over starting states
            (prob, path) = lViterbi[j]
            prob *= multiplier[j][i]
            if prob > probMax:
                pathMax = path + [i]
                probMax = prob
        nViterbi.append((probMax, pathMax))
    return nViterbi
\end{verbatim}
}


\end{document}